%% file: main.tex
\documentclass{report}
\usepackage[T1]{fontenc}
\usepackage{textcomp}
\newcommand{\signaturepage}{}
\usepackage{subfiles}
\usepackage{suthesis-2e}

\usepackage{pdfpages}
\usepackage{textgreek}
\usepackage{float}
\usepackage{tikz-cd}
\usepackage{amsmath}
\usepackage{amsfonts}
\usepackage{amsthm}
\usepackage{amssymb}
\usepackage{mathtools}
\usepackage{placeins}
\usepackage{hyperref}
\usepackage{bm}
\usepackage{xcolor}
\usepackage{varwidth}
\usepackage{dsfont}
\usepackage[frozencache,cachedir=.]{minted}

\usepackage{array}
\usepackage[thinlines]{easytable}
\usepackage{thmtools}
\usepackage{empheq}
\usepackage{enumerate}
\usepackage[shortlabels]{enumitem}
\usepackage{varwidth}
\usepackage{tasks}

\usepackage{blindtext}
\usepackage{algorithm}
\usepackage{algpseudocode}
\usepackage{graphicx}
\usepackage{caption}
\usepackage{subcaption}
\usepackage{braket}

\newtheorem{problem}{Problem}

\newtheorem{theorem}{Theorem}[section]

\newtheorem{definition}{Definition}
\DeclarePairedDelimiter{\parens}{\lparen}{\rparen}
\DeclarePairedDelimiter{\abs}{\lvert}{\rvert}
\DeclarePairedDelimiter{\norm}{\lVert}{\rVert}

\DeclarePairedDelimiter{\floor}{\lfloor}{\rfloor}
\DeclareMathOperator{\Tr}{Tr}

\dept{Physics}

\begin{document}
\title{Ray Optics Approach to Holography}
\author{Andrii Torchylo}
\principaladviser{Jason Hogan}
\firstreader{Joonhee Choi}
\copyrightfalse
\tablespagefalse

\beforepreface
\prefacesection{Abstract}
Retrieving the phase of a complex-valued field from the measurements of its amplitude is a crucial problem with a wide range of applications in microscopy and ultracold atomic physics. In particular, obtaining an accurate and efficient solution to this problem is a key step in shaping laser beams for trapping atoms in optical tweezer arrays and applying high-fidelity entangling gates on a neutral atom quantum computer. Current approaches to this problem fail to converge on the optimal solution due to a phenomenon known as vortex formation. In this work, we present an efficient optimization algorithm using Optimal Transport. Our approach completely bypasses the creation of phase vortices and allows for a state-of-the-art solution both in terms of accuracy and efficiency. Furthermore, we show a deep theoretical connection between the Optimal Transport plan and the ray-optics limit of the Wigner distribution of the unknown complex-valued field, and show that our method can be used to retrieve the phase-space transformation of any unknown quadratic phase system. Finally, we reinterpret this problem in the modern quantum learning framework. The techniques we develop provide both useful intuition and practical tools for advancing the frontiers of phase retrieval and laser beam shaping.

\prefacesection{Preface}
This thesis summarizes the work on the problem of phase retrieval during my research as an undergraduate in Jason Hogan's lab at Stanford. I started working on this problem during the summer of 2023 and continued until the spring of 2025. My work in the summer of 2023 was supported by funding from the Stanford Undergraduate Research Program.

\prefacesection{Acknowledgments}
First, I want to thank Jason Hogan and Hunter Swan for the amazing three years of research. This work was the result of a very close collaboration, and it certainly would not be possible without either of them. Jason, you made me the physicist that I am today. You are a very kind person and a very passionate teacher, and I could not wish for a better mentor in my career. I will forever remember many instances of ``just one very last thing" conversations that would continue until late in the night. These conversations inspired to try new ideas and look for new connections. Hunter, you taught me how to think rigorously about the physics problems in the lab and how to convert hand-wavy explanations into concrete mathematical proofs. I admire your dedication and passion for both physics and math, and I learned so so so much from you.

I also acknowledge my thesis reader, Joonhee Choi, for his genuine interest in my work and support throughout my career at Stanford. Also, I want to acknowledge Rafe Mazzeo for the valuable conversation regarding functional analysis, and Robert Huang for helping with the quantum learning interpretation of the problem. In addition, I would like to acknowledge my close friend John Wang for working with me on deep learning aspects of the problem, and Lucas Tellez for helping me understand the ins and outs of convex optimization. Finally, I would like to acknowledge my close friend and collaborator, Dan-Stefan Eniceicu, for continuous support throughout this project (and beyond).

In addition, I want to acknowledge my close friends and family for their love and support. Thank you to my girlfriend Kylie and all my friends at Stanford: Igor,  Claire, Lucas, Dan, John Bailey, Eric, Ursula, Patrick, Garin, Avery, Sisely, Hannah, Elly, and many more. Also, thank you to my sister Tanya, who is the best sister in the world.

Finally, I want to say thank you to everyone in the Ukrainian military for protecting the safety and well-being of my family and my country. 

\afterpreface

\subfile{sections/chapter1.tex}

\subfile{sections/chapter2.tex}
\subfile{sections/chapter3.tex}

\subfile{sections/chapter4.tex}
\subfile{sections/chapter5.tex}
\subfile{sections/conclusions.tex}
\subfile{appendix/A.tex}

\subfile{appendix/B.tex} 
\subfile{appendix/C.tex}
\subfile{appendix/D.tex}

\bibliographystyle{unsrt}
\bibliography{main}

\end{document}

%% file: sections/chapter1.tex
\chapter{Introduction}
\label{chapter:introduction}

\section{Spatial Control of Laser Light}
\label{section:spatial-control-of-laser-light}

The careful engineering and control of laser light is essential for interacting with ultracold atoms, performing high-resolution microscopy, and studying the optical properties of biomaterials \cite{Pasienski:08, varnavides2024iterativephaseretrievalalgorithms, Zhou2020}. The arbitrary manipulation of laser light intensity, also known as beam shaping, is an enabling technology for modern atomic physics experiments, which often involve creating Bose-Einstein condensates (BEC) and trapping atoms in optical lattices \cite{BoseEinsteinLasers, Lewenstein01032007}. Beam shaping is also important for matter-wave lensing, a technique that allows for cooling down atoms to picokelvin temperatures \cite{Kovachy_2015}.

A recent rise of quantum computing using trapped ions and neutral atoms further increased the interest of the community in accurate and efficient beam shaping \cite{manetsch2024tweezerarray6100highly, Zhang_2024, Bernien2017, Ebadi2021}. In this field, people are interested in creating arrays of optical traps that hold atoms for the purpose of processing quantum information and performing quantum computation. The quality of such devices is contingent on creating an optical tweezer array that is both accurate (the desired intensity pattern is as close as possible to the target intensity) and efficient (utilizing most of the last light power). Furthermore, the entangling gate fidelity of the neutral-atom quantum computer is limited by beam inhomogeneity, which can be improved by better spatial control of the laser light \cite{fidelityPaper}.

This thesis describes our progress towards understanding this problem and the rich mathematical structure that underlies it. This field has many elegant connections between seemingly far removed disciplines such as signal processing, optimal transport, and even quantum learning. By carefully investigating these connections, we have arrived at powerful techniques that are pushing the state-of-the-art in accurate and efficient beam shaping.

\section{Beam Shaping and Beam Estimation}
\label{section:beam-shaping-and-beam-estimation}

For typical beam-shaping applications, we often want to characterize the laser light as precisely as possible. In wave optics theory, this amounts to knowing the complex-valued electric field in a single plane of an optical setup. Then, if we know the medium of propagation, we can predict the electric field in any other plane using Helmholtz or paraxial propagation. However, there is a caveat. Measurement of the complex-valued electric field is extremely difficult experimentally.

Denote the complex-valued electric field at the plane $z=z_0$ as follows:
\begin{equation}
    \vec{E}(x,y)=g(x,y)e^{i \phi(x,y)}\hat{p}(x,y)
\end{equation}
where $g(x,y)$ is called the amplitude, $\phi(x,y)$ is the phase, and $\hat{p}(x,y)$ is the position-dependent polarization. We will assume a uniform polarization and will only talk about the scalar-valued electric field from now on. In an experimental setting, it is relatively easy to measure the $g(x,y)$ by placing a camera at the plane $z=z_0$, while measuring the phase $\phi(x,y)$ is generally much more difficult — it is possible to use Shack-Hartmann sensor to do the measurement, but the resulting resolution of the phase map will be much lower than that of the amplitude \cite{Akondi:21}.

To overcome this challenge, it is possible to infer the electric field phase from measurements of the amplitude alone across different planes. As an example, suppose that in addition to the amplitude $g$ at the plane $z=z_0$, we acquired the amplitude $G$ of the same laser light at $z=z_0+d$ where $d \to \infty$. In appropriate dimensionless units, these quantities are related via the Fourier transform (as we will show in the Chapter \ref{chapter:wave-optics-theory}):
\begin{equation}
\label{eq:fourier-relationship}
    G(\mu,\nu)=|\mathcal{F}[g(u,v)e^{i\phi(u,v)}](\mu,\nu)|
\end{equation}
Thus, the goal becomes to find a phase $\phi(u,v)$ such that the predicted output amplitude is as close as possible to $G(\mu,\nu)$. More formally, we have the following optimization problem (adopted from \cite{swan2024highfidelityholographicbeamshaping}):
\begin{problem} 
\label{prob:phase-generation}
Given input beam modulus and target output beam modulus $g,G:\mathbb{R}^2\rightarrow\mathbb{R}_{\geq 0}$ with $\norm{g}_2 = \norm{G}_2$, find a phase function $\phi:\mathbb{R}^2\rightarrow\mathbb{R}$ minimizing 
\begin{equation}
d\left(G(\mu,\nu) , \abs*{\mathcal{F}\!\left[g(u,v)\, e^{2\pi i \, \phi(u,v)}\right]\!(\mu,\nu)}\right),
\end{equation}
where $d$ is some chosen distance function between two images. 
\end{problem}
A solution to the problem \ref{prob:phase-generation} could be used to reshape a laser light into an arbitrary pattern using a device called the spatial light modulator (SLM) \footnote{Or many other similar phase modulating gadgets such as phase light modulator (PLM), metalenses, etc.}. Given an input beam with a known amplitude $g(u,v)$, we can use the SLM to imprint the computed phase onto the beam, which will effectively reshape the laser beam into intensity $G(\mu,\nu)$ in the Fourier plane. Notice that we do not need to have planes separated by $d\to\infty$: we can use a lens with focal length $f$ to bring the Fourier plane to $d=2f$ (see Figure 1).

\begin{figure}[t]
\centering
\includegraphics[width=0.65\columnwidth]{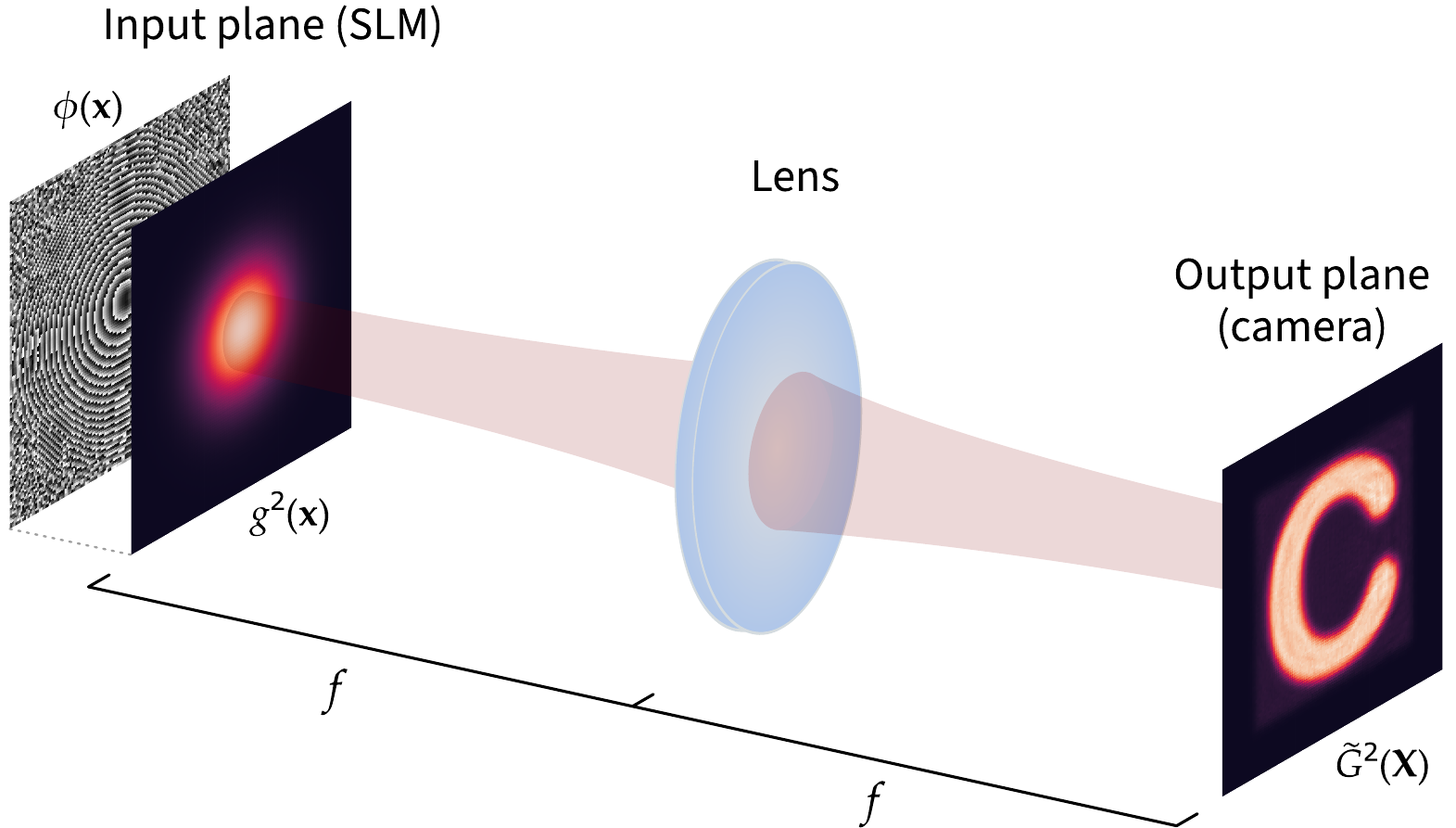}
\caption[Model optical system]{Model optical system. An input laser beam with intensity $g^2(\vec{x})$ is reflected off an SLM with applied phase $\phi(\vec{x})$, passes through a lens of focal length $f$ at distance $f$ from the SLM, and is then imaged on the output (camera) plane at distance $f$ from the lens, with output intensity $\tilde{G}^2(\vec{X})$.}
\label{fig:setup}
\end{figure}

Of course, sometimes we do not know the amplitude of the input beam $g(u,v)$ incident on the SLM and its intrinsic phase $\psi(u,v)$, making it difficult to perform the reshaping task. In this case, we can use the SLM to apply a family of phases $\{\phi_j\}_{j=1}^{n}$ and measure the corresponding amplitudes on the output plane $\{G_j\}_{j=1}^{n}$ to infer the original beam $g(u,v)$ and its intrinsic phase $\psi(u,v)$. Although the applied phases can be arbitrary, we restrict our attention to a family of quadratic phases $\phi_j(u,v)= -2\pi  (u^2+v^2)/2r^2_j$ for some curvatures $r_j\in \mathbb{R}$. We will see in Chapter \ref{chapter:ray-optics-reduction-with-optimal-transport} that using these quadratic phases is equivalent to placing a thin lens in the input plane, which effectively implements a fractional Fourier transform by some angle that depends on $r_j$. We formulate our problem as follows:

\begin{problem}
\label{prob:beam-estimation}
Given coefficients $r_j\in \mathbb{R}$ and corresponding amplitudes measurement $G_j:\mathbb{R}^2\rightarrow\mathbb{R}_{\geq 0}$ with $\norm[\big]{G_j}_2 = 1$, $j=1,\dots,n$, find $g:\mathbb{R}^2\rightarrow\mathbb{R}_{\geq 0}$, $\psi:\mathbb{R}^2\rightarrow\mathbb{R}$ minimizing 
\begin{equation}
\sum_j  d\left(G_j (\mu,\nu), \left|\mathcal{F}\!\left[g(u,v)\, e^{2\pi i \left(\psi(u,v) - {(u^2 +v^2)}/{ 2r_j^2}\right)}\right]\!(\mu,\nu)\right|\right),
\end{equation}
where $d$ is a chosen distance function.
\end{problem}

We will refer to the problem \ref{prob:beam-estimation} as \textit{beam estimation} and the problem \ref{prob:phase-generation} as \textit{phase generation}. In practice, we usually want to solve problem \ref{prob:beam-estimation} first to characterize the beam, and then use the solution to problem \ref{prob:phase-generation} to reshape the laser into the desired intensity pattern.

\section{Past Work}
\label{section:past-work}
There has been a lot of work on the problem of \textit{phase retrieval}\footnote{which is a more common name for the problem of phase generation} from both purely theoretical and algorithmic perspectives. 

From an algorithmic side, the most common approach goes by the name of the Gerchberg-Saxton (GS) algorithm or the Iterative Fourier Transform Algorithm (IFTA) \cite{gerchberg1972practical}. This approach works fairly well in practice, but it does not have any guarantees of convergence \cite{noll2021alternatingprojectionsapplicationsgerchbergsaxton}. In fact, failure to converge is associated with the formation of so-called phase vortices, which stagnate convergence in intensity loss (\ref{eq:Lint-def}). There are methods such as Mixed Region Amplitude Freedom (MRAF) algorithm \cite{Pasienski:08} and Cost Function Minimization (CFM) \cite{Harte:14}, which improve the smoothness of the retrieved phase solutions, but do not eliminate the issue of vortex formation. 

The failure to convergence can be associated with the complicated non-convex landscape of the optimization problem \ref{prob:phase-generation}. In fact, from a purely mathematical perspective, the problem is ill-posed, due to the phenomenon known as transversality \cite{wellPosedness}.

In this work, we show that one can use the stationary phase approximation to approximate the problem of phase generation to the problem of optimal transport, which is a convex problem that has a guaranteed solution and can be solved in polynomial time. We show that this reduction is equivalent to taking a ray-optics limit of the problem. 

We also reformulate the problem of phase generation as the problem of Wigner distribution estimation and show how the optimal transport plan arises as the ray-optic limit of the Wigner distribution. Furthermore, we show that one can retrieve the phase-space transformation of any optical system that corresponds to a linear canonical transform (i.e., has a real-valued ABCD matrix) by using only two projective measurements.

The idea of using optimal transport in the context of ray optics is not new. People have made concrete theoretical connections between the ray-optics theory and optimal transport \cite{Gutierrez2023}, and used it for caustic design \cite{caustic1, caustic2, caustic3, caustic4, caustic5}. However, to the best of our knowledge, the only application of ray optics in holography was the geometric beam shaping in 1D \cite{dickey2014laser}. Thus, our work is unique in the way that it bridges the gap between holography and ray-optics theory. Furthermore, it establishes a very concrete way of understanding Wigner distributions in the ray-optics limit.

\section{Thesis Outline}
\label{section:thesis-outlint}

The thesis has the following structure. Chapter \ref{chapter:wave-optics-theory} is a self-contained introduction to wave optics theory using the language of fractional Fourier transforms and Wigner distributions. Much of this chapter follows the book \cite{ozaktas2001fractional}, the authors of which pioneered the use of fractional Fourier transforms in optics. Chapter \ref{chapter:ray-optics-reduction-with-optimal-transport} is mostly original research that uses optimal transport to solve the ray-optics approximation of the problems \ref{prob:phase-generation} and \ref{prob:beam-estimation}, which we published in \cite{swan2024highfidelityholographicbeamshaping}. In addition, we prove entirely new, unpublished results, generalizing our work to fractional Fourier transforms and establishing connections to Wigner distributions. In Chapter \ref{chapter:algorithms}, we describe the landscape of algorithmic approaches to both problems and present our current state-of-the-art solution. Additionally, we reduce the $\mathcal{O}(n^4)$ memory constraint of the original algorithm proposed in \cite{swan2024highfidelityholographicbeamshaping} to $\mathcal{O}(n^2)$, allowing the generation of high-resolution holograms with optimal transport. In Chapter \ref{chapter:from-phase-holography-to-quantum-state-tomography}, we establish a deep theoretical connection to the theory of quantum learning, proposing a new avenue for holography research.

%% file: sections/chapter2.tex
\chapter{Wave Optics Theory}
\label{chapter:wave-optics-theory}

The goal of this chapter is to introduce the reader to the relevant concepts from wave optics theory, Wigner distributions, and fractional Fourier transforms. Much of this chapter closely follows the book \cite{ozaktas2001fractional}, but many interesting facts are also derived independently. 

\section{Conventions}
\label{section:conventions}

Throughout this thesis, we will refer to the electric field as a ``signal" to highlight the fact that this theory can be applied to any complex-valued function in an abstract separable Hilbert space. The small variables $x,y,z$ (usually) correspond to the position variables (units of length), and the Greek variables $\sigma_x,\sigma_y,\sigma_z$ are the spatial frequencies (units of 1/length). We will also use $u$, $v$, $w$ for dimensionless position variables and $\mu$, $\nu$, and $\eta$ for their dimensionless conjugate variables. Notice that for every physical system that we will describe, there will be a conversion factor $s$ (unit of length) that relates dimensional and dimensionless variables:
\begin{align}
    x/s = u && y/s=v && z/s = w \\
    \sigma_xs=\mu && \sigma_ys=\nu && \sigma_zs=\eta
\end{align}
This is the same notational convention as in \cite{ozaktas2001fractional}, and, in my humble opinion, it is one of the best ones I have seen. To learn more about conventions used in this thesis, see Appendix \ref{chapter:fourier-analysis}.

\section{Paraxial Wave}
\label{section:paraxial-wave}

The classical wave equation is a very useful tool that helps us to understand light propagation through a medium. However, very often, we are interested in a special case where the planar waves mostly propagate along a well-defined axis, call it the z-axis, and do not deviate too much from $z$. This is the case for lasers, for example, where most of the light travels in a straight line with small off-axis variations. The goal of this section is to derive an approximation for the scalar wave equation, \cite{siegman1986lasers}
\begin{equation}
\label{eq:classical-wave-eq}
\frac{\partial^2}{\partial x^2}E + \frac{\partial^2}{\partial y^2}E + \frac{\partial^2}{\partial z^2}E + k^2 E=0
\end{equation}
where $E(x,y,z)$ is the time-independent solution, while time-dependence is captured by a ``wiggle factor" $\exp(i\omega t)$. To proceed, we define an envelope field $\tilde{E}(x,y,z)$, by factoring out an exponent from the original field $E$:
\begin{equation}
\label{eq:u-definition}
    E(x, y, z)=\tilde{E}(x, y, z)\exp(-ikz)
\end{equation}
Plugging in (\ref{eq:u-definition}) into (\ref{eq:classical-wave-eq}) and simplifying leads to:
\begin{equation}
    \frac{\partial^2}{\partial x^2}\tilde{E} + \frac{\partial^2}{\partial y^2}\tilde{E} + \frac{\partial^2}{\partial z^2}\tilde{E} = 2ki\frac{\partial \tilde{E}}{\partial z}
\end{equation}
If $\tilde{E}$ does not change rapidly in z direction, we can use paraxial wave approximation $\left|\frac{\partial^2\tilde{E}}{\partial z^2}\right| \ll \left|2k\frac{\partial \tilde{E}}{\partial z}\right|$ or $\left|\frac{\partial^2 \tilde{E}}{\partial x^2}\right|$ or $\left|\frac{\partial^2 \tilde{E}}{\partial y^2}\right|$. This approximation allows us to neglect the term $\frac{\partial^2 \tilde{E}}{\partial z^2}$, which leads to the paraxial wave equation:

\begin{equation}
\label{eq:paraxial} 
    \frac{\partial^2}{\partial x^2} \tilde{E} + \frac{\partial^2}{\partial y^2} \tilde{E} = 2ki\frac{\partial \tilde{E}}{\partial z}
\end{equation}

\subsection{Paraxial Propagation}
\label{subsection:paraxial-propagation}

The paraxial wave equation (\ref{eq:paraxial}) is exactly Schrödinger's equation in 2D with 0 potential once we change variables from $z$ to $t$. We know that the eigenstates of this equation are plane waves. This inspires a solution strategy — decompose the function $u$ into plane waves, and derive an effective propagation equation for the plane waves.

We start by taking the Fourier transform of $\tilde{E}(x, y, z)$ with respect to variables $x$ and $y$, which we will call $G(\sigma_x, \sigma_y, z)$. Note that both $G$ and $\tilde{E}$ are functions of the same $z$ variable. Using the inverse Fourier transform, we can write $\tilde{E}$ in terms of $G$ and plug it back into the paraxial wave equation (\ref{eq:paraxial}):
\begin{multline}
    \left(\frac{\partial^2}{\partial x^2} + \frac{\partial^2}{\partial y^2}\right)\iint G(\sigma_x, \sigma_y, z)\exp(i 2\pi (\sigma_x x + \sigma_y y))d\sigma_xd\sigma_y\\
    = 2ki\frac{\partial}{\partial z}\iint G(\sigma_x, \sigma_y, z)\exp(i 2\pi (\sigma_x x + \sigma_y y))d\sigma_xd\sigma_y
\end{multline}
Differentiating under integral sign, operators $\frac{\partial^2}{\partial x^2}$ and  $\frac{\partial^2}{\partial y^2}$ pulls down $(i2\pi \sigma_x)^2$ and $(i2\pi \sigma_y)^2$ terms from the exponents respectively. Now the derivative $\frac{\partial}{\partial z}$ will just operate on the function $G$. We obtain:
\begin{multline}
    \left((i2\pi \sigma_x)^2 + (i2\pi \sigma_y)^2\right)\iint G(\sigma_x, \sigma_y, z)\exp(i 2\pi (\sigma_x x + \sigma_y y))d\sigma_xd\sigma_y \\
    = 2ki\iint\frac{\partial}{\partial z} G(\sigma_x, \sigma_y, z)\exp(i 2\pi (\sigma_x x + \sigma_y y))d\sigma_xd\sigma_y
\end{multline}
Rearranging the terms:
\begin{equation}
    \iint\left[4\pi^2(\sigma_x^2 + \sigma_y^2) G(\sigma_x, \sigma_y, z) + 2ki \frac{\partial}{\partial z} G(\sigma_x, \sigma_y, z)\right]\exp(i 2\pi (\sigma_x x + \sigma_y y))d\sigma_xd\sigma_y = 0
\end{equation}
Since the collection $\{\exp(i 2\pi (\sigma_x x + \sigma_y y)): \sigma_x, \sigma_y \in \mathbb{R}\}$ forms a basis for the $L^2(\mathbb{R}^2)$ space, we conclude that coefficients next to each basis vector must be $0$.
\begin{equation}
    4\pi^2(\sigma_x^2 + \sigma_y^2) G(\sigma_x, \sigma_y, z) + 2ki \frac{\partial}{\partial z} G(\sigma_x, \sigma_y, z) = 0 \quad \forall \sigma_x, \sigma_y
\end{equation}
From which follows:
\begin{equation}
    \frac{\partial}{\partial z} G(\sigma_x, \sigma_y, z) = i\frac{4\pi^2(\sigma_x^2 + \sigma_y^2)}{2k}G(\sigma_x, \sigma_y, z)  \quad \forall \sigma_x, \sigma_y
\end{equation}
This differential equation has an easy solution:
\begin{equation}
G(\sigma_x, \sigma_y, z) = G(\sigma_x, \sigma_y, 0) \exp\left(i\frac{4\pi^2(\sigma_x^2 + \sigma_y^2)}{2k}z\right)  \quad \forall \sigma_x, \sigma_y
\end{equation}
Or in terms of original function $\tilde{E}$:
\begin{equation}
\label{eq:_propagator}
 \mathcal{F}[\tilde{E}(x, y, z)] = \mathcal{F}[\tilde{E}(x, y, 0)]\exp\left(i\frac{4\pi^2(\sigma_x^2 + \sigma_y^2)}{2k}z\right) 
\end{equation}
This gives a recipe for propagating the paraxial wave equation \ref{eq:_propagator} — decompose the envelope field $u(x,y,0)$ into the Fourier modes, multiply each by an appropriate exponential factor, and take the inverse Fourier transform, to find the field $u(x,y,0)$.

Alternatively, we can propagate our wave fully in the position space. We will use a convolution theorem (property \ref{FT1} in Appendix \ref{chapter:fourier-analysis}). First, we compute the inverse Fourier transform of the exponential term:
\begin{multline}
\mathcal{F}^{-1}\left[\exp\left(i\frac{4\pi^2(\sigma_x^2 + \sigma_y^2)}{2k}z\right)\right]=\\
= \int \int \exp\left(i\frac{4\pi^2(\sigma_x^2 + \sigma_y^2)}{2k}z\right) \exp(i 2\pi (\sigma_x x + \sigma_y y)) d\sigma_x d\sigma_y\\
= \frac{(1/2+i/2)^2k}{\pi z}\exp\left(\frac{-ik(x^2 + y^2)}{2z}\right) = \frac{ik}{2\pi z}\exp\left(\frac{-ik(x^2 + y^2)}{2z}\right)
\end{multline}
Now we can rewrite equation (\ref{eq:_propagator}) as:
\begin{equation}
 \mathcal{F}[\tilde{E}(x, y, z)] = \mathcal{F}[\tilde{E}(x, y, 0)]\mathcal{F}\left[\frac{ik}{2\pi z}\exp\left(-\frac{ik(x^2+y^2)}{2z}\right)\right]
\end{equation}
Using the convolution theorem \ref{FT1}:
\begin{align}
\tilde{E}(x, y, z) &= \tilde{E}(x, y, 0) \circledast \frac{ik}{2\pi z}\exp\left(-\frac{ik(x^2+y^2)}{2z}\right) \\
&= \frac{ik}{2\pi z} \int \int \tilde{E}(x', y', 0) \exp\left(-\frac{ik((x-x')^2+(y-y')^2)}{2z}\right)dx' dy'
\end{align}
The last equation is also known as a \textit{Huygen's integral} in a paraxial limit. So, to summarize this section, we derived two important results. Given the field at $z = z_0$ we can compute the field at an arbitrary location $z = z_0 + L$ in either position space or Fourier space using one of the following formulas. 
\begin{equation}
\label{eq:propagator}
 \mathcal{F}\{\tilde{E}(x, y, z)\} = \mathcal{F}\{\tilde{E}(x, y, z_0)\}\exp\left(i\frac{4\pi^2(\nu_x^2 + \nu_y^2)}{2k}L\right) 
\end{equation}

\begin{equation}
\label{eq:convolution}
  \tilde{u}(x, y, z) = \frac{ik}{2\pi L} \int dx'\int dy'\tilde{u}(x', y', z_0) \exp\left(-\frac{ik}{2L}\left((x-x')^2+(y-y')^2\right)\right)
\end{equation}

\subsection{Gaussian Beam Propagation}
\label{subsection:gaussian-beam-propagation}

As an application of the equations above, we will describe the Gaussian beam propagation in free space and later generalize this result to a propagation through an optical system of lenses.

Consider an electric field at location $(x, y, 0)$ of the following form:
\begin{equation}
    \tilde{E}(x, y, 0) = E_0 \exp{\left(-\frac{x^2+y^2}{w_0^2}\right)}
\end{equation}
Notice that at $\tilde{E}(x, y, 0)=E(x, y, 0)$ since $\exp(-ikz)$ is unity at $z=0$. The parameter $w_0$ is called the waist of a Gaussian beam, which defines a contour where intensity drops by $e^{-2}$ of its maximum value. This is because intensity $I \propto E^2$, thus: 
\begin{equation}
    \frac{I(w_0)}{I(0)} = \exp{\left(-2\frac{w_0^2}{w_0^2}\right)} = e^{-2} 
\end{equation}
 We want to know how this field propagates along z due to the Paraxial wave equation (\ref{eq:paraxial}). To do this, we will use our convolution formula (\ref{eq:convolution}):
\begin{align}
    \tilde{E}(x, y, z) &= \frac{ikE_0}{2\pi z} \int dx'\int dy'\exp{\left(-\frac{x'^2+y'^2}{w_0^2}\right)} \exp\left(-\frac{ik}{2z}\left((x-x')^2+(y-y')^2\right)\right)\\
    &= \frac{ikE_0}{2\pi z} \int dx'\exp{\left(-\frac{x'^2}{w_0^2}\right)}\exp\left(-\frac{ik}{2z}(x-x')^2\right) \int (...) dy' \\
    &= E_0\frac{kw_0^2}{kw_0^2 - 2 iz} \exp\left(-\frac{k(x^2+y^2)}{kw_0^2-2iz}\right) \\
    &= E_0\frac{kw_0^2(kw_0^2 + 2iz)}{k^2w_0^4 + 4z^2} \exp\left(-\frac{k^2w_0^2(x^2+y^2)}{k^2w_0^4+4z^2}\right)\exp\left(-i\frac{2zk(x^2+y^2)}{k^2w_0^4+4z^2}\right)
\end{align}
To simplify this, let's define the following substitutions (to be physically interpreted later):
\begin{align}
     z_R = \frac{w_0^2 k}{2} && w(z) = w_0 \sqrt{1 + \frac{z}{z_R}} \\
     R(z) = z\left(1 + \frac{z_R^2}{z^2}\right) && \psi(z) = \arctan\left(\frac{z}{z_R}\right)
\end{align}
Then, after simplifying, we obtain the usual Gaussian beam formula:\footnote{This is a full expression for the time-independent electric field, including. $\exp(ikz)$ factor}
\begin{equation}
\label{eq:paraxial-helmholtz} 
    E(x, y, z) = E_0 \frac{w_0}{w(z)}\exp\left(-\frac{(x^2+y^2)}{w(z)^2}\right)\exp\left(-i\left(\frac{k(x^2+y^2)}{2R(z)} - \psi(z) + kz \right)\right)
\end{equation}
Looking at this equation, our substitutions can be interpreted physically. $w(z)$ represents the waist of the Gaussian at the distance $z$ from the origin. $z_R$ is called a Rayleigh range, which is a distance where $w(z) = \sqrt{2} w_0$. $R(z)$ is the radius of curvature, which characterizes the curvature of the wavefront. Finally, $\psi(z)$ is the Gouy phase shift that a Gaussian beam undergoes throughout propagation. 

\section{Hermite Gaussian Basis}
\label{section:hermite-gaussian-basis}

\begin{figure}[!ht]
\centering
\includegraphics[width=0.85\columnwidth]{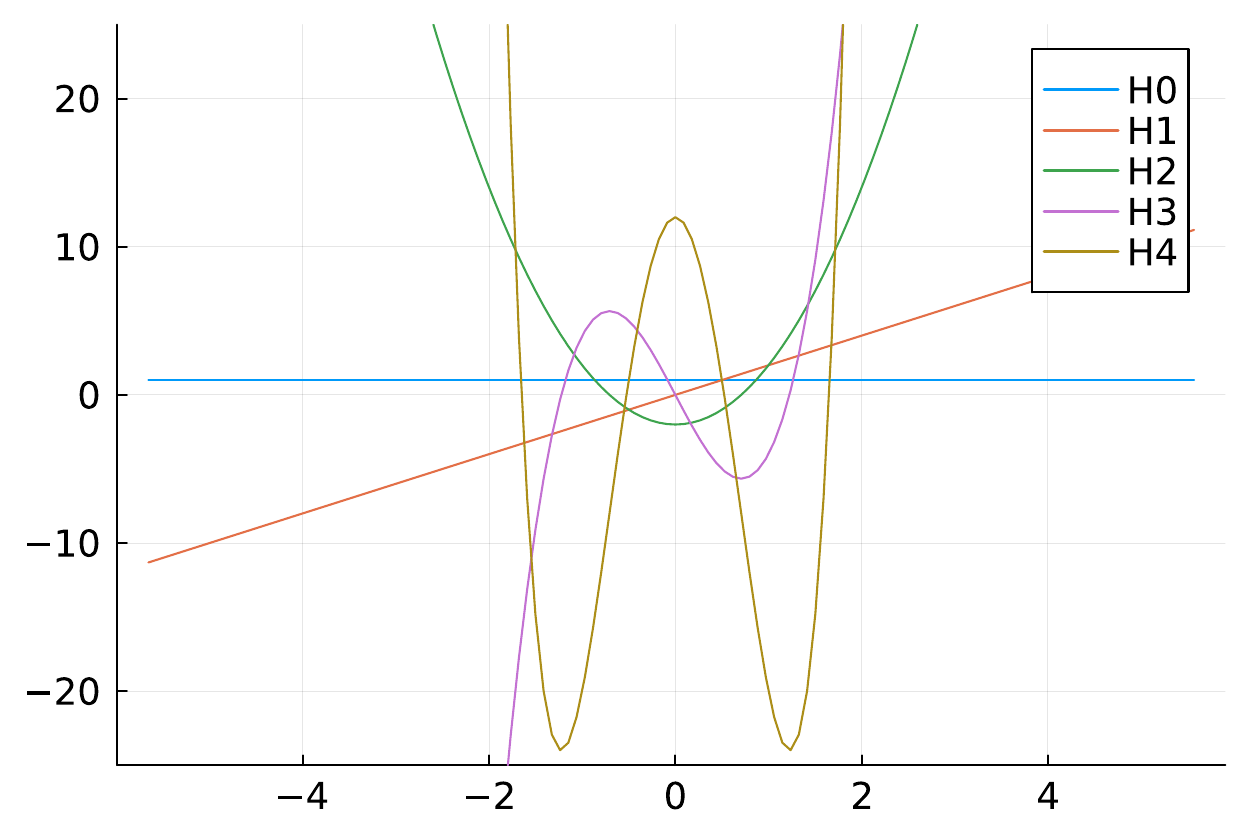}
\caption{First five Hermite polynomials $H_n(u)$ }
\label{fig:hermite-poly}
\end{figure}

\begin{figure}[!ht]
\centering
\includegraphics[width=0.85\columnwidth]{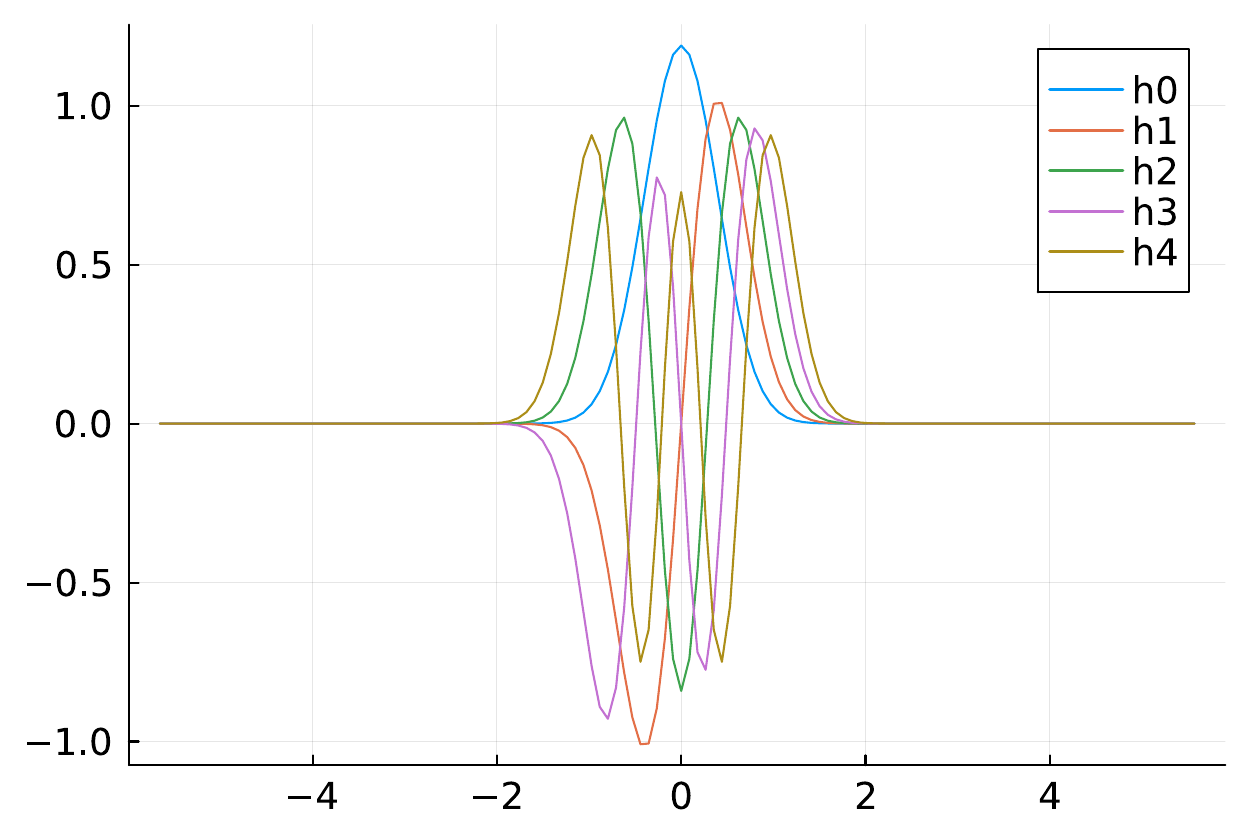}
\caption{First five Hermite-Gaussian functions $h_n(u)$}
\label{fig:hermite-gauss}
\end{figure}

Gaussian beam (\ref{eq:paraxial-helmholtz}) is only one possible solution to the paraxial wave equation. It turns out there is a more general family of solutions, described in terms of Hermite-Gaussian functions \cite{ozaktas2001fractional}:
\begin{multline}
\label{eq:hermite-paraxial} 
    E_{l,m}(x, y, z) = \frac{1}{w(z)}h_l\left(\frac{x}{w(z)}\right)h_m \left(\frac{y}{w(z)}\right) \exp\left(-i\left(\frac{k(x^2+y^2)}{2R(z)} - (l+m+1)\psi(z) + kz \right)\right)
\end{multline}
where $h_l$ and $h_m$ are $l$-th and $m$-th Hermite-Gaussian functions:
\begin{align}
\label{eq:hermite-functions}
    h_n=A_n H_n(\sqrt{2\pi}u)\exp(-\pi u^2) && A_n=\frac{2^{1/4}}{\sqrt{2^n n!}}
\end{align}
defined in terms of the Hermite polynomials:
\begin{equation}
    H_n(u) = (-1)^n e^{u^2}\frac{d^n}{d u^n}e^{-u^2}
\end{equation}
The first few Hermite polynomials and associated Hermite-Gaussian functions are given in the Table \ref{tab:hermite-polynomials} and Figures \ref{fig:hermite-poly} and \ref{fig:hermite-gauss}. Notice that they satisfy the recurrence relation $H_{n+1}(u) =2uH_n(u)-2nH_{n-1}$, which makes them easy to generate.
\begin{table}[ht]
\centering
\begin{tabular}{c|l|l}
$n$ &  $H_n(u)$ & $h_n(u)$\ \\
\hline
0 & \(  1 \) & \( 2^{1/4} e^{-\pi  u^2} \)\\
1 & \( 2u \) & \( 2^{3/4}\sqrt\pi u e^{-\pi  u^2}  \) \\
2 & \( 4u^2 - 2 \) & \( 2^{-1/4}\left(4\pi u^2-1\right)e^{-\pi  u^2}  \) \\
3 & \( 8u^3 - 12u \) & \( \frac{2^{3/4}\sqrt{\pi}}{\sqrt{3}} \left(4\pi u^3-3u\right) e^{-\pi  u^2}  \)\\
4 & \( 16u^4 - 48u^2 + 12 \) & \(\frac{1}{2^{5/4} \sqrt{3}} \left(16\pi^2u^4-24\pi u^2+3\right) e^{-\pi  u^2} \) \\
\end{tabular}
\caption{First five Hermite-Gaussian polynomials \( H_n(u) \)}
\label{tab:hermite-polynomials}
\end{table}

\subsection{Hermite-Gaussian Properties}
\label{subsection:hermite-gaussian-properties}

\subsubsection{Complete Basis of $L^2(\mathbb{R})$}
Notice that each $h_l$ is square-integrable, because it is suppressed by $e^{-\pi u^2}$, so it decays sufficiently fast. They have norm 1, according to the standard inner product of $L^2(\mathbb{R})$, and they are orthogonal to each other. In fact, one can view Hermite polynomials as a Gram-Schmidt orthogonalization of the basis $\{1, x, x^2,...\}$. So, $(h_l)_{l=0}^\infty$ is an orthonormal sequence in $L^2(\mathbb{R})$. To see that it is complete, it suffices to show that kernel $K_m(u)$:
\begin{equation}
    K_m(u)=\frac{1}{m+1}\sum_{j=0}^m\sum_{n=0}^jh_n(u)
\end{equation}
converges to a delta function in weak topology, i.e. $K_m(u)\rightharpoonup\delta(u)$ as $m\to\infty$, which is indeed true \cite{Young_1988}. Thus, set of all Hermite-Gaussian $(h_l)_{l=0}^\infty$ functions forms a complete basis of $L^2(\mathbb{R})$.

\subsubsection{Complete Basis for Paraxial Propagation}
We can generalize this completeness result even further. As it turns out, the family of functions $E_{l,m}$ described in (\ref{eq:hermite-paraxial}) forms a complete basis for the solutions to the paraxial wave equation (\ref{eq:paraxial}). This gives us yet another way to propagate a paraxial wave (akin to equations (\ref{eq:propagator}) and (\ref{eq:convolution})). We start by considering the plane $z=0$ where the electric field $E(x,y,z)$ is collimated (i.e., it has a flat phase). Then we can decompose:
\begin{align}
    E(x,y,0)&=\sum_{l=0}^\infty\sum_{m=0}^\infty C_{lm} \frac{1}{w_0}\psi_l\left(\frac{x}{w_0}\right)\psi_m\left(\frac{y}{w_0}\right)\\
    C_{lm}&=\iint\frac{1}{w_0}\psi_l\left(\frac{x}{w_0}\right)\psi_m\left(\frac{y}{w_0}\right) E(x,y,0)dxdy
\end{align}
Then we can easily propagate the electric field to an arbitrary plane using eq. (\ref{eq:hermite-paraxial}):
\begin{align}
    E(x,y,z)=\sum_{l=0}^\infty\sum_{m=0}^\infty C_{lm} \frac{1}{w(z)}\exp\left(-\frac{(x^2+y^2)}{w(z)^2}\right)\exp\left(-i\left(\frac{k(x^2+y^2)}{2R(z)} -  (l+m+1)\psi(z) + kz \right)\right)
\end{align}

\subsubsection{Eigenfunction Properties}
It is a well-known fact that Hermite-Gaussian functions are eigenstates of the Quantum Harmonic Oscillator (QHO). Equivalently, we can say that $h_n$ solves the following differential equation \cite{ozaktas2001fractional}:
\begin{equation}
    \frac{d^2h_n(u)}{du^2}+4\pi^2\left(\frac{2n+1}{2\pi}-u^2\right)h_n(u)=0
\end{equation}
which we can rewrite as an eigenvalue equation:
\begin{equation}
\label{eq:qho-eigenvalues}
    \left(-\frac{d^2}{du^2}+4\pi^2u^2\right)h_n(u)=2\pi(2n+1)h_n(u)
\end{equation}
where the operator $\left(-\frac{d^2}{du^2}+4\pi^2u^2\right)$ can be viewed as a dimensionless QHO Hamiltonian, $h_n(u)$ is the $n$-th eigenfucntion with eigenvalue $2\pi(2n+1)$. 
Another important property is that $h_n$ is the eigenfunction of the continuous Fourier transform (\ref{eq:ft}) as shown in \cite{ozaktas2001fractional}:
\begin{equation}
\label{eq:fourier-eigen}
    \mathcal{F}[h_n](u)=e^{-in\pi/2}h_n(u)
\end{equation}
This makes Hermite-Gaussians an especially convenient choice of basis for the problem, since it simultaneously diagonalizes both QHO and the Fourier transform operator $\mathcal{F}$. In particular, we can easily compute a Fourier transform by decomposing any function $f\in L^2(\mathbb{R})$ into the Hermite-Gaussians $(h_l)_{l=0}^\infty$.
\begin{equation}
    f = \sum_{n=0}^\infty a_nh_n \implies\mathcal{F}[f]=\sum_{n=0}^\infty a_n\mathcal{F}[h_n]= \sum_{n=0}^\infty a_n e^{-in\pi/2} h_n
\end{equation}
where we used the linearity of the Fourier transform and the eigenvalue property (\ref{eq:fourier-eigen}).

\section{Fractional Fourier Transformation}
\label{section:fractional-fourier-transformation}

There is a natural generalization of the Fourier transformation, which can be obtained using the eigenfunction property (\ref{eq:fourier-eigen}):

\begin{definition}
\label{def:frft-hermite}
Let $f = \sum_{n=0}^\infty a_nh_n$ be any function in $L^2(\mathbb{R})$, and let $\alpha\in[0,2 \pi)$. The fractional Fourier transform (FrFT) of angle $\alpha$, is a function $\mathcal{F_\alpha}[f]\in L^2(\mathbb{R})$ defined as:

\begin{equation}
\label{eq:frft-defintion}
    \mathcal{F_\alpha}[f] =\sum_{n=0}^{\infty}a_n e^{-in\alpha}h_n
\end{equation}
where again $h_n$ are n-th Hermite-Gaussians defined in (\ref{eq:hermite-functions}).
\end{definition}
 
In particular, when $\alpha =\pi/2$, $\mathcal{F}_{\pi/2}$ coincides with our regular notion of the Fourier transform $\mathcal{F}$. A trivial consequence of this definition is that Hermite-Gaussians $h_n$ are still eigenfunctions of $\mathcal{F}_\alpha$  operator, with the eigenvalue $e^{-in\alpha}$. So, one way of understanding the operator $\mathcal{F}_\alpha$ is to think of it as a propagator similar to the paraxial wave equation. Once we decompose a function $f$ into Hermite Gaussian modes, the fractional Fourier transform evolves each coefficient of the eigendecomposition $a_n$ by rotating it in the complex plane by an angle $n\alpha$, i.e., $a_n\mapsto a_ne^{-in\alpha}$.

Another way to view a fractional Fourier transform $\mathcal{F}_\alpha$ is to treat it as a Fourier transform $\mathcal{F}$ applied $\alpha/(\pi/2)$ times. Let's elaborate on this interpretation when $\alpha$ is an integer multiple of $\pi/2$.

Consider the simplest case of composing the Fourier transform twice, which corresponds to $\alpha=\pi$.
\begin{equation}
    \mathcal{F}_\pi=\mathcal{F}^2=\mathcal{F}\circ\mathcal{F}
\end{equation}
The first operator takes a function $f\in L^2(\mathbb{R})$ to its Fourier conjugate $F\in L^2(\mathbb{R})$ where by definition:
\begin{equation}
    F(\mu) =\mathcal{F}[f](\mu)= \int_{-\infty}^\infty f(u)\\e^{-i2\pi \mu u}du
\end{equation}
Now, applying the Fourier transform one more time, we obtain:
\begin{equation}
     g(u) =\mathcal{F}^2[f](u)=\mathcal{F}[F](u)=\int_{-\infty}^\infty F(\mu)\\e^{-i2\pi \mu u}d\mu
\end{equation}
Notice that this is almost the inverse Fourier transform of $F(\mu)$ — except we have a negative sign in the exponent. Grouping this negative sign with $u$ we get that $g(-u)$ is the inverse Fourier transform of $F(\mu)$. But $F(\mu)$ is a Fourier transform of $f(u)$, which means that $g(u)=f(-u)$. In other words, $\mathcal{F}^2$ is just a parity operator $\mathcal{P}$, i.e., it switches $u\mapsto-u$.

Now, $\mathcal{F}^0$ is equivalent to applying a Fourier transform 0 times, so it is just an identity operator $\mathds{1}$. Similarly, $\mathcal{F}^1=\mathcal{F}$ is just the normal Fourier transform. Now what about $\mathcal{F}^3$? Notice the following observation
\begin{equation}
\mathcal{F}^4=\mathcal{F}^3\circ\mathcal{F}=\mathcal{F}^2\circ\mathcal{F}^2=\mathcal{P}\circ\mathcal{P}=\mathds{1}
\end{equation}
where we used our previous result that $\mathcal{F}^2=\mathcal{P}$. So, if $\mathcal{F}^3\circ\mathcal{F}=\mathcal{F}\circ\mathcal{F}^3=\mathds{1}$ then it must be that $\mathcal{F}^3=\mathcal{F}^{-1}$ is the inverse Fourier transform. Notice that there is no point in computing arbitrary powers beyond 4, because $\mathcal{F}^{n+4}=\mathcal{F}^n \quad\forall n\in\mathbb{Z}$, using our previous observation that $\mathcal{F}^4=\mathds{1}$. 

\begin{figure}[!ht]
\centering
\includegraphics[width=0.85\columnwidth]{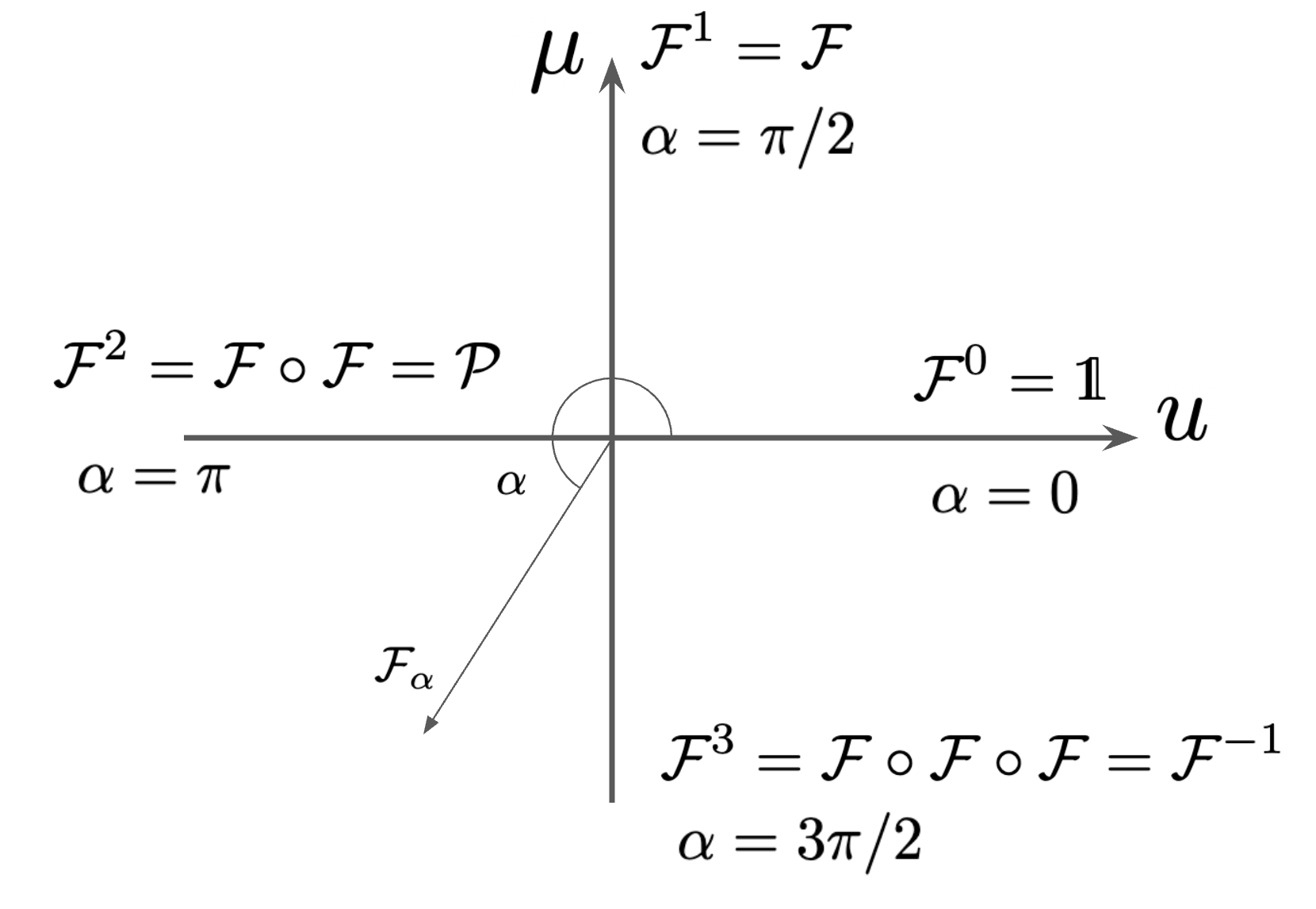}
\caption{Pictorial representation of a fractional Fourier transform}
\label{fig:fractional-frft}
\end{figure}

To summarize, we have found that for integer powers $\mathcal{F}^n$, ``lives" on a circle (see Figure \ref{fig:fractional-frft}). In fact, we can think of this circle as a phase space representation of the function. The Fourier transform $\mathcal{F}^1$ can be thought of as a change of basis from the position domain to the momentum domain, while $\mathcal{F}^2=\mathcal{P}$ is a change of basis within the position basis — it takes $u\mapsto-u$. In general, a fractional Fourier transform of angle $\alpha$ represents a function in a mixed position and momentum basis. More concretely, we can have the following equivalent definition of the fractional Fourier transform:
\begin{definition}
    Let $f(x)\in L^2(\mathbb{R})$ be a function expressed in position basis, then the \textit{fractional Fourier transform} of angle $\alpha\notin \pi \mathbb{Z}$ can be written as:
    \begin{equation}
    \mathcal{F}_\alpha[f](\mu) = \sqrt{1 - i \cot \alpha} \exp\left(-i\pi \mu^2\cot\alpha\right)\int_{-\infty}^{\infty} duf(u) 
    \exp\left(
    -i2\pi \left(u\mu \csc \alpha -\frac{u^2}{2}\cot \alpha \right)
    \right)
    \end{equation}
Furthermore, if $\alpha=0$, we define $\mathcal{F}_\alpha=\mathds{1}$, and if $\alpha=\pi$ then we define $\mathcal{F}_\alpha=-\mathds{1}$
\end{definition}
This definition is very similar to our usual Fourier Transform on $L^2(\mathbb{R})$ (\ref{eq:ft}), except for an additional curvature phase factor of $\frac{u^2}{2}\cot \alpha$ in the definition. We will see that this can be implemented physically by adding an extra lens to our optical setup.

By construction, we see that $\mathcal{F}_\alpha$ agrees with our previous analysis when $\alpha=n\pi/2$ (see that $\csc\alpha=1$ and $\cot\pi/2=0$). Verifying that the two definitions agree when $\alpha\neq \pi/2$ requires performing some integrals of Hermite Gaussian functions, and is left as an exercise. Importantly, the new definition explicitly reveals the ``mixed" position/momentum basis we described above. To see this, define a basis $(e_\mu^{(\alpha)})_{\mu\in\mathbb{R}}$ where each element is the following function:
\begin{equation}
\label{eq:fractional-fourier-basis}
    e^{(\alpha)}_\mu(u)=\exp\left(-2\pi i\left(\frac{\mu^2}{2}\cot\alpha -\mu u\csc\alpha +\frac{\mu^2}{2}\cot\alpha\right)\right)
\end{equation}
Then whenever $\alpha\neq\pi\mathbb{Z}$ we see that $\mathcal{F}_\alpha[f](\mu)=\braket{e_\mu^{(\alpha)}|f}$. In other words, $\mathcal{F}_\alpha$ simply changes position basis into the basis $(e_\mu^{(\alpha)})_{\mu\in\mathbb{R}}$, which is another complete orthonormal basis for $L^2(\mathbb{R})$ \cite{ozaktas2001fractional}.

To demystify the fractional Fourier transform even further, we directly relate it to the unitary evolution under the Quantum Harmonic Oscillator Hamiltonian.
\begin{definition}
    Let $\mathcal{H}$ be a Quantum Harmonic Oscillator Hamiltonian, which we define by its action on a function $f(u)$, i.e.,  $\mathcal{H}f(u)=-\frac{d^2}{du^2}f(u)+4\pi^2u f(u)$. Then we define the fractional Fourier transform to be an associated unitary operator applied for time $t_\alpha=\alpha/4\pi$:
    \begin{equation}
        \mathcal{F_\alpha}=U(t_\alpha)=e^{-i\mathcal{H}t_\alpha}
    \end{equation} 
\end{definition}
To see that this definition is indeed equivalent to Definition \ref{def:frft-hermite}, consider its action on the $n$-th Hermite Gaussian, $\ket{n}$. As previously discussed, $h_n(u)=\braket{u|n}$ is an eigenfucntion of the QHO with the eigenvalue $E_n=2\pi(2n + 1)$. It immediately follows that:
\begin{equation}
    U(t_\alpha)\ket{n}=e^{-iE_n t_\alpha}\ket{n}=e^{-i2\pi(2n+1)\alpha/4\pi}\ket{n}=e^{-i\alpha/2}e^{-in\alpha}\ket{n}
\end{equation}
which is exactly the action of the $\mathcal{F_\alpha}$ in the Defintion \ref{def:frft-hermite} up to an overall phase factor $e^{-i\alpha/2}$.

\section{Wigner Distribution}
\label{section:wigner-distribution}

After we defined several orthonormal bases for $L^2(\mathbb{R})$, we have the ability to represent a given complex-valued signal $f\in L^2(\mathbb{R})$ in a few different ways. We can talk about the position basis representation of the function $f(u)$, which formally corresponds to the inner product $\braket{\delta_u|f}$  where $\delta_u$ is 0 everywhere except at $x$.\footnote{Technically $\delta_u\in \mathcal{D}(\mathbb{R})$ and $\delta_u\not\in L^2(\mathbb{R})$. We call such objects \textit{distributions}.} Similarly, we can represent the same function in the Fourier domain, which is nothing but $F(\mu) =\braket{e_\mu|f}$ where $e_\mu(u)=e^{-i2\mu u}$. Finally, we can use the fractional Fourier basis \ref{eq:fractional-fourier-basis}, which gives us yet another representation $f_\alpha(\mu)=\braket{e_\mu^{(\alpha)}|f}$. 

Each representation is valuable in its own right as it emphasizes important features of the signal. Depending on the choice of the basis, we are either using the position space, momentum space, or, in the case of fractional Fourier transform, a combination of the two. It turns out that there is a way to represent a function simultaneously in terms of position \textit{and} momentum space, and the correct mathematical object in this case is called \textit{ Wigner Distribution}.

\newpage

\begin{definition}
    Let $f\in L^2(\mathbb{R})$ be a function that has a position basis representation $f(u)$ i.e., we can view $f:\mathbb{R}\to\mathbb{C}$.  The Wigner Distribution of the function $f$ is a function $W_f:\mathbb{R}^2\to\mathbb{R}$ defined as:
    \begin{equation}
    \label{eq:wigner-def}
        W_f(u, \mu)=\int f(u+u'/2)f^*(u-u'/2)e^{-i2\pi u'\mu}du'
    \end{equation}
\end{definition}
Roughly speaking, $W_f(u, \mu)$ can be treated as a phase space energy density function that shows how much energy is concentrated in a $\Delta u \Delta \mu$ chunk of phase space area. This is only a rough intuition, because $W_f(u, \mu)$ can take negative values due to interference effects, which is why we call it a \textit{quasi-probability distribution} instead of a usual non-negative probability distribution. 

It is always true that:
\begin{align}
    |f(u)|^2&=\int W_f(u,\mu)d\mu\\
    |\mathcal{F}[f](\mu)|^2&=\int W_f(u,\mu)du
\end{align}
so the marginals of the Wigner distribution are exactly the intensity of the function in the position domain and the momentum domain. Furthermore, if we integrate $W_f$ over both $u$ and $\mu$ we obtain the $L^2(\mathbb{R})$ norm of the function (i.e., the total energy):
\begin{equation}
    ||f||_{L^2}^2=\int |f(u)|^2dx=\int W_f(u,\mu)du d\mu=\int W_f(u,\mu) d\mu du=\int |\mathcal{F}[f](\mu)|^2d\mu=||\mathcal{F}[f]||_{L^2}^2
\end{equation}
In other words, interchanging the order of integration $dud\mu\mapsto d\mu du$ manifests in Parseval's identity.

Wigner distribution $W_f$ contains all of the information about the signal $f$, but it is encoded in a different way. It is possible to retrieve the signal $f$ (up to a constant phasor $e^{i\phi_0}$) from the Wigner distribution $W_f$ using these identities:
\begin{align}
\label{eq:signal-from-wigner}
    &f(u+u'/2)f^*(u-u'/2)=\int W_f(u,\mu)e^{i2\pi\mu u'}d\mu\\
    &f(u)=\frac{1}{f^*(0)}\int W_f(u/2,\mu)e^{i2\pi\mu u}d\mu
\end{align}
In particular, this implies that there is one-to-one relationship between the complex-valued functions $f\in L^2(\mathbb{R})$ modded by an equivalence relation $f\sim g$ if $f(u)=g(u)e^{i\phi_0}$ for some $\phi_0\in\mathbb{R}$, and the set of all Wigner distributions, which is a subset of $L^2(\mathbb{R}^2)$. Importantly, as we will see not every function in $L^2(\mathbb{R}^2)$ is a Wigner distribution, and we will give an alternative description of the set of all Wigner distributions using Laguerre Gaussians.  

\newpage

\subsection{Examples of Wigner Distribution}
\label{subsection:examples-of-wigner-distributions}

It will be instructive to take a look at a Wigner distribution for several important signals (\ref{tab:wigner-examples}).\footnote{Table adopted from p.70 of \cite{ozaktas2001fractional}.}. In the table below $\text{rect}(u)$ is $1$ if $u\in[-0.5, 0.5]$ and $0$ otherwise, and $\text{sinc}(u)=\text{sin}(\pi u)/(\pi u)$.

\begin{table}[ht]
\centering
\begin{tabular}{c|l|l}
 & $f(u)$ & $W_f(u,\mu)$ \\
\hline
1 & \(  \exp(i2\pi u \xi) \) & \( \delta(\mu-\xi) \)\\
2 & \(  \delta(u-\xi)\) & \( \delta(u-\xi)  \) \\
3 & \(  \exp(i\pi(\chi u^2+2\xi u+\zeta)) \) & \( \delta(\mu-\chi u-\xi)\) \\
4 & \(  (2\chi)^{1/4}\exp(-\pi\chi u^2) \) & \( 2\exp(-2\pi(\chi u^2+\mu^2/\chi)) \)\\
5 & \(  \text{rect}(u) \) & \( 2(1-|2u|)\text{rect}(u)\text{sinc}(2(1-|2u|)\mu) \) \\
\end{tabular}
\caption{Wigner distribution of some common signals}
\label{tab:wigner-examples}
\end{table}

\begin{figure}[!ht]
\centering
\includegraphics[width=\columnwidth]{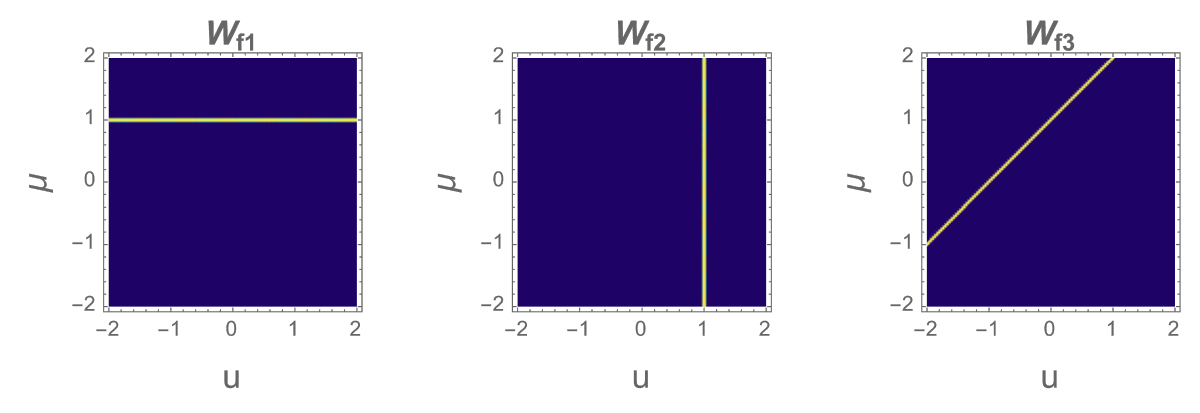}
\caption[Wigner distributions of some simple signals]{Wigner distributions of $\exp(i2\pi u \xi)$ (left), $\delta(u-\xi)$ middle, and $\exp(i\pi(\chi u^2+2\xi u+\zeta))$ (right). The values $\xi$ and $\chi$ are set to 1.}
\label{fig:wigner-deltas}
\end{figure}

Let's interpret some of the signals. The Wigner distribution of $\exp(i2\pi u \xi)$, $\delta(u-\xi)$, and $\exp(i\pi(\chi u^2+2\xi u+\zeta))$ are just straight lines through the phase space (see Figure \ref{fig:wigner-deltas}). For instance, the signal $\exp(i2\pi u \xi)$ corresponds to a line centered at the momentum variable $\mu=\xi$. In fact, a single point on a Wigner distribution can be loosely thought of as a ray at position $u$ with momentum $\mu$, which is related to the angle of the ray propagation via $\mu=\sin(\theta)s/\lambda$ where $\lambda$ is the wavelength, $s$ is the scale factor, and $\theta$ is the angle measured from the $u$ axis (in paraxial limit $\sin\theta\approx\theta$. So, with this interpretation, the Wigner distribution of $\exp(i2\pi u \xi)$ is precisely a bundle of rays on $u\in(-\infty, \infty)$ all propagating in the same direction $\mu=\theta s/\lambda$.

\begin{figure}[!ht]
\centering
\includegraphics[width=0.50\columnwidth]{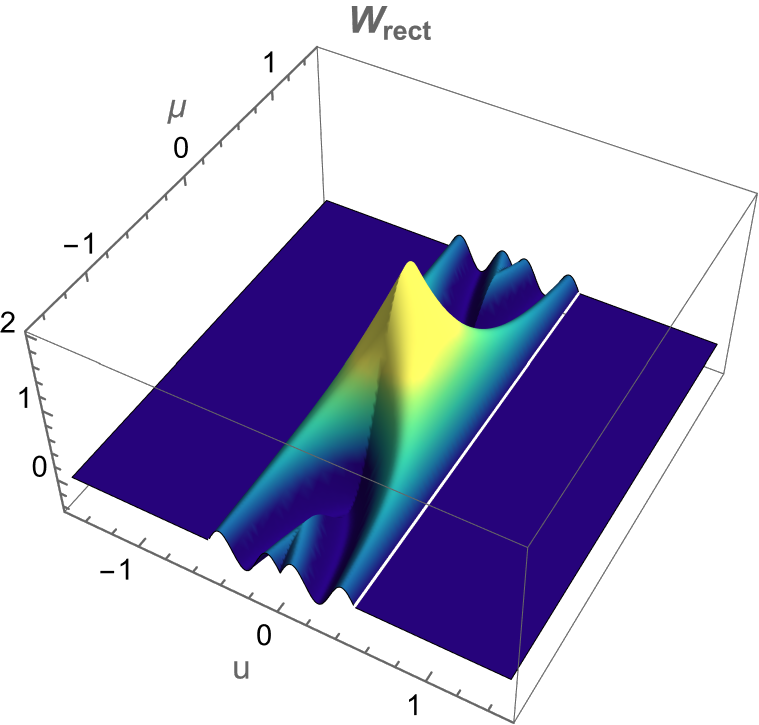}
\caption[Wigner distribution of $\text{rect(u)}$]{Wigner distribution of $\text{rect(u)}$.}
\label{fig:wigner-rect}
\end{figure}

Now, let's consider a Wigner distribution of the $\text{rect}(u)$ function (see Figure \ref{fig:wigner-rect}). As you can see, this distribution has support only at $u\in[-0.5, 0.5]$ (same as the support of $\text{rect}(u)$), but the support in $\mu$ direction is unbounded. In fact, this Wigner distribution achieves negative values, which could not be interpreted using the ray optics picture we presented above. We will refer to a Wigner distribution with negative values as \textit{non-classical}, while a \textit{classical} Wigner distribution will be strictly positive.

\subsection{Hermite-Gaussian Decomposition}
\label{subsection:hermite-gaussian-decomposition}

Recall that Hermite Gaussian basis $(h_l)_{l=0}^\infty$ is a complete orthonormal basis for $L^2(\mathbb{R}^2)$. So, we can express any function as $f=\sum_{n=0}^\infty C_n f_n$. This allows us to express the Wigner distribution as follows:
\begin{align}
\label{eq:wigner-hermite-decompostion}
    W_f(u,\mu)&=\int\sum_{n=0}^\infty C_n h_n(u+u'/2) \sum_{m=0}^\infty C_m^*h_m(u-u'/2)e^{-i2\pi u'\mu}du'\\
    &=\sum_{n=0}^\infty \sum_{m=0}^\infty C_n C_m^*\int h_n(u+u'/2)h_m(u-u'/2)e^{-i2\pi u'\mu}du'\\
    &=\sum_{n=0}^\infty \sum_{m=0}^\infty C_n C_m^*W_{h_n,h_m}(u,\mu)
\end{align}
where the last line is a definition of $W_{h_n,h_m}(u,\mu)$, which is sometimes called cross-Wigner distribution. It turns out that this term can be computed analytically for all $n$ and $m$ in terms of the associated Laguerre polynomials \cite{VanValkenburgh_2008}. 

\begin{equation}
\label{eq:wigner-laguerre-formula}
W_{h_n,h_m}(u,\mu) = 2 (-1)^n (4\pi)^{(m-n)/2} \sqrt{\frac{n!}{m!}}Z^{m-n} L_n^{m-n}(4\pi R^2) e^{-2\pi R^2}
\end{equation}
where we defined $Z=u+i\mu$ is a position on a the phase space and $R = \sqrt{u^2+\mu^2}$ is a distance from the origin. Also, $L_m^{(k)}$ is the associated Laguerre polynomial, which can be defined using the Rodrigues formula.
\begin{equation}
L_{k}^{(\alpha)}(u)
  = \frac{1}{k!}\,
    u^{-\alpha}\,
    e^{u}\,
    \frac{d^{\,k}}{du^{k}}
    \!\Bigl(e^{-u}\,u^{k+\alpha}\Bigr),
\qquad
\alpha > -1,\; k = 0,1,2,\dots
\end{equation}

This invites a very nice interpretation of the Wigner Distribution — once we decompose the signal $f$ into Hermite-Gaussian modes with coefficients $C_n$, the Wigner distribution becomes a linear combination of the associated Laguerre polynomials. 

There are a lot of nice symmetries in the expression \ref{eq:wigner-laguerre-formula} that highlight our intuition about the phase space of the quantum harmonic oscillator. One interesting consequence is that the cross-Wigner distribution is conjugate symmetric and integrates to the Kronecker delta:
\begin{align}
    W_{h_n,h_m}(u,\mu)=[W_{h_m,h_n}(u,\mu)]^* && \iint W_{h_n,h_m}(u,\mu)dud\mu = \delta_{n,m}
\end{align}
which should be reminiscent of the inner product in the Hilbert space, except of course $W_{h_n,h_m}$ is a function and not a scalar. Also notice that when $n=m$ we get:
\begin{equation}
    W_{h_n}=W_{h_n,h_n}=2(-1)^nL_n(4\pi R^2)e^{-2\pi R^2}
\end{equation}

\begin{figure}[!ht]
\centering
\includegraphics[width=0.80\columnwidth]{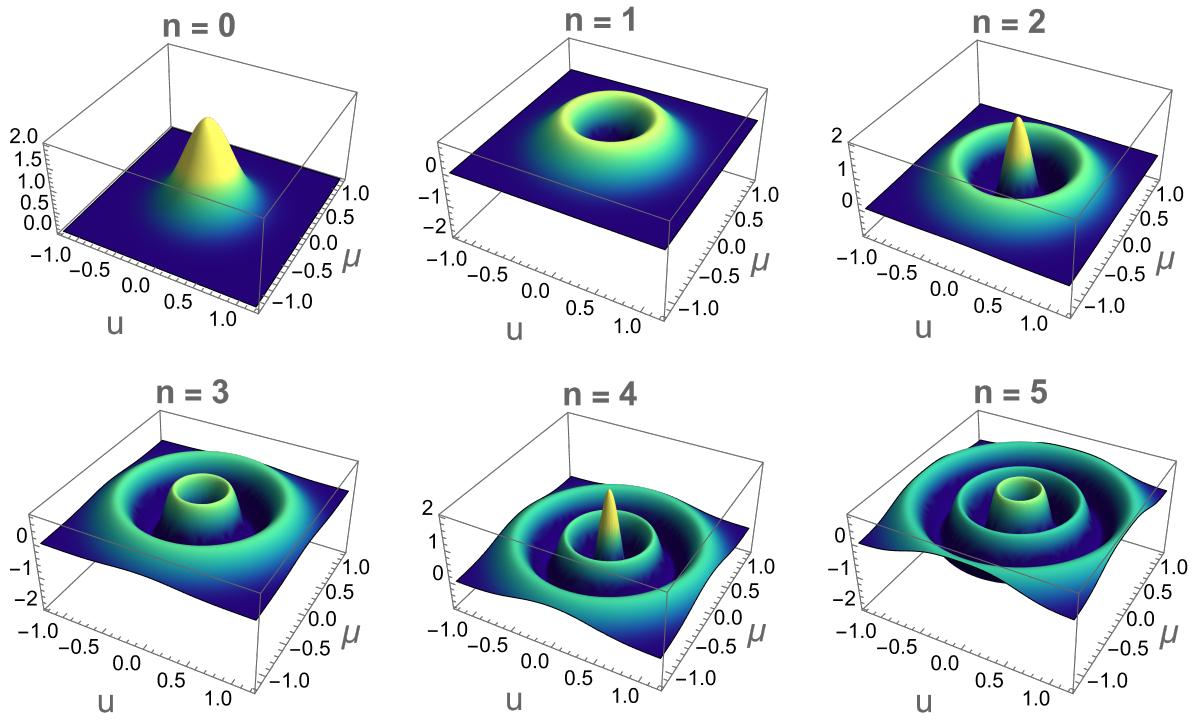}
\caption{Wigner distribution of first five Hermite Gaussians}
\label{fig:wigner-hermite}
\end{figure}

This means that the Wigner distribution of the $n$-th Hermite Gaussian (see Figure \ref{fig:wigner-hermite}) is proportional to the $n$-th Laguerre Gaussian. This is exactly what we expect from the phase-space representations of the eigenfunction of the Quantum Harmonic Oscillator. Since the unitary action of the QHO can be thought of as a rotation on the phase space (i.e., a fractional Fourier transformation), the eigenfunctions do not change their functional form under this evolution, so they must be radially symmetric.

Notice also that the Wigner distribution of the Hermite Gaussians could be negative whenever $n\geq1$ since $L_n(4\pi R^2)$ admits $n$ zeros in $R$. This was not the case for the Wigner distribution of ``classical" signals such as a bundle of rays. This is perhaps not surprising since the eigenstates of the harmonic oscillator rarely appear in classical physics. In fact, their classical ``counterpart" is known as a coherent state, which is a linear combination of the following form:
\begin{equation}
    f_\alpha(u)=\sum_{n=0}^\infty \frac{\alpha^n}{\sqrt{n!}}e^{-|\alpha|^2/2}h_n(u)
\end{equation}
where $\alpha\in\mathbb{C}$ is a single parameter of a coherent state that has a nice physical interpretation in the case of the quantum state of the electromagnetic field — $|\alpha|^2$ is related to the mean photon number and $\text{Arg}(\alpha)$ is the phase of the coherent light source. We can compute the Wigner distribution of the coherent state using the Wigner-Hermite decomposition formula \ref{eq:wigner-hermite-decompostion} \footnote{Alternatively, this computation can be performed using displacement operator formalism.}:
\begin{align*}
    W_{f_\alpha}(u,\mu)&=\sum_{n=0}^\infty \sum_{m=0}^\infty \frac{\alpha^n}{\sqrt{n!}}e^{-|\alpha|^2/2}\frac{\alpha^m}{\sqrt{m!}}e^{-|\alpha|^2/2} W_{h_n,h_m}(u,\mu)\\
    &=2e^{-|\alpha|^2}e^{-2\pi R^2}\sum_{n=0}^\infty \sum_{m=0}^\infty\frac{\alpha^{n}(\alpha^*)^{m}}{\sqrt{n!m!}}\sqrt{\frac{n!}{m!}} (-1)^n (4\pi)^{(m-n)/2} Z^{m-n} L_n^{m-n}(4\pi R^2) \\
    &=2e^{-|\alpha|^2}e^{-2\pi R^2}\sum_{n=0}^\infty \sum_{m=0}^\infty\frac{\alpha^{n}(\alpha^*)^{m}}{m!} (-1)^n (4\pi)^{(m-n)/2} Z^{m-n} L_n^{m-n}(4\pi R^2)\\
    &=2e^{-|\alpha|^{2}}e^{-2\pi R^{2}} \sum_{k=0}^{\infty}\sum_{n=0}^{\infty} \frac{\alpha^{k}\,|\alpha|^{2n}}{(n+k)!}(-1)^{n}(4\pi)^{k/2}Z^{k} L_{n}^{k}(4\pi R^2) 
\end{align*}
Next, we can use the generating function of the associated Laguerre polynomials to perform the summation over $n$:
\begin{equation}
    \sum_{n=0}^{\infty} t^{n}\,L_{n}^{k}(z)
   = \frac{\exp\!\bigl[-z\,t/(1-t)\bigr]}{(1-t)^{k+1}}
   \qquad |t|<1
\end{equation}
Skipping the remaining algebraic manipulations, we obtain the final result:
\begin{equation}
    W_{f_\alpha}(u,\mu) = 2\exp\Bigl[-2\pi\bigl((u-u_{\alpha})^{2}+(\mu-\mu_{\alpha})^{2}\bigr)\Bigr]
\end{equation}
where $u_\alpha=\text{Re}[\alpha]/\sqrt{\pi}$ and $\mu_\alpha=\text{Im}[\alpha]/\sqrt{\pi}$. 
So, we get that the Wigner distribution of the coherent state is simply a Gaussian centered at $u_\alpha, \mu_\alpha$ with the standard deviations $\frac{1}{2\sqrt{\pi}}$. The Wigner distribution is strictly positive, which highlights the classical nature of the coherent state.

\subsection{Important Properties}
\label{subsection:important-properties}

Another important property of the Wigner distribution is that it plays nicely with the fractional Fourier transform. Let $f\in L^2(\mathbb{R})$ be some signal with the associated Wigner distribution $W_f$. Suppose further that we compute a fractional Fourier transform of the signal $f_\alpha=\mathcal{F}_\alpha[f]$. What will be the Wigner distribution of the new signal $f_\alpha$?

It turns out that the answer is simply a clockwise rotation in a phase space by an angle $\alpha$. To be precise, define an operator $R_\alpha$ that rotates phase space coordinates counter-clockwise by angle $\alpha$. Then, $W_{f_\alpha}$ can be written as:
\begin{equation}
    W_{f_\alpha}(u,\mu)=W_f(R_\alpha^{-1}(u,\mu))=W_f(u\cos\alpha-\mu\sin \alpha, u\sin\alpha+\mu\cos\alpha)
\end{equation}
Furthermore, we can obtain the magnitude of the $f_\alpha$ by integrating out the second coordinate. 
\begin{equation}
    |f_\alpha(u)|^2=\int W_{f_\alpha}(u,\mu)d\mu
\end{equation}
which is exactly the Radon Transform, $\mathcal{RDN}_\alpha$, of a two-dimensional distribution $(\mathcal{RDN}_{\alpha} W_f)(u)$. 
\begin{equation}
\label{eq:radon-transform}
    (\mathcal{RDN}_{\alpha} W_f)(u)
  \;=\;
  \int_{-\infty}^{\infty}
  \bigl(W \circ R_{-\alpha}\bigr)(u,\mu)\,d\mu
  \;=\;
  \int_{-\infty}^{\infty}
  W\!\bigl(u\cos\alpha - \mu\sin\alpha,\;u\sin\alpha + \mu\cos\alpha\bigr)\,d\mu
\end{equation}

This means that we can obtain $|f_\alpha(u)|^2$ in two different ways — either performing direct integration over $\mu$ of $W_{f_\alpha}(u,\mu)$ or by projecting the original $W(u,\mu)$ onto an axis, which makes an angle $\alpha$ with the $u$ axis. The first method can be compactly written as $\mathcal{RDN}_0(W_{f_\alpha})$ while the other is simply $\mathcal{RDN}_\alpha(W_f)$. This all can be summarized in the following commuting diagram (\ref{fig:wigner-frft-commute}). 

\begin{figure}[!ht]
\begin{center}
\begin{tikzcd}[row sep=huge, column sep=huge]
  f
    \arrow[r,"W"]
    \arrow[d,"\mathcal{F_\alpha}"']
  &
  W_f
    \arrow[r,"\mathcal{RDN}_0"]
    \arrow[d,"\mathcal{R}"]
    \arrow[dr,"\mathcal{RDN}_\alpha"]
  &
  {|f(u)|^{2}}
  \\
  f_\alpha
    \arrow[r,"W"']
  &
  W_{f_\alpha}
    \arrow[r,"\mathcal{RDN}_0"']
  &
  {|f_\alpha(u)|^{2}}
\end{tikzcd}
\caption[Commutative diagram between $W_f$, $\mathcal{F}_\alpha$, and $\mathcal{RDN}_\alpha$]{Relationship between the Wigner distribution, fractional Fourier Transform, and Radon transform.}
\label{fig:wigner-frft-commute}
\end{center}
\end{figure}
In a sense, $W_f$ and $W_{f_{\alpha}}$ are the same functions up to the coordinate rotation. Similarly, $f$ and $f_\alpha$ are the same functions up to the basis rotation. In fact, it is pretty easy to show that this leads to the following equivalent definition of the Wigner distribution:
\begin{definition}
    Let $f\in L^2(\mathbb{R})$ be a function that has a position basis representation $f(u)$ i.e., we can view $f:\mathbb{R}\to\mathbb{C}$.  The Wigner Distribution of the function $f$ is a function $W_f:\mathbb{R}^2\to\mathbb{R}$ defined as:
    \begin{equation}
    \label{eq:wigner-def2}
        W_f(u, \mu)=\int f_\alpha(u+u'/2)f_\alpha^*(u-u'/2)e^{i2\pi u'\mu}du'
    \end{equation}
    where $f_\alpha = \mathcal{F}_\alpha[f]$ is the fractional Fourier transformation of $f$ for any angle $\alpha\in(0,\pi)$.
\end{definition}
An easy way to see that this is true is to notice that the conjugation in $f_\alpha^*$ term flips $i\mapsto-i$ in the FrFT kernel, effectively canceling the kernel from $f_\alpha$. We end up with a triple integral, but two of these integrals become delta functions, which eventually reduces us to the original definition.

\section{Higher Dimension Generalization}
\label{section:higher-dimension-generalization}

Our definitions for the fractional Fourier transform and the Wigner distribution can be trivially extended to functions of two (and more) variables. To do so, consider a function $f(u,v)\in L^2(\mathbb{R}^2)$. We will denote the conjugate variables as $\mu$ and $\nu$. In Appendix \ref{chapter:fourier-analysis}, we defined the Fourier transform of such function as:
\begin{equation}
      F(\mu, \nu) = \mathcal{F}[f](\mu,\nu) =  \int_{-\infty}^{+\infty}\int_{-\infty}^{+\infty} f(u, v)\exp(-i 2\pi (u \mu + v \nu)dudv  
\end{equation}
The analogous expression generalizes Definition 2 of the FrFT:
\begin{align}
\mathcal{F}_\alpha[f](\mu,\nu) &\equiv \int_{-\infty}^{\infty}\int_{-\infty}^{\infty}  f(u,v) K_\alpha(u,\mu)K_\alpha(v,\nu) du dv\\
&K_\alpha(u,\mu)\equiv\sqrt{1 - i \cot \alpha} \exp\left(
-i2\pi \left(-\frac{\mu^2}{2}\cot\alpha+u\mu \csc \alpha -\frac{u^2}{2}\cot \alpha \right)
\right)
\end{align}
Notice that in principle it is even possible to rotate $u$ and $v$ coordinates by different angles $\alpha_1$ and $\alpha_2$, but we are not going to complicate our lives this far. A similar line of reasoning applies to the Wigner distribution, which now becomes a 4-dimensional object $W_f(u,v,\mu,\nu)$.
\begin{equation}
    W_f(u, v, \mu, \nu)=\int f(u+u'/2, v+v'/2)f^*(u-u'/2, v-v'/2)e^{-i2\pi( u'\mu+v'\nu)}du'dv'
\end{equation}
We can also extend all of our ideas with the Hermite-Gaussian basis to 2 dimensions. Since $(h_n(u))_{n=0}^\infty$ forms a complete basis for $L^2(\mathbb{R})$, a product basis $(h_n(u)h_m(v))_{n,m=0}^\infty$ can be used for $L^2(\mathbb{R}^2)$. In this case, we can decompose any function $f(x,y)$ as follows:
\begin{align}
    f(u,v)=\sum_{n=0}^\infty\sum_{m=0}^\infty C_{n,m}h_n(u)h_m(v) && C_{n,m}=\braket{h_n(u)h_m(v)|f(u,v)}
\end{align}
The last idea will turn out to be very useful in later chapters. Generalizing to $n$-dimensions is as simple as adding more integral (summation) signs.

\section{Optical Implementation}
\label{section:optical-implementation}

So far, our discussion has been mainly focused on theoretical objects that describe the propagation of waves. Of course, we would not be talking about these objects if they did not matter in the experimental settings. In this section, we will show how one could experimentally implement a fractional Fourier transform and explain what measurement information is accessible to the experimenter. Our discussion will closely follow the Chapter 8 from \cite{ozaktas2001fractional}.

\subsection{Linear Canonical Transforms}
\label{subsection:linear-canonical-transforms}

The fractional Fourier transform is just a particular example of a more general class of transformations that we can perform with common optical elements such as lenses, sections of free space, and quadratic graded-index medium. The most general allowed transformation goes by the name of a linear canonical transformation, and it is a map that takes a function $f$ to a function $g$ via the following integral:
\begin{align}
\label{eq:1d-linear-canonical}
\hat{g}(x)&=\int\hat{K}(x,x')\hat{f}(x')dx'\\
\hat{K}(x,x')&\equiv \sqrt{\hat{\beta}}e^{-i\pi/4}\exp\left[i\pi\left(\hat{\alpha}x^2-2\hat{\beta}xx'+\hat{\gamma}x'^2\right)\right]
\end{align}
where $\hat{\alpha}$, $\hat{\beta}$, and $\hat{\gamma} \in\mathbb{R}$ are three independent parameters that define the linear canonical transform. A two-dimensional generalization is very straightforward. 
\begin{align}
\label{eq:2d-linear-canonical}
\hat{g}(x,y)&=\iint\hat{K}(x,x')\hat{K}(y,y')\hat{f}(x',y')dx'dy'
\end{align}
Notice also that we use hats to denote that the function (or parameter) has dimensional units. We will use a scale parameter $s$ (units of length) to go from dimensionless coordinates $u,v$ to physical coordinates $x$ and $y$ that have units of length. We define:
\begin{align}
\label{eq:dimensionless-rescaling}
    f(u,v)&\equiv f(x/s,y/s)\equiv s \hat{f}(x,y)\\
    F(U,V)&\equiv F(s\sigma_x, s\sigma_y)\equiv s^{-1}\hat{F}(\sigma_x,\sigma_y) 
\end{align}
Similarly, for the parameters of the transformation:
\begin{align}
    \hat{\alpha}\equiv\alpha/s^2 && \hat{\beta}\equiv\beta/s^2 &&\hat{\gamma}\equiv\gamma/s^2
\end{align}
See Section \ref{sec:dimensionless-form} in Appendix \ref{chapter:fourier-analysis} for more details and rationale behind these definitions. It is possible to relate the parameters of the linear canonical transform to the ABCD ray matrix:
\begin{equation}
  \begin{bmatrix}
    \hat{A} & \hat{B} \\
    \hat{C} & \hat{D}
\end{bmatrix}
\;=\;
\begin{bmatrix}
    \hat{\gamma}/\hat{\beta} & 1/\hat{\beta} \\
    -\hat{\beta}+\hat{\alpha}\hat{\gamma}/\hat{\beta} & \hat{\alpha}/\hat{\beta}
\end{bmatrix}
\end{equation}
where as expected we have $\hat{A}\hat{D}-\hat{B}\hat{C}=1$. Once again, the dimensionless version of the matrix can be obtained via
\begin{align}
    \hat{A}\equiv A && \hat{B}\equiv s^2B && \hat{C}\equiv C/s^2 && \hat{D}\equiv D
\end{align}
In fact, one can rewrite the kernel of the linear canonical transform using the ABCD matrix:
\begin{equation}
\label{eq:linear-canonical-ABCD}
    \hat{h}(x,x')=\sqrt{\frac{1}{\hat{B}}}e^{-i\pi/4}\exp\left[\frac{i\pi}{\hat{B}}(\hat{D}x^2-2xx'+\hat{A}x'^2)\right]
\end{equation}
where the parameter $\hat{C}$ does not appear in the expression above. This is because it was eliminated using the determinant condition $\hat{A}\hat{C}-\hat{B}\hat{D}=1$. This makes sense because linear canonical transforms have three independent degrees of freedom, so one of the parameters in the ABCD matrix is redundant.

We can use this connection to the ABCD matrices to understand the effect of the general linear canonical transform on a Wigner distribution. Recall that in ray optics, the ABCD matrix describes the coordinate transformation between the input and the output ray to the optical system. Concretely, if the input ray has position $x_1$ and propagates in the direction $\sigma_{x_1}\equiv\theta_1/\lambda$, then on the output it will become:
\begin{equation}
  \begin{bmatrix}
    x_2 \\
    \sigma_{x_2}
\end{bmatrix}
\;=\;
\begin{bmatrix}
    \hat{A} & \hat{B} \\
    \hat{C} & \hat{D}
\end{bmatrix}
\begin{bmatrix}
x_1\\
\sigma_{x_1}
\end{bmatrix}
\end{equation}
So, at least in the ray-optics limit, the linear canonical transformation is just a change of coordinates. This is actually true in wave optics as well, as one can formally prove that \cite{ozaktas2001fractional}
\begin{equation}
\label{eq:wigner-through-linear-canonical}
    \hat{W}_{\hat{g}}(x,\sigma_x)=\hat{W}_{\hat{f}}(\hat{D}x-\hat{B}\sigma_x,-\hat{C}x+\hat{A}\sigma_x)
\end{equation}
where $\hat{g}(x)$ is the output of the linear canonical transform applied to a function $\hat{f}(x)$ and $\hat{A}, \hat{B},\hat{C}$ and $\hat{D}$ are the associated parameters of the transformation.

\subsection{Optical Components}
\label{subsection:optical-components}

Now, we will describe the relevant optical components through the language of the linear canonical transforms. For each optical component, we will demonstrate the transformation kernel, associated ABCD matrix, and explain its effect on the Wigner distribution.

\subsubsection{Free Space Propagation}
\begin{figure}[!ht]
\centering
\includegraphics[width=1\columnwidth]{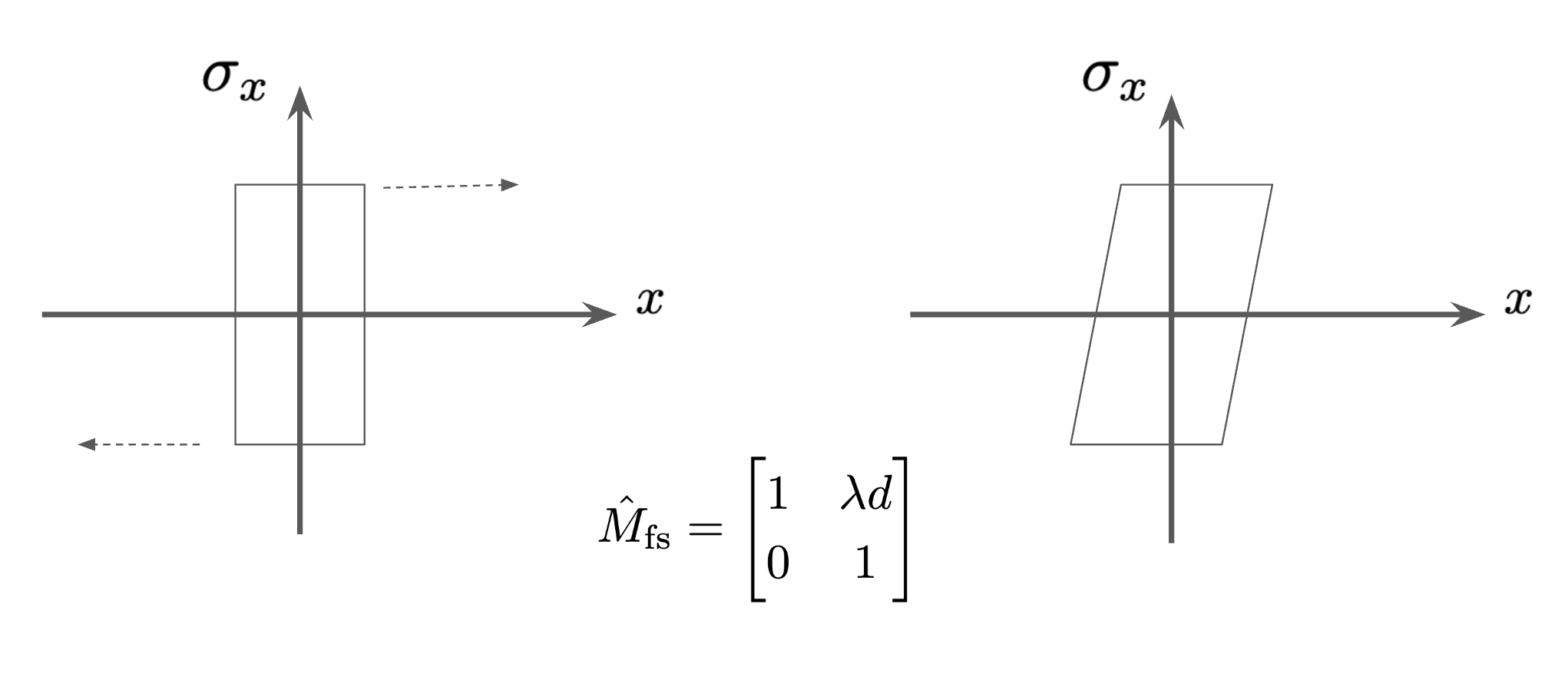}
\caption{Effect of the free space propagation on the Wigner distribution}
\label{fig:shear-fs-transform}
\end{figure}

Suppose you propagate a laser of wavelength $\lambda$ through a section of free space of length $d$. Then the associated linear canonical transform is:
\begin{align}
\label{eq:free-space-canonical}
\hat{g}(x) &=\int\hat{K}_{\text{fs}}(x,x')\hat{f}(x')dx'\\
\hat{K}_{\text{fs}}(x,x')&\equiv e^{-i\pi/4}\sqrt\frac{1}{\lambda d}e^{i\pi(x-x')^2/\lambda d}
\end{align}
where the associated ABCD matrix is:
\begin{equation}
  \hat{M}_{\text{fs}}=\begin{bmatrix}
    1 & \lambda d \\
   0 & 1
\end{bmatrix}
\end{equation}
This has the effect of shearing the Wigner distribution in the $x$ direction (see Figure \ref{fig:shear-fs-transform}).

\subsubsection{Thin Lens}
\label{subsection:thin-lens}

\begin{figure}[!ht]
\centering
\includegraphics[width=1\columnwidth]{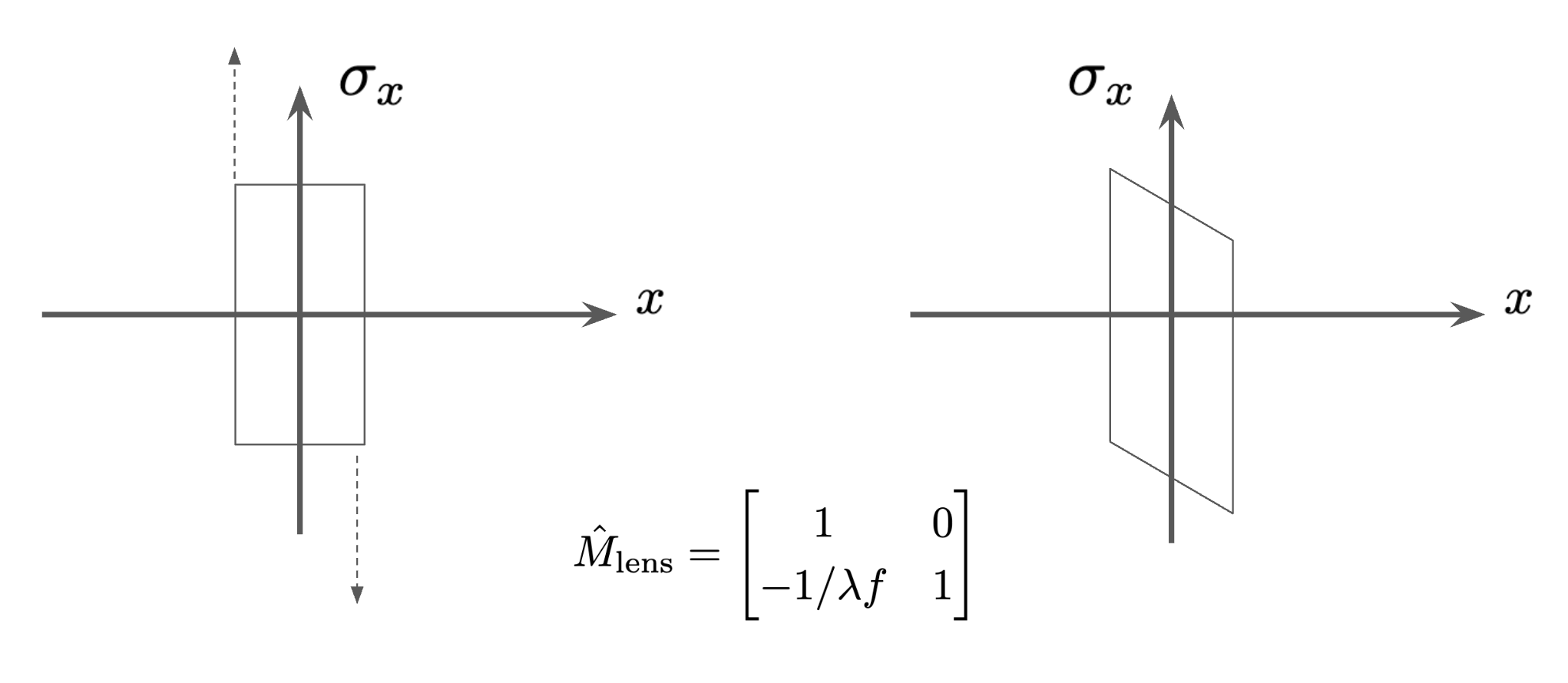}
\caption{Effect of the propagation through the lens on the Wigner distribution}
\label{fig:shear-lens-transform}
\end{figure}

Now, imagine that your light encounters a thin lens \footnote{We are assuming a paraxial limit and that lens has no aberrations.} with a focal distance $f$ (positive $f$ corresponds to a convex lens). The associated linear canonical transform is just a multiplication by a quadratic phase factor:
\begin{align}
\label{eq:thin-lens-canonical}
\hat{g}(x) &=\int\hat{K}_{\text{lens}}(x,x')\hat{f}(x')dx'\\
\hat{K}_{\text{lens}}(x,x')&\equiv e^{-i\pi x^2/{\lambda}f}\delta(x-x')
\end{align}
The ABCD matrix is given by:
\begin{equation}
  \hat{M}_{\text{lens}}=
\begin{bmatrix}
    1 & 0 \\
   -1/\lambda f & 1
\end{bmatrix}
\end{equation}
This has the effect of shearing the Wigner distribution in the $\sigma_x$ direction (see Figure \ref{fig:shear-lens-transform}).

\subsubsection{Quadratic graded-index medium}
\begin{figure}[!ht]
\centering
\includegraphics[width=1\columnwidth]{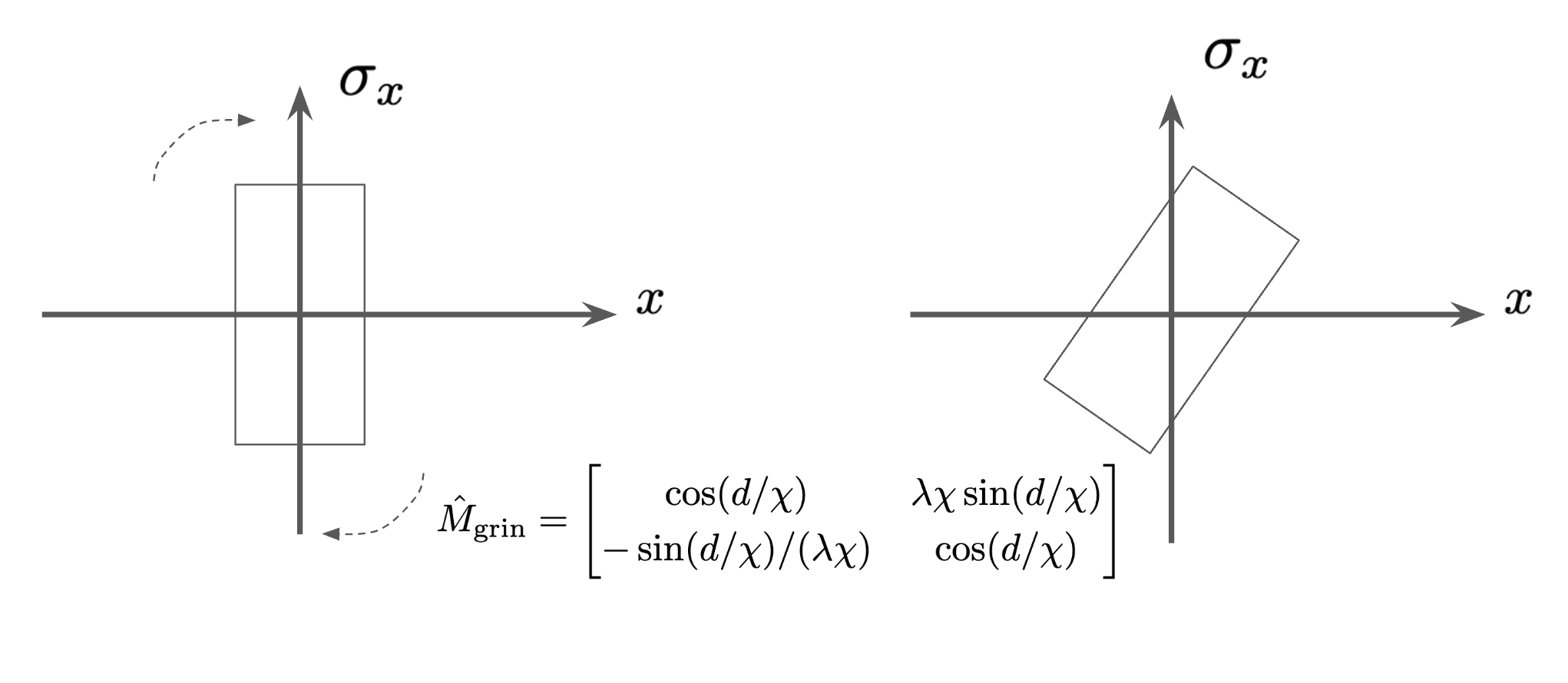}
\caption[Effect of the propagation through GRIN medium on the Wigner distribution]{Effect of the propagation through the quadratic-graded index medium on the Wigner distribution}
\label{fig:grin-transform}
\end{figure}

Now, suppose that we propagate distance $d$ through the material whose index of refraction changes with $x$:
\begin{equation}
    n^2(x)=n_0^2[1-(x/\chi)^2]
\end{equation}
where $\chi$ in the units of length is the parameter of the medium that characterizes how quickly the index of refraction changes, as we deviate from the main optical axis $x=0$. In 2D we can have:
\begin{equation}
    n^2(x,t)=n_0^2[1-(x/\chi_x)^2-(y/\chi_y)^2]
\end{equation}
but we will only focus on a special case when $\chi_x=\chi_y$. Despite not being as common as lenses and free space in the lab, quadratic graded-index materials (also known as GRIN) are commercially available and can be purchased from e.g. \href{https://www.thorlabs.com/newgrouppage9.cfm?objectgroup_id=11167}{Thorlabs}. One can show (see Chapter 7 from \cite{ozaktas2001fractional}) that quadratic graded-index corresponds to the following linear canonical transform:
\begin{align}
\label{eq:grin-canonical}
\hat{g}(x) &=\int\hat{K}_{\text{grin}}(x,x')\hat{f}(x')dx'\\
\hat{K}_{\text{grin}}(x,x')&\equiv 
\begin{cases}
\sqrt{\frac{1-i\cot (d/\chi)}{\lambda\chi}}e^{-id/2\chi} e^{\frac{i\pi}{\lambda\chi}(\cot(d/\chi) x^2-2\csc(d/\chi) x x'+\cot(d/\chi) x'^2)} \quad &d\neq n\pi\chi \\ 
e^{-id/2\chi}\delta(x-x')\quad &d=2n\pi\chi\\ 
e^{-id/2\chi}\delta(x+x')\quad &d=(2n\pm1)\pi\chi
\end{cases}
\end{align}
which looks almost exactly like Definition 2 of the fractional Fourier transform — more on this later. The associated ABCD matrix is
\begin{equation}
  \hat{M}_{\text{grin}}=\begin{bmatrix}
    \cos (d/\chi) & \lambda\chi \sin(d/\chi) \\
    -\sin (d/\chi)/(\lambda\chi) & \cos(d/\chi)
\end{bmatrix}
\end{equation}
which is exactly a rotation of the Wigner Distribution by an angle of $\alpha = d/\chi$ if $\lambda\chi =1$ (more generally it is a rotation with scaling) (see \ref{fig:grin-transform}).

\subsubsection{Spherical Reference Surface}
\begin{figure}[!ht]
\centering
\includegraphics[width=0.5\columnwidth]{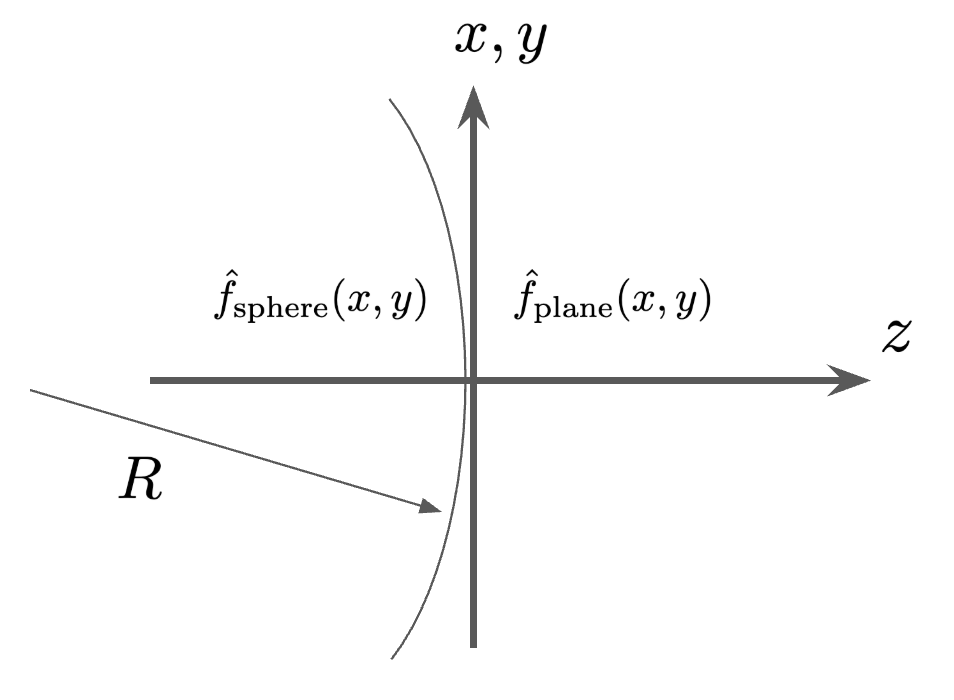}
\caption{Spherical reference surface}
\label{fig:spherical-ref}
\end{figure}

Although this is not strictly an optical element, it is very useful sometimes to switch between the planar and spherical surfaces to represent the same signal $\hat{f}$ (see Figure \ref{fig:spherical-ref}).
\begin{equation}
    \hat{f}_{\text{plane}}(x,y)=\hat{f}_{\text{sphere}}(x,y)\exp[i\pi(x^2+y^2)/\lambda R]
\end{equation}
This has the effect of multiplying the signal by the phase, which depends on the radius of curvature of the chosen surface. In the paraxial limit, we can model this as a thin lens, so all of the formulas from the Subsection \ref{subsection:thin-lens}  apply here too.

\subsection{Optical Systems}
\label{subsection:optical-systems}

\begin{figure}[!ht]
\centering
\includegraphics[width=0.5\columnwidth]{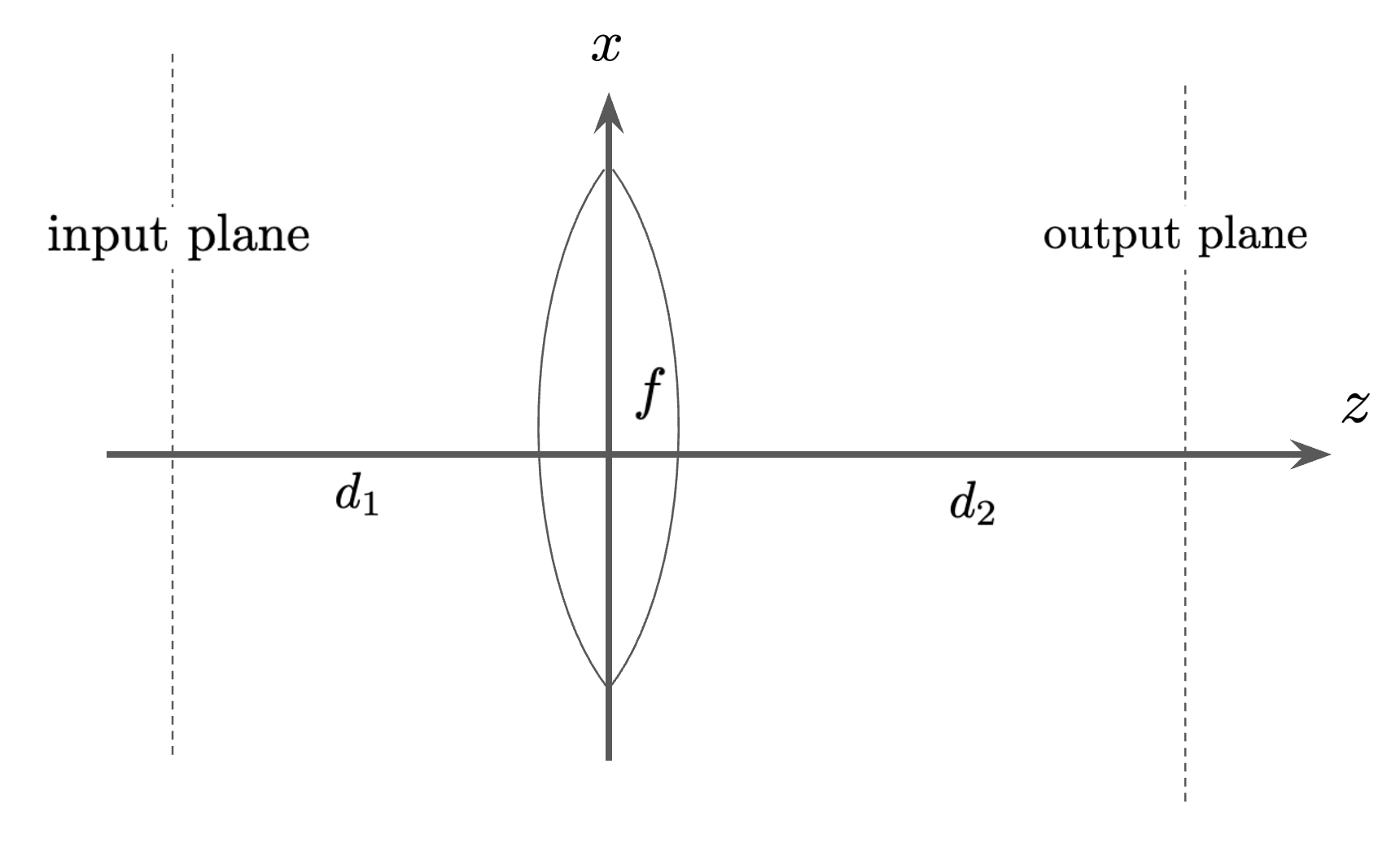}
\caption[Space-lens-space optical system]{Optical system composed of a section of free space, followed by a lens, followed by another section of free space.}
\label{fig:space-lens-space}
\end{figure}

Suppose now we arrange a system that combines a section of free space of distance $d_1$ followed by a thin lens with focal distance $f$, and another section of free space of distance $d_2$ (see Figure \ref{fig:space-lens-space}).

Let's figure out the effect of this system on the input signal $\hat{f}(x)$. One way to do this would be to perform tedious integration through each optical component, which is what I did when I just started doing research in this area. It was a great learning experience, but thankfully, now I know a better way. Given that any linear canonical transform is parametrized by the associated ABCD matrix, we can simply perform two quick 2x2 matrix multiplications to find the matrix of the whole system. This method fully determines the linear canonical transform (eq. \ref{eq:linear-canonical-ABCD}).
\begin{equation}
\hat{M}_{\text{system}}=\hat{M}_{\text{fs2}}\hat{M}_{\text{lens}}\hat{M}_{\text{fs1}}
\end{equation}
Plugging in expressions for each optical component, we get:
\begin{equation}
\label{eq:ABCD-general-space-lens-space}
\begin{bmatrix}
    \hat{A} & \hat{B} \\
   \hat{C} & \hat{D}
\end{bmatrix}=
\begin{bmatrix}
    1 & \lambda d_1 \\
   0 & 1
\end{bmatrix}
\begin{bmatrix}
    1 & 0 \\
   -1/\lambda f & 1
\end{bmatrix}
\begin{bmatrix}
    1 & \lambda d_2 \\
   0 & 1
\end{bmatrix}=
\begin{bmatrix}
    1-\frac{d_1}{f} & \lambda  \left(d_1 + d_2 - d_1d_2/f \right) \\
 -\frac{1}{f \lambda } & 1-\frac{d_2}{f}  
\end{bmatrix}
\end{equation}
This system corresponds to the following general linear canonical transform:
\begin{equation}
    \hat{h}(x,x')=\frac{e^{-i\pi/4}}{\sqrt{\lambda\left(d_1+d_2- d_1d_2/f\right)}}\exp\left[\frac{i\pi((1-d_2/f)x^2-2xx'+(1-d_1/f)x'^2)}{\lambda\left(d_1+d_2- d_1d_2/f\right)}\right]
\end{equation}
The effect on the Wigner distribution can be inferred from the formula \ref{eq:wigner-through-linear-canonical}. Now we consider several special cases of the equation above.

\subsubsection{Imaging System implementation}
Consider a special case when 
\begin{equation}
    \frac{1}{f}=\frac{1}{d_1}+\frac{1}{d_2}
\end{equation}
Notice that in this case, the ray matrix simplifies to
\begin{equation}
\begin{bmatrix}
    \hat{A} & \hat{B} \\
   \hat{C} & \hat{D}
\end{bmatrix}=
\begin{bmatrix}
    -d_2/d_1 & 0 \\
    -1/\lambda f & -d_1/d_2
\end{bmatrix}\equiv
\begin{bmatrix}
    M & 0 \\
    M/\lambda R & 1/M
\end{bmatrix}
\end{equation}
where in the last line we defined $M\equiv-d_2/d_1$ and $R=fd_2/d_1$. This corresponds\footnote{A mathematically inclined reader might be troubled that we are dividing by 0 in the equation \ref{eq:linear-canonical-ABCD} when $\hat{B}=0$. The correct way to derive the kernel above is to do this computation in the Fourier domain, apply the convolution theorem and take the limit as defocusing parameter goes to 0.} to the following linear canonical transformation:
\begin{equation}
    \hat{h}(x,x')=e^{i\pi x^2/\lambda R}\sqrt{M}\delta(x-Mx')
\end{equation}
which allows for an easy interpretation of what this system is doing to the input signal $f$.
\begin{equation}
    \hat{g}(x)=\int\hat{h}(x,x')\hat{f}(x)=e^{i\pi x^2/\lambda R} \sqrt{M} f(Mx)
\end{equation}
As we can see, the output signal at point $x$ corresponds to the input signal at point $Mx$, where $M$ is the negative value that signifies the \textit{magnification} of the image. If the absolute value $|M|<1$, then the image on the output plane is smaller than the image on the input, and vice versa if $|M|>1$. In fact, if $f>0$ (i.e., the lens is convex), we always expect the image to be inverted, which should be reminiscent of the fractional Fourier transform by the angle $\pi$, which corresponds to the parity operator. 

Also, notice that the final image acquires an additional quadratic phase characterized by the \textit{radius of curvature} $R=fd_2/d_1$. We can deal with it in several ways. We can construct our system in such a way that $R$ is very large, so this phase is negligible, or introduce another lens to correct for it, or just use a spherical reference surface that eliminates this phase shift mathematically. 

\subsubsection{Fourier Transform implementation}

Let us consider a different limit of the original expression when $d_1=d_2=f$. Then our matrix simplifies significantly to:
\begin{equation}
\begin{bmatrix}
    \hat{A} & \hat{B} \\
   \hat{C} & \hat{D}
\end{bmatrix}=
\begin{bmatrix}
    0 & \lambda f \\
 -1/f \lambda  & 0 
\end{bmatrix}
\end{equation}
This corresponds to the following linear canonical transformation
\begin{equation}
\label{eq:foruier-transform}
    \hat{h}(x,x')=\sqrt{\frac{1}{\lambda f}}e^{-i\pi/4}\exp\left[\frac{i\pi}{\lambda f}(-2xx')\right]
\end{equation}
Then the output of this system is simply a rescaled Fourier transform of the input:
\begin{equation}
    \hat{g}(x)=\frac{e^{-i\pi/4}}{\sqrt{\lambda f}}\int \hat{f}(x')\exp{(-2\pi i x'(x/\lambda f))}dx'=\frac{e^{-i\pi/4}}{\sqrt{\lambda f}} \mathcal{F}[\hat{f}](x/\lambda f)
\end{equation}
We can make this even better by expressing everything in dimensionless variables and using the scale parameter $s=\sqrt{\lambda f}$. We obtain:
\begin{equation}
    g(u)=e^{-i\pi/4} \mathcal{F}[f](u)
\end{equation}
So, this is exactly the Fourier transform of dimensionless functions up to a non-essential constant $e^{-i\pi/4}$, which never matters in practice. 

A natural generalization of this is the case when $d_1=f$ but $d_2$ is free to vary (we will call it $d$). Then we will have an additional quadratic phase factor in our kernel:

\begin{equation}
    \exp\left[\frac{i\pi}{\lambda}\left(\frac{f-d}{f^2}\right)x^2\right]\equiv\exp\left[\frac{i\pi}{\lambda R}x^2\right]
\end{equation}
where we once again defined $R=(f-d)/f^2$ to be the radius of curvature of the wavefront. 

\subsection{Fractional Fourier Transform Implementations}
\label{subsection:Fractional-Fourier-Transform-Implementations}

Below, we describe several equivalent ways to physically implement the fractional Fourier transform in the lab. We will see that any optical element can be viewed as a fractional Fourier transformation (with magnification and additional quadratic phase factor).

\subsubsection{Quadratic Graded-index Medium}

The most natural implementation of the fractional Fourier transform is by using the quadratic graded-index medium. Then propagating by distance $d$ will perform a Fourier rotation by the angle $\alpha=d/\chi$. This means that the signal at a distance $d=\chi\pi/2$ is exactly the Fourier transform, then at the distance $d=\chi\pi$ we get a parity-flipped signal, etc.

Notice also that this system has a natural choice of the scale factor $s=\sqrt{\chi\lambda}$, and with this scale factor, the relationship between the dimensionless input and the output is precisely a fractional Fourier transform:
\begin{align}
    g(u)&=\mathcal{F_\alpha}[f](u) 
\end{align}
where as usual $f(u)\equiv f(x/s)\equiv s^{1/2}\hat{f}(x)$. To recover the final dimensional output, we  apply $\hat{g}(x)\equiv s^{-1/2}g(x/s)\equiv s^{-1/2}g(u)$. 

\subsubsection{Space Lens Space Implementation}
The most natural way to obtain a fractional Fourier transform using the space lens space optical system described above is to pick $d$ and $f$ according to the following relationship:
\begin{align}
    d&=\frac{s^2}{\lambda}\tan(\alpha/2)\\
    f&=\frac{s^2}{\lambda}\csc(\alpha)
\end{align}
where $s$ is the resulting scale factor. For any $\alpha\neq n\pi$ there are infinitely many solutions $(d, f, s)$ to the constraints above, so it is always possible to find a configuration that would implement $\mathcal{F}_\alpha$ on the dimensionless functions $f(u)\equiv f(x/s)=s^{1/2}\hat{f}(x)$ and $g(u)\equiv g(x/s)=s^{1/2}\hat{g}(x)$.

\subsubsection{Free Space Implementation}
It turns out that even if we do not place any optics, we will observe a fractional Fourier transformation of the signal. Recall that a signal propagating a distance $d$ through free space will undergo the following transformation:

\begin{align}
\hat{g}(x) &=e^{-i\pi/4}\sqrt\frac{1}{\lambda d}\int e^{i\pi(x-x')^2/\lambda d}\hat{f}(x')dx'
\end{align}
One can relate this system to the fractional Fourier transform by making the following identification:
\begin{align}
    \tan(\alpha)=\frac{\lambda d}{s^2}&& M=\sqrt{1+\left(\frac{\lambda d}{s^2}\right)^2}&& R=\frac{s^4+\lambda^2d^2}{d}
\end{align}

So, at a distance $d$ from the input plane, we will observe a fractional Fourier transform $\mathcal{F}_\alpha$ of the original signal. The transformed signal will live on a spherical reference surface with radius $R$ and will be magnified by $M$. Alternatively, we can say that the image on a planar surface will have a quadratic phase factor set by the radius of curvature $R$. As $d\to \infty$, the angle $\alpha \to \pi/2$, and we recover the usual Fraunhofer diffraction equation.

\subsubsection{General Implementation}

Recall that the general class of linear canonical transforms is characterized by three variables $\hat{\alpha}, \hat{\beta}$, and $\hat{\gamma}$, while the fractional Fourier transform is a class of transformations indexed by one parameter $\alpha$. It turns out that if we allow for magnification $M$ and additional radius of curvature parameter $R$ we will recover all possible linear canonical transforms.

The mapping between the two transformations is given by the following equations:
\begin{align}
    \hat{\alpha} =\frac{\cot(\alpha)}{s^2M^2}+\frac{1}{\lambda R} &&
    \hat{\beta} =\frac{\csc(\alpha)}{s^2M} &&
    \hat{\gamma} = \frac{\cot{(\alpha)}}{s^2}
\end{align}
Alternatively, we can relate any ABCD matrix to a fractional Fourier transformation plus magnification and radius of curvature $R$.
\begin{align}
    &\hat{A}=M\cos(\alpha) &&
    &\hat{B}=s^2M\sin(\alpha) \\
    &\hat{C}=-\frac{\sin(\alpha)}{s^2M}+\frac{M\cos(\alpha)}{\lambda R} &&
    &\hat{D}= \cos(\alpha)/M+\frac{s^2M\sin(\alpha)}{\lambda R} 
\end{align}
which can be inverted to obtain $\alpha$, $M$ and $R$:
\begin{align}
\label{eq:ABCD-to-FrFT}
    \tan(\alpha)=\frac{1}{s^2}\frac{\hat{B}}{\hat{A}} &&
    M=\sqrt{\hat{A}^2+(\hat{B}/s^2)^2} && 
    \frac{1}{\lambda R}&=\frac{1}{s^4}\frac{\hat{B}/\hat{A}}{\hat{A}^2+(\hat{B}/s^2)^2}+\frac{\hat{C}}{\hat{A}}
\end{align}
So, in summary, any optical element that can be described by a real-valued ABCD matrix implements a fractional Fourier transform of some angle $\alpha$, which will appear at a spherical surface of radius $R$, magnified by $M$. As long as we keep track of the relevant data and only apply the fractional Fourier transform to the dimensionless functions, we can use the equations above to predict the electric field at any plane in the optical system!

\subsection{Camera projection}
\label{subsection:camera-projection}

Lastly, we will briefly mention the mathematical interpretation of placing a camera in the optical setup. In general, a camera implements an intensity measurement of the planar surface:
\begin{equation}
    I_z(x,y)\propto |f(x,y,z)|^2=f(x,y,z)f^*(x,y,z)
\end{equation}
which erases all of the phase information. In particular, if the signal naturally lives on a spherical surface with radius of curvature $R$, the phase term $\exp[i\pi(x^2+y^2)/\lambda R]$ will not be visible for the camera.

One way to get more information from the camera images is to use the fractional Fourier transformation. First of all, notice that we can view the camera projection as a Radon transform of the angle $\alpha=0$ of the Wigner distribution $W_f$, because we have that:
\begin{equation}
    (\mathcal{RDN}_{0} W_f)(x)=|f(x)|^2 
\end{equation}
Recall, also, that we can use pretty much any optical system to implement the fractional Fourier transformation to obtain a new signal $f_\alpha=\mathcal{F_\alpha}[f]$ and a new Wigner distribution $W_{f_\alpha}$. Now, if we take a picture, we will get:
\begin{equation}
    (\mathcal{RDN}_{0} W_{f_\alpha})(x)=|f_\alpha(x)|^2
\end{equation}
but on the other hand this is equivalent to $(\mathcal{RDN}_{\alpha} W_{f})$ — i.e., projecting the original Wigner distribution $W_f$ onto an axis tilted by angle $\alpha$. 
By the Fourier slicing theorem, each such image corresponds to a single slice through the Fourier transform of the Wigner distribution. \footnote {For $\hat{W}_f(x,\sigma_x)$ we need to Fourier transform both variables. For a general $f:\mathbb{R}^n\to\mathbb{C}$, we need to apply the $n$-dimensional Fourier transform to use the Fourier slicing theorem.} So, if we collect enough such slices, we can reconstruct the entire Wigner distribution, which will give us the corresponding complex-valued signal $f$. This mathematical idea has been implemented by \cite{Akondi:21} and \cite{ANCORA201881} to experimentally measure the Wigner distribution of a light-field.

%% file: sections/chapter3.tex
\chapter{Ray Optics Reduction with Optimal Transport}
\label{chapter:ray-optics-reduction-with-optimal-transport}

In this chapter, we will use the theory developed in Chapter 2 to reformulate the problem of phase generation and beam estimation. We will see that there is a natural way to view these problems through the language of fractional Fourier transforms and Wigner distributions. Next, we will use the stationary phase approximation to obtain the ray-optics limit of both problems. We will establish a deep theoretical connection to the theory of optimal transport, and explain how the transport plan naturally arises as a ray-optics limit of a Wigner distribution.

\section{Problem reformulation}
\label{section:problem-reformulation}

Equipped with the wave optics theory from the previous chapter, we can reinterpret the problem of phase generation \ref{prob:phase-generation} and beam estimation \ref{prob:beam-estimation}.

\subsection{Wigner Phase Generation}
\label{subsection:wigner-phase-generation}

\begin{figure}[!ht]
\centering
\includegraphics[width=\textwidth]{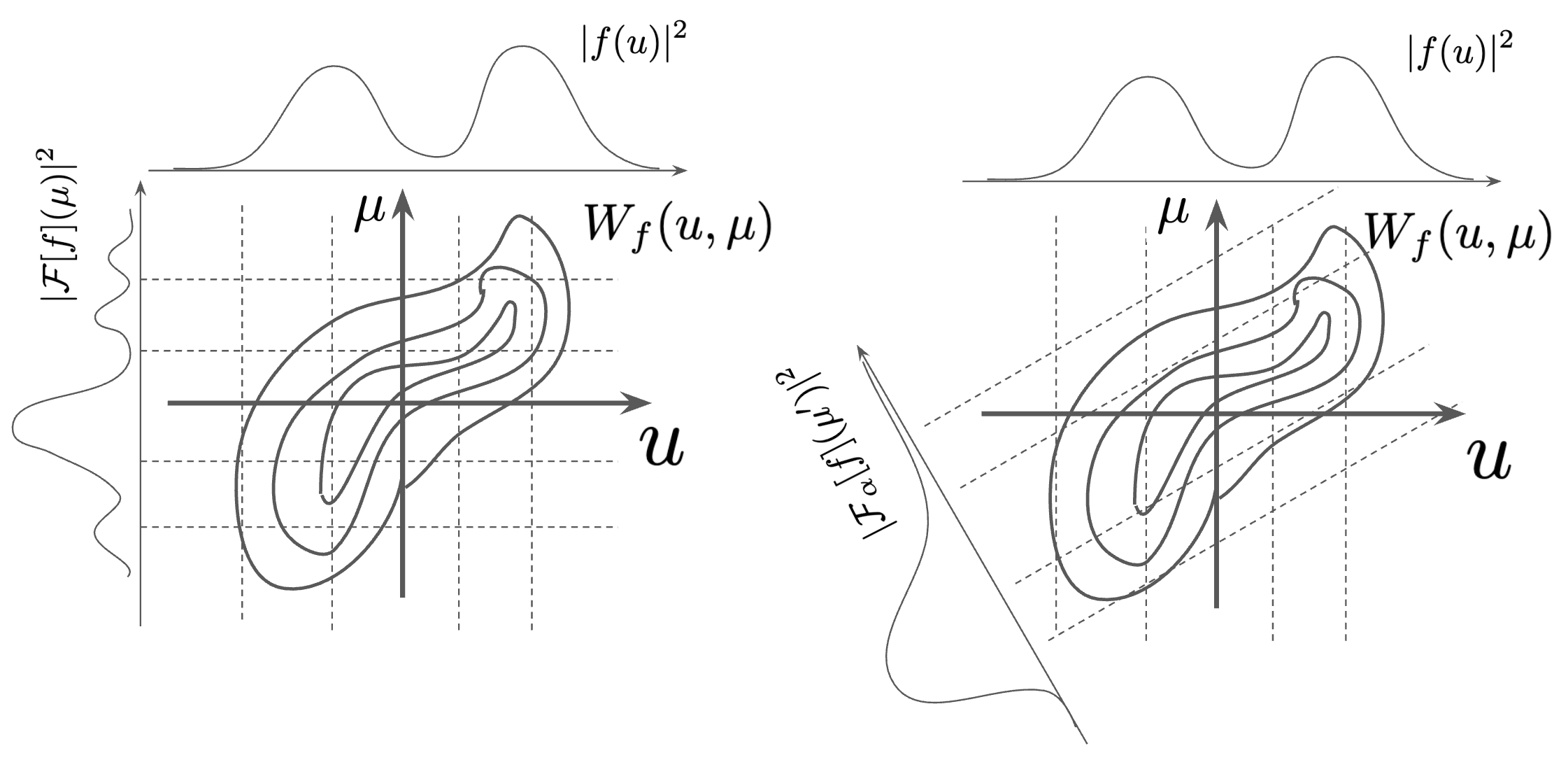}
\caption[Wigner distribution $W_f(u,\mu)$ and its marginal projections]{Schematic representation of the Wigner distribution $W_f(u,\mu)$ and its marginal projections. The left figure shows marginal projections $\mathcal{RDN}_0(W)$ and $\mathcal{RDN}_{\pi/2}(W)$. The right figure shows marginal projections $\mathcal{RDN}_0(W)$ and $\mathcal{RDN}_\alpha(W)$ for some arbitrary angle $\alpha$.}
\label{fig:WignerMarginalProjections}
\end{figure}

Consider the setting of the problem of phase generation in 1D. We are given access to the beam moduli at the input and the output $g, G:\mathbb{R}\to \mathbb{R}_+$, and our task is to find a phase $\phi:\mathbb{R}\to \mathbb{R}$ minimizing 
\begin{equation}
d\left(G(\mu) , \abs*{\mathcal{F}\!\left[g(u)\, e^{i \phi(u)}\right]\!(\mu)}\right)
\label{eq:prob-1-metric}
\end{equation}
Consider the case when this problem can be solved exactly, meaning that there exists a phase $\phi$ such that the cost is 0 \footnote{We know that this is not possible for all distributions $g$ and $G$. A trivial counterexample in the continuous setting is when both $g$ and $G$ are compact.}. Then we can consider a Wigner distribution $W_f(u,\mu)$ of the retrieved signal $f(u)=g(u)e^{i\phi(u)}$. If we truly found the right signal $f$ then we must have:
\begin{align}
    g^2(u)&=|f(u)|^2=\int W_f(u,\mu)d\mu \\
    G^2(\mu)&=|\mathcal{F}[f](\mu)|^2=\int W_f(u,\mu)du
\end{align}
In other words, the retrieved signal has a Wigner distribution whose marginals match the required constraints $g(u)$ and $G(\mu)$ (see Figure \ref{fig:WignerMarginalProjections}). Generalizing this to 2 dimensions and allowing some error in the marginals allows us to reformulate the problem of phase generation using the Wigner distribution.

\begin{problem} 
\label{prob:wigner-phase-generation}
Given input beam modulus and target output beam modulus $g,G:\mathbb{R}^2\rightarrow\mathbb{R}_{\geq 0}$ with $\norm{g}_2 = \norm{G}_2$, find a Wigner distribution $W:\mathbb{R}^4\rightarrow\mathbb{R}$ minimizing 
\begin{equation}
d\left(G(\mu,\nu), \tilde{G}(\mu,\nu)\right)+d\left(g(u,v), \tilde{g}(u,v)\right),
\end{equation}
where $d$ is some chosen distance function between two images and $\tilde{g}$ and $\tilde{G}$ are projections of $W$:
\begin{equation}
    \tilde{g}(u,v)=\iint W(u,v,\mu,\nu)d\mu d\nu
\end{equation}
\begin{equation}
    \tilde{G}(\mu,\nu)=\iint W(u,v,\mu,\nu)dudv
\end{equation}
\end{problem}

Implicit in this problem is the condition that a function $W(u,v,\mu,\nu)$ has to be a Wigner distribution. As we showed in Chapter \ref{chapter:ray-optics-reduction-with-optimal-transport}, Section \ref{section:wigner-distribution}, not every function $W:\mathbb{R}^4\to\mathbb{R}$ corresponds to a Wigner distribution, but every function $f:\mathbb{R}^2\to\mathbb{C}$ has an associated Wigner distribution $W_f$. See \cite{ozaktas2001fractional} for the exact conditions that the function has to satisfy to be a Wigner distribution.

It is clear on physical grounds that solutions to Problem 3 can be used to derive approximate solutions to Problem 1, and vice versa, though the equivalence is not perfect. Problem 3 can be thought of as a relaxation of Problem 1 which allows the input intensity to vary slightly. In the problem above, once we optimize over the space of all possible Wigner distributions, we can retrieve a final signal $f(u,v)$ using equation (\ref{eq:signal-from-wigner}). Then we can recover the phase $\phi(u,v)=\text{Arg}(f(u,v))$. The reverse direction also holds. Once we find the phase $\phi(u,v)$ we know the full complex-valued signal $f(u,v)=g(u,v)\exp(i\phi(u,v))$ from which we can easily obtain the optimal Wigner distribution $W_f(u,v,\mu,\nu)$.

\subsection{Wigner Beam Estimation}
\label{subsection:wigner-beam-estimation}

\begin{figure}[!ht]
\centering
\includegraphics[width=0.7\textwidth]{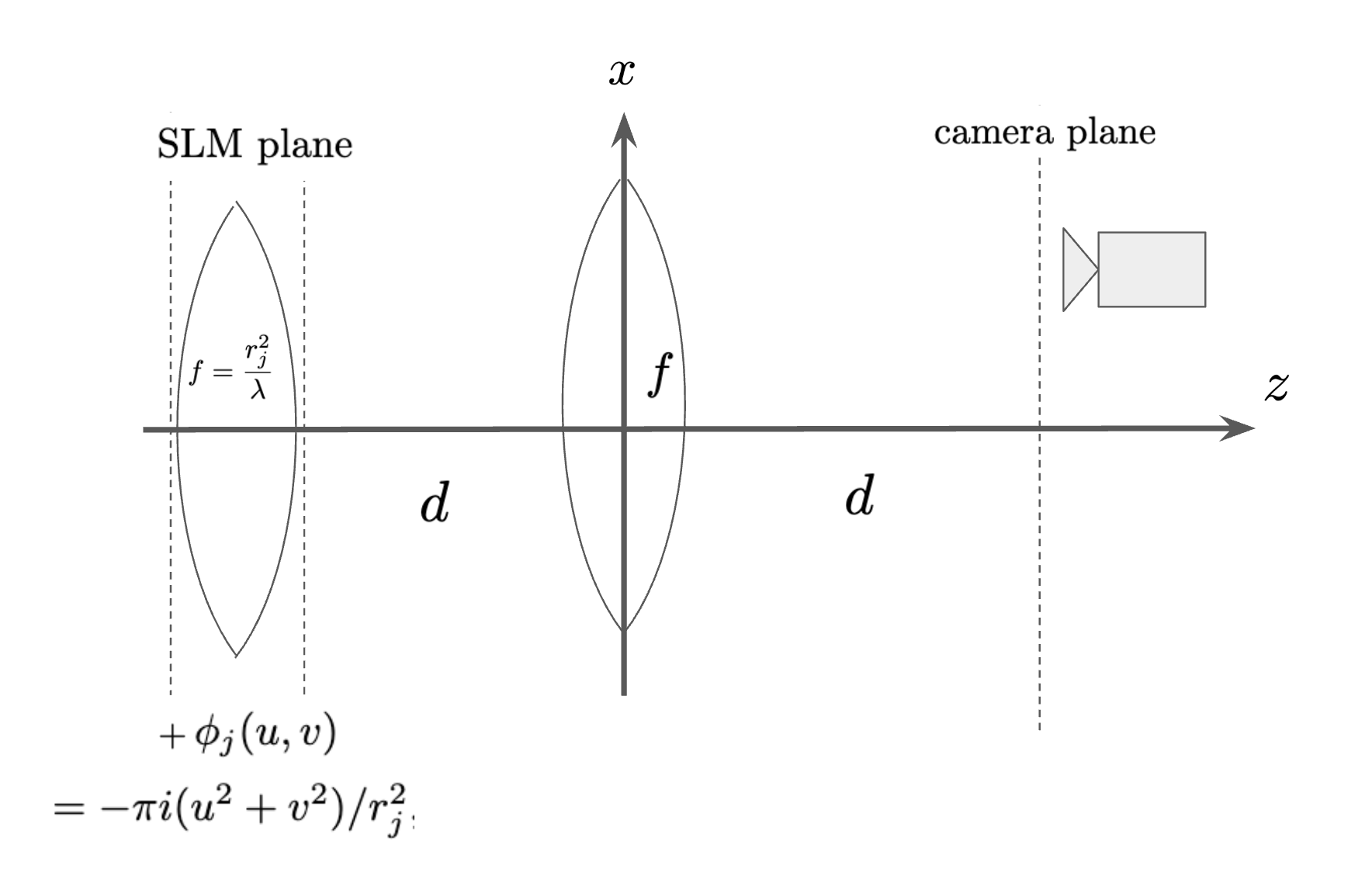}
\caption[Optical set-up for the beam estimation]{Optical set-up for the beam estimation. If the SLM on the input plane applies a quadratic phase shift $\phi_j(u,v)=-\pi i (u^2+v^2)/r_j^2$, we can interpret it as a lens of focal distance $f=-1/\lambda f_j$.}
\label{fig:phase-diversity-set-up}
\end{figure}

Let us find a similar re-interpretaion for the problem of beam estimation. In this setting, we do not know the input beam amplitude $g(u,v)$, but we can apply a family of phases $\{\phi_j\}_{j=1}^{n}$ on the input plane and measure the corresponding amplitudes on the output plane $\{G_j\}_{j=1}^{n}$ (see Figure \ref{fig:phase-diversity-set-up}) to infer the input beam $g(u,v)$ and its intrinsic phase $\psi(u,v)$. In other words, we want to find a full complex-valued beam $f:\mathbb{R}^2\to\mathbb{C}$ from $n$ amplitude measurements. 

If we restrict our family to quadratic phases $\phi_j(u,v)=-2\pi(u^2+v^2)/2r_j^2$ for some curvatures $r_j\in \mathbb{R}$, then we obtain a clear interpretation of this set-up in terms of the fractional Fourier transformation. The combined system implements the following ABCD matrix:

\begin{equation}
\label{eq:ABCD-to-FrFT-phase-diversity}
\begin{bmatrix}
    \hat{A} & \hat{B} \\
   \hat{C} & \hat{D}
\end{bmatrix}=
\begin{bmatrix}
    0 & \lambda f \\
    -1/\lambda f & 0
\end{bmatrix}
\begin{bmatrix}
    1 & 0 \\
    -1/r_j^2 & 1
\end{bmatrix}=\begin{bmatrix}
    -\lambda f /r_j^2 & \lambda f \\
    -1/\lambda f & 0
\end{bmatrix}
\end{equation}
We can relate this matrix to a fractional Fourier transform parameter $\alpha$, magnification $M$, and radius of curvature $R$ using equations \ref{eq:ABCD-to-FrFT}:
\begin{align}
\label{eq:alphaMR-phase-diversity}
\alpha_j = \arctan\!\left(-\frac{r_{j}^{2}}{s^{2}}\right)
&&
M = |\lambda f|\,
     \sqrt{\frac{1}{r_{j}^{4}}+\frac{1}{s^{4}}}
&&
R = \frac{(\lambda f)^{2}\!\left(s^{4}+r_{j}^{4}\right)}
          {r_{j}^{2}s^{4}}.
\end{align}
where we are free to choose any scale factor $s$ to make our life easier. Notice also that because we are making an intensity measurement, the spherical phase factor $R$ does not matter for our purposes. Thus, we can see that our optical set-up implements a magnified version of the fractional Fourier transform of angle $\alpha_j$ defined above. Each time we apply a quadratic phase with radius of curvature $r_j$, the measured modulus $G_j$ is the magnitude of the fractional Fourier transformation by angle $\alpha_j$ (up to magnification)\footnote{Below we absorb the magnification effect into $G_j$ fix an overall scale factor $s$ for all $j$.}.
\begin{equation}
    |G_j(\mu,\nu)|^2=|\mathcal{F}_{\alpha_j}[f](\mu,\nu)|^2
\end{equation}
which can be rewritten in terms of the Wigner distribution using the commutative diagram \ref{fig:wigner-frft-commute}:
\begin{equation}
    |G_j(\mu,\nu)|^2=(\mathcal{RDN}_0W_{f_{\alpha_j}})(\mu,\nu)=(\mathcal{RDN}_{\alpha_j}W_{f})(\mu,\nu)
\end{equation}
So, the diversity image is nothing but a Radon projection onto a tilted plane (see Figure \ref{fig:WignerMarginalProjections}). This interpretation inspires the following problem, which is closely related to the beam estimation problem:

\begin{problem}
\label{prob:wigner-beam-estimation}
Given angles $\alpha_j\in \mathbb{R}$ and corresponding diversity image moduli $G_j:\mathbb{R}^2\rightarrow\mathbb{R}_{\geq 0}$ with $\norm[\big]{G_j}_2 = 1$, $j=1,\dots,n$, find a Wigner distribution $W:\mathbb{R}^4\to\mathbb{R}$ minimizing 
\begin{equation}
\sum_j  d\left(G_j (\mu,\nu), (\mathcal{RDN}_{\alpha_j}W_f)(
\mu,\nu)\right)
\end{equation}
where $d$ is a chosen distance function.
\end{problem}

As we can see, we can think of both problems \ref{prob:wigner-phase-generation} and \ref{prob:wigner-beam-estimation} as estimating Wigner distributions given 2 (or more) measurements of its marginals. This should be reminiscent of a quantum learning problem, and we will make this connection precise in the Chapter 5. For now, we will focus on finding an approximate solution to the above problems.

\section{Ray Optics Limit of Phase Generation}
\label{section:ray-optics-limit-of-phase-generation}

Suppose once again that we have a perfect solution to the Problem \ref{prob:phase-generation} (or equivalently it solves \ref{prob:wigner-phase-generation})\footnote{It is easy to show that a solution to the Problem 1 with a 0 loss is also a solution to the Problem 2 with a 0 loss, and vice-a-verse.}. This means that there exists a phase $\phi$ such that:
\begin{equation}
    \left|\mathcal{F}[g(u,v)e^{i\phi(u,v)}](\mu,\nu)\right|=G(\mu,\nu)
\end{equation}
In other words, $\phi(u,v)$ satisfies an integral equation:
\begin{equation}
\label{eq:phase-interference-integral}
    \left|\iint g(u,v)e^{i\phi(u,v)} e^{-i2\pi (u\mu +v\nu)}du dv\right|=G(\mu,\nu)
\end{equation}
It is extremely difficult to solve for a function $\phi(u,v)$ using the equation above. Intuitively, one needs to figure out the exact interference pattern that produced the desired amplitude $G(\mu,\nu)$. Solving the interference phenomena is generally computationally difficult, so we will pursue an idea of reducing this problem to its ray-optics equivalent, which should be much more tractable.

\subsection{Stationary Phase Approximation}
\label{subsection:stationary-phase-approximation}

An incredibly useful tool for approximating highly oscillatory integrals is known as the Stationary Phase Approximation (SPA). This method is often used to retrieve the classical limit of the Feynman path integrals. We will use it to obtain the ray-optics limit of the integral equation (\ref{eq:phase-interference-integral}).

The basic idea is the observation that a rapidly changing phase averages the contributions to the integral. So we can approximate the value of a highly-oscillating integral by points where $\nabla \phi=0$.
\begin{theorem}
    Let $f:\mathbb{R}^2\to\mathbb{C}$ be a function that can be written as $f(u,v)=g(u,v)e^{ik\psi(u,v)}$ where $g:\mathbb{R}^2\to\mathbb{R}_+$ is the amplitude, $\psi:\mathbb{R}^2\to\mathbb{R}$ is the phase, and $k\in\mathbb{R}_+$ is some scale factor. We further assume that either $g$ is compactly supported or it exponentially decays. Then in the limit $k\rightarrow \infty$ the following integral can be approximated by:
    \begin{equation}
        I=\iint f(u,v)dudv= \sum_{(u_0,v_0)\in S}(2\pi/k)g(u_0,v_0)e^{ik\psi(u_0,v_0)}\frac{e^{\frac{i\pi}{4}\text{sgn}(\nabla^2(\psi(u_0,v_0)))}}{\sqrt{\det(\nabla^2(\psi(u_0,v_0))}}+o(k^{-1})
    \end{equation}
    where $S$ is the set of points $(u_0,v_0)$ where $\nabla \psi$ vanishes, and $\text{sgn}$ refers to the signature of the Hessian matrix $\nabla^2\psi$ i.e. the number of positive eigenvalues minus the number of negative eigenvalues.
\end{theorem}
To prove this theorem, it suffices to expand the phase up to the second order around each point $(u_0,v_0)$ where the phase gradient vanishes $\nabla\psi|_{(u_0,v_0)}=0$. Then the phase can be approximated as follows:
\begin{equation}
    \psi(u,v) \approx \psi(u_0,v_0)+1/2[u-u_0, v-v_0]\nabla^2({\psi}(u_0,v_0))\begin{bmatrix}u-u_0 \\v-v_0\end{bmatrix}
\end{equation}
Then, we also approximate the amplitude $g(u,v)$ up to the zeroth order $g(u,v)\approx g(u_0,v_0)$. After these approximations, the integral becomes a simple complex-valued Gaussian integral, which is easy to compute analytically. For the details of the proof, we refer the reader to \cite{BleisteinHandelsman1986}.

\subsection{Monge-Ampere PDE}
\label{subsection:monge-ampere-PDE}

Let us apply the Stationary Phase Approximation to the integral (\ref{eq:phase-interference-integral}). Specifically, we consider a set of points where the combined phase factor $\psi(u,v)=\phi(u,v)-2\pi(\mu u + \nu v)$ has a vanishing gradient:
\begin{equation}
    S=\Bigl\{(u_0,v_0)\in\mathbb{R}^2\;\Bigl|\; \nabla\psi|_{(u_0,v_0)}=0\Bigr\}
\end{equation}
where the set condition can be rewritten as:
\begin{equation}
\label{eq:inverse-of-T}
   \nabla\phi|_{(u_0,v_0)}=2\pi\begin{bmatrix}\mu \\ \nu\end{bmatrix} 
\end{equation}
Fix some $\mu$ and $\nu$. For an arbitrary phase function $\phi$ there could be many points $(u_0,v_0)$ such that $\nabla\phi|_{(u_0,v_0)}=2\pi[\mu \quad \nu]^T$, which makes it difficult to proceed with our approximation. Thus, we will make additional assumptions that $\phi$ is a \textit{strictly convex (or strictly concave) and twice differentiable function}.

With the following assumption in place, one can easily see that $\psi(u,v)$ is also a strictly convex or a strictly concave function for a fixed $\mu$ and $\nu$. To see that, observe that $\nabla^2\psi=\nabla^2\phi$. Therefore, for each choice of $(\mu,\nu)$ there exists at most one unique point $(u_0, v_0)\in\mathbb{R}^2$ such that $\nabla\phi_{(u_0,v_0)}=2\pi[\mu\quad \nu]^T$. Therefore, we can define a function $T:\mathbb{R}^2\to\mathbb{R}^2$ that maps $(\mu,\nu)$ into the corresponding point $(u_0,v_0)$.

Now, recall the integral equation we are trying to solve:
\begin{equation}
    \left|\iint g(u,v)e^{i\phi(u,v)} e^{-i2\pi (u\mu+v\nu)}du dv\right|=G(\mu,\nu)
\end{equation}
Once we fix $\mu,\nu$ on the output, the set $S$ becomes a singleton $\{T(\mu,\nu)\}$ and we get the following approximation:
\begin{equation}
    G(\mu,\nu)\approx\frac{2\pi g(T(\mu,\nu))}{\sqrt{\det\nabla^2\phi(T(\mu,\nu))}}
\end{equation}
We can use the defining relationship of the function $T$ to switch variables:
\begin{equation}
    G(\nabla\phi(u,v)/2\pi)\approx\frac{2\pi g(u,v)}{\sqrt{\det\nabla^2\phi(u,v)}}
\end{equation}
Rescaling the function $\phi$ by $2\pi$ and squaring both sides we get:
\begin{equation}
\label{eq:monge-ampere}
    G^2(\nabla\phi(u,v))\det\nabla^2\phi(u,v)=g^2(u,v)
\end{equation}
This is a second-order, nonlinear, partial differential equation for an unknown function $\phi:\mathbb{R}^2\to\mathbb{R}^2$, which is known as a \textit{Monge-Ampere PDE}. To formulate it rigorously, we also need to provide the boundary conditions. Let $\Omega$ and $\Omega^*\subset\mathbb{R}^2$ be the support of the functions $g$ and $G$, respectively, then the Monge-Ampere PDE with the so-called ``second boundary value'' can be formulated as follows:
\begin{equation}
\label{eq:monge-ampere-bv}
\left\{
\begin{aligned}
G^{2}\!\bigl(\nabla\phi(u,v)\bigr)\det\nabla^{2}\phi(u,v) &= g^{2}(u,v)
&& \text{for all }(u,v)\in\Omega\\[6pt]
\nabla\phi(\Omega) &= \Omega^{\!*}\\[6pt]
\phi(u_{0},v_{0}) &= 0
&& \text{for some fixed }(u_{0},v_{0})\in\Omega
\end{aligned}
\right.
\end{equation}
The second condition here ensures that $\nabla\phi$ is a bijection between the support regions of the input and output distributions. The third condition simply fixes the overall offset of the function $\phi$ since clearly $\phi(u,v)+C$ is also a solution to the Monge-Ampere PDE.

The above boundary value problem is known to admit a unique solution under mild regularity and convexity assumptions on the functions $g$ and $G$ \cite{Caffarelli1990, Urbas1997}. In Chapter \ref{chapter:algorithms}, Section \ref{section:optimal-transport}, we will see that solving this boundary value problem is actually a linear program, which can be solved in polynomial time.

\subsection{Validity of the Approximation} 
\label{subsection:validity-of-the-approximaiton}

Any time an approximation is introduced, one would like to know how good it is, and how to improve it if it is not good enough. The accuracy of the stationary phase approximation will be analyzed presently, but before doing so it is worthwhile to pause and examine some qualitative features of the approximation. \footnote{All credit for this section goes to Hunter Swan} 

Firstly, it is important to understand that we ultimately do not require the solution obtained via the stationary phase approximation to be tremendously accurate, because this solution is for us only a starting point for solving the phase generation problem.  Our intention is to simply use the approximation to seed an iterative solver algorithm, and the only requirement of the approximate solution is thus that it be near enough to the globally optimal solution so that our refinement procedure does not get trapped in local extrema.  For this reason it is not especially important that our approximate solutions have, for instance, a low loss with respect to some metric. 

Secondly, it will turn out that the error of the stationary phase approximation in deriving the Monge Ampere equation is controlled by a dimensionless parameter $\beta := \frac{2\pi R_{in}R_{out}}{f\lambda}$, where $R_{in}, R_{out}$ are the characteristic feature sizes of the input and output beams.  The approximation becomes accurate for $\beta \gg 1$.  This is precisely the limit in which diffraction effects are insignificant \cite{dickey2014laser}, and thus we interpret the result as simply stating that we must operate in the ray optics limit for our approximation to be valid. To attain this, or to make the approximation better, we may use any combination of: making the size of the input beam larger; making the size of the target beam larger; using a smaller wavelength $\lambda$; or using a smaller focal length $f$. 

Thirdly, in practice, all of the aforementioned parameters which affect $\beta$ are typically fixed by experimental constraints, and we thus often have no control over the quality of the approximation.  The quality of the approximation thus depends on whether our application demands diffraction-limited performance or not.  Typically smooth light fields do not demand diffraction-limited performance, while e.g. spot arrays do.  Our methods are thus more useful in the former case.

To analyze the error in the stationary phase approximation, we will briefly reinsert units in our expression for the optical propagator.  Let $s_{in}$ and $s_{out}$ be length scales for the input and output planes, respectively, such that $s_{in}s_{out} = f\lambda$.  Then let us write $R_j = r_j s_j$, where $j\in\{\text{in},\text{out}\}$, $R_j$ is again the dimensionful characteristic feature size of the beam in the input or output plane, and $r_j$ is a corresponding dimensionless characteristic size.  Let $g(u,v)$, $G(\mu,\nu)$, and $\phi(u,v)$ satisfy the Monge Ampere equation (\ref{eq:monge-ampere}).  Then the rescaled (but still dimensionless) functions $\hat g(u,v) \equiv g(u/r_{in},v/r_{in})/r_{in}$, $\hat G(\mu,\nu) \equiv G(\mu/r_{out},\nu/r_{out})/r_{out}$, and $\hat \phi(u,v) \equiv r_{in}r_{out}\phi(u/r_{in},v/r_{in})$ also satisfy the Monge Ampere equation: 

\begin{align*}
\hat G^2\left(\nabla\left(\hat\phi(u,v)\right)\right)\det\nabla^2\left(\hat\phi(u,v)\right) & = \frac{1}{r_{out}^2} G^2\left(\frac{1}{r_{out}}\left(\frac{r_{in}r_{out}}{r_{in}}\nabla\phi\right)\left(\frac{u}{r_{in}},\frac{v}{r_{in}}\right)\right) \\
& \qquad\qquad\times\det\!\left[\frac{r_{in}r_{out}}{r_{in}^2}\left(\nabla^2\phi\right)\left(\frac{u}{r_{in}},\frac{v}{r_{in}}\right)\right]\\
& = \frac{1}{r_{out}^2} G^2\left(\left(\nabla\phi\right)\left(\frac{u}{r_{in}},\frac{v}{r_{in}}\right)\right)\frac{r_{out}^2}{r_{in}^2}\det\left[\left(\nabla^2\phi\right)\left(\frac{u}{r_{in}},\frac{v}{r_{in}}\right)\right] \\
& = \frac{1}{r_{in}^2} G^2(\nabla\phi)\det\nabla^2\phi \Big|_{(u/r_{in},v/r_{in})} \\
& = \frac{1}{r_{in}^2} g^2\left(\frac{u}{r_{in}},\frac{v}{r_{in}}\right) \\
& = \hat g^2(u,v)
\end{align*}

We now examine the wave-optical propagation of the beam $\hat g e^{i 2\pi \hat \phi}$, evaluated at the point $\nabla\hat\phi|_{(u_0,v_0)} = r_{out}\left(\nabla\phi\right)(u_0/r_{in},v_0/r_{in})$ for some fixed point $(u_0,v_0)$ in the input domain.  Letting $\mu_0, \nu_0$ denote the components of $\nabla\hat\phi|_{(u_0,v_0)}/r_{out}$, we have
\begin{align*}
\left|\mathcal{F}\left[\hat g e^{i2\pi \hat\phi}\right]\left(r_{out}\mu_0,r_{out}\nu_0\right)\right|
& =\left|\iint \hat g(u,v) e^{i2\pi \hat \phi(u,v)} e^{-i2\pi r_{out} (u\mu_0+v\nu_0)}dudv\right| \\
& = \Bigg| \iint \frac{1}{r_{in}} g\left(\frac{u}{r_{in}},\frac{v}{r_{in}}\right) \times \\
& \qquad \exp\left(i2\pi\left[r_{in}r_{out}\phi\left(\frac{u}{r_{in}},\frac{v}{r_{in}}\right) - r_{out}\left(u\mu_0 + v \nu_0\right)\right]\right) dudv \Bigg| \\
(u,v\rightarrow r_{in}u,r_{in}u)\qquad & = r_{in}\left| \iint g(u,v) \exp\left( i\left(2\pi r_{in}r_{out}\right) \left(\phi(u,v) - u\mu_0 - v\nu_0\right)\right) dudv \right| \\
(\text{by SPA}) \qquad\qquad & = \frac{2\pi r_{in}g(u_0/r_{in},v_0/r_{in})}{2\pi r_{in}r_{out}\sqrt{\det\nabla^2\phi(u_0/r_{in},v_0/r_{in})}} + r_{in}o\left(\frac{1}{2\pi r_{in}r_{out}}\right) \\ 
& = \hat G(\nabla \hat \phi(u_0,v_0)) + r_{in}o\left(\frac{1}{2\pi r_{in}r_{out}}\right)
\end{align*}

This shows that in the limit $r_{in}r_{out}\rightarrow \infty$ the output field generated by input beam $\hat g e^{i2 \pi \hat\phi}$ approaches the scaled target beam $\hat G$, with error term regulated by $\beta \equiv 2\pi r_{in}r_{out}$.  Note that the magnitude of $\hat G$ has scaling $\frac{1}{r_{out}} = \frac{r_{in}}{r_{in}r_{out}}$, so even though $\hat G$ may itself be small for large $r_{out}$, the error term vanishes faster. 

Inserting dimensionful parameters into the expression for $\beta$:
\begin{equation}
\beta = 2\pi \frac{R_{in}}{s_{in}}\frac{R_{out}}{s_{out}} = \frac{2\pi R_{in}R_{out}}{f\lambda}
\end{equation}
Having $\beta\gg 1$ is precisely the condition for the ray optics limit to be valid \cite{dickey2014laser}. 

To recapitulate, a scaled solution $\phi$ to the Monge Ampere equation (\ref{eq:monge-ampere}) yields an input beam $ge^{i2\pi\phi}$ which well approximates the target output beam $G$ in a scaling limit where:
\begin{itemize}
\item The input beam $g(u)$ is rescaled to $g(u/r_{in},v/r_{in})/r_{in}$.
\item The output beam $G(\mu)$ is rescaled to $G(\mu/r_{out},\nu/r_{out})/r_{out}$.
\item The phase $\phi$ is rescaled to $r_{in}r_{out} \phi(u/r_{in},v/r_{in})$.
\item The point of comparison $(\mu_0,\nu_0) = \nabla\phi(u_0,v_0)$ between the propagated beam and target beam is scaled commensurately with the output beam, $(\mu_0,\nu_0)\rightarrow (r_{out}\mu_0,r_{out}\nu_0)$.
\item The parameter $\beta = 2\pi r_{in}r_{out}\rightarrow \infty$. 
\end{itemize}

What this means in practice is that in order for the ray optics limit to yield good results for phase generation, the product of the feature sizes in the input and output planes must be much larger than $f\lambda$.  

\subsection{Optimal Transport}
\label{subsection:optimal-transport}

It turns out that the Monge-Ampere PDE is deeply connected to the mathematical theory of optimal transport (OT), which is a very well-studied problem with a lot of open-source solvers \cite{OptimalTransport.jl, POT}. In this section, we will provide a basic overview of the theory of optimal transport \footnote{Explanation taken from the paper \cite{swan2024highfidelityholographicbeamshaping} that I co-authored.} needed to solve the problem above. We refer the interested reader to \cite{Villani2009, Villani2021} for the detailed introduction to the subject.

The basic problem of OT is to find a way of rearranging one
probability density $\mu(x)$ into another $\nu(y)$ that optimizes some
cost $c(x, y)$ for the rearrangement process. For example, we may
think of $\mu(x)$ as the height of a pile of sand, $\nu(y)$ as the depth of a nearby hole, and $c(x, y)$ as the cost to move sand from position $x$ to fill a hole at position $y$. OT seeks to find a way to move sand into the hole with minimal total cost, encapsulated in a function $\gamma(x)$ called \textit{the transport map} which indicates where to send the sand at location $x$, and which minimizes $\int c(x,\gamma(x))\mu(x)dx$.

In more rigorous terms, OT problems are defined on Radon spaces $X$ and $Y$, and the cost function $c:X\times Y\to[0,\infty)$ is a Borel-measurable function. Given probability measures $\mu$ on $X$ and $\nu$ on $Y$, we seek to find a transport map $\gamma:X\to Y$ that minimizes $\int_X c(x,\gamma(x))d\mu(x)$ subject to the condition that $\gamma_*(\mu)=\nu$. 
This should be reminiscent of the boundary value problem (\ref{eq:monge-ampere-bv}). The pushforward condition is exactly the same, while the optimality condition is encoded in a slightly different language. 

The key fact needed from OT theory \cite{brenier1987decomposition, brenier1991polar, de2014monge} is that in the special case where the probabilities $\mu$ and $\nu$ have domain $\mathbb{R}^n$ and are well-behaved, and where the cost function is $c(x, y) =||x-y||^2_2$, an optimal transport map $\gamma$ exists and is the gradient of some scalar function $\phi: \mathbb{R}^n \to \mathbb{R}$, where $\phi$ satisfies:
\begin{equation}
    \nu(\nabla\phi(x))\det\nabla^2\phi(x)=\mu(x)
\end{equation}
This is exactly the Monge-Ampere equation \ref{eq:monge-ampere} with $\nu=G^2$ and $\mu=g^2$. Solving the optimal transport problem with these distributions and the above quadratic cost thus exactly solves the boundary-value problem \ref{eq:monge-ampere-bv} and yields an approximate solution to the phase generation problem.

\subsection{Kantorovich Relaxation of OT}
\label{subsection:kantorovich-relaxation-of-ot}

\begin{figure}[!ht]
\centering
\includegraphics[width=0.80\textwidth]{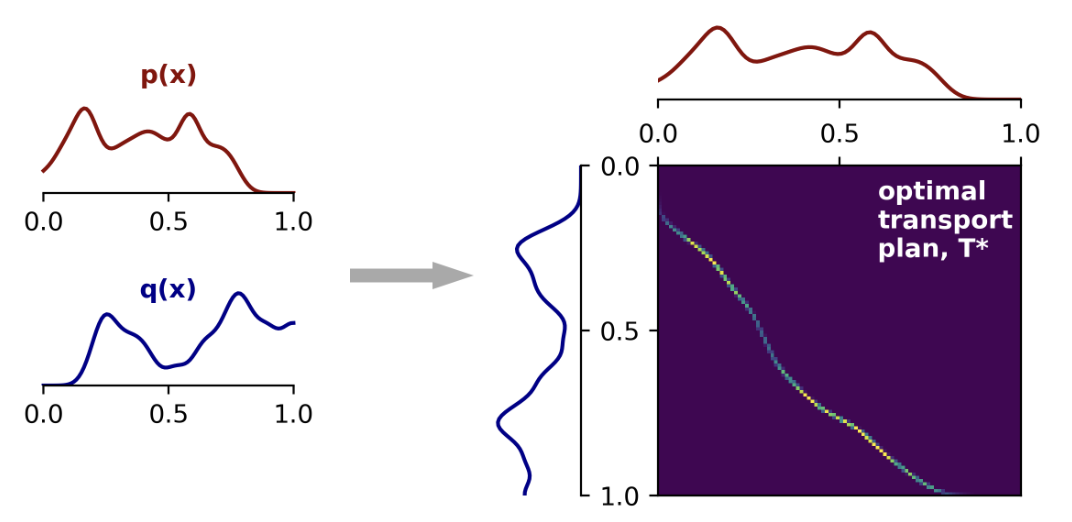}
\caption[Optimal transport plan]{Input distributions $p(x)$, $q(x)$, and the optimal transport plan $T^*$ (in our notation $\Gamma$).}
\label{fig:ot-kantorovich-relaxation}
\end{figure}

Analytical solutions to the above OT problem only exist in the 1-dimensional case, where we can use clever integration to find the transport plan $\gamma$. For higher-dimensional spaces, it becomes incredibly difficult to solve for the transport map $\gamma$ directly. Leonid Kantorovich showed that if one promotes a transport map $\gamma$ to a distribution on a product space $\Gamma:X\times Y\to[0,\infty)$, called a \textit{transport plan} (see Figure \ref{fig:ot-kantorovich-relaxation}\footnote{Courtesy of the image to this \href{https://alexhwilliams.info/itsneuronalblog/2020/10/09/optimal-transport/}{Optimal Transport (blog post)}.}), then the optimization problem becomes much more tractable, and in the discrete cases reduces to a linear program. This invention later earned him a Nobel Prize in economics.

The ``trick" is to optimize $\int\Gamma(x,y)c(x,y)dxdy$ over all possible transport plans $\Gamma(x,y)$ subject to conditions that $\int\Gamma(x,y)dxdy=1$, $\int\Gamma(x,y)dy=\mu(x)$, and $\int\Gamma(x,y)dx=\nu(y)$. In the case of the quadratic cost function, and well-behaved inputs $g$ and $G$, we can recover the transport map from a transport plan via:
\begin{equation}
\label{eq:retrieve-gamma}
    \gamma(x)=\frac{1}{\mu(x)}\int y\Gamma(x,y)dy
\end{equation}
Recall that $\gamma(x)=\nabla\phi(x)$, which allows us to obtain the phase $\phi$:
\begin{equation}
\label{eq:phase-from-monge}
    \phi=\int_{C_x} \gamma(s)\cdot ds
\end{equation}
where $C_x$ is any path from a reference point $x_0$ to $x$.

Notice that this optimization problem is very similar to the problem of Wigner phase generation \ref{prob:wigner-phase-generation}, where we optimize over the space of all Wigner distributions that match marginal constraints. In fact, we will later show that the transport plan $\Gamma$ is the ray-optics limit of the Wigner distribution.

\section{Fractional Fourier Transform Generalization}
\label{section:fractional-fourier-transform-generalization}

Above, we have shown that if the input and output planes are related by the Fourier transform:
\begin{equation}
    \left|\mathcal{F}[g(u,v)e^{i\phi(u,v)}](\mu,\nu)\right|=G(\mu,\nu)
\end{equation}
We can approximate the equation above with the Monge-Ampere boundary value problem \ref{eq:monge-ampere-bv}, which in turn can be solved efficiently using Optimal Transport. The natural question arises if this is also possible for a fractional Fourier transform by the angle $\alpha$. The answer is yes.

Suppose the input and the output of the optical system are related by a fractional Fourier transform of angle $\alpha\in(0,\pi/2)$. Then we seek to find the phase $\phi$ that would satisfy the following integral equation:
\begin{equation}
    G(\mu,\nu)=\left|\mathcal{F_\alpha}[g(u,v)e^{i\phi(u,v)}](\mu,\nu)\right|=\left|\iint g(u,v)e^{i\phi(u,v)}K_\alpha(u,\mu)K_\alpha(v,\nu) du dv \right|
\end{equation}
where as before:
\begin{equation}
    K_\alpha(u,\mu)\equiv\sqrt{1 - i \cot \alpha} \exp\left(
-i2\pi \left(-\frac{\mu^2}{2}\cot\alpha+u\mu \csc \alpha -\frac{u^2}{2}\cot \alpha \right)
\right)
\end{equation}
Because of the absolute value sign, we can drop the quadratic phase factor $\exp(-i2\pi(-\frac{\mu^2}{2}\cot{\alpha}))$. So, the combined prefactor in front of the integral becomes\footnote{Notice that the case of $\alpha=0$ as usual requires some special care.} $|1 - i \cot \alpha|=|1/\sin\alpha|$. 
\begin{equation}
    |G(\mu,\nu)|=\frac{1}{|\sin\alpha|}\left|\iint g(u,v)e^{i\phi(u,v)-i2\pi((u\mu+v\nu)\csc\alpha +i\pi(u^2+v^2)\cot\alpha}dudv \right|
\end{equation}
where we can combine phases into a combined phase function $\psi$:
\begin{equation}
    \psi(u,v)\equiv\phi(u,v)-2\pi((u\mu+v\nu)\csc\alpha +\pi(u^2+v^2)\cot\alpha
\end{equation}
Next, we will assume that $\phi$ is twice differentiable and strictly convex \footnote{If $\phi$ is strictly concave, the argument does not work, but in practice this restriction does not matter.}. Computing the Hessians yields:
\begin{equation}
    \nabla^2\psi=\nabla^2\phi +2\pi \mathds{1} \cot{\alpha}
\end{equation}
where $\mathds{1}$ is a $2\times 2$ identity matrix. Looking at these matrices in the eigen-basis basis of $\nabla^2\phi$ we see that $\nabla^2\psi$ has strictly positive eigenvalues since $\cot{\alpha}>0$ when $\alpha\in(0,\pi/2)$. Therefore, $\psi$ is a strictly convex function for a fixed $\mu$ and $\nu$, since its Hessian has positive eigenvalues. Thus, for each output coordinate $\mu, \nu$, there exists a unique point $(u_0,v_0)$ such that $\nabla\psi(u_0,v_0)=0$. This is equivalent to the following condition:
\begin{equation}
    \nabla \phi (u_0,v_0)-2\pi \csc\alpha \begin{bmatrix}\mu\\\nu\end{bmatrix} +2\pi \cot\alpha\begin{bmatrix}u_0\\v_0\end{bmatrix}=\begin{bmatrix}0\\0\end{bmatrix}
\end{equation}
which can be rewritten as:
\begin{equation}
    \begin{bmatrix}\mu\\\nu\end{bmatrix} =\frac{\sin \alpha}{2\pi}\nabla\phi(u_0,v_0)+\cos\alpha \begin{bmatrix}u_0\\v_0\end{bmatrix}
\end{equation}
Once again, we consider the function $T:\mathbb{R}^2\to\mathbb{R}^2$ that maps a pair of points $(\mu,\nu)$ to corresponding points $(u_0,v_0)$. The equation above can be thought of as the inverse of $T$, analogous to equation (\ref{eq:inverse-of-T}) in the case of the Fourier transform.
Now, performing the Stationary Phase Approximation, we obtain the equation:
\begin{equation}
    G(\mu,\nu)=\frac{2\pi}{|\sin\alpha|}\frac{g(T(\mu,\nu))}{\sqrt{\det(\nabla^2\phi(T(\mu,\nu))+2\pi\text{I}\cot\alpha)}}
\end{equation}
Now, we change variables to $(u,v)$ which gives us:
\begin{equation}
\label{eq:SPA-to-FrFT}
    G\left(\frac{\sin \alpha}{2\pi}\nabla\phi(u,v)+\cos\alpha \begin{bmatrix}u\\v\end{bmatrix}\right)=\frac{2\pi}{|\sin\alpha|}\frac{g(u,v)}{\sqrt{\det(\nabla^2\phi(u,v)+2\pi\mathds{1}\cot\alpha)}}
\end{equation}
To simplify this further, we change the phase variable:
\begin{equation}
\label{eq:varphi-phase}
    \varphi(u,v)=\frac{\sin\alpha}{2\pi}\phi(u,v)-\frac{\cos\alpha}{2}(u^2+v^2)
\end{equation}
which has the following gradient and the Hessian:
\begin{align}
    \nabla \varphi(u,v)&=\frac{\sin\alpha}{2\pi}\nabla\phi(u,v)-\cos\alpha\begin{bmatrix}u\\v\end{bmatrix}\\
    \nabla^2\varphi(u,v)&=\frac{\sin\alpha}{2\pi}\nabla^2\phi(u,v)-\cos\alpha \mathds{1}
\end{align}
With the following substitution, the equation becomes:
\begin{equation}
G\left(\nabla\varphi(u,v)\right)=\frac{g(u,v)}{\sqrt{\det(\nabla^2\varphi(u,v))}}
\end{equation}
This is exactly our old Monge-Ampere equation \ref{eq:monge-ampere} for an unknown phase $\varphi(u,v)$. Therefore, we can use all of the machinery described above to efficiently compute $\varphi(u,v)$. An important caution is that one needs to remember to invert equation \ref{eq:varphi-phase} to recover the original phase $\phi$:
\begin{equation}
    \phi(u,v)=\frac{2\pi}{\sin\alpha}\varphi(u,v) + \cot\alpha\pi(x^2 +y^2)
\end{equation}
Notice that the quadratic phase factor is identical to the one that appears in the phase diversity imaging used for the beam estimation problem. To see that, compare the equation above with equations (\ref{eq:ABCD-to-FrFT-phase-diversity}) and (\ref{eq:alphaMR-phase-diversity}).

\section{Ray Optics Limit of the Wigner Distribution}
\label{section:ray-optics-limit-of-the-wigner-distribution}

As we saw, there is a beautiful connection between the theory of optimal transport and the problem of phase generation. In the next chapter, we will later leverage this connection to obtain the state-of-the-art numerical solutions to the problem of phase generation and beam estimation. 

It turns out that optimal transport is not just a convenient computational tool, but it also has a deep connection to the Wigner distribution. Next, we will show that the optimal transport plan naturally arises as the zero-order approximation to the Wigner distribution, and demonstrate that it perfectly retrieves the effect of any unknown linear canonical transform on a collimated beam.

\subsection{Zero-order Term}
\label{subsection:zero-order-term}

To motivate this even further, consider the Wigner distribution $W_f(u,\mu)$ of some signal $f(u)=g(u)e^{i 2\pi \phi(u)}$, with modulus $g$ and convex phase $\phi$ \footnote{Here we are using a slightly different convention for the phase by having an explicit factor of $2\pi$ upfront. With this convention $\phi$ is measured in cycles as opposed to radians.}. We will assume that the signal is normalized, i.e. $||f||_2=1$. Notice that both $W_f(u,\mu)$ and $\Gamma(u,\mu)$\footnote{we switch here the probabilistic variables $x,y$ to our wave optics variables $u,\mu$. Notice that $\mu$ is no longer used for the output probability distribution.} satisfy a lot of common properties:
\begin{align}
    \iint W_f(u,\mu)dud\mu=1 && \iint \Gamma(u,\mu)dud\mu=1 \\
    \int W_f(u,\mu)d\mu=|f(u)|^2 && \int \Gamma(u,\mu)d\mu=|f(u)|^2\\
    \int W_f(u,\mu)du=|F(\mu)|^2 && \int \Gamma(u,\mu)du=|F(\mu)|^2\\
    \int \mu W_f(u,\mu)d\mu=g(u)^2\phi'(u) && \int \mu \Gamma(u,\mu)d\mu=g(u)^2\phi'(u)
\end{align}
with the exception that $\Gamma$ is strictly positive, while $W_f$ could be negative. As we discussed earlier, negative values of the Wigner distribution correspond to the interference effects, so we expect to obtain the optimal transport $\Gamma$ in the ray-optics limit of the Wigner distribution. 

We start by expanding the definition of the Wigner distribution:
\begin{align}
    W_f(u,\mu)&=\int f(u+u'/2)f^*(u-u'/2)e^{-i2\pi u'\mu}du'\\
    &=\int g(u+u'/2)g(u-u'/2)e^{i2\pi\phi(u+u'/2)-i2\pi\phi(u-u'/2)-i2\pi u'\mu}du'
\end{align}
Expand $\phi(u+u'/2)$ and $\phi(u-u'/2)$ in Taylor series up to the second order:
\begin{align}
\label{eq:phi-quadratic}
\phi(u+u'/2)&\approx\phi(u)+\phi'(u)\frac{u'}{2}+\frac{\phi''(u)}{2!}\left(\frac{u'}{2}\right)^2\\
    \phi(u-u'/2)&\approx\phi(u)-\phi'(u)\frac{u'}{2}+\frac{\phi''(u)}{2!}\left(\frac{u'}{2}\right)^2
\end{align}
Suppose also that we expand $g$ up to zeroth order around $u_0$:
\begin{align}
\label{eq:amplitude-taylor}
    g(u+u'/2)&\approx g(u) \\
    g(u-u'/2)&\approx g(u)
\end{align}
With the following approximations, the Wigner distribution becomes:
\begin{align}
    W^{(0)}_f(u,\mu)&\equiv \int g(u)^2 e^{i 2\pi u'\phi'(u) - i2\pi \mu u'} du'\\
    &= g(u)^2\int e^{i2\pi u' (\phi'(u) - \mu)} du'\\
    &= g(u)^2\delta(\phi'(u)-\mu)
\end{align}
Let's verify that the marginal property is still satisfied. Fix a variable $\mu$, and find a unique point $u$ such that $\nabla\phi(u)=\mu$. Integrating against $\mu$ we get:
\begin{equation}
    \int W^{(0)}_f(u,\mu)d\mu=g^2(u)\int \delta(\phi'(u)-\mu)d\mu=g^2(u)
\end{equation}
Similarly, integration against $u$:
\begin{equation}
    \int W^{(0)}_f(u,\mu)d\mu=\int g^2(u)\delta(\phi'(u)-\mu)du=\frac{g^2((\phi')^{-1}(\mu))}{\phi''((\phi')^{-1}(u))}
\end{equation}
Recall that in the Monge-Ampere formulation of the problem \ref{eq:monge-ampere-bv}, this exactly corresponds to $G^2(\mu)$.

In the Kantorovich relaxation of OT, the transport plan $\Gamma$ is given by \footnote{As long as $g^2$ and $G^2$ satisfy some mild regularity conditions \cite{Villani2009, Villani2021}.}:
\begin{equation}
    \Gamma=g(u)^2\delta(\phi'(u)-\mu)
\end{equation}
but this is exactly the zeroth-order approximation to the Wigner distribution we found above.
\begin{equation}
    W^{(0)}_f=\Gamma
\end{equation}
Notice that $W^{(0)}_f$ is a compact, and everywhere positive distribution. So, we can interpret this object naturally as a bundle of rays at positions $u$ and propagating in direction $\mu=\phi'(u)$. In the 2D generalization of this, which is straightforward but tedious, we obtain a bundle of rays parametrized by positions $u,v$ and propagating in the direction $\nabla\phi(u,v)$. 

This is exactly the ray-optics limit of the problem. Recall that at the input plane $z=0$, we can decompose any electric field $E(x,y,z)$ into its Fourier modes $A_{\vec{k}}(x)=\exp(i\phi_{\vec{k}}(\vec{x}))$ indexed by wave vectors $\vec{k}$, where $\phi_{\vec{k}}(\vec{x})=\vec{k} \cdot \vec{x}$. Then, the $\nabla\phi_{\vec{k}}$ exactly corresponds to the wave vector $\vec{k}$, and can be interpreted as the direction of the propagating ray. So, our approach to the problem of phase generation \ref{prob:phase-generation}, which is equivalent to the problem of the Wigner phase generation \ref{prob:wigner-phase-generation}, can be interpreted as finding the best ray bundle approximation that will match the constraints.

\subsection{Higher-order Terms}
\label{subsection:higher-order-terms}

A very important clarification regarding our previous result is that $W^{(0)}_f(u,\mu)=\Gamma(u,\mu)$ is \textit{not} our proposed solution to the problem of Wigner phase generation \ref{prob:wigner-phase-generation}. Recall that in this problem, our inputs are $g^2(u)$ and $G^2(\mu)$, which correspond to a Wigner distribution of some unknown signal $W_f$, and we are tasked to find the closest Wigner distribution $\tilde{W}$ that matches these constraints. 

Our optimal transport algorithm produces an approximated phase $\phi^{(0)}$ (using \ref{eq:phase-from-monge}), which we later use to find the corresponding complex-valued signal $\psi$ and the desired Wigner distribution:
\begin{equation}
    \psi(u)=g(u)e^{i\phi_0(u)}\implies \tilde{W}= W_{\psi}
\end{equation}
To understand how $W_{\psi}$ is different from $W^{(0)}_f$, let's expand the amplitude up to the second order:
\begin{align}
    g(u+u'/2)&=g(u) + g'(u)\frac{u'}{2}+\frac{g''(u)}{2!}\left(\frac{u'}{2}\right)^2+\mathcal{O}(u'^3)\\
    g(u-u'/2)&= g(u) -g'(u)\frac{u'}{2}+\frac{g''(u)}{2!}\left(\frac{u'}{2}\right)^2+\mathcal{O}(u'^3)
\end{align}
Then, grouping terms together:
\begin{equation}
    g(u+u'/2)g(u-u'/2)=g^2(u)+(g(u)g''(u)-g'(u)^2)\left(\frac{u'}{2}\right)^2+\mathcal{O}(u'^3)
\end{equation}
As before, we will assume a quadratic Taylor approximation for the phase (see eq. \ref{eq:phi-quadratic}). This results in the following approximation to $W_\psi$: 
\begin{align}
    W_{\psi}(u,\mu)&=\int \psi(u+u'/2)\psi^*(u-u'/2)e^{i2\pi u'(\phi'(u) - \mu)}du' \\
    &=\int g(u+u'/2)g(u-u'/2)e^{i2\pi u'(\phi'(u) - \mu)}du'\\
    &=\int \left( g^2(u)+(g(u)g''(u)-g'(u)^2)u'^2/4+\mathcal{O}(u'^3) \right)e^{i2\pi u'(\phi'(u) - \mu)}du'\\
    &= W^{(0)}_f+ \frac{1}{4}\left(g(u)g''(u)-g'(u)^2\right)\int u'^2 e^{i2\pi u'(\phi'(u) - \mu)}du' +\text{h.o.t}\\
    &=W^{(0)}_f + \frac{1}{4}\left(g(u)g''(u)-g'(u)^2\right)\delta''(\phi'(u) - \mu) +\text{h.o.t}\\
    &\equiv W^{(0)}_f + W^{(2)}_f + \text{h.o.t}
\end{align}
where the last line is the definition of $W^{(2)}_f$, which can be thought of as a second-order Taylor approximation of the unknown $W_f$ (notice that the first-order contributions $\mathcal{O}(u')$ vanish).

Let's form some intuition about the $W^{(2)}_f$ term of this expansion. If we imagine discretizing everything on a grid of $N$ points, we can think of $\delta(x)$ as the $N\times N$ identity matrix $\mathds{1}$. Derivative of the $\delta$ function can be thought of as taking the following limit \footnote{With the caution that these are not functions but distributions.}:
\begin{align}
    \delta'(u)&=\lim_{h\to0}\frac{\delta(u+h)-\delta(u)}{h}\\
    \delta''(u)&=\lim_{h\to0}\frac{\delta'(u+h)-\delta'(u)}{h}=\lim_{h\to0}\frac{\delta(u+2h)-2\delta(u+h)+\delta(u)}{h^2}
\end{align}
So, if $\delta(u)$ corresponds to the identity matrix $\mathds{1}$, then $\delta''(u)$ corresponds to the following banded matrix $L_\Delta$:
\begin{equation}
    L_\Delta \;=\; \frac{1}{h^{2}}
\begin{pmatrix}
-2 &  1 &  0 & \cdots & 0 & 0\\
 1 & -2 &  1 & \ddots &   & 0\\
 0 &  1 & -2 & \ddots & \ddots & \vdots\\
\vdots & \ddots & \ddots & \ddots & 1 & 0\\
 0 &   & \ddots & 1 & -2 & 1\\
 0 & 0 & \cdots & 0 & 1 & -2
\end{pmatrix},
\qquad
\end{equation}
where $h=\Delta u$ is the discretization of the grid. Notice that $L_\Delta$ is related to the 1D Laplacian, and it effectively ``smears" out the thin transport plan. We will see that it also adds additional stripes to the transport plan because of the interference with amplitude terms. 
As a result, we conclude that $W_f^{(0)}$ corresponds to the thin transport plan $\Gamma$, while $W_f^{(2)}$ adds additional stripes, and smears out the distribution. The next higher-order terms are related to the higher-order derivatives of the $\delta$ functions and have a similar effect on the Wigner distribution. 

\subsection{Numerical Experiments}
\label{subsection:numerical-experiments}

Next, we perform a simple numerical experiment to demonstrate the ideas above. As we shall see, our retrieved Wigner distribution $W_\psi$ looks almost identical to the ground truth $W_f$ when the phase is close to quadratic. Once we introduce a cubic term, we will see that $W_\psi$ starts to deviate from $W_f$.

We will work on a natural lattice $L_0$ with $N=128$ points. Thus, both $u$ and $\mu$ grids are discretized with $\Delta u=\Delta \mu = 1/\sqrt{128}$ and length $L_u=L_\mu=\sqrt{128}\approx11.3$ dimensionless units (see Section \ref{section:coordinate-lattices} in Appendix \ref{chapter:numerical-implementation} for more on lattices). We consider the following function $f:\mathbb{R}\to\mathbb{C}$:
\begin{align}
    f(u)&=\text{rect}(u/\sigma)e^{i(\alpha u^2+\beta u^3)}
\end{align}
where $\alpha, \beta,\sigma\in \mathbb{R}_+$ are some fixed parameters, and $\text{rect}(u)$ is 1 if $u\in[-0.5,0.5]$ and $0$ otherwise.

We evaluate $f$ on the lattice $L_0$ to get an array $f_i$, and normalize it using \ref{eq:unitary-normalization}. Then we define:
\begin{align}
    g_i&=|f_i|^2\\
    G_i&=|\mathcal{F}[f]_i|^2
\end{align}
where $\mathcal{F}$ here refers to a shifted discrete Fourier transform (see Section \ref{subsection:visualizing-shifted-dft} of Appendix \ref{chapter:numerical-implementation} for the details). Using optimal transport between distributions $g^2$ and $G^2$ we find our convex phase $\phi_0$ from which we deduce the complex-valued function $\psi(u)=g(u) e^{i\phi_0(u)}$, and $W_\psi$. We also compute the predicted marginal:
\begin{equation}
    \tilde{G}_i=|\mathcal{F}[\psi]_i|^2
\end{equation}

For each choice of parameters $\alpha, \beta, \sigma$, we compare the target marginal $G^2$ to the predicted marginal $\tilde{G}^2$ and compare Wigner distributions. Our experiments confirm that whenever cubic terms are small $\beta\ll\alpha$, the retrieved Wigner distribution is an excellent approximation to the ground truth.

\subsubsection{Flat Phase Experiment $\alpha=0$, $\beta=0$}

\begin{figure}[!ht]
\centering
\includegraphics[width=0.80\textwidth]{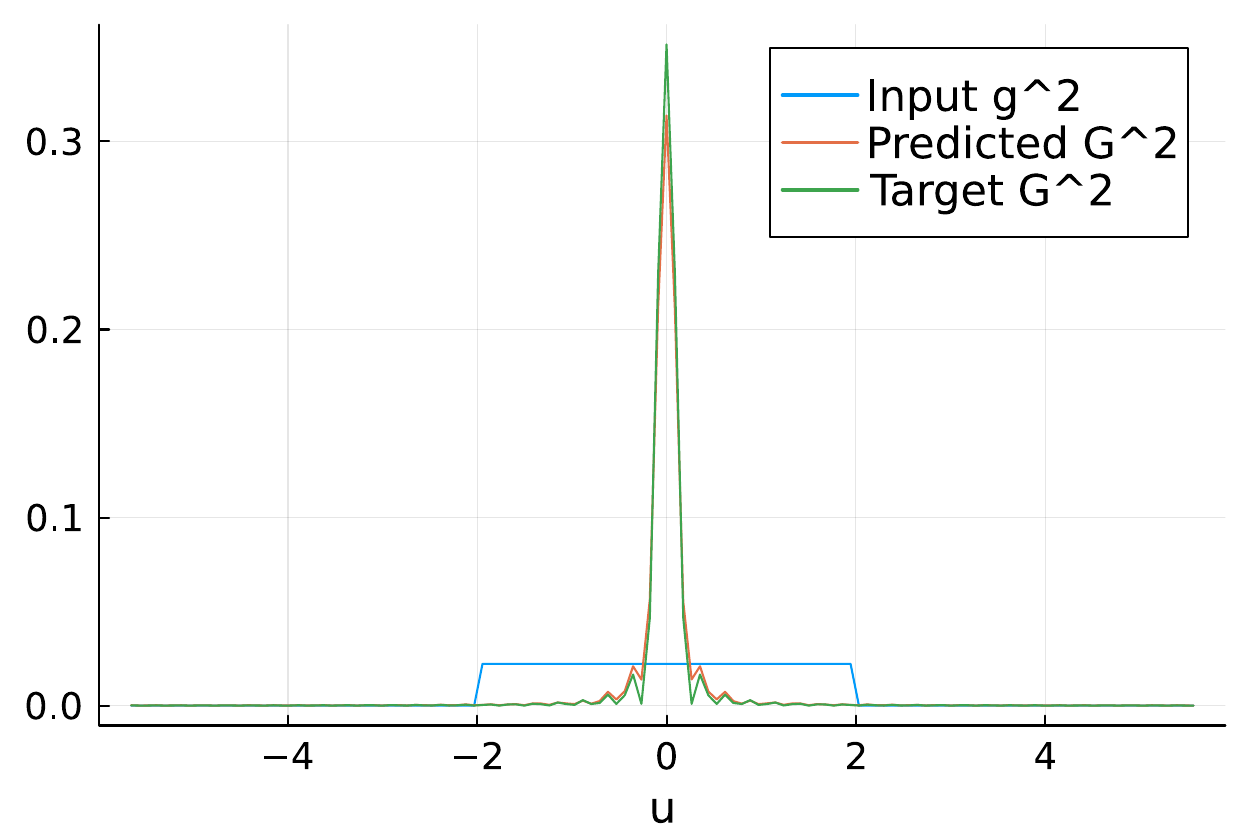}
\caption{Marginal intensities for $\alpha=0$, $\beta=0$, $\sigma=2$}
\label{fig:40}
\end{figure}

\begin{figure}[!ht]
\centering
\includegraphics[width=\textwidth]{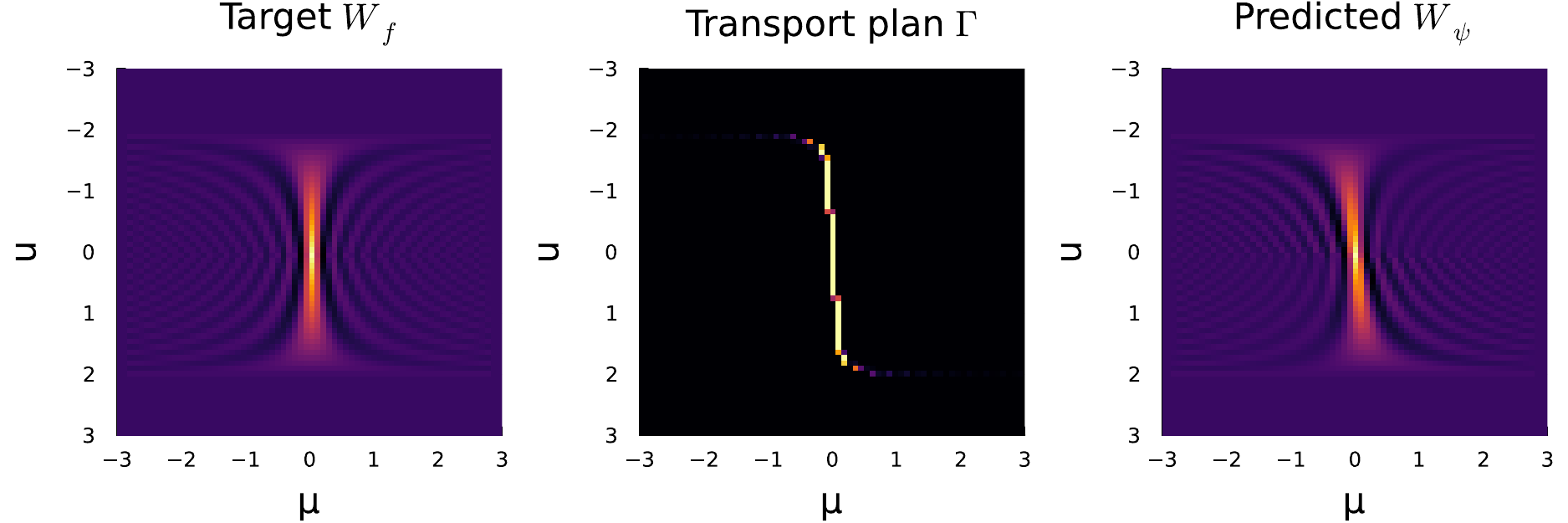}
\caption[Wigner estimation for $\alpha=0$, $\beta=0$, $\sigma=2$]{Target Wigner distribution $W_f$ (left), transport plan $\Gamma=W^{(0)}_f$ (middle), and retrieved Wigner distribution $W_\psi$ (right) for $\alpha=0$, $\beta=0$, $\sigma=2$}
\label{fig:41}
\end{figure}

For this experiment, we fix $\alpha=0$, $\beta=0$, and $\sigma=2$. The target is a square modulus of the Fourier transform of $\text{rect}(u)$ function, which is exactly $(\sin(2\pi \sigma \mu)/(2\pi\sigma\mu))^2$. The target Wigner distribution on the plot (\ref{fig:41}) is just a discretization of the continuous Wigner distribution (see Figure \ref{fig:wigner-rect}). 
As we can see from Figures \ref{fig:40} and \ref{fig:41}, the predicted Wigner distribution matches the desired Wigner distribution almost perfectly with the transport plan $W^{(0)}=\Gamma$ correctly highlighting the line of the biggest support of the distribution. The only caveat here is that the optimal transport assumes a strict convexity (or concavity), which results in a slight tilt of the main axis of the predicted distribution

\subsubsection{Quadratic Phase Experiment $\alpha=2$, $\beta=0$}
\begin{figure}[!ht]
\centering
\includegraphics[width=0.80\textwidth]{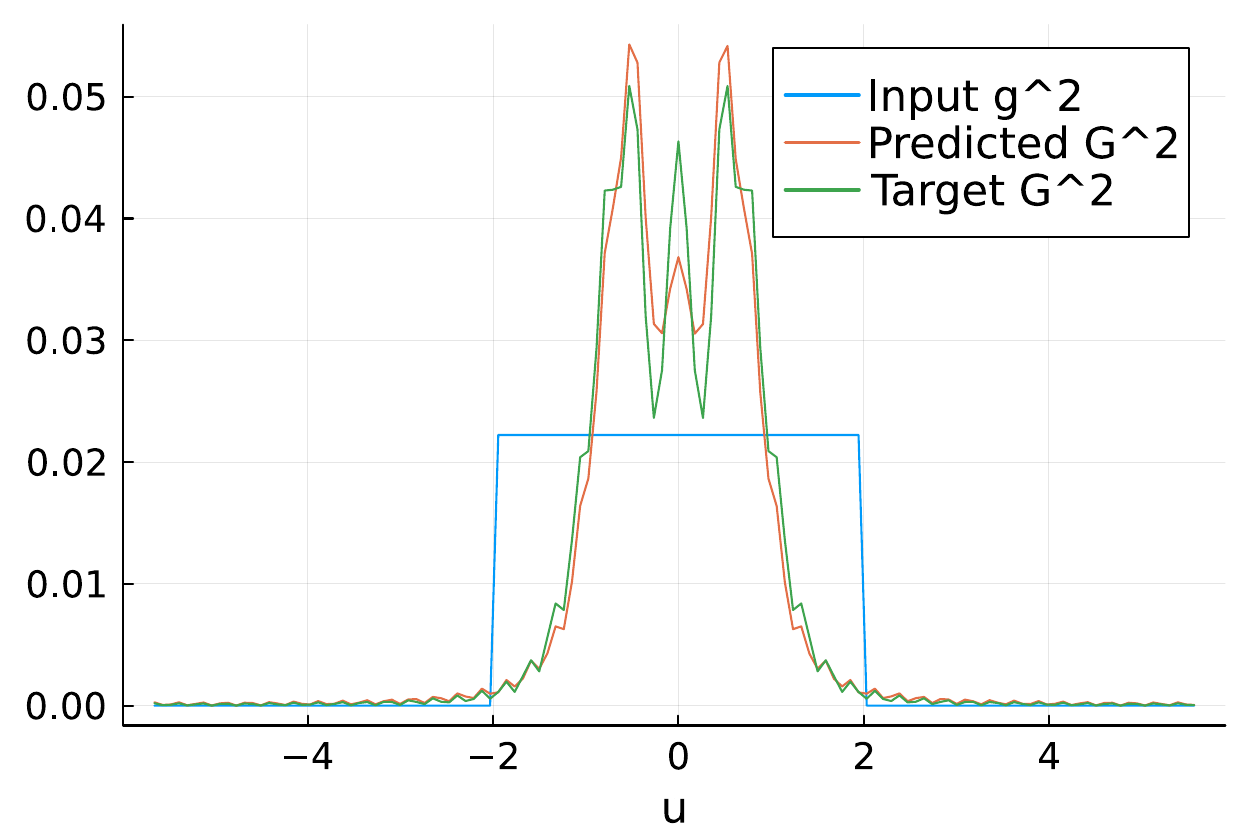}
\caption{Marginal intensities for $\alpha=2$, $\beta=0$, $\sigma=2$}
\label{fig:42}
\end{figure}

\begin{figure}[!ht]
\centering
\includegraphics[width=\textwidth]{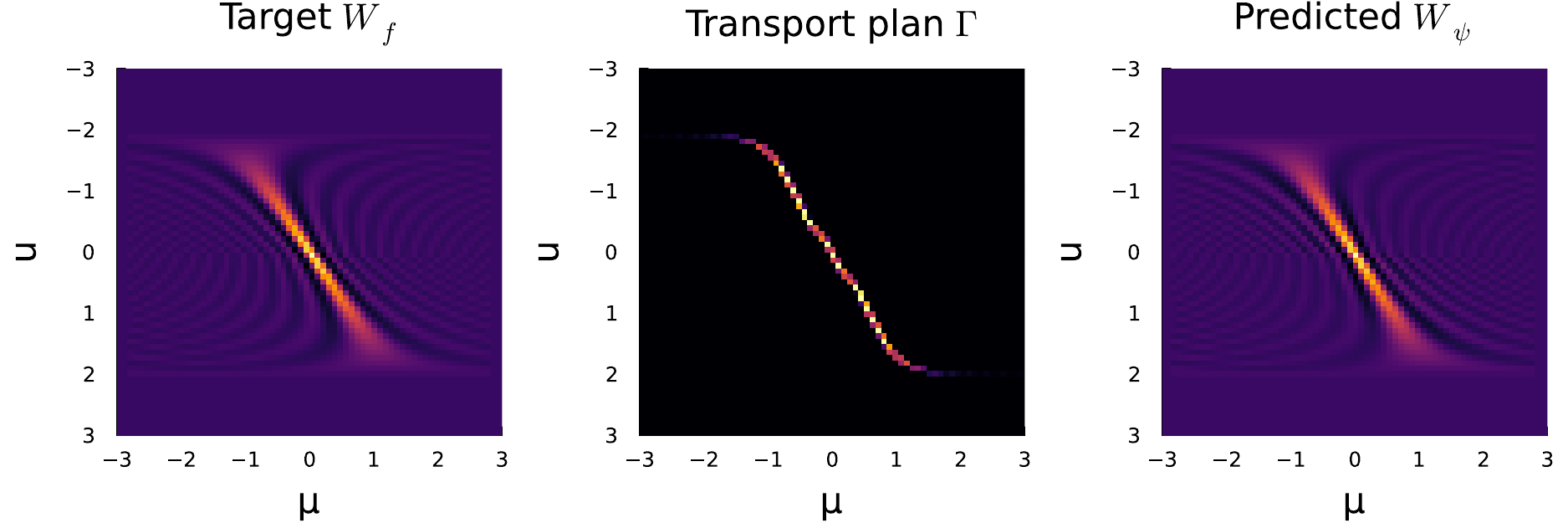}
\caption[Wigner estimation for $\alpha=2$, $\beta=0$, $\sigma=2$]{Target Wigner distribution $W_f$ (left), transport plan $\Gamma=W^{(0)}_f$ (middle), and retrieved Wigner distribution $W_\psi$ (right) for $\alpha=2$, $\beta=0$, $\sigma=2$}
\label{fig:43}
\end{figure}

For this experiment, we fix $\alpha=2$, $\beta=0$, and $\sigma=2$. This is equivalent to placing a lens in our optical set-up, or as we argued before, taking a fractional Fourier transform of the input distribution. This corresponds to a shearing transformation in the phase space. 
As we can see from Figures \ref{fig:42} and \ref{fig:43}, our approach recovers the wigner distribution almost perfectly, with the transport plan correctly highlighting the line of the biggest support of the distribution.

\subsubsection{Cubic Phase Experiment $\alpha=2$, $\beta=-2$}
\begin{figure}[!ht]
\centering
\includegraphics[width=0.80\textwidth]{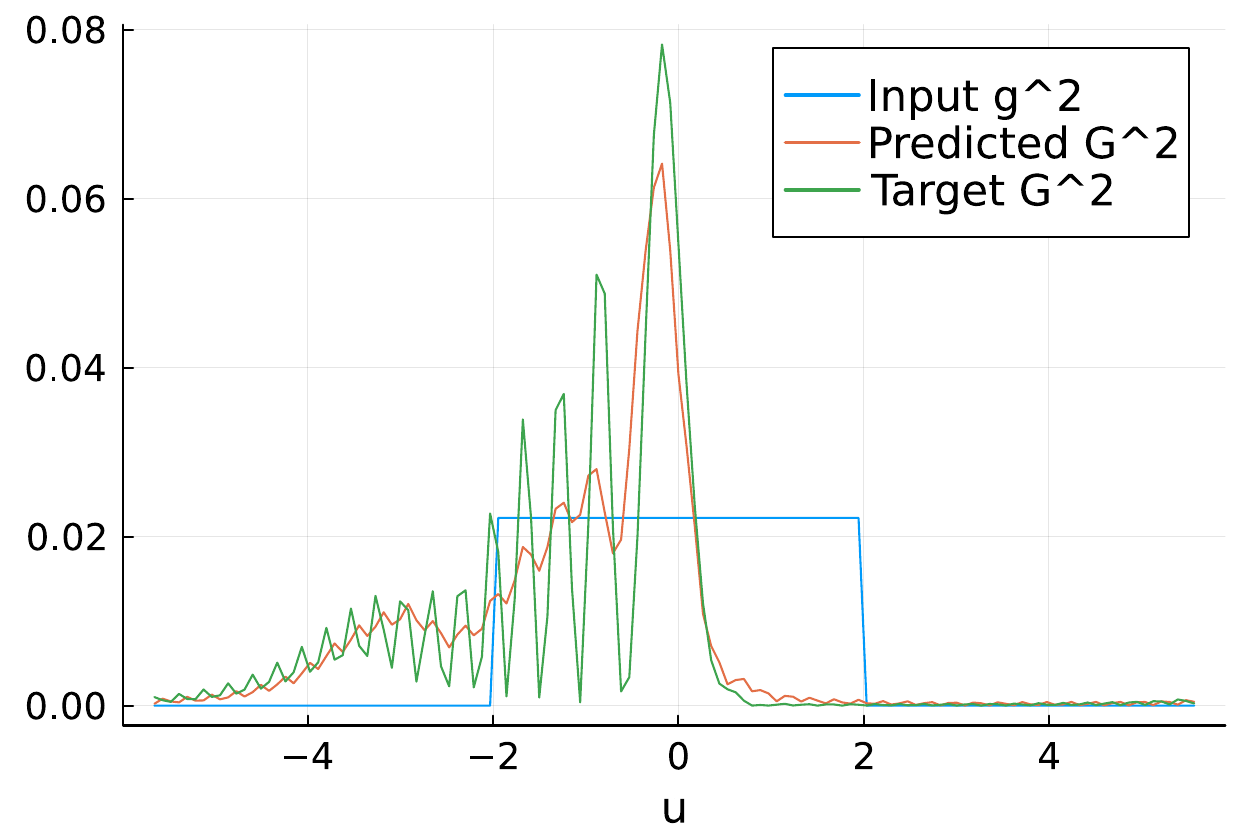}
\caption{Marginal intensities for $\alpha=2$, $\beta=-2$, $\sigma=2$}
\label{fig:44}
\end{figure}

\begin{figure}[!ht]
\centering
\includegraphics[width=\textwidth]{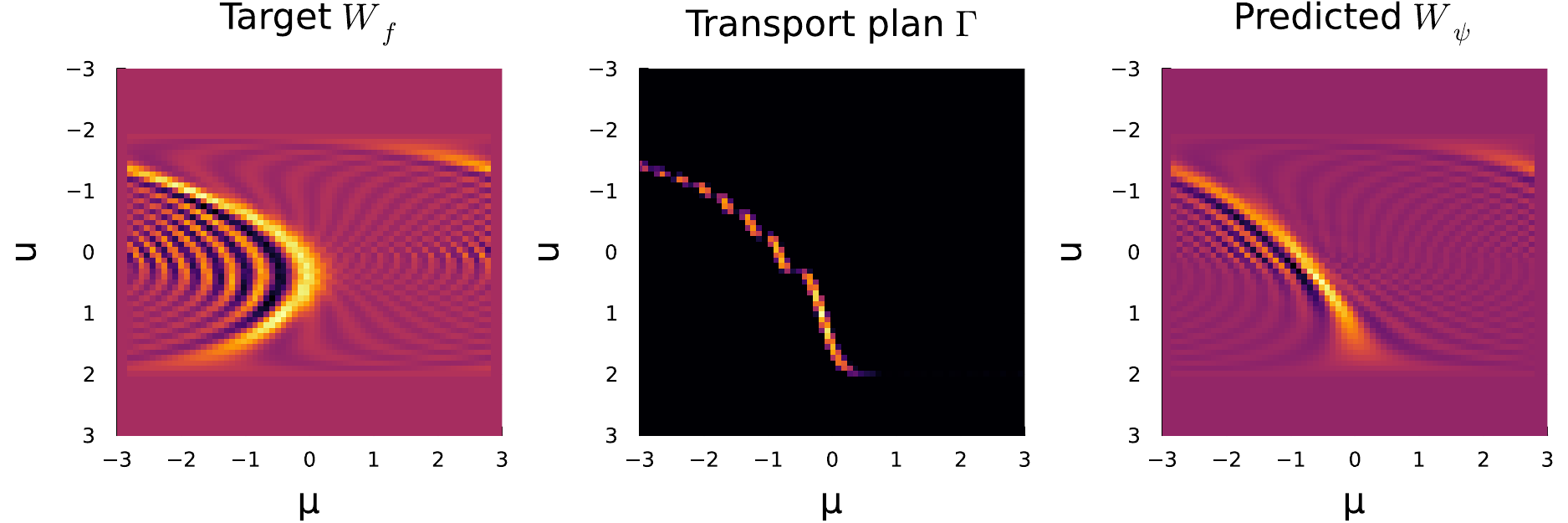}
\caption[Wigner estimation for $\alpha=2$, $\beta=-2$, $\sigma=2$]{Target Wigner distribution $W_f$ (left), transport plan $\Gamma=W^{(0)}_f$ (middle), and retrieved Wigner distribution $W_\psi$ (right) for $\alpha=2$, $\beta=-2$, $\sigma=2$}
\label{fig:45}
\end{figure}

For this experiment, we fix $\alpha=2$, $\beta=-2$, and $\sigma=2$. We note that this choice of the phase is neither convex nor concave and cannot be approximated well by a quadratic. Furthermore, one can show that adding a cubic phase \textit{does not} correspond to any linear canonical transform. So, this transformation is not something one usually sees in an optics lab. In other words, the cubic phase and the target Wigner distribution can be classified as \textit{non-classical}. Nevertheless, our approach does a decent job of retrieving the most important features of the distribution, by approximating it with the closest linear-canonical transformation (\ref{fig:44} and \ref{fig:45})

\subsubsection{Parameter Sweep Experiment}
\begin{figure}[!ht]
\centering
\includegraphics[width=\textwidth]{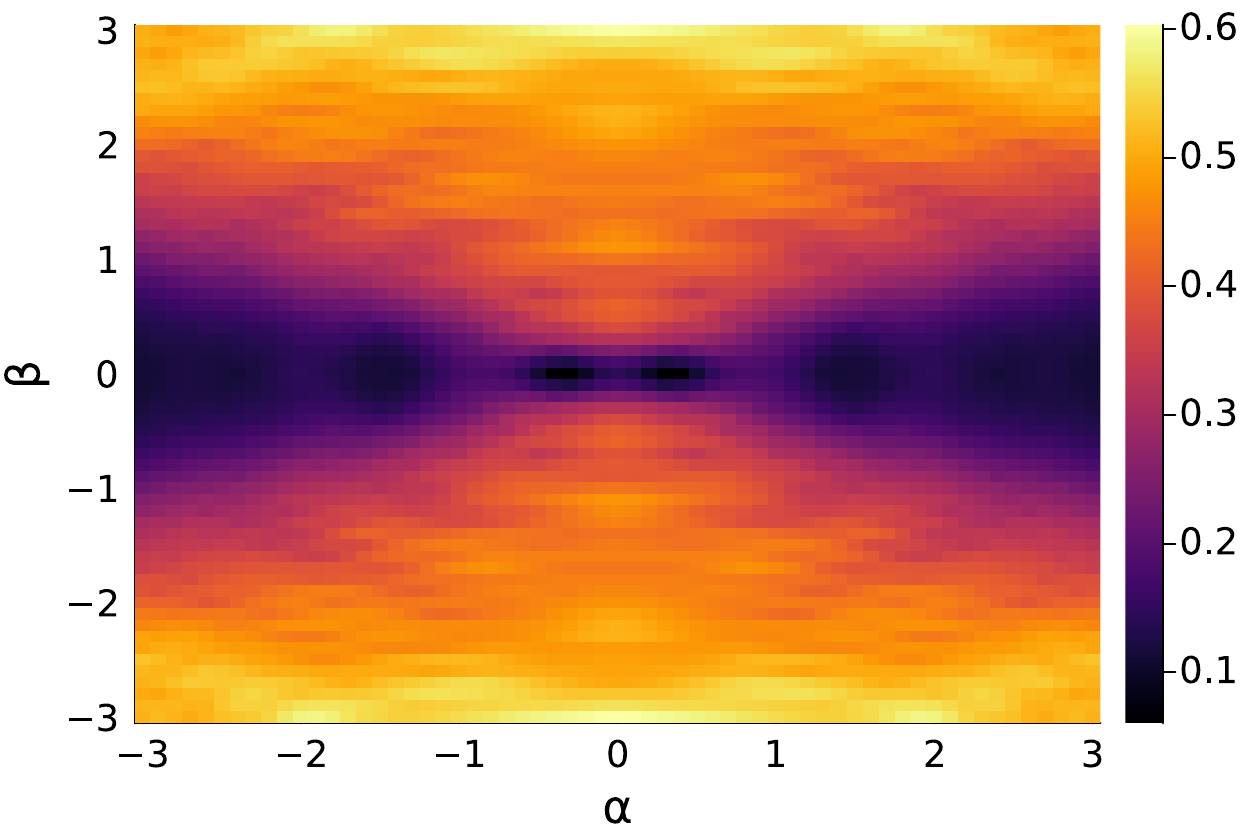}
\caption[Parameter sweep experiment]{Intensity loss between the target and predicted distributions for $\alpha$ and $\beta$ in a range of $[-3,3]$ with a step of $0.1$. We can see a clear increase in loss as we increase the cubic $\beta$ term, but it is relatively stable when we vary $\alpha$}
\label{fig:48}
\end{figure}

In this experiment, we sweep through the range of parameters $\alpha$ and $\beta$, recording the intensity loss (\ref{eq:Lint-def}) between the predicted $\tilde{G}$ and the target marginal $G$. The range of both parameters $\alpha$ and $\beta$ is from $-3$ to $3$ in steps of $0.1$. Figure \ref{fig:48} highlights the observation above that our approximation suffers from non-classical cubic contributions to the phase.

\subsection{Retrieving Linear Canonical Transforms}
\label{subsection:retrieving-linear-canonical-transforms}

The experiment above shows evidence that this method can be used to approximate the Wigner distribution of any $f:\Omega\to\mathbb{C}$, where $\Omega$ is a compact subset of $\mathbb{R}^2$, as long as the phase is strictly convex or strictly concave. In particular, if we multiply the modulus $g(u,v)$ by some quadratic phase, we can always retrieve the Wigner distribution of the combined signal. 

We can use this idea to estimate the effect of any optical system on the Wigner distribution that can be characterized by the real ABCD matrix (so-called quadratic-phase systems). Recall from Chapter 2 that any ABCD matrix corresponds to the following linear canonical transform:

\begin{equation}
    \hat{h}(x,x')=\sqrt{\frac{1}{\hat{B}}}e^{-i\pi/4}\exp\left[\frac{i\pi}{\hat{B}}(\hat{D}x^2-2xx'+\hat{A}x'^2)\right]
\end{equation}

Suppose next that the input to the system is a well-collimated beam i.e. $\hat{f}(x,y)=g(x,y)e^{i\phi(x,y)}$ and phase $\phi(x,y)=\phi_0$ is constant. Then applying this system to $\hat{f}$ we get the output signal $\hat{F}$, which can be computed as follows:
\begin{align}
    \hat{F}(x,y)&=\iint \hat{f}(x',y')\hat{h}(x,x')\hat{h}(y,y')dx' dy'\\
    &=\frac{e^{i\varphi(x,y)}} {\hat{B}}\iint \hat{g}(x', y')e^{i2\pi \psi(x',y')}dx'dy'
\end{align}
Where we grouped phases into:
\begin{align}
    \psi(x',y')&\equiv \frac{\hat{A}}{2\hat{B}}\left(x'^2+y'^2-2(xx'+yy')\right) \\
    \varphi(x,y)&\equiv \phi_0-\frac{\pi}{2}+\pi(x^2+y^2)\frac{\hat{D}}{\hat{B}}
\end{align}
Now the modulus at the input is $g(x,y)$, by definition, and the modulus of the output is defined as:
\begin{equation}
    G(x,y)=\frac{1}{\hat{B}}\left|\iint \hat{f}(x', y')e^{i2\pi \psi(x',y')}dx'dy' \right|
\end{equation}
where the phase factor $\varphi(x,y)$ disappears after taking the absolute value. 

Using our optimal transport solution between $G^2$ and $g^2$ we will get a direct estimate of $\psi$, which corresponds to learning $\hat{A}/2\hat{B}$. Now, if we apply a similar idea to phase-diversity imaging, we can obtain additional amplitude $G_j$ from which we can infer $2$ more coefficients of the ABCD matrix. We only need 3 coefficients to figure out everything about the linear canonical transformation system. So, our intuition is that you need maybe 1 or 2 additional diversity images. We leave exact implementations of this idea as future directions for this work.

%% file: sections/chapter4.tex
\chapter{Algorithms}
\label{chapter:algorithms}

Both the phase generation (\ref{prob:phase-generation}) and the beam estimation (\ref{prob:beam-estimation}) are non-convex optimization problems. For general amplitudes, ``perfect" solutions might not exist. \footnote{By ``perfect" we mean solutions with zero loss. It is easy to show that such solutions do not generally exist (e.g. if input \& output beam amplitudes have compact support)}. Even if we define a loss that captures the reconstructed error and try to optimize over the phase of the beam, there is a high chance that the optimization algorithm will get stuck in a local minimum. This is due in large part to the fact that a phase solution is only defined mod $2\pi\mathbb{Z}$ at each point, which makes the loss landscape extremely tricky to optimize and gives rise to vortex formation — an unwanted artifact that stagnates convergence of most iterative algorithms for phase retrieval. 

There are several, now standard, approaches to phase generation in the literature. The most common is known as a Gerchberg-Saxton (GS) algorithm or the Iterative Fourier Transform Algorithm (IFTA) \cite{gerchberg1972practical}, which uses a method of iterated projections to optimize for the phase. Unfortunately, this algorithm suffers from vortex formation and tends to converge on suboptimal solutions. In \cite{Pasienski:08}, the GS algorithm was improved to the Mixed Region Amplitude Freedom (MRAF) algorithm, which opened a possibility for creating high-accuracy holograms at the expense of wasting some of the laser power. Finally, there are several approaches known as Cost Function Minimization (CFM) \cite{Harte:14}, which use the conjugate gradient method to optimize the phase. However, the authors of this paper state that this method is ``inadequate for large continuous patterns due to the emergence of optical vortices within the trapping region during calculation. These vortices are characterised by a sudden drop in intensity coinciding with a local phase winding by a multiple of $2\pi$, and they arise because they can be initially beneficial to cost function reduction."

All of the algorithms above start the optimization from some initial guess for the phase and then iteratively update it until some convergence objective is reached. The initializations are usually quite simple — either a random phase, a constant phase, or the argument of the inverse Fourier transform of the target amplitude. As it turns out, the exact choice of the initial phase is extremely important for the convergence of the algorithm, and if chosen correctly, it can bypass any vortex formation.

In this chapter, we will first quickly review the standard approaches to solving the problem and numerically demonstrate their drawbacks. Next, we will show how to obtain an initial phase guess using a convex relaxation of the problem, which significantly improves the convergence of any of the above methods. 

\section{Method of Iterated Projections}
\label{section:method-of-iterated-projections}

Suppose you have some general Hilbert space $H$ and two closed subsets $A$ and $B$ such that $A\cap B\neq \emptyset$. These sets are called \textit{constraint sets} and our job is to find any element $x\in A\cap B$. In other words, an element $x\in H$ that satisfies both constraints. To do so, we are given access to a starting point $x_0\in H$ and two projection functions $P_A: H \to A$ and $P_B: H \to B$. These functions have a property that they project onto the closest point in the set (this point might not be unique). With these assumptions a simple algorithm is to do the following:

\begin{algorithm}[!ht]
\caption{Method of Iterated Projections}
\label{alg:mip}
\begin{algorithmic}[1]
\State $x \gets x_0$ \Comment{initialization}
\State $i \gets 0$
\While{$i \leq N$}
\State $x \gets P_A(x)$ \Comment{Project onto A}
\State $x \gets P_B(x)$ \Comment{Project onto B}
\State $i = i + 1$
\EndWhile
\end{algorithmic}
\end{algorithm}

\begin{figure}[htbp]
    \centering
    \begin{subfigure}[b]{0.45\textwidth}
        \centering
        \includegraphics[width=\textwidth]{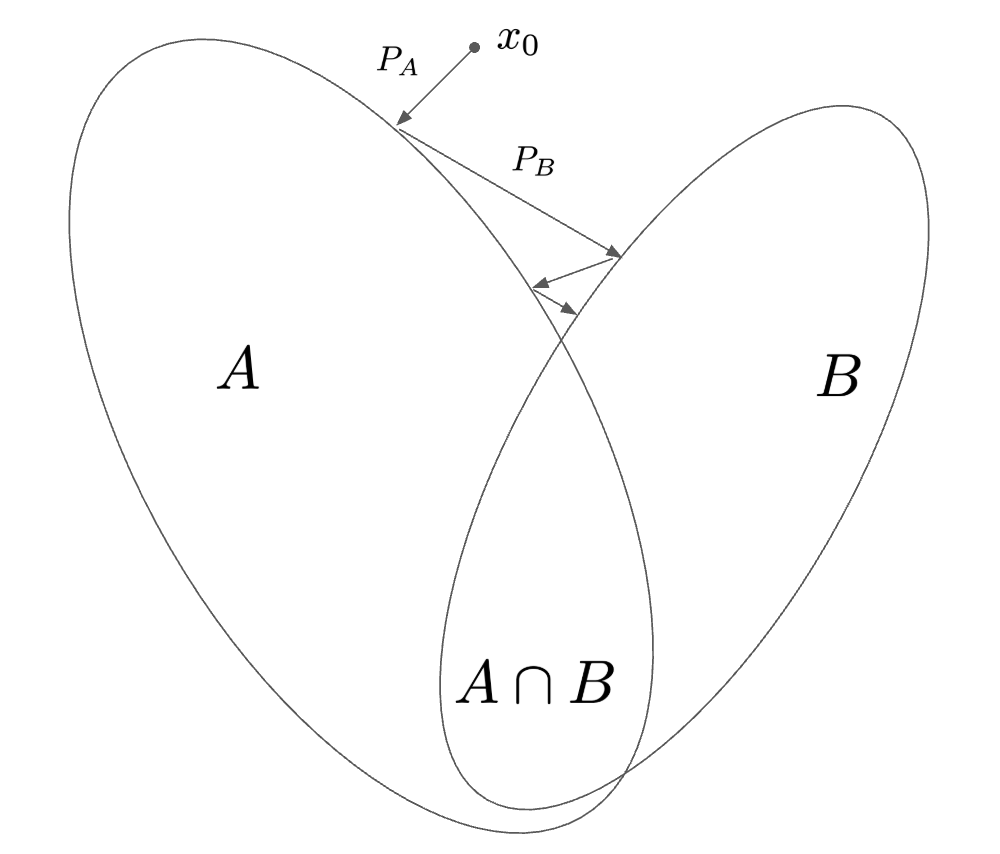}
        \caption{Convergence with convex sets}
    \end{subfigure}
    \hfill
    \begin{subfigure}[b]{0.45\textwidth}
        \centering
        \includegraphics[width=\textwidth]{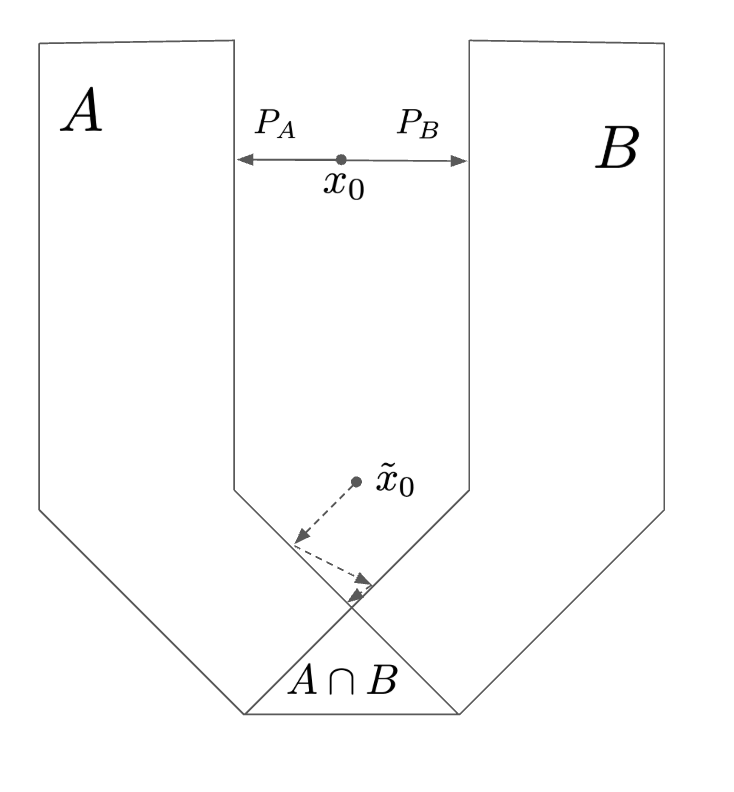}
        \caption{Convergence with non-convex sets}
    \end{subfigure}
    \caption[Convex and non-convex method of the iterated projections]{Convergence of the method of iterated projections depending on the geometry of sets $A$ and $B$. The panel (a) demonstrates that convergence is guaranteed for any starting point $x_0$. The  panel (b) shows that depending on the starting position, the algorithm might (or might not) converge}
    \label{fig:conv-iterated}
\end{figure}

Without any guarantees on the geometry of the sets $A$ and $B$, the convergence is not guaranteed (see Figure \ref{fig:conv-iterated} for a counterexample). However, if we assume that sets $A$ and $B$ are \textit{convex} (meaning that $\forall x,y\in A, \alpha\in[0,1]\implies \alpha x+(1-\alpha)y\in A$) we have a guarantee that the algorithm will converge, as proven by Cheney and Goldstein in \cite{proximitymapsConvex}. Meaning that $x_n$, the point we obtain after applying $n$ iterations, converges to a point $y\in A \cap B$:
\begin{equation}
    \lim_{n\to\infty} ||x_n-y||=0
\end{equation}
where $||z||=\sqrt{\braket{z|z}}$ is the norm induced by the inner product of the Hilbert space. 

A similar proof can be used in the case when the sets $A$ and $B$ are convex but non-intersecting ($A\cap B = \emptyset$). In this case, the algorithm will eventually iterate between points $x_i\in A$ and $x_{i+1}\in B$ that are closest to each other. In other words, it will find a pair of closest points between two convex non-intersecting sets.

Moreover, convexity also guarantees that the projection maps are well-defined, meaning that there exists a unique point in a set A (or set B) closest to $x$. This is due to the closest point property of the Hilbert space, which is only true for non-empty, closed, convex sets (see Theorem 3.8 in \cite{Young_1988}).  

If we consider non-convex sets, however, there is no guarantee of convergence. Then the initialization steps become very important, and if we start ``close" to the intersection $A\cap B$, then we have a better hope of converging. This will be a basis for our intuition for the rest of this chapter.

\section{Gerchberg-Saxton}
\label{section:gerchberg-saxton}

Let us consider a particular example of the method of iterated maps known as the Gerchberg-Saxton (GS) algorithm (sometimes called IFTA). The general idea is that we want to find a function that satisfies amplitude constraints at the input plane and at the output plane, and use a method of iterated maps to find such a function.

For this example, we consider a space of functions $L^2(D)$ defined in (\ref{eq:l2d-def}) of Appendix \ref{chapter:fourier-analysis}. Recall that this is a Hilbert space with the norm:
\begin{equation}
    \braket{f|g}=\iint_D f^*(u,v)g(u,v)dudv
\end{equation}

Suppose now that we want to find a function $f\in L^2(D)$ that has amplitude $g^2:D\to\mathbb{R}_{\geq0}$ at the input plane, and an amplitude $G^2:D\to\mathbb{R}_{\geq0}$ at the Fourier plane. These conditions can be viewed as constraint sets:
\begin{align}
\label{eq:GS-constraint-sets}
    A &= \{f\in L^2(D):|f|^2=g^2\} \\
    B &= \{f\in L^2(D):|\mathcal{F}[f]|^2=G^2\}
\end{align}
In order to define projections onto these sets, we first decompose our function into polar coordinates $f(u,v)=r(u,v)e^{i\phi(u,v)}$. To project onto a set $A$, we simply replace the current amplitude $r(u,v)$ with the target amplitude $g(u,v)$ while keeping the phase. Similarly, for the Fourier domain, we take the Fourier transform, replace the amplitude, and then take the inverse Fourier transform. 
\begin{align}
\label{eq:gs-projections}
    P_A(f)&=g\frac{f}{|f|}  \\
    P_B(f)&=\mathcal{F}^{-1}\left[G\frac{\mathcal{F}[f]}{|\mathcal{F}[f]|}\right]
\end{align}
Where $|f|$ is the amplitude in the input domain ($r(u,v)$), and $|\mathcal{F}[f]|$ is the amplitude in the Fourier domain. With the projections operations in place, we can summarize the Gerchberg-Saxton algorithm as follows:
\begin{algorithm}[!ht]
\label{alg:gs-alg}
\caption{Gerchberg-Saxton Algorithm}
\begin{algorithmic}[1]
\State $\phi \gets \text{Angle}\left(\mathcal{F}^{-1}[G]\right)$ \Comment{Phase initialization}
\State $i \gets 0$
\While{$i \leq N$}
\State $f_2 \gets \mathcal{F}[g e^{i\phi}]$ \Comment{Estimate Fourier plane field}
\State $\phi \gets \text{Angle}(f_2)$ \Comment{Discard amplitude}
\State $f_1 \gets \mathcal{F}^{-1}[G e^{i\phi}]$ \Comment{Estimate input plane field}
\State $\phi \gets \text{Angle}(f_1)$ \Comment{Discard amplitude}
\EndWhile
\end{algorithmic}
\end{algorithm}

This is just a particular example of the algorithm \ref{alg:mip} where lines 4,5 implement $P_A$ and lines 6,7 implement $P_B$. The phase initialization is performed by taking the inverse Fourier transform of the target amplitude. Another common choice is to initialize the phase to a constant value or start from a random matrix.

\subsection{Convergence Properties}
\label{subsection:convergence-properties}

Let us consider the geometry of the constraint sets. To see that they are not convex, pick $f_1$ and $f_2 \in A$, and $\alpha\in[0,1]$. Without loss of generality, we can write $f_1=ge^{i\phi_1}$, $f_2=ge^{i\phi_2}$. Notice that:
\begin{align}
    |\alpha f_1+(1-\alpha)f_2|^2&=\alpha^2 |f_1|^2+(1-\alpha)^2 |f_2|^2+2\alpha(1-\alpha)\text{Re}[f_1f^*_2] \\
    &=\alpha^2 g^2+(1-\alpha)^2g^2+2\alpha(1-\alpha)g^2\cos(\phi_1-\phi_2)
\end{align}
which is clearly not convex (similar reasoning applies to $B$).
Because sets are not convex, we are not guaranteed to converge using the method of the iterated projections. Moreover, the projections are not unique — there could be multiple points in the constraint set equidistant from $f$\footnote{This happens if and only if the amplitude of f is zero at some point where g is non-zero, in which case any phase gives an equidistant point.}. That said, GS projections (\ref{eq:gs-projections}) are optimal. 
\begin{align}
    ||f-P_A(f)||^2_{L^2}&=||f-\frac{fg}{|f|}||^2_{L^2}\\
    &=||\frac{f}{|f|}(|f|-g)||^2_{L^2}\\
    &=\int_D\left|\frac{f}{|f|}(|f|-g)\right|^2\\
    &=\int_D\left||f|-g\right|^2
\end{align}
Now, if we consider an arbitrary member of the constraint set $A$, call it $\tilde{f}=ge^{i\psi}$, we see that, by reverse triangle inequality:
\begin{align}
    ||f-\tilde{f}||^2_{L^2}&=\int_D |f-ge^{i\psi}|^2\\
    &\geq\int_D\left||f|-|g|\right|^2= ||f-P_A(f)||^2_{L^2}
\end{align}
So, $P_A$ projects onto the closest point in the constraint set $A$. The same argument applies to the projection $P_B$. 

Recent convergence analysis \cite{noll2021alternatingprojectionsapplicationsgerchbergsaxton} of the Gerchberg-Saxton algorithm has shown that the algorithm can fail to converge even if the solution is feasible (i.e. $A\cap B\neq\emptyset)$. In most cases, L2 loss between the target image and the reconstructed image stagnates at a suboptimal value. However, the convergence is guaranteed under the following 3 conditions: \footnote{These conditions were summarized from the paper \cite{noll2021alternatingprojectionsapplicationsgerchbergsaxton} by GPT-o3.}

\begin{enumerate}
    \item \textbf{Angle Condition}:
    There exists a uniform lower bound on the angle between consecutive projection steps. Formally, the projections from one set to the other must not become too aligned (i.e., the angle between projection directions does not shrink to zero too fast). This ensures consistent progress toward convergence and avoids ``grazing" behavior where the iterates slide along the surface without approaching a solution.

    \item \textbf{Hölder Regularity}:
    This condition controls how the two sets interact geometrically near the point of convergence. It ensures that the distance between the sets grows in a controlled (Hölder continuous) manner near the solution. It prevents situations where the sets become too flat or tangential to each other, which could hinder convergence.

    \item \textbf{Saturated Gaps}:
    This guarantees that when a point is near the target set (e.g., in the frequency domain), its projection lands in a predictable neighborhood of the other set (e.g., in the spatial domain). It ensures that proximity in one domain results in proximity in the other, thereby preventing erratic or divergent projection behavior near the solution.
\end{enumerate}

Intuitively speaking, most of these conditions can be met if we start ``close" to the intersection of two sets. In practice, we see an order of magnitude improvement in L2 loss if we initialize the phase using the convex relaxation of the problem \cite{swan2024highfidelityholographicbeamshaping}.

\subsection{Phase vortices}
\label{subsection:phase-vortices}

\begin{figure}[!ht]
\centering
\includegraphics[width=0.67\columnwidth]{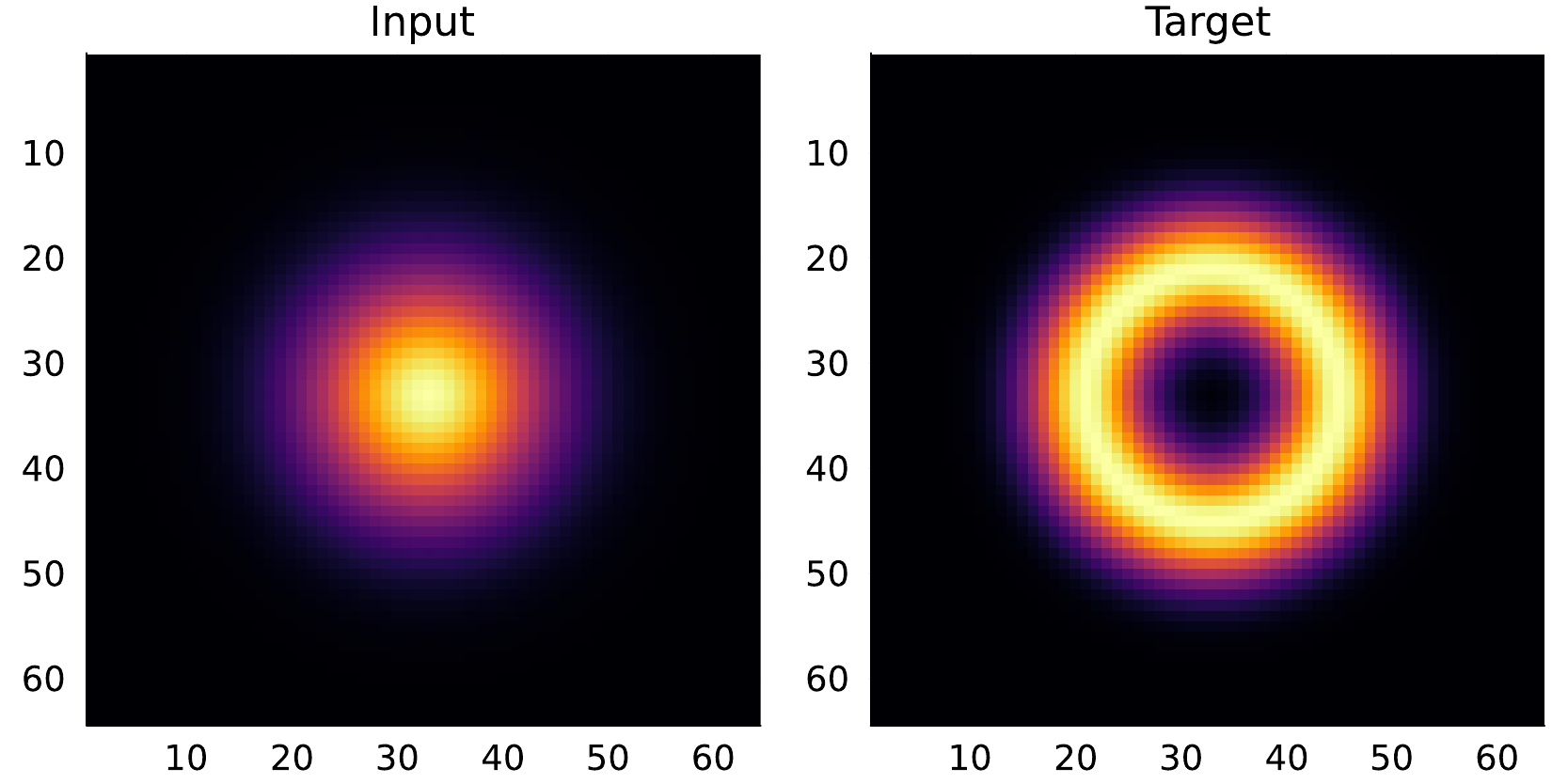}
\caption[Gaussian input and ring output]{Input is a Gaussian intensity pattern with $\sigma=1.0$ unit. The target intensity is a Gaussian ring of radius $R=2.5$ and standard deviation of $\sigma=1.0$ unit.}
\label{fig:input_output}
\end{figure}

Our numerical experiments show that the stagnation of the convergence in the GS algorithm is associated with the formation of the phase vortices. To demonstrate this, we create an example of input and output intensities, discretized on a 64x64 grid (see Figure \ref{fig:input_output}). We are working with a natural lattice (see Subsection \ref{subsection:natural-lattice}  in Appendix \ref{chapter:numerical-implementation}) — discretization is set to $\Delta u=\Delta v=1/\sqrt{64}$ and the total window size is $L_u=L_v=\sqrt{64}= 8$ dimensionless units wide.

We initialize the phase to all zeros and run 500 iterations of Gerchberg-Saxton while keeping track of the intensity loss (see eq. \ref{eq:Lint-def} in Appendix \ref{chapter:numerical-implementation}) between the reconstructed image and the objective.

\begin{figure}[!ht]
\centering
\includegraphics[width=0.65\columnwidth]{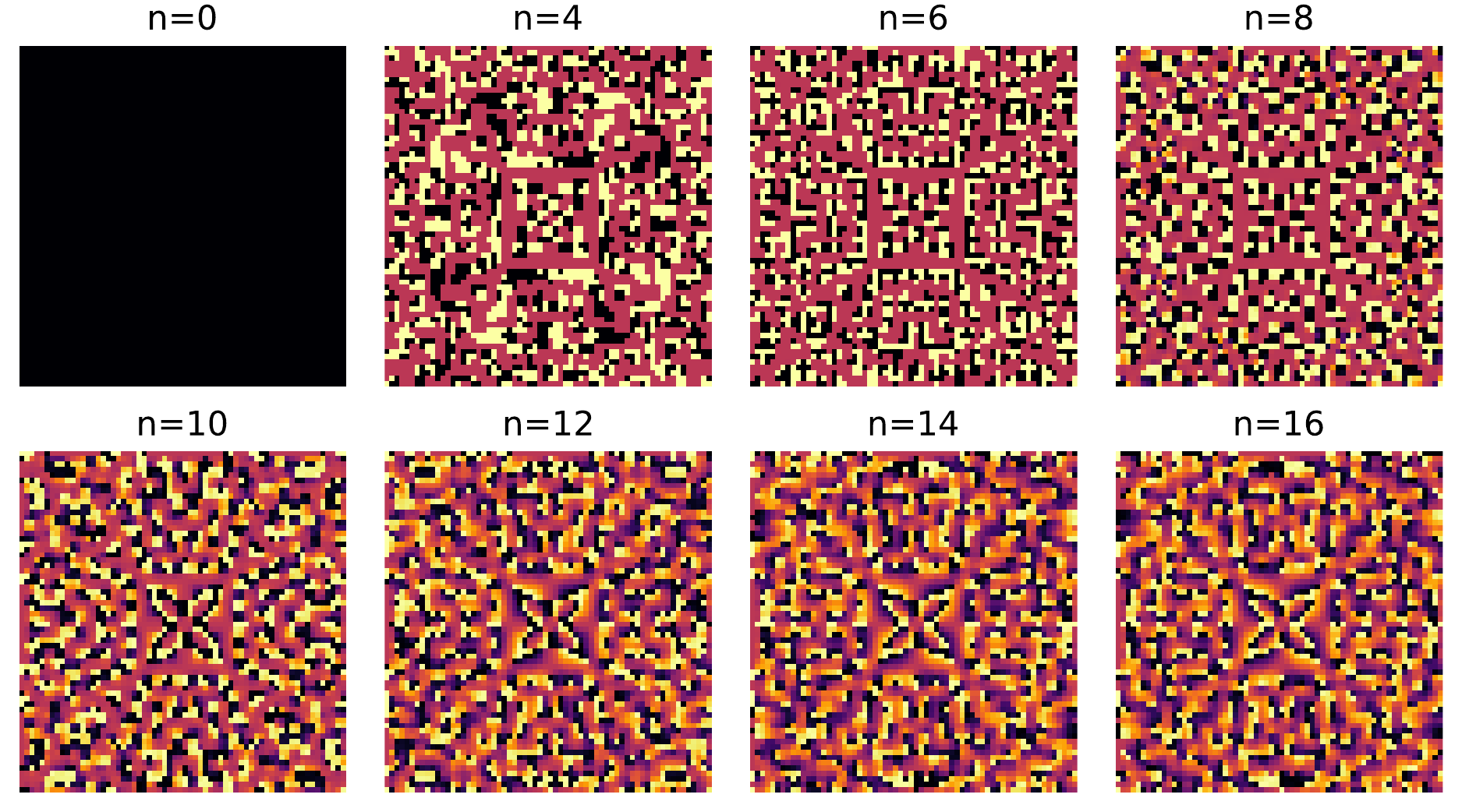}
\caption[First 16 iterations of the GS algorithm]{Optimized phase over first 16 iterations. We start with a flat phase of all zeros, and after ~10 iterations, phase vortices start to form and they persist for all remaining iterations.}
\label{fig:vortex_formation}
\end{figure}

\begin{figure}[!ht]
\centering
\includegraphics[width=0.55\columnwidth]{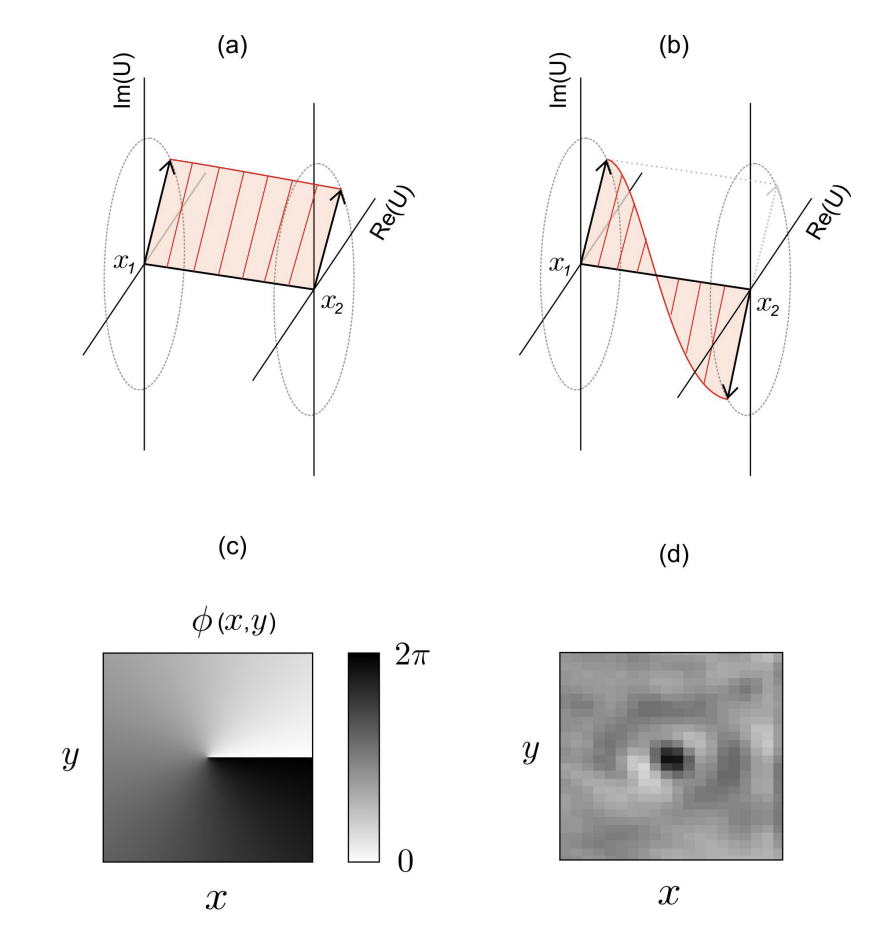}
\caption[Vortex formation and its effect on intensity]{Visual diagram depicting vortex formation (courtesy to \cite{quantumEngineeringThesis}). (a) Two adjacent points $x_1$ and $x_2$ in the focal plane with the same amplitude and phase. The physical light field interpolates between the two points and maintains approximately the same amplitude. (b) Two adjacent points with opposite phase, where the amplitude of the interpolated light field crosses zero. (c) A phase singularity in the light field is a point where the phase assumes all values between 0 and $2\pi$ in an arbitrarily small area around the singularity. (d) Measured intensity at the location of a phase singularity, showing a dip in the intensity, called a speckle. The speckle size is on the order of the diffraction limit.}
\label{fig:vortex_diagram}
\end{figure}

As you can see from Figure \ref{fig:vortex_formation}, after a couple of iterations, the phase of the algorithm roughly stabilizes around a solution that has a lot of branch cuts, with so-called phase vortices — places where phase wraps by $2\pi$. Each vortex has a property that if you go around it in a circle, the phase will continuously change from $-\pi$ to $\pi$ and then jump to $-\pi$ again (see Figure \ref{fig:vortex_diagram}). One way to understand these vortices is to think of them as locations where the curl of the gradient of the phase is nonzero. Recall that in a classical (ray-optics) limit, the light propagates along the ray direction $\nabla \phi$. Thus, phase vortices are an example of a non-classical effect, far-removed from the ray-optics limit.

\newpage

\newpage
 
\begin{figure}[!ht]
\centering
\includegraphics[width=0.70\columnwidth]{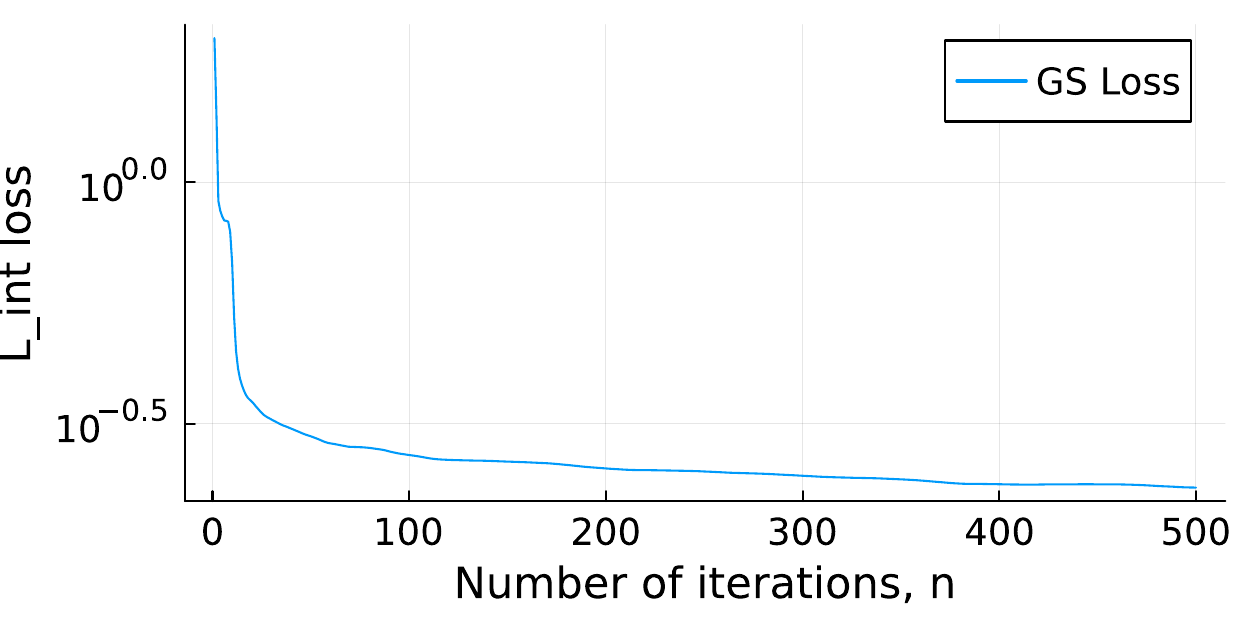}
\caption{Intensity Loss as defined in (\ref{eq:Lint-def}) as a function of number of iterations.}
\label{fig:L2convergence}
\end{figure}

\begin{figure}[!ht]
\centering
\includegraphics[width=0.70\columnwidth]{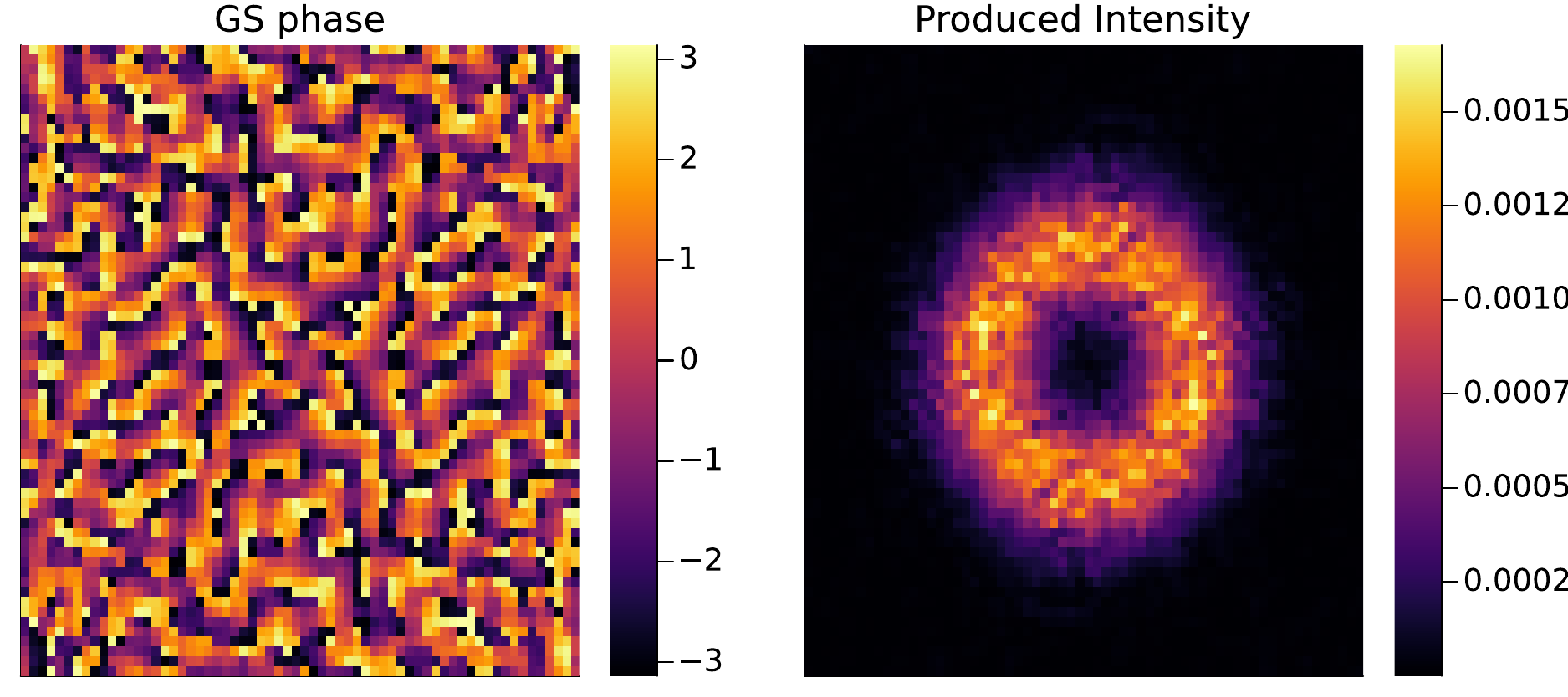}
\caption[Gerchberg-Saxton solution with vortecies]{The phase solution at iteration 500 of the Gerchberg-Saxton algorithm. The left panel shows many places where ``phase vortices" appear — places where phase wraps around the circle in $2\pi$. Each phase vortex corresponds to a black dot (0-intensity spot) on the output image (right panel).}
\label{fig:finalimage}
\end{figure}

In practice, these phase vortices limit the convergence of the Gerchberg-Saxton algorithm by introducing too many degrees of freedom to the space of the solutions, which inevitably leads to the GS algorithm being stuck in local minima. Moreover, phase vortices, when propagated to the Fourier plane, result in 0-intensity dots on the produced image (see Figure \ref{fig:finalimage}). These dots are the main contribution to the intensity loss, as we can see from (see Figure \ref{fig:L2convergence}). As the iteration progresses, GS is unable to remove the phase vortices, but instead just moves them around until the lowest possible loss is achieved.

We will come back to this example later and show that by initializing the Gerchberg-Saxton algorithm with the appropriate phase, no vortices will form, and we will get a solution with a much lower loss. Notice also that such phase vortices are a common unwanted artifact for other iterative solvers as well, such as MRAF \cite{Pasienski:08} and conjugate gradients CFM \cite{Harte:14}.

\section{MRAF Algorithm}
\label{section:MRAF-algorithm}

\begin{figure}[!ht]
\centering
\includegraphics[width=0.70\columnwidth]{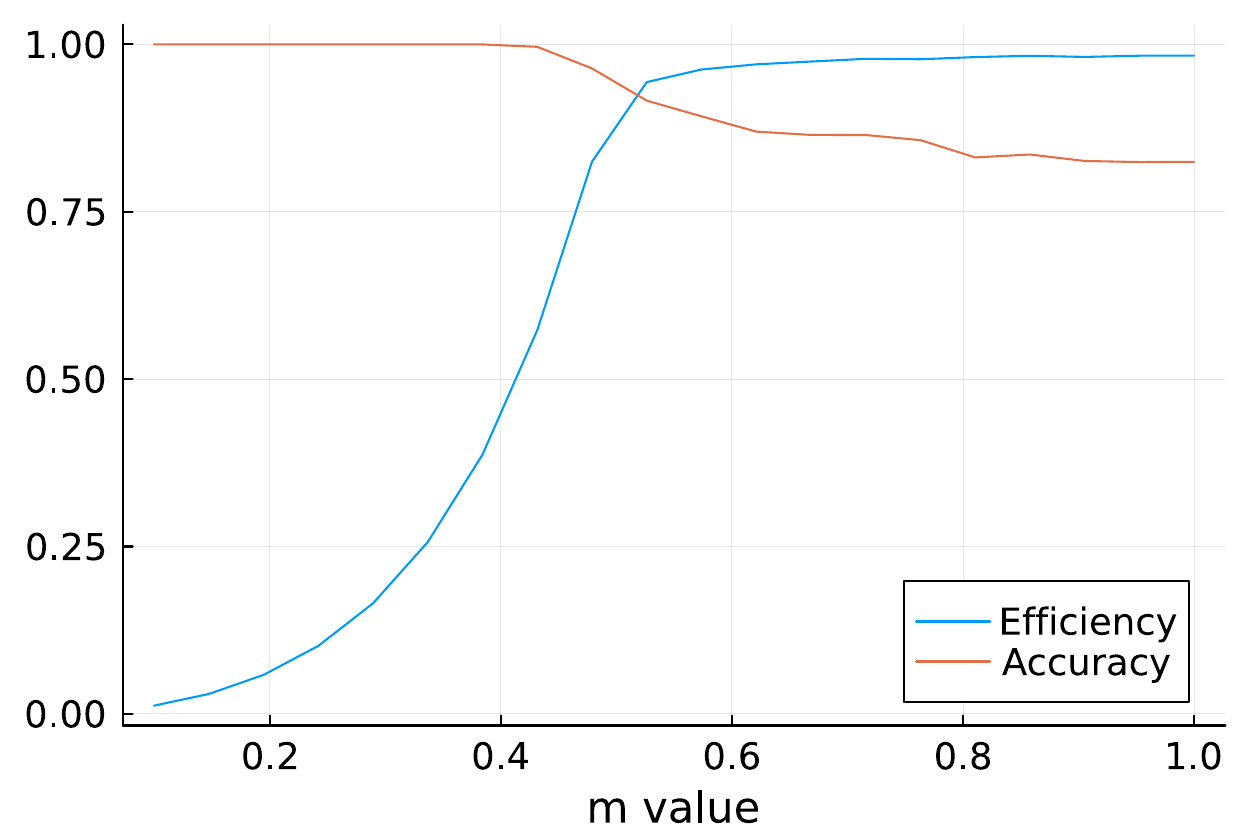}
\caption[Accuracy vs. Efficiency trade-off]{Accuracy vs. Efficiency trade-off. By controlling the value of the parameter $m$, we can change how much light is sent into the signal region. By losing most of the laser power in the noise region, we can obtain arbitrarily high accuracy in the signal region.}
\end{figure}

\begin{figure}[!ht]
\centering
\includegraphics[width=0.65\columnwidth]{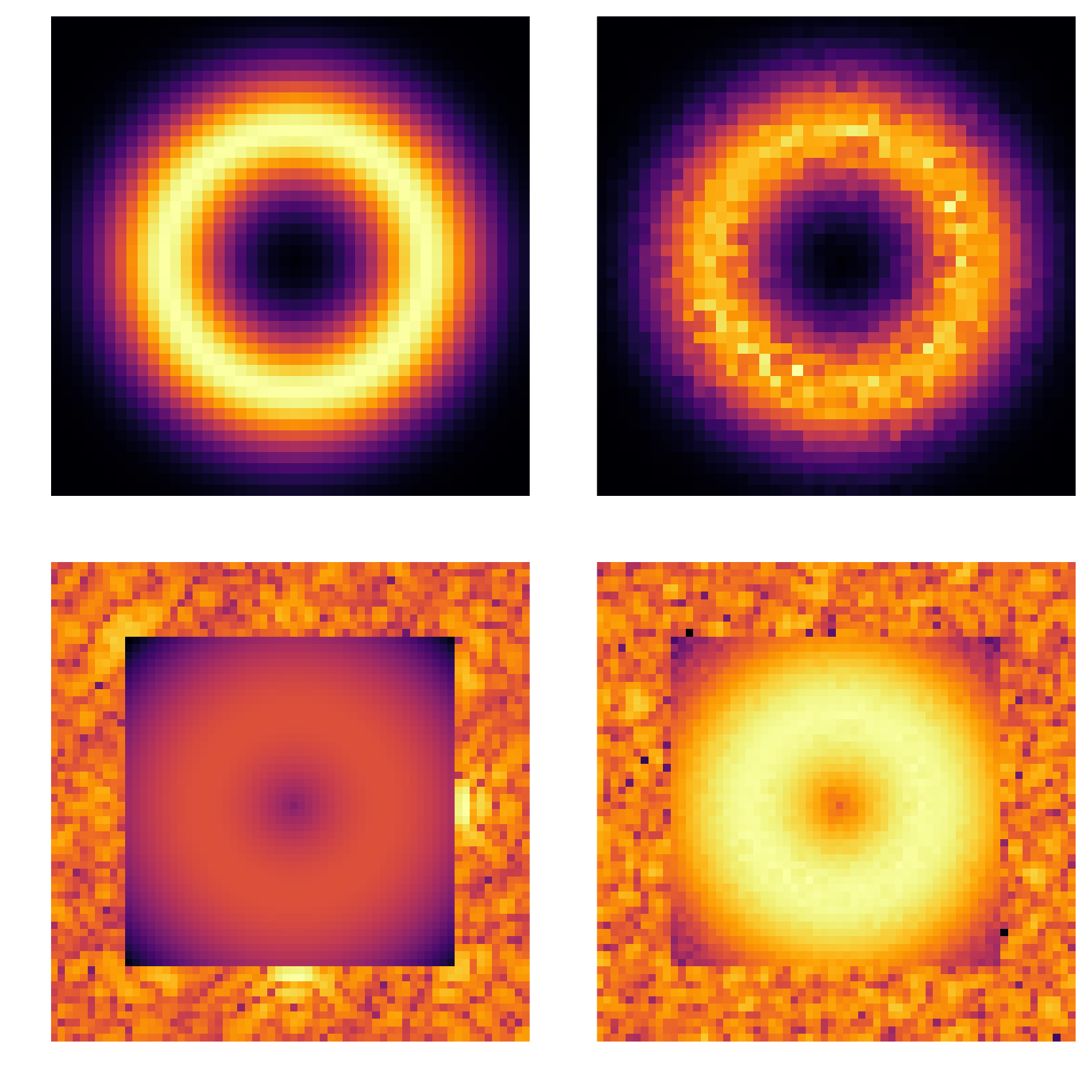}
\caption[MRAF solutions]{MRAF solutions. The top left is the m-value of 0.1, and the top right is 0.5 — both are cropped around the signal region. The bottom left and the bottom right are corresponding full images with the logarithm applied to make both regions visible.}
\label{fig:mraf-solution}
\end{figure}
As we saw in Subsection \ref{subsection:phase-vortices}, the phase vortices limit the accuracy of the produced intensity patterns. One can wonder if there is a way to obtain higher-accuracy intensity patterns, and what is the fundamental limit on the accuracy of reconstruction for an arbitrary phase hologram. As was shown in \cite{Pasienski:08}, we can create much higher accuracy intensity patterns if we are willing to sacrifice the efficiency of the laser power. In fact, there exists a natural trade-off between the accuracy and the efficiency of the hologram.

The idea proposed in the paper was to relax the number of constraints by dividing the target image into the signal region (SR) and the noise region (NR). The signal region is used to constrain the produced intensity, while in the noise region, we allow the intensity and the phase of the light to do whatever. More precisely, we modify the Gerchberg-Saxton constraints to be:
\begin{align}
    A &= \{f\in l^2([N]^2):|f|^2=g^2\} \\
    B &= \{f\in l^2([N]^2):|\mathcal{F}[f]|^2_{jk}=G^2_{jk}\quad\forall (j,k)\in SR\}
\end{align}
So, we just require the $SR$ subset of the output plane to match the desired intensity $G^2$. Now we have to modify the projection operators.
\begin{align}
    P_A(f)=g\frac{f}{|f|}  &&
    P_B(f)=\mathcal{F}^{-1}\left[m\left(G\frac{\mathcal{F}[f]}{|\mathcal{F}[f]|}\right)\Big|_{SR} + (1-m)\mathcal{F}[f]\Big|_{NR}\right]
\end{align}
where $(\cdot)\Big|_{SR}$ means that we only select elements of the array corresponding to the signal region. The constant $m\in[0,1]$ here is a so-called mixing parameter, and it controls how much light will be sent into the signal region, while $1-m$ is the fraction of light that will be sent into the noise region. We can sweep through the value of the parameter $m$ and record the efficiency and accuracy of the produced hologram. Efficiency here refers to the fraction of the intensity in the SR, and accuracy is 1 - $L_{int}$ but only inside the SR (see Appendix \ref{chapter:fourier-analysis} for details).

This experiment demonstrates a clear trade-off between the accuracy and the efficiency of the MRAF algorithm (and laser beam shaping in general). Low values of $m$ will lead to low efficiency but highly accurate solutions, while high values of $m$ will result in low accuracy but high efficiency. Notice also that MRAF still produces phase vortices during the optimization (see Figure \ref{fig:mraf-solution}). For the high values of $m$, these vortices exist in both SR and NR, while for the smaller values of $m$, they are pushed away (along with the majority of laser power) into the noise region.

\section{CFM and Conjugate Gradients}
\label{section:cfm-and-conjugate-gradients}

The overview of the approaches to the phase generation would be incomplete without the cost function minimization approach \cite{Harte:14}. We start by defining a cost function:
\begin{align}
\label{eq:loss-func}
    C(\phi)&=\sum_{n,k}\left(G_{n,m}^2-\tilde{G}_{n,m}^2\right)^2 \\ 
    \tilde{G}_{n,m}&\equiv\sum_{j,k}W_{n,j} g_{j,k}\exp(i\phi_{jk})W_{m,k}
\end{align}
where $g^2$ and $G^2$ are input and target intensities, respectively, and $W$ is a shifted DFT matrix defined in Appendix \ref{chapter:numerical-implementation}, eq. (\ref{eq:shifted-dft}). This loss function is similar to $L_\text{int}$ from Appendix \ref{chapter:numerical-implementation}, eq. (\ref{eq:Lint-def}) in the sense that it is agnostic to the phase of the beam in the output plane. The only difference is that this loss function uses the L2 norm between the intensities.\footnote{Technically L2 norm of the intensities will be the $\sqrt{C(\phi)}$, but we can use the fact that $\text{argmin}_\phi C(\phi)=\text{argmin}_\phi \sqrt{C(\phi)}$.} This is a common choice in a non-convex optimization, as it provides better convergence using gradient descent methods.

One can imagine performing backpropagation through the loss function (\ref{eq:loss-func}) while using $\phi$ as a parameter. In this approach, we will start with some initialization $\phi_0(x,y)$ and then do $N$ iterations of gradient descent to find the optimal phase.

We argue that the naive gradient descent on the loss function \ref{eq:loss-func} is suboptimal, as it suffers from the formation of phase vortices similarly to GS. The intuition for this is that this loss landscape is closely related to the non-convex geometry of the GS constraint sets \ref{eq:GS-constraint-sets}. More concretely, if we start at some unknown phase $\phi_0$, each gradient descent step picks the closest point on the constraint set $B$ (notice that constraint $A$ is always satisfied) and makes a small step towards this point, where the size of the step is controlled by the learning rate parameter. This is not any better than the projection $P_B$, which projects directly onto $B$ in a single step. Furthermore, the proximity in the phases $\phi$ does not necessarily imply the proximity in the loss because \ref{eq:loss-func} is extremely non-local due to the Fourier transform. Therefore, the naive gradient descent approach will usually perform worse than the GS algorithm.

The paper \cite{Harte:14} has two main advantages with respect to the naive gradient descent approach.
\begin{itemize}
    \item The conjugate gradient method, as opposed to classic gradient descent (or GS method), adjusts the optimization direction according to the history of the optimization steps during the optimization. After the first step, the conjugate gradient direction will be quite different from the GS projection direction.
    \item Augmenting the loss function with additional terms that penalize large localized deviations or actively enforce smoothing over the four nearest-neighbor pixels, which helps to reduce vortex formation for small parts of the target intensity.
    
\end{itemize}
With these two modifications, the loss landscape (and the descent procedure) of \cite{Harte:14} is very different from the GS algorithm or the naive backpropagation through the loss \ref{eq:loss-func}, so it is hard to assess optimality of this algorithm using the theoretical understanding above.

This method is still prone to vortex formation when the desired pattern is a large continuous intensity \cite{Harte:14}, which could be an indication of sub-optimality of the proposed optimization algorithm (at least in that specific case). Furthermore, the initialization step in  \cite{Harte:14} can be improved significantly using optimal transport, as we will see later in this chapter.

\section{Other Deep Learning Methods}
\label{section:other-deep-learning-methods}

\begin{figure}[!ht]
    \centering
    \includegraphics[width=0.60\columnwidth]{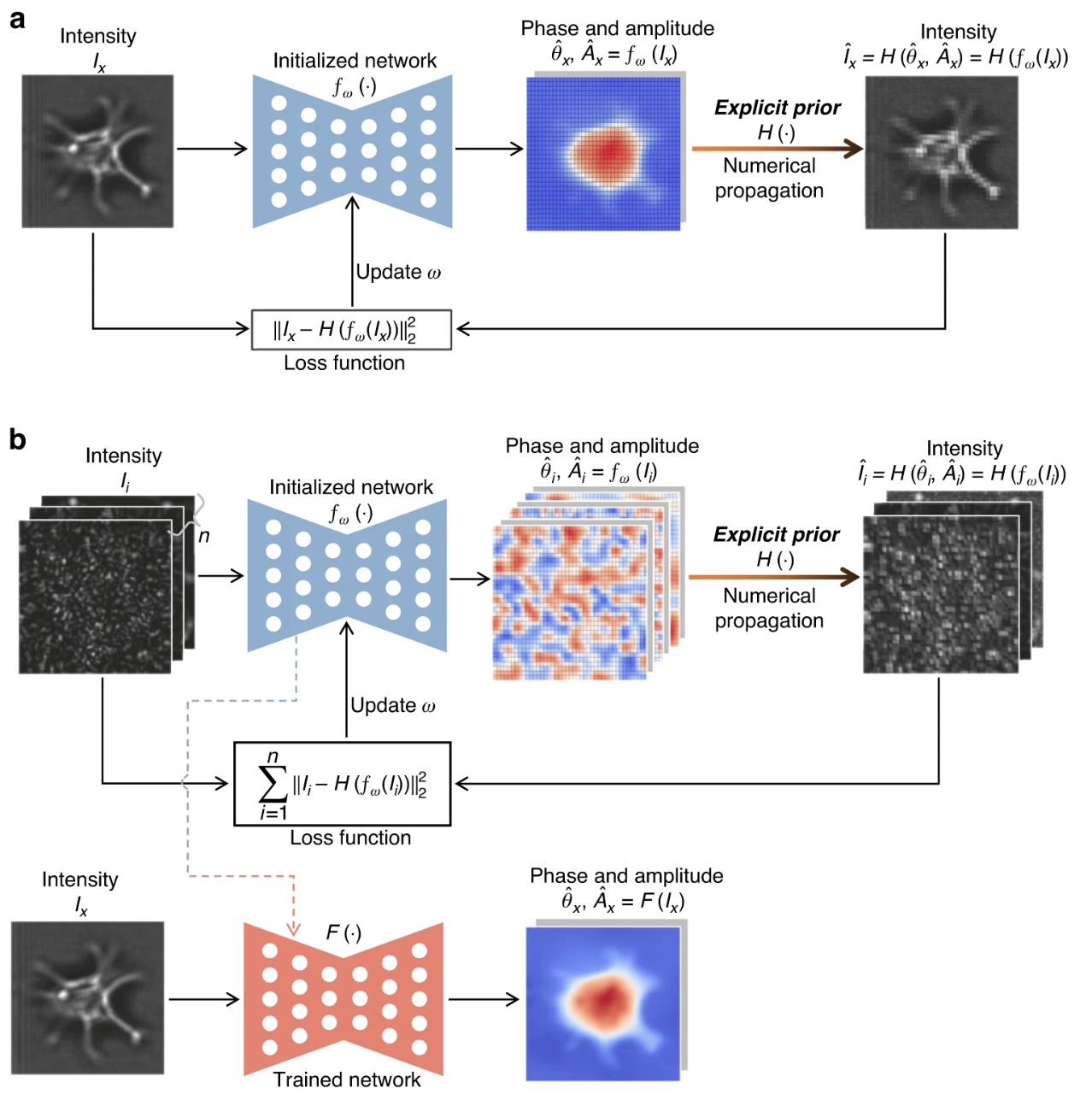}
    \caption[Overview of Deep Learning Based Approaches]{Overview of Deep Learning Based Approaches. 1a shows approach (1) and 1b shows approach (2). Figure taken from \cite{pr_survey}.}
    \label{fig:dl-overview}
\end{figure}

Of course, one should not be too quick to discard any neural network approach, since they are quite common in literature (see this review paper \cite{pr_survey}). At the end of the day, we are trying to predict an $n\times n$ phase image $\hat{\phi}$, given two $n\times n$ intensity images $g^2$ and $G^2$, so a convolutional neural network such as UNet \cite{physen_net} seems like a perfect candidate. In fact, the best deep learning solution, PhysenNet, leverages a UNet architecture with 4 downsampling layers, a bottleneck, and 4 upsampling layers \cite{physen_net}\footnote{Just to be clear, our OT solution outperforms this by an order of magnitude (could be a little less, depending on the exact choice of the distributions).}

Most neural networks can be placed into one of two main approaches: (1) an untrained, iterative scheme and (2) a data-driven, trained scheme \cite{physen_net}. In the approach (1), the neural network takes a single pair of intensities $G^2, g^2$ as the input, and outputs a phase prediction $\hat\phi$. The neural network is optimized to predict the best possible hologram $\hat\phi$. In the approach (2), the neural networks are trained on the paired dataset of inputs $(G^2_i, g_i^2)$ and corresponding target outputs $\phi_i$. The goal is to optimize a function that will map $(G^2_i, g_i^2)\to\phi_i$.

While approach (2) is the typical setup for most machine learning approaches, we are interested in approach (1), where the neural network acts as an iterative optimizer. This approach does not require a training dataset and always\footnote{One can argue that approach (1) is equivalent to overfitting on a single item in the dataset, which in deep learning literature is considered significantly easier then learning a a generalized map as in approach (2)} provides more accurate phase predictions because it iterates continuously over the same input target pair, while the trained network must predict the target phase with one forward pass.

One could argue that the neural network approach (1) will necessarily suffer from the same issues as the GS algorithm and the gradient descent methods presented above. To understand this mathematically, consider the schematic diagram \ref{fig:deep-learnin-approach}. We start with an initial set of parameters $\omega\in\mathbb{R}^M$, where $M$ could be an arbitrary large number. We use a neural network to map $\omega$ along with the fixed $G^2$ and $g^2$ into our phase estimate $\hat{\phi}(\omega)$, which is an $n\times n$ image. Finally, we compute the loss using equation \ref{eq:loss-func}, and propagate the gradients back to $\omega$.

\begin{figure}[!ht]
\begin{center}
\begin{tikzcd}[row sep=huge, column sep=huge]
\omega \arrow[r,"\text{NN}"] &
\tilde{\phi}(\omega) \arrow[r,"L_{\text{int}}"] &
l\!\bigl(\tilde{\phi}(\omega)\bigr)
\end{tikzcd}
\caption[General deep learning approach]{Deep learning approach when one starts from the unknown weights $\omega\in\mathbb{R}^M$ and outputs a phase guess $\tilde{\phi}\in\mathbb{R}^N$ and corresponding loss $l(\tilde{\phi}(\omega))$. This loss is later used to optimize the weights of the neural network. Throughout the training, the pair of intensities $g^2$ and $G^2$ remains fixed.}
\label{fig:deep-learnin-approach}
\end{center}
\end{figure}

The hope is that by choosing $M$ large enough, we can use high-dimensional optimization and the universal approximation theorem of neural networks to find the optimal phase. One could make the argument that by choosing $M\gg n^2$, it is possible to avoid local minima of the loss, but this turns out to be \textit{false}. 

The reason why this is not true is because the optimization procedure is \textit{bottlenecked} by the mapping $\hat\phi(\omega) \to l(\hat\phi(\omega))$, which is the right-hand side of diagram \ref{fig:deep-learnin-approach}. To see that, suppose you get stuck in the local minima $\phi_0$ of the loss function (e.g., randomly initialized weights $\omega$ are likely to output such $\phi_0$ at the first step of the optimization). Then, backpropagation will find a small step $\omega\mapsto\omega+\Delta\omega$, which will map into a new phase prediction $\hat\phi(\omega+\Delta \omega)$. This new phase output should be relatively close to the original prediction $\hat\phi(\omega)$ since the mapping $\omega\to\phi$ is close to continuous, but we know that the $\hat\phi(\omega)$ is the local minima of the loss $l$. Thus, any small perturbation will only increase the loss.

Therefore, we conclude that the neural network approach (1) is just as ill-posed as the GS problem (if not worse). Our extensive numerical experiments indeed confirm this result. Even with the very nice initialization, the neural network approach (1) is unable to outperform our best method, which we will present next. See our work with neural networks amended to the end of this thesis.

\section{Optimal Transport}
\label{section:optimal-transport}

Now, we present our state-of-the-art solution to the problem of phase generation \ref{prob:phase-generation}, using the theoretical connection to the problem of optimal transport that we developed in the previous chapter. We later generalize this to the problem of beam estimation \ref{prob:beam-estimation}. The optimal transport phase solutions have several important benefits and are complementary to many of the existing approaches:
\begin{itemize}
    \item \textbf{Best initialization}: our method solves once and for all the question of what phase one should start with as the initial input to their iterative algorithm. We argue that for smooth potentials, the answer should always be the optimal transport phase, which comes from the ray-optics limit of the problem.
    \item \textbf{Convex, unwrapped phase}: our method is guaranteed to return a smooth, convex, and unwrapped phase (not modded by $2\pi$), which will never contain any vortices. Furthermore, if we use this phase to seed any of the iterative algorithms such as GS or MRAF, we observe no vortex formation and an order of magnitude improvement in terms of the intensity loss.
    \item \textbf{Efficiency and accuracy}: unlike MRAF, Optimal Transport approach does not sacrifice any of the laser power. If one wishes to sacrifice $m$-fraction of the laser light to generate better accuracy holograms, it is very easy to ``plug in" our solution into MRAF or a similar algorithm with a signal and noise regions. According to our experiments, the combination of OT and MRAF gives us a state-of-the-art in terms of \textit{both} efficiency and accuracy.
    \item \textbf{Hyper parameters}: our method has a single hyperparameter $\epsilon$, which in general we will set to $10^{-3}$ or $10^{-4}$ depending on the intensities. So, it does not require any tuning. 
    \item \textbf{Physical interpretation}: it is elegantly linked to retrieving the ray-optics limit of the Wigner distribution, which could be very useful in its own right.
\end{itemize}

\begin{figure}[!ht]
\centering
\includegraphics[width=0.70\columnwidth]{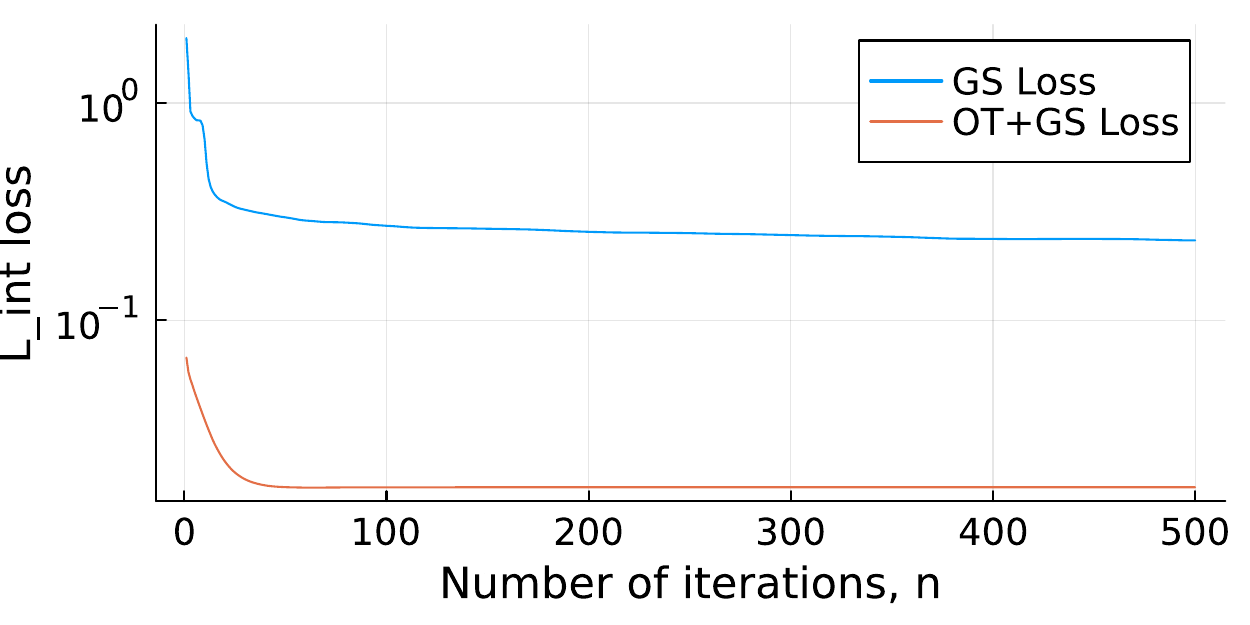}
\caption[GS vs OT loss comparison]{The optimization loss of the GS algorithm if we start from iFFT phase (blue line), or the optimal transport phase (red line). The optimal transport (unrefined) solution is the value of the red curve at $n=0$, which is already strictly better than the naive GS solution}
\label{fig:GSOT-convergence}
\end{figure}

\begin{figure}[!ht]
\centering
\includegraphics[width=0.65\columnwidth]{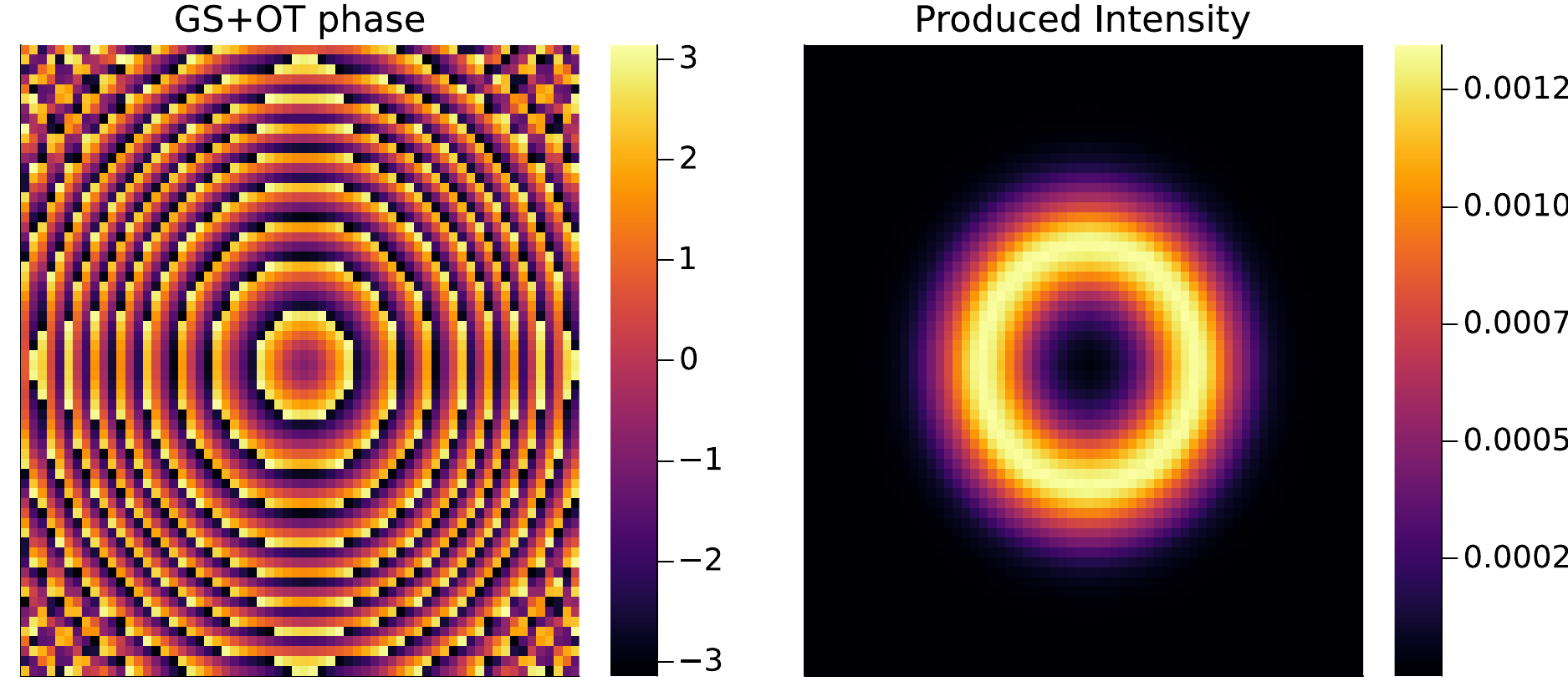}
\caption[GS+OT solution to phase generation]{Final phase, produced from OT initialization and GS refinement (left) and the corresponding target intensity image (right). We see that GS+OT method yields no visible vertices in the region where the input intensity distribution $g^2$ has a significant support.}
\label{fig:GSOT-phase}
\end{figure}

Thus, our state-of-the-art solution\footnote{We open-source all of our algorithmic work at \cite{SLMTools}} is very simple — extract an initial phase guess $\phi_0$ using optimal transport, which is a well-posed convex problem, and then plug it into GS, MRAF, or a similar algorithm for a slight refinement.

\begin{figure}[!ht]
    \centering
    \includegraphics[width=0.9\textwidth]{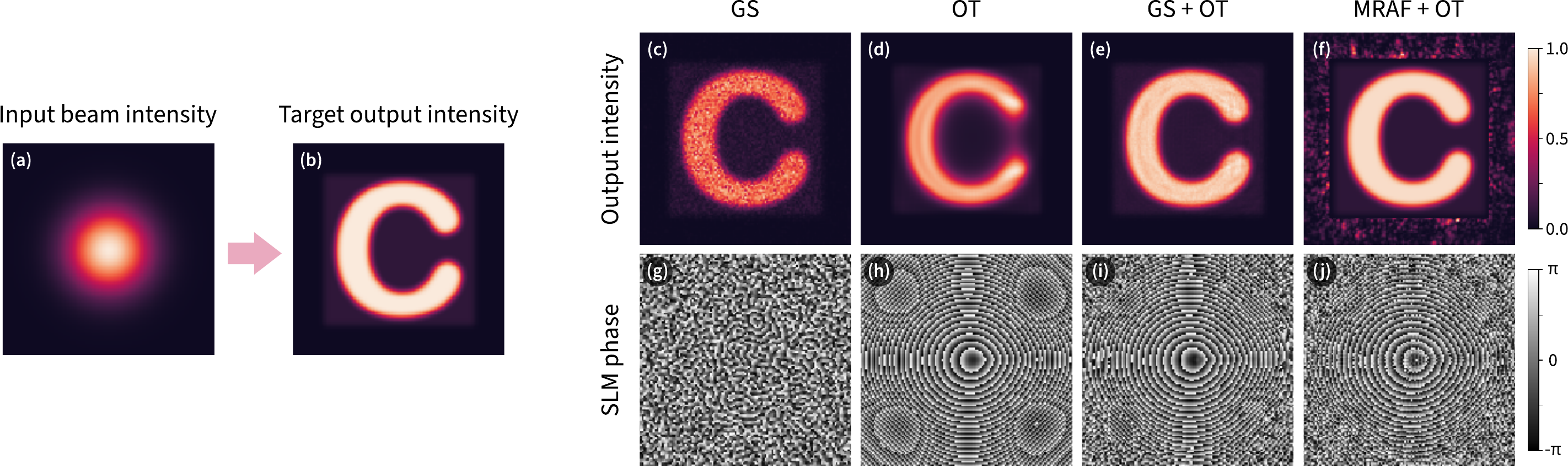}
    \caption[Comparison of phases and output beams from various phase generation algorithms]{Comparison of phases and output beams from various phase generation algorithms.  All images are $128 \times 128$ pixels.  (a) is the input beam intensity.  (b) is the target output beam intensity. (c-f) are output intensities realized by the phase displayed immediately below. (g) is from GS initialized with a random phase, with RMS error $\epsilon = 13.9\%$ and efficiency $\eta = 99.13\%$. (h) is from OT; $\epsilon = 14.3\%$, $\eta = 99.96\%$. (i) is from GS initialized by OT; $\epsilon =2.58\%$, $\eta = 99.91\%$. (j) is MRAF initialized by OT; $\epsilon = 5.95\times10^{-16}$, $\eta = 85.15\%$.  All iterative algorithms were run for 10,000 iterations.  The MRAF hyperparameter was set by hand to 0.48. A centered $96\times 96$ pixel box was used as the MRAF signal region and the region for computing all efficiencies $\eta$.}
    \label{fig:pg-comparison}
\end{figure}

To motivate the reader further, we show that our method drastically improves the quality of the hologram (see Figures \ref{fig:GSOT-convergence} and \ref{fig:GSOT-phase}) for a simple numerical experiment that we have started in Subsection \ref{subsection:phase-vortices} of this chapter. Also, take a look at the Figure \ref{fig:pg-comparison} from our paper \cite{swan2024highfidelityholographicbeamshaping}. We refer the reader to the supplementary materials of \cite{swan2024highfidelityholographicbeamshaping} for a comprehensive test of this method on a wide range of input/output intensities, and a quantitative proof of its advantages on various metrics.

\subsection{Memory Bottleneck}
\label{subsection:memory-bottleneck}

To obtain the optimal transport solution, we do the following. We start by defining a natural lattice $L=(L^{(x)},L^{(y)})$ that lives on $n\times n$ points (see Section \ref{section:coordinate-lattices} of Appendix \ref{chapter:numerical-implementation} for more explanations regarding lattices), and discretize the input and the target intensities $g^2$ and $G^2$ using $L$\footnote{We can easily use different lattices for input and output distributions, which works perfectly fine with the proposed algorithm. The only reason why we use the same lattice is for exposition clarity.}. This gives us discretized arrays, which we call $g^2_{jk}$ and $G^2_{JK}$ (capital index will correspond to the output plane). 

Next, we define a cost matrix $C_{jkJK}$, which will encode the quadratic cost of transportation needed for the theoretical link between Monge-Ampere PDE and Optimal Transport \cite{de2014monge}.

\begin{equation}
\label{eq:c-4-tensor}
C_{jkJK}= \left(L^{(x)}_{j} - L^{(x)}_{J}\right)^2 + \left(L^{(y)}_k-L^{(y)}_K\right)^2,
\end{equation}

Notice that in general $C_{jkJK}$ is a 4-tensor that has $n^4$ real-valued (non-zero) entries. Storing such an object in memory requires gigabytes of RAM for $n>10^3$, which was one of the biggest bottlenecks of the algorithm that my collaborators and I proposed in \cite{swan2024highfidelityholographicbeamshaping}. In this work, we reduce this $\mathcal{O}(n^4)$ memory constraint to $\mathcal{O}(n^2)$, allowing for a quick and efficient computation of holograms with 1024x1024 pixels (and possibly more).

Once discretized objects $g^2, G^2$, and $C$ are defined, we can use pretty much any optimal transport solver such \cite{POT} or \cite{OptimalTransport.jl} to find the transport plan, and the corresponding phase using equations \ref{eq:retrieve-gamma} and \ref{eq:phase-from-monge}. Our implementation is \textit{open-source} and can be found in \cite{SLMTools}. 

The rest of this subsection will be focused on how to use the structure of the problem to reduce the memory constraint, while the nitty-gritty details of the algorithm can be found in my previous paper \cite{swan2024highfidelityholographicbeamshaping} and the corresponding supplementary materials. 

I wrote this section presenting the material in the way that I best understand it, but a more methodological introduction to these topics can be found in the book \cite{peyré2020computationaloptimaltransport}.

\subsection{Discrete Optimal Transport}
\label{subsection:discrete-optimal-transport}

In order to relax the memory constraint, we need to carefully understand the optimization problem at hand. So, we start from the rigorous formulation of the discrete optimal transport problem.

Suppose we are given discretized probability distributions $\mu \in \mathbb{R}^N$ and $\nu \in \mathbb{R}^N$. One can think of these as ``flattened" versions of $g^2_{jk}$ and $G^2_{JK}$ distributions, so $N=n^2$. In practice, we will keep $g^2_{jk}$ and $G^2_{JK}$ as $n\times n$ matrices, because it will be useful for our final algorithm. By definition, we assume that $\mu_i\geq0$, $\nu_i\geq0$ and $\sum \mu_i=1$ and $\sum \nu_i=1$. We are also given a positive cost matrix $C\in \mathbb{R}^{N\times N}$, such that $C_{ij}\geq0$. 

We can formulate the Kantorovich relaxation of the optimal transport as the following optimization problem\footnote{Notice that this is a natural discretization of the continuous Kantorovich relaxation of optimal transport described in Chapter 2.}:
\begin{align}
\min_{\Gamma} \quad & \sum_{i,j}C_{ij}\Gamma_{ij}\\
\textrm{s.t.} \quad & \sum_j\Gamma_{ij}=\mu_i \nonumber \quad\sum_i\Gamma_{ij}=\nu_j \nonumber \quad \Gamma_{ij} \geq 0
\end{align}
where $\Gamma\in\mathbb{R}^{N\times N}$ is the optimization variable. Here, the optimization function, equality, and inequality constraints are all affine, so the problem is convex. Specifically, it is a linear program (LP) \cite{boyd2004convex}. So, we can use any convex optimization software to find $\Gamma$. The only issue is that $N=n^2$ here could be rather large, which makes it very difficult to store the transport plan $\Gamma$ and the cost matrix $C$ in memory, since both have $N^2=n^4$ coefficients.

\subsection{Change of Basis?}
\label{subsection:change-of-basis}

Looking at the optimization problem (\ref{eq:primal-OT}), it seems hopeless to avoid storing the entire $\Gamma$ in memory during optimization, at least that is what we all thought for a while.

A breakthrough discussion happened when Jason Hogan suggested changing the basis in the problem \ref{eq:primal-OT} from the pixel coordinates to a better-suited basis, such as the Hermite-Gaussian basis or Laguerre basis. Together, we worked out a theorem \ref{theorem:basis-change} that shows that this is indeed possible. The only downside was the fact that the optimal transport in the new basis acquired additional $\mathcal{O}(N^2)$ constraints, which made the optimization very slow.

This line of reasoning inspired several important threads, such as considering the dual formulation of the optimal transport problem and the famous entropic relaxation, which uses the Sinkhorn-Knopp algorithm. These threads were extremely useful as they eventually allowed us to reduce the memory constraint from $\mathcal{O}(n^4)$ to $\mathcal{O}(n^2)$.

\subsection{Reformulations of Optimal Transport}
\label{subsection:reformulations-of-optimal-transport}

\subsubsection{Kantorovich Dual OT}
There are several optimization problems closely related to \ref{eq:primal-OT}. First, and foremost, one can try to solve the Kantorovich dual problem \ref{eq:dual-OT} (see Appendix \ref{chapter:primer-on-convex-optimization} for the detailed explanation):
\begin{align}
\max_{\phi\in\mathbb{R}^N,\;\psi\in\mathbb{R}^N}\quad
	&\sum_{i}\phi_i\mu_i \;+\;\sum_{j}\psi_j\nu_j \\[2pt]
\text{s.t.}\quad
	&\phi_i+\psi_j \;\le\; C_{ij},
	\qquad\forall i,j.\nonumber
\end{align}
where we are now optimizing over the dual variables $\phi,\psi\in\mathbb{R}^N$. This approach is nice because we do not have to store the transport plan $\Gamma\in \mathbb{R}^{N\times N}$ at any point of the optimization. Unfortunately, we have to enforce additional $N^2$ constraints of the form $\phi_i+\psi_j \;\le\; C_{ij}$, which would slow down the algorithm considerably. It turns out that we can do much better.

\subsubsection{Entropy Regularized OT}
In entropic relaxation of optimal transport (proposed in \cite{cuturi}), there is an additional term $\epsilon\Omega(\Gamma)$ added to the loss function where $\Omega(\Gamma)=\sum_{ij}\Gamma_{ij}\log(\Gamma_{ij})$ is the negative entropy of the transport plan. Notice that entropy has an implicit constraint of $\Gamma_{ij}\geq0$ because of the logarithm. Thus, the problem becomes:
\begin{align}
\label{eq:entropic-primal}
\min_{\Gamma} \quad & \sum_{i,j}C_{ij}\Gamma_{ij} +\epsilon \sum_{ij}\Gamma_{ij}\log(\Gamma_{ij})\\
\textrm{s.t.} \quad & \sum_j\Gamma_{ij}=\mu_i \nonumber \quad\sum_i\Gamma_{ij}=\nu_j \nonumber 
\end{align}
where $\epsilon$ is the regularization parameter, controlling how much smearing of the transport plan we can allow. If $\epsilon\to0$, we recover the original unregularized optimal transport problem. This is the only hyperparameter that one needs to choose in our algorithm, and we will usually set it to $10^{-3}$. In practice, the entropy regularized optimal transport works better with the noisy data as it provides a more stable convergence. In particular, our paper \cite{swan2024highfidelityholographicbeamshaping} uses entropic regularization for all of our experiments. 

Of course, the entropic regularization of optimal transport does not solve our memory issue, as it still uses $\Gamma \in \mathbb{R}^{N\times N}$ as the optimization variable. Let us consider the dual problem:

\begin{align}
\label{eq:entropic-dual}
\max_{\phi,\psi} \quad & \sum_i\phi_i\mu_i + \sum_i\psi_i\nu_i -\sum_{i,j}\epsilon \exp((\phi_i+\psi_j - \epsilon -C_{i,j})/\epsilon)
\end{align}

Notice that all of the constraints (implicit and explicit) have been removed in the process of computing the dual. This is due to a very subtle reason that entropic regularization makes the constraint $\Gamma_{ij}$ implicit, i.e. it has a smaller \textit{feasibility domain} (see definition \ref{eq:feasibility-domain} in Appendix \ref{chapter:primer-on-convex-optimization})

We are left with the optimization problem that uses two $\mathcal{O}(N)=\mathcal{O}(n^2)$ variables, as opposed to a single $\mathcal{O}(N^2)=\mathcal{O}(n^4)$ variable $\Gamma$, and a smooth, convex, differentiable loss function that we need to maximize. This is pretty much as good as it gets in terms of optimization problems, but once again, we can do even better by leveraging a connection to the following mathematical result. 

\subsection{Sinkhorn theorem}
\label{subsection:sinkhorn-theorem}

\begin{definition}
    We call an $n \times n$ positive matrix $A$ \textit{doubly stochastic} if its rows and columns add up to 1, i.e., $\sum_i A_{ij}=1$ and $\sum_j A_{ij}=1$
\end{definition}
\begin{theorem}
\label{theo:sinkhorn}
    \textbf{(Sinkhorn)} If $A$ is an $n \times n$ matrix with strictly positive elements, then there exist diagonal matrices $D_1$ and $D_2$ with strictly positive diagonal elements such that $D_1 A D_2$ is doubly stochastic. The matrices $D_1$ and $D_2$ are unique modulo multiplying the first matrix by a positive number and dividing the second one by the same number.
\end{theorem}
This result was proven by Sinkhorn in 1972 \cite{10.1214/aoms/1177703591}. To apply this theorem to our problem, we need a slightly more general setting, where the marginal sums of the $A$ matrix are matching strictly positive vectors $\mu$ and $\nu$, instead of being constant 1. This is known as the generalized Sinkhorn theorem.
\begin{theorem}
\label{theo:sinkhorn-general}\textbf{(Generalized Sinkhorn Theorem)}
 Let $\mu\in \mathbb{R}^N$ and $\nu\in \mathbb{R}^N$ be any strictly positive vectors such that $\sum\mu_i=1$ and  $\sum\nu_i=1$. Then, there exist diagonal matrices $D_1$ and $D_2$ with strictly positive diagonal elements such that $D_1 A D_2$ has marginals of $\mu$ and $\nu$, i.e. $\sum_j(D_1 A D_2)_{ij}=\mu_i$ and $\sum_i(D_1 A D_2)_{ij}=\nu_j$
\end{theorem}
The proof of this results follows from a result proven by Tverberg in 1976 \cite{TVERBERG1976674} (see also the theorem 4.5 in a more modern review of the subject \cite{idel2016reviewmatrixscalingsinkhorns}).

The theorem \ref{theo:sinkhorn-general} guarantees existence and uniqueness (up to a trivial rescaling) to the problem of matrix renormalization with known marginal constraints. As we will show, this problem is equivalent to solving the entropy regularized optimal transport problem \ref{eq:entropic-primal} (or equivalently its dual \ref{eq:entropic-dual}). Furthermore, this reformulation leads to a simple iterative algorithm, which is currently the state-of-the-art solution in terms of time complexity for the entropic regularized OT \cite{cuturi, complexitySinkhorn}.

\subsection{Sinkhorn-Knopp Algorithm}
\label{subsection:sinkhorn-knopp algorithm}

Suppose you are given a positive matrix $A\in\mathbb{R}^{N\times N}$ and two positive marginal constraints $\mu, \nu \in \mathbb{R}^N$. By the theorem \ref{theo:sinkhorn-general} there exist unique (up to rescaling) positive diagonal matrices $D_1$ and $D_2$ such that the matrix $D_1AD_2$ has the desired marginals $\mu$ and $\nu$. But how do we find these matrices?

The answer is once again the method of iterated projections (\ref{alg:mip}). Define the space of $\Omega=\mathbb{R}_{+}^N\times\mathbb{R}_{+}^N$, where any element $(u,v)\in\Omega$ corresponds to the pair of positive diagonal matrices:
\begin{align}
    D_1 = \text{diag}(u_1,u_2,\dots,u_N) && D_2=\text{diag}(v_1,v_2,\dots,v_N)
\end{align}
We define the following constraint sets:
\begin{align}
    C_1&=\{(u, v)\in\Omega: \sum_j u_iA_{ij}v_j=\mu_i\}\\
    C_2&=\{(u, v)\in\Omega: \sum_iu_iA_{ij}v_j=\nu_j\}
\end{align}
Now we need to define projections $P_{C_1}:\Omega\to C_1$ and $P_{C_2}:\Omega\to C_2$ onto these constraint sets:
\begin{align}
    P_{C_1}((u,v))=(u',v) \quad \text{where}\quad u'_i\equiv u_i\mu_i/\sum_j u_iA_{ij}v_j\\
    P_{C_2}((u,v))=(u,v') \quad \text{where}\quad v'_j\equiv v_j\nu_j/\sum_i u_iA_{ij}v_j
\end{align}
By initializing $(u,v)$ to constant ones, and iteratively applying $P_{C_1}$ and $P_{C_2}$ we obtain the famous Sinkhorn-Knopp algorithm:

\begin{algorithm}[!ht]
\caption{Sinkhorn-Knopp algorithm}
\label{alg:sinkhorn-knopp}
\begin{algorithmic}[1]
\State $u \gets \mathds{1}_N$ \Comment{Initialize scaling vectors}
\State $v \gets \mathds{1}_N$
\State $D_1 \gets \mathrm{diag}(u)$
\State $D_2 \gets \mathrm{diag}(v)$
\For{$i = 1$ \textbf{to} max\_iter}
    \State $v \gets \left( v^\top \oslash \sum(D_1 A D_2, \text{axis}=1) \right)^\top$ \Comment{Normalize columns}
    \State $v \gets v \odot \nu$ \Comment{Enforce marginal $\nu$}
    \State $D_2 \gets \mathrm{diag}(v)$
    \State $u \gets u \oslash \sum(D_1 A D_2, \text{axis}=2)$ \Comment{Normalize rows}
    \State $u \gets u \odot \mu$ \Comment{Enforce marginal $\mu$}
    \State $D_1 \gets \mathrm{diag}(u)$
\EndFor
\State \Return $u, v$
\end{algorithmic}
\end{algorithm}
\noindent where $\odot$ and $\oslash$ are the element-wise multiplication and division, respectively. It might seem like we arrived once again at the GS-like iteration algorithm, which is a bit ironic. However, there is one very important difference between the Sinkhorn-Knopp (SK) algorithm and the Gerchberg-Saxton (GS) algorithm: the geometry of the optimization landscape. The SK algorithm optimizes between two closed, convex, intersecting sets $C_1$ and $C_2$ (intersecting due to the Sinkhorn theorem \ref{theo:sinkhorn}), while the GS algorithm works with two closed, nonconvex, possibly nonintersecting subsets $A$ and $B$ \ref{eq:GS-constraint-sets}. Think of this distinction as the left-hand side and the right-hand side of Figure \ref{fig:conv-iterated}. The nice geometry of the constraint sets in the SK algorithm is exactly what guarantees optimal convergence properties \cite{peyré2020computationaloptimaltransport}.

In the next subsection, we show how to reduce the entropy regularized optimal transport problem to the matrix renormalization problem, which we can efficiently solve using the Sinkhorn-Knopp algorithm above.

\subsection{Efficient OT Solution}

Investigating carefully our derivation of the entropic dual \ref{eq:sinkhorn-OT-dual} in Appendix \ref{chapter:primer-on-convex-optimization}, we notice that minimizing the Lagrangian of the problem results in the equation, which directly links the optimal transport plan $\Gamma$ to the dual variables $\psi,\phi$. 
\begin{equation}
    \Gamma_{i,j} = \exp((\phi_i+\psi_j - \epsilon -C_{i,j})/\epsilon)\quad \forall i,j
\end{equation}
Normally, this equation is used to eliminate the primal variable $\Gamma$ and rewrite the optimization problem in terms of the dual variables $\phi$ and $\psi$, but here we will use it map our problem to the setting of the generalized Sinkhorn theorem \ref{theo:sinkhorn-general}. Rewriting this condition just slightly, we obtain:
\begin{equation}
    \Gamma_{i,j} = \exp(\phi_i/\epsilon- 1/2)
    \exp( -C_{i,j}/\epsilon)\exp(\psi_j/\epsilon- 1/2)
\end{equation}
Now, let's make the following substitutions:
\begin{align}
    A_{ij}&=\exp(-C_{i,j}/\epsilon)\\
    D_1&=\text{diag}(\exp(\phi_1/\epsilon-1/2),\dots,\exp(\phi_n/\epsilon-1/2)) \label{eq:d1tophi}\\
    D_2&=\text{diag}(\exp(\psi_1/\epsilon-1/2),\dots,\exp(\psi_n/\epsilon-1/2)) \label{eq:d2topsi}
\end{align}
from which we obtain (in matrix notation):
\begin{equation}
\label{eq:gamma-to-dad}
    \Gamma = D_1AD_2
\end{equation}
Notice that the transport plan $\Gamma$ always has to satisfy marginal constraints $\mu$ and $\nu$. Furthermore, according to our derivation in Appendix \ref{chapter:primer-on-convex-optimization} the \textit{optimal} transport plan can be factored into $D_1AD_2$ via equation \ref{eq:gamma-to-dad}, where $D_1$ and $D_2$ are related to the \textit{optimal} dual variables $\phi$ and $\nu$ via equations \ref{eq:d1tophi} and \ref{eq:d2topsi}. Furthermore, this decomposition is unique (up to trivial rescaling), according to the generalized Sinkhorn theorem \ref{theo:sinkhorn-general}. Therefore, we can use Sinkhorn-Knopp theorem to find $D_1$ and $D_2$, thereby solving the entropy regularized optimal transport problem.

Notice that if one so desires, it is possible to retrieve the transport plan $\Gamma$ using:
\begin{equation}
    \Gamma=D_1AD_2 \quad \text{where} \quad D_1=\mathrm{diag}(u) \quad \text{and}\quad D_2=\text{diag(v)}
\end{equation}
But this is not required for the sake of optimization. In fact, the algorithm above naturally is $\mathcal{O}(N)$ in memory, as opposed to the primal problem, which is $\mathcal{O}(N^2)$. In the next section, we will show that we can generalize our entire phase generation pipeline without ever computing the memory-heavy transport plan $\Gamma$.

\begin{figure}[!ht]
\centering
\includegraphics[width=0.5\columnwidth]{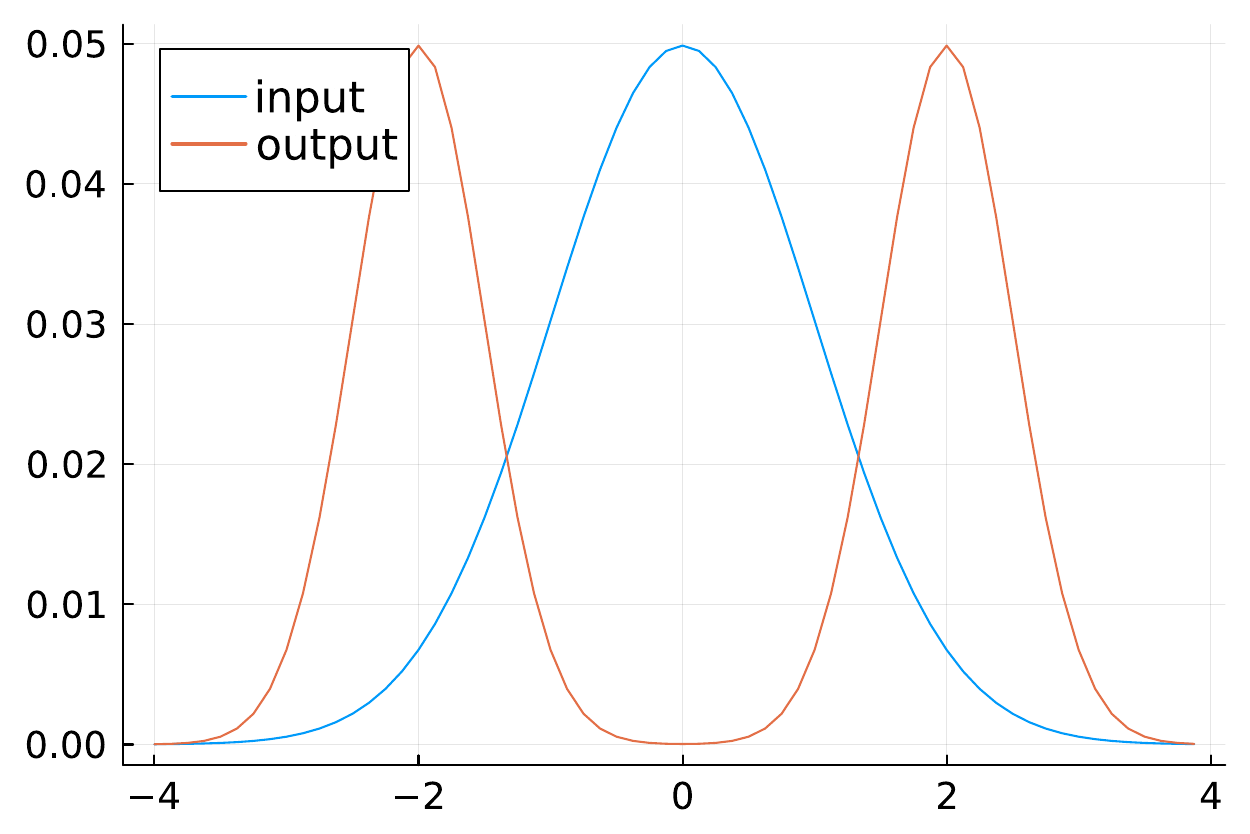}
\caption{Input and output distributions for the experiment with the Sinkhorn-Knopp algorithm.}
\label{fig:SK-input-output}
\end{figure}

Let's demonstrate the working of the algorithm on a quick numerical example. We consider two distributions $\mu, \nu \in \mathbb{R}^{64}$ that live on a natural lattice $L$ with $64$ points. $\mu$ is a Gaussian with a standard deviation of $1$, while $\nu$ is a sum of two Gaussians with standard deviations of $1/2$ centered at $-2$ and $2$, respectively (see Figure \ref{fig:SK-input-output}).

\begin{figure}[!ht]
\centering
\includegraphics[width=\columnwidth]{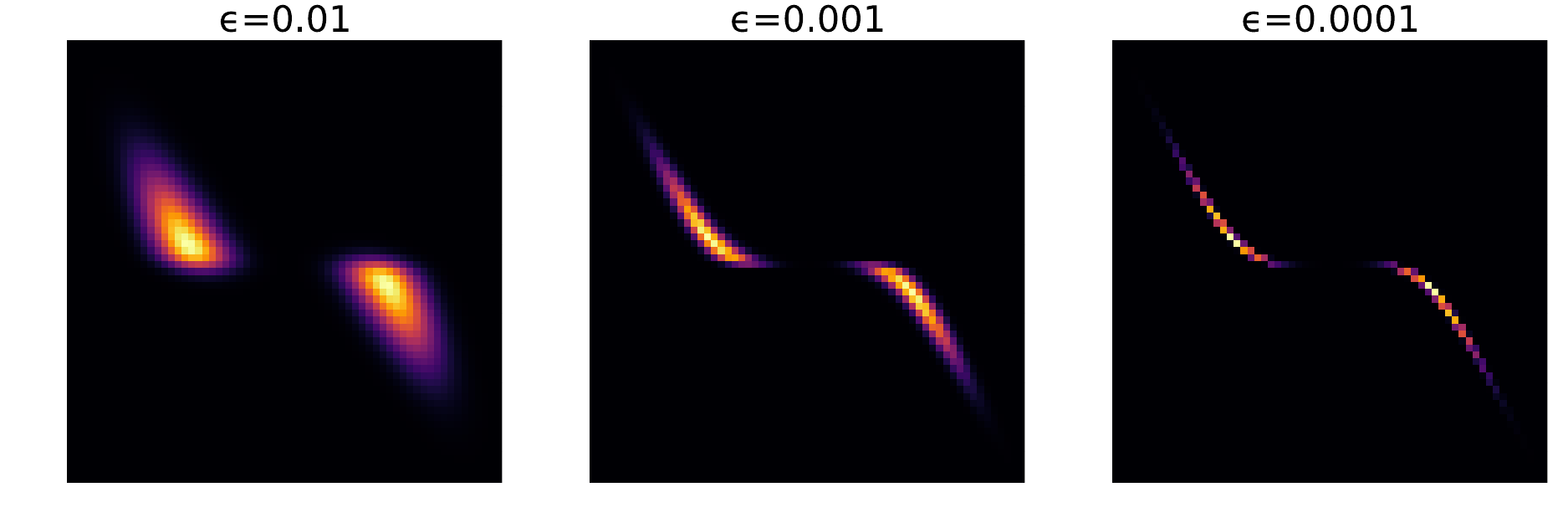}
\caption[Optimal transport using the Sinkhorn-Knopp algorithm for various values of $\epsilon$]{Optimal transport using the Sinkhorn-Knopp algorithm for various values of $\epsilon$. The input is the left marginal, and the output is the top marginal. We can see that in the limit of $\epsilon\to0$, we recover the unregularized optimal transport plan}
\label{fig:SK-epsilon-sweep}
\end{figure}

We can see in Figure \ref{fig:SK-epsilon-sweep} that decreasing the parameter $\epsilon$ in Algorithm \ref{alg:sinkhorn-knopp} we obtain a thinner, less entropic, transport plan. In the limit of $\epsilon\to0$, we recover the solution to the unregularized optimal transport problem \cite{cuturi}.

\subsection{Efficient 2D Implementation}
\label{subsection:efficient-2d-implementation}

Now, let's generalize the algorithm above to 2D distributions. Now, the dual variables $\phi$ and $\psi$ will be $n\times n$ matrices. The cost $C$ and the positive matrix $A$ now become a rank-4 tensor $n\times n\times n\times n$ of real coefficients, which we defined in \ref{eq:c-4-tensor}. To simplify the notation, we will  define $X_j\equiv L^{(x)}_{j}$ and $Y_k\equiv L^{(y)}_{k}$, and assume the natural lattice with even $n$, i.e. both $j$ and $k$ go over the set of integers $\{-n/2,\dots,n/2-1\}$, which correspond to the range of $[-\frac{\sqrt{n}}{2},\frac{\sqrt{n}}{2}-\frac{1}{\sqrt{n}}]$ units. We can rewrite $C$ and $A$ tensors as:
\begin{align}
\label{eq:C-A-def}
C_{jkJK}&= \left(X_{j} - X_{J}\right)^2 + \left(Y_k-Y_K\right)^2\\
A_{jkJK}&= \exp(-C_{jkJK}/\epsilon)\\
&=\exp{\left(-\left(X_j-X_J\right)^2/\epsilon\right)}\exp{\left(-\left(Y_k-Y_K\right)^2/\epsilon\right)}
\end{align}
Notably, storing the full tensors for $C$ and $A$ in memory is not practical, because it will result in a $\mathcal{O}(n^4)$ memory algorithm. The key is to recognize that $C$ and $A$ tensors have a lot of symmetries, which allows us to encode them in a memory efficient way. Looking at \ref{eq:C-A-def}, we can see that both $C$ and $A$ are shift-invariant. 
\begin{align}
    &C_{j+q,k+r,J+q,K+r}=C_{j,k,J,K} \\
    &A_{j+q,k+r,J+q,K+r}=A_{j,k,J,K} \quad \forall{q,r\in\mathbb{Z}}
\end{align}
So, instead of storing the original 4D tensor $A_{j,k,J,K}$, we can just set last two indices to $0$, and store the remaining 2D matrix:
\begin{equation}
K_{j,k}\equiv A_{j,k,0,0}=\exp{\left(-\left(X_j^2+Y_k^2\right)/\epsilon\right)}
\end{equation}
So, $K$ matrix is a discretized centered  Gaussian matrix with the width controlled by the parameter $\epsilon$. Thus, if we fix the last two indecies of $J$ and $K$ of the $A$ tensor, we can interpret it as a shifted $K$ matrix, i.e. 
\begin{equation}
    A_{j,k,J,K}=A_{j-J,k-K,0,0}=K_{j-J,k-K}
\end{equation}

Next, we will show that there is an elegant way of expressing steps (7) and (10) in algorithm \ref{alg:2d-sinkhorn-knopp} using a convolution with a $K$ matrix. Then, we will compute the final phase gradients using a similar convolution trick.

Consider the marginal sums of the transport plan $\Gamma =D_1A D_2$ :
\begin{align}    \sum_{J,K}u_{j,k}A_{j,k,J,K}v_{J,K}&=u_{j,k}\sum_{J,K}A_{j,k,J,K}v_{J,K}\\
    &=u_{j,k}\sum_{J,K}K_{j-J,k-K}v_{J,K}\\
    &=u_{j,k}(K\circledast v)_{j,k}
\end{align}
where $\circledast$ is a discrete convolution operation. Similarly for the other marginal:
\begin{align}
\sum_{j,k}u_{j,k}A_{j,k,J,K}v_{J,K}&=v_{J,K}\sum_{j,k}u_{j,k}A_{j,k,J,K}\\
    &=v_{J,K}\sum_{j,k}u_{j,k}K_{j-J,k-K}\\
    &=v_{J,K}(u\circledast K)_{J,K}
\end{align}  
This naturally leads to the following modification of the Sinkhorn-Knopp algorithm:

\begin{algorithm}[!ht]
\caption{2D Sinkhorn-Knopp algorithm}
\label{alg:2d-sinkhorn-knopp}
\begin{algorithmic}[1]
\State $K_{jk} \gets \exp(-(X_j^2+Y_k^2)/\epsilon)$ \Comment{Form a convolution kernel}
\State $u \gets \mathds{1}_{N\times N}$ \Comment{Initialize scaling vectors}
\State $v \gets \mathds{1}_{N\times N}$
\For{$i = 1$ \textbf{to} max\_iter}
    \State $v'\gets u\odot (K\circledast v)$
    \State $v \gets v \oslash v'$ \Comment{Normalize columns}
    \State $v \gets v \odot \nu$ \Comment{Enforce marginal $\nu$}
    \State $u'\gets (u\circledast K)\odot v$
    \State $u \gets u \oslash u'$ \Comment{Normalize rows}
    \State $u \gets u \odot \mu$ \Comment{Enforce marginal $\mu$}
\EndFor
\State \Return $u, v$
\end{algorithmic}
\end{algorithm}
\noindent where once again $\odot$ is the element-wise multiplication, $\oslash$ is the element-wise division, and $\circledast$ is the convolution operation. Next, we show that we can use the convolution trick to efficiently compute the phase gradients. Recall, the gradient of the phase is related to the fist moments of $\Gamma$, which can be derived from (\ref{eq:retrieve-gamma}).\footnote{by discretizing the integral with a lattice, and generalizing the problem to 2 dimensions.}

\begin{align}
    \left(\frac{\partial \phi}{\partial x}\right)_{jk}&=\frac{1}{\mu_{jk}}\left(\sum_{JK} \Gamma_{jkJK}X_J\right) \\
    \left(\frac{\partial \phi}{\partial y}\right)_{jk}&=\frac{1}{\mu_{jk}}\left(\sum_{JK} \Gamma_{jkJK}Y_K\right)
\end{align}
Recall that $\Gamma=D_1 AD_2$, so we can write:
\begin{align}
\sum_{JK} \Gamma_{jkJK}X_J &= \sum_{JK}u_{jk} A_{jkJK}v_{JK}X_J\\
&=u_{jk}\sum_{JK} K_{j-J,k-K}v_{JK}X'_{JK}\qquad\qquad X'_{JK}\equiv X_J\\
&=u_{jk}(K\circledast(v\odot X'))_{jk}
\end{align}
where we defined $X'_{JK}$ matrix by copying $X_J$ along the second axis. Repeating the arguments above, we get the following compact expressions for the phase gradients:
\begin{align}
   \frac{\partial \phi}{\partial x}=u\odot(K\circledast(v\odot X'))\oslash\mu \\
    \frac{\partial \phi}{\partial y}=u\odot(K\circledast(v\odot Y'))\oslash\mu
\end{align}
and the following algorithm \ref{alg:phase-gradient}:
\begin{algorithm}[!ht]
\caption{2D Phase Gradient Algorithm}
\label{alg:phase-gradient}
\begin{algorithmic}[1]
\State $X'_{JK} \gets X_J$ 
\State $Y'_{JK} \gets Y_K$
\State $\nabla\phi_x \gets u\odot(K\circledast(v\odot X'))\oslash \mu$
\State $\nabla\phi_y \gets u\odot(K\circledast(v\odot Y'))\oslash\mu$ 
\State \Return $\nabla\phi_x$, $\nabla\phi_y$
\end{algorithmic}
\end{algorithm}

Our implementation of these algorithms can be found in Appendix \ref{chapter:numerical-implementation}, Section \ref{section:code-examples}, and our open-source package \cite{SLMTools}.

\subsection{Convolution implementation}
\label{subsection:convolution-implementation}

Both algorithms \ref{alg:2d-sinkhorn-knopp} and \ref{alg:phase-gradient} require a numerically stable way to compute the convolution with the Gaussian kernel $K$. Our goal is to have an algorithm that would work for small values of $\epsilon$, since we recover the original non-regularized OT solution in the limit $\epsilon\to0$. Generally speaking, working with small values of $\epsilon$ in the original 1D Sinkhorn-Knopp algorithm is difficult, and requires careful algorithmic design \cite{Schmitzer2019, ThibaultEtAl2017, Berman2020, ThorntonCuturi2023}. In the case of 2D, additionally, we have to worry about the numerical stability of the convolution operation, because for small $\epsilon$, the convolution kernel $K$ becomes sharply peaked. 

Notice that in the case of $\epsilon\to0$, gaussian kernel $K$ approaches the delta function. So, it is tempting to approximate steps 5 and 8 of the algorithm \ref{alg:2d-sinkhorn-knopp} using the convolution with the delta function (i.e. the identity operation). This, unfortunately, will not work, because there will be no updates for $u$ and $v$ variables throughout the iterations of the algorithm. Thus, we see that non-zero width of $K$ is essential for ``mixing" of $u$ and $v$ variables.

There are many ways of implementing a Gaussian convolution operation, and we will mostly focus on two most obvious approaches \textit{Fourier multiplication method} and \textit{padded convolution approach}. The first method relies on the convolution theorem \ref{FT1}, which allows us to multiply matrices in the Fourier domain instead of performing the convolution:
\begin{equation}
    K\circledast v=\mathcal{F}^{-1}[\mathcal{F}[K]\odot \mathcal{F}[u]]
\end{equation}
Notice that it is recommended to pre-compute $\mathcal{F}[K]$ at the beginning of the algorithm, to avoid unnecessary computation at each iteration. This Fourier approach works down to $\epsilon=0.003$ in practice for the input/output distribution described above. 

An alternative approach is to perform the discrete convolution operation directly. The issue here is that the output of a convolution of an $N\times N$ matrix with the $M\times M$ kernel is $(N+M-1)\times(N+M-1)$ in size. So, in order to make sure that $u$ has the same dimensionality at each iteration, we implement the following trick. We define the kernel $K$ over a lattice with $(2N-1)\times(2N-1)$ points, which results in $(3N-2)\times (3N-2)$ output of the convolution, and then we center crop $N \times N$ window, which is our updated $u'$. This implementation has a similar numerical stability to our Fourier method, which was around $\epsilon=0.0025$ for our input/output distributions.

Finally, we refer the reader to \cite{ipol.2013.87}, which is an excellent survey of the existing gaussian convolution algorithms. Because of the limited time, we haven't explored the full landscape of these algorithms, so we leave the experiments with the convolution algorithms as a future direction to this work.

\subsection{Complexity Analysis}
\label{subsection:complexity-analysis}

In conjunction, algorithms \ref{alg:2d-sinkhorn-knopp} and \ref{alg:phase-gradient}, and a simple phase integration step from \cite{swan2024highfidelityholographicbeamshaping} allows us to compute the ray-optics solution to the problem of phase generation while only keeping $\mathcal{O}(n^2)$ data in memory. This is a huge improvement compared to the $\mathcal{O}(n^4)$ memory constraint of the algorithm we originally proposed in \cite{swan2024highfidelityholographicbeamshaping}.

The time complexity of the original Sinkhorn-Knopp algorithm is proven to be $\mathcal{O}(N^2/\epsilon^2)$, where $N$ is the dimensionality of the input $\mu$ (and the output $\nu$) and $\epsilon$ is the regularization parameter \cite{complexitySinkhorn}. The main computational bottleneck of the SK algorithm \ref{alg:sinkhorn-knopp} is the matrix-vector multiplication, which takes $\mathcal{O}(N^2)$ if implemented naively. 

In the 2D case, we have $N=n^2$, so a naive Sinkhorn-Knopp algorithm will have a runtime of $\mathcal{O}(n^4/\epsilon^2)$. However, we replace the matrix-vector multiplication with a convolution operation, which has a runtime of $\mathcal{O}(n^2\log(n)/\epsilon^2)$. So, our final runtime is a modest $\mathcal{O}(n^2 \log(n)/\epsilon^2)$

Just to summarize, our convolution trick allows us to reduce the memory from $\mathcal{O}(n^4)$ to $\mathcal{O}(n^2)$ and the run-time from $\mathcal{O}(n^4/\epsilon^2)$ to $\mathcal{O}(n^2\log (n)/\epsilon^2)$, which gives us the ability to compute the ray-optics solution to the problem with a computational effort comparable to a few round of the Gerchberg-Saxton algoirthm. All of the algorithms use operations that are natively supported on most GPUs, allowing us to obtain a very fast and efficient solution to the problem of phase generation.

Unfortunately, we were not the first ones to invent the convolution trick. As we later discovered, this exact idea was presented by Gabriel Peyré during the seminar of the Kantorovich Initiative \cite{peyreKickoff2023}. In fact, Marco Cuturi (who invented Sinkhorn) and Gabriel Peyré have a book about computational Optimal Transport \cite{peyré2020computationaloptimaltransport} published in 2020 that describes this convolution idea. Nevertheless, Hunter and I were very happy to rediscover this important result, and extend it to a full phase generation pipeline!

\newpage
\section{Beam Estimation Algorithms}
\label{section:beam-estimation-algorithms}

In the problem of beam estimation, we are estimating the unknown complex-valued beam, $f(u,v)=g(u,v)e^{i2\pi\psi(u,v)}$, incident upon our phase-modulating device. To do so, we apply a family of quadratic phases $\phi_j(u,v)=-\pi i(u^2+v^2)/r_j^2$ for $j=1,2,...,m$, and record the corresponding beam moduli $G_j$, where as we shown in Chapter 3:
\begin{equation}
    |G_j(\mu,\nu)|^2=|\mathcal{F}_{\alpha_j}[f](\mu,\nu)|^2
\end{equation}
where $\alpha_j$ is related to the curvature of the quadratic phase $r_j$ via equation \ref{eq:alphaMR-phase-diversity}. The task of the beam estimation \ref{prob:beam-estimation} is to find the closest beam that matches required constraints $G_j$. 
Because the input to this problem are $m$ images $G_j$ as opposed to two images in the case of the problem of phase generation \ref{prob:phase-generation}, we need to slightly generalize the method of iterated projections.

\subsection{Method of Iterated Projections with m constraints}
\label{subsection:method-of-iterated-projections-with-m-constraints}

We start by defining constraint sets $B_j\subset\Omega$, where $\Omega=L^2(\mathbb{R}^2)$ is the space of all square-integrable functions.
\begin{equation}
    B_j=\{f\in L^2(\mathbb{R}^2): |\mathcal{F}_{\alpha_j}[f]|^2=G_j^2\}
\end{equation}
The projection onto a constraint set $B_j$ can be written as:
\begin{equation}
    P_{B_j}(f)=\mathcal{F}^{-1}_{\alpha_j}\left[G_j \frac{\mathcal{F}_{\alpha_j}[f]}{|\mathcal{F}_{\alpha_j}[f]|}\right]
\end{equation}
which is a very natural generalization of the typical GS projections \ref{eq:gs-projections}. Next, we define a product constraint set $C=B_1\times B_2\times\dots\times B_m\subset\Omega^m$ and the following diagonal constraint set:
\begin{equation}
    D=\{(f,f,\dots,f)\in\Omega^m: f\in\Omega\}
\end{equation}

We claim that an element $\omega\in\Omega^m$ in the intersection of $C\cap D$ will give us a function $f$ that satisfies all constraint sets $B_j$. To see that, notice that since $\omega \in D$, we get that $\omega=(f,f,\dots,f)$ for some $f\in\Omega$. Now, since $w\in C$, we get that $f\in B_j$ for all $j$, as required. The reverse direction is also true. If there exists a function $f\in\bigcap_{j=1}^mB_j$, then $\omega\equiv(f,f,\dots,f)\in D$ and $\omega\in C$. Therefore, we conclude that:
\begin{equation}
    C\cap D=\bigcap_{j=1}^mB_j
\end{equation}

One can also define projections $P_C$ and $P_D$, as follows.
\begin{align}
    P_C(f_1, f_2, \dots, f_m)&=(P_{B_1}(f_1), P_{B_2}(f_2), \dots, P_{B_m}(f_m))\\
    P_D(f_1, f_2, \dots, f_m)&=\left(\frac{1}{m}\sum_{j=1}^mf_i, \frac{1}{m}\sum_{j=1}^mf_i, \dots, \frac{1}{m}\sum_{j=1}^mf_i\right)
\end{align}

These projections allow us to apply the regular algorithm for the  method of iterated maps (alg. \ref{alg:mip}) to the constraints $C$ and $D$, which optimizes for the function $f$ that satisfies all of the constraints, thereby solving the beam estimation problem.

The remaining of this section will investigate the quality of the beam estimation solution depending on the number of acquired diversity images $m$.

\subsection{One-shot Beam Estimation}
\label{subsection:one-shot-beam-estimation}

Generally speaking, one image is insufficient to uniquely determine the complex-valued beam, because in the case of $m=1$ the problem of beam estimation \ref{prob:beam-estimation} is under-constrained. 

But suppose we only want a rough estimate of the beam. How much information can one extract from a single diversity image $G_j$? 
\begin{equation}
    G_j(\mu,\nu)=|\mathcal{F}_{\alpha_j}[f](\mu,\nu)|
\end{equation}
where we derived before that $\alpha_j=\arctan(-r_j^2/s^2)$ and $s$ was chosen according to the equation \ref{eq:alphaMR-phase-diversity} in such a way that magnification factor $M=1$. 

As we showed in Chapter 3, applying the stationary phase approximation (SPA) to $G_j$ yields:
\begin{equation}
\label{eq:SPA-to-Gj}
G_j\left(\nabla\varphi(u,v)\right)\approx\frac{g(u,v)}{\sqrt{\det(\nabla^2\varphi(u,v))}}
\end{equation}
where $\varphi$ is a new variable defined as:
\begin{equation}
\varphi(u,v)=\sin(\alpha_j)\psi(u,v)-\frac{\cos\alpha_j}{2}(u^2+v^2)
\end{equation}
We can invert equation \ref{eq:SPA-to-Gj} to obtain our approximate beam if we assume that the intrinsic phase has no curvature $\nabla^2\psi(u,v)=0$, which means that $\nabla^2{\varphi}=-\cos(\alpha_j)\mathds{1}$. Then we get:
\begin{equation}
    g(u,v)\approx G_j(\nabla\varphi(u,v))\cos(\alpha_j)
\end{equation}
If we further assume that the intrinsic phase is flat i.e. $\nabla\psi(u,v)=0$, then:
\begin{equation}
    g(u,v)\approx G_j(u\cos{\alpha_j}, v\cos\alpha_j)\cos(\alpha_j)
\end{equation}
Once again, if we only have one diversity image, we can't infer much about the intrinsic phase of the beam $\psi$, but we get a one-shot beam estimate of its amplitude $g$. 

Generalizing the error analysis from Subsection \ref{subsection:validity-of-the-approximaiton}, we find that the error of this estimate scales as $o\left(\frac{\sin\alpha_j}{r_{in}r_{out}}\right)$, where $r_{in}$, and $r_{out}$ are the characteristic sizes of the input and output beams in dimensionless units. Practically, this means that the best choice of the applied phase for the one-shot beam estimation is the one that has the steepest phase $r_j\to0$. In this case, the optical system \ref{fig:phase-diversity-set-up} approaches the identity.

\subsection{Two-shot Beam Estimation with Optimal Transport}
\label{subsection:two-shot-beam-estimation-with-optimal-transport}

Assume now that we collect two diversity images $G_1$ and $G_2$, corresponding to some FrFT angles $\alpha_1$ and $\alpha_2$. 
\begin{align}
    G_1(\mu,\nu) &=|\mathcal{F}_{\alpha_1}[f](\mu,\nu)|\\
    G_2(\mu,\nu) &= |\mathcal{F}_{\alpha_2}[f](\mu,\nu)|
\end{align}
By the  composition property of the FrFT:
\begin{equation}
    \mathcal{F}_{\alpha_1+\alpha_2}[f]=\mathcal{F}_{\alpha_2-\alpha_1}[\mathcal{F}_{\alpha_1}[f]]=\mathcal{F}_{\beta}[f_{\alpha_1}]
\end{equation}
where we defined $f_{\alpha_1}=\mathcal{F}_{\alpha_1}[f]$ and angle $\beta\equiv \alpha_2-\alpha_1$. Furthermore notice that:
\begin{align}
    G_1&=|f_{\alpha_1}|\label{eq:beam-moduli-g1}\\
    G_2&=|\mathcal{F}_\beta[f_{\alpha_1}]|
\end{align}
So, we need to find an unknown complex-valued beam $f_{\alpha_1}$ given it's amplitude $G_1$, and the amplitude of the fractional Fourier transform of $f_{\alpha_1}$ by the known angle $\beta$, which we call $G_2$. This is exactly the setting of the theorem we proved in Chapter \ref{chapter:ray-optics-reduction-with-optimal-transport}, Section \ref{section:fractional-fourier-transform-generalization}. 

Just to recap, this problem can be reduced to the following Monge-Ampere PDE for the unkown phase $\phi(u,v)=\text{Arg}[f_{\alpha_1}]$:
\begin{align}
G_2\left(\nabla\varphi(u,v)\right)&=\frac{G_1(u,v)}{\sqrt{\det(\nabla^2\varphi(u,v))}}+o (\sin(\beta)^{-1})\label{eq:G2-beta}\\
\varphi(u,v)&\equiv\sin(\beta)\phi(u,v)-\frac{\cos\beta}{2}(u^2+v^2) \label{eq:varphi-beta}
\end{align}
which can be solved very efficiently using the entropic regularization of Optimal Transport that we described above. Once the Optimal Transport phase $\varphi$ is obtained, we need to remember to invert the equation \ref{eq:varphi-beta} to find the phase $\phi$ of the complex-valued signal $f_{\alpha_1}$.
\begin{equation}
    \phi(u,v)=\frac{\varphi(u,v)}{\sin\beta} +\frac{u^2+v^2}{2\tan{\beta}}
\end{equation}
Finally, we can get the unknown beam estimate $f$ by inverting equation \ref{eq:beam-moduli-g1}:
\begin{equation}
    f=\mathcal{F}^{-1}_{\alpha_1}[G_1e^{i2\pi\phi}]
\end{equation}
which is our best two-shot-estimate for the problem of beam estimation. 

As before, the error of this term scales as $o\left(\frac{\sin\beta}{r_{in}r_{out}}\right)$. So, we want our $\beta$ to be as close as possible to $\pi/2$. To do so, we can collect one diversity image $G_1$ by applying no phase $r_j=\infty$, and $G_2$ by applying a very steep phase $r_j\to0$. In practice, however, there is a limit to how small $r_j$ can be applied set by the discretization of the SLM screen. 

\subsection{Metrics and Performance}
\label{subsection:metrics-and-performance}

In order to quantify\footnote{this section is adapted from our paper \cite{swan2024highfidelityholographicbeamshaping}} beam estimation performance in terms of experimentally accessible quantities, we define an error metric $\delta$ for a beam estimate $\parens[\big]{g(u,v),\psi(u,v)}$ by 
\begin{equation}
\delta \coloneq \sqrt{\frac{1}{m}\sum_{j=1}^m \norm*{ G_j^2(u,v) - \abs*{\mathcal{F}_{\alpha_j}\left[g(u,v)e^{i2\pi\psi(u,v)}\right]}^2 }_2^2}\,.
\label{eq:pd-error-metric}
\end{equation}
In words, $\delta$ is the $L^2$ distance between the measured diversity image $G_j^2$ and that predicted by the beam estimate, averaged in quadrature over all diversity images. 

\begin{figure}[!ht]
    \centering
    \includegraphics[width=\columnwidth]{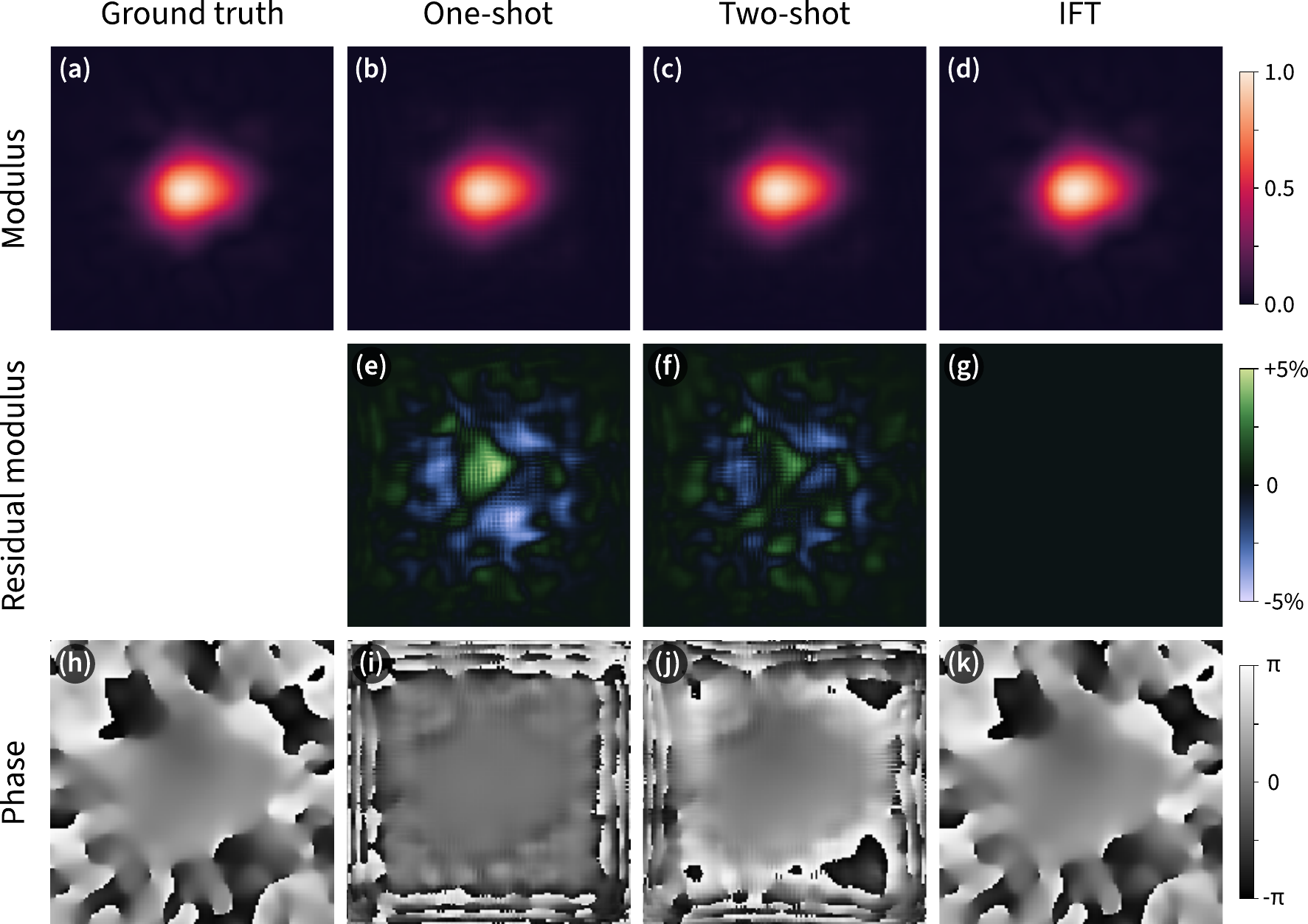}
    \caption[Beam estimates using various phase diversity algorithms]{Beam estimates using various phase diversity algorithms.  The top row (a-d) is the beam modulus.  The middle row (e-g) is the residual modulus, i.e.\ the difference between the modulus of the estimate and that of the ground truth.  The bottom row (h-k) is the phase.  (a,h) Ground truth beam. (b,e,i) One-shot beam estimate with diversity coefficient $1/r^2=1.5$ ($\delta = 0.02$). (c,f,j) Two-shot beam estimate with diversity coefficients $1/r_j^2=1.5$ and $1/r_k^2 = 0.1$ ($\delta = 0.005$). (d,g,k) IFT estimate with 15 diversity images, $1/r^2=0.1,0.2,\dots,1.5$, and 1000 iterations ($\delta = 3.3\times 10^{-17}$). For visual comparison, a global phase has been chosen for each image such that the local phase in the center of the image is $0$.}
    \label{fig:beam-estimates-hunter}
\end{figure}

We test performance of beam estimation algorithms on a simulated input beam generated by summing Hermite-Gaussian modes with random amplitudes (see Supplement of \cite{swan2024highfidelityholographicbeamshaping}). Figure \ref{fig:beam-estimates-hunter} shows a comparison of the ground truth input beam and the estimate of modulus and phase produced by each of the above algorithms.  The one-shot (with $1/r_j^2=1.5$) and two-shot (with $1/r_j^2 = 1.5$, $1/r_k^2 = 0.1$) estimates have error metrics $\delta = 0.02$ and $\delta = 0.005$, respectively.\footnote{Notice a difference in the definitions of the diversity images between this work and \cite{swan2024highfidelityholographicbeamshaping}. In this work we use radius of curvature $r_j$, while the paper \cite{swan2024highfidelityholographicbeamshaping} uses parameter $\alpha_j=1/r_j^2$, which is not the fractional Fourier angle.}

\begin{figure}[t]
\centering
\includegraphics[width=0.8\columnwidth]{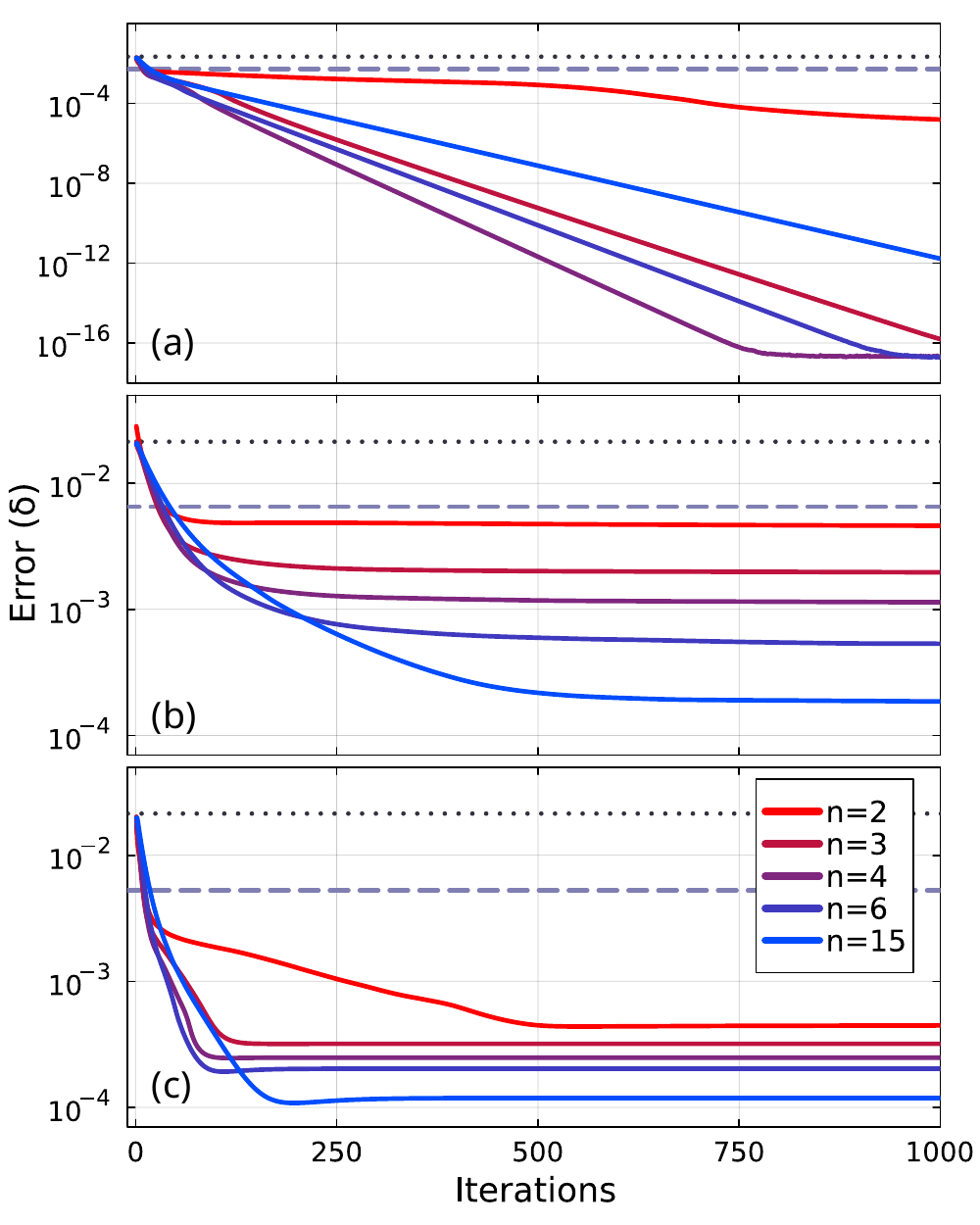}
\caption[Beam estimation error metrics]{Beam estimation error metrics, (a) in absence of noise, (b) with additive noise, and (c) with Poissonian shot noise. Dotted and dashed lines indicate $\delta$ for the one-shot and two-shot algorithms, respectively. Solid lines show $\delta$ vs. the number of iterations of the IFT algorithm with different numbers $n$ of diversity images.  Coefficients $1/r^2$ in each case are as follows. $n=2: 1/r^2 \in \{0.1, 1.5\}$. $n=3: 1/r^2\in \{0.1,0.8,1.5\}$. $n=4: 1/r^2 \in \{0.1,0.6,1.0,1.5\}$. $n=6: 1/r^2 \in \{0.1,0.4,0.7,0.9,1.2,1.5\}$. $n=15: 1/r^2 \in \{0.1,0.2,\dots,1.5\}$.  In (b), we approximate a 16-bit camera with up to two dark counts per pixel by adding to each diversity image pixel $G^2_{j,LM}$ a random value in the range $[0,2^{-15}\times \max_{PQ} G^2_{j,PQ}]$.  In (c), we approximate shot noise for a 16-bit camera by letting each diversity image pixel value be a Poissonian random variable with mean $2^{16}\times G^2_{j,LM}/\max_{PQ}G^2_{j,PQ}$, where $G^2_{j,LM}$ is the corresponding noiseless pixel value.  In all cases, $\delta$ is computed using all 15 noiseless diversity images.
}
\label{fig:pd-convergence}
\end{figure}

In absence of noise, we find that the method of iterated projections algorithm described above converges to within machine precision of ground truth (modulo a global phase) when at least 3 diversity images are used.  The rate of convergence depends on the range of diversity phase coefficients $\alpha_j$ and the number of diversity images.  Using more diversity images does not always lead to more rapid convergence. The rate of convergence is shown in \ref{fig:pd-convergence}~(a).  

In the presence of image noise, the IFT phase diversity algorithm no longer exactly reproduces the ground truth solution.  Instead, the error metric stagnates at a level which depends on the magnitude of the noise and the number of diversity images used.  In \ref{fig:pd-convergence}~(b,c) we show the performance of the same three algorithms in the presence of two models of noise.  In computing the error metric in these cases, we use the uncorrupted images $G^2_j$, since this provides a better measure of how close the estimated beam is to the ground truth.

As future research directions of the beam estimation problem, it is interesting to investigate the following ideas. We have some evidence that the deceleration of IFT convergence when many diversity images are used [see \ref{fig:pd-convergence}~(a)] can be understood in the fractional Fourier domain as an effect of oversampling of low spatial frequencies.  Applying some form of high-pass filtering may alleviate this effect and lead to better convergence.  Additionally, it is interesting to investigate the performance of phase diversity under more realistic noise models in an SLM system.

%% file: sections/chapter5.tex
\chapter{From Phase Holography to Quantum State Tomography}
\label{chapter:from-phase-holography-to-quantum-state-tomography}

Most of the approaches to phase generation or beam estimation problems frame the phase of the beam as an unknown quantity and the amplitudes as constraints. As we saw before, optimizing for phase directly tends to create unwanted phase vortices, which stagnate the convergence of most algorithms. These artifacts arise partially due to the fact that the optimization variable lives in a space of real $N\times N$ matrices that are modded by $2\pi$, which can be succinctly written as $(S^1)^{N\times N}$ where $S^1$ is the circle group $S^1\cong\mathbb{R}/2\pi\mathbb{Z}$. The reason why the Optimal Transport does not run into this issue is because it optimizes over the variable $\nabla \phi$, from which we can extract the phase that is not modded by $2\pi$.

In this chapter, we present a completely new approach to the problem of phase generation, which uses Hermite-Gaussian decomposition of the beam and treats complex-valued coefficients in the basis expansion as the optimization variable, which (hopefully)\footnote{this claim requires more numerical experiments} does not create vortices. We also establish a deep connection to the theory of Quantum Learning.

\section{Hermite-Gaussian Phase Generation in 1D}
\label{section:hermite-gaussian-phase-generation-in-1d}

Assume we are trying to learn everything about a complex-valued function $f\in L^2(\mathbb{R})$, given intensity measurements $g^2, G^2\in L^2(\mathbb{R})$ that are normalized to unity $||g||_{L_2}=||G||_{L_2}=1$. Where, as usual, we define $g^2(u)=|f(u)|^2$ and $G^2(u)=|\mathcal{F}[f]|(u)^2$. 
Let us consider a Hermite-Gaussian basis (defined in Section \ref{section:hermite-gaussian-basis}, eq. \ref{eq:hermite-functions}), which we copy here for convenience:
\begin{align}
    h_n=A_n H_n(\sqrt{2\pi}u)\exp(-\pi u^2) && A_n=\frac{2^{1/4}}{\sqrt{2^n n!}}
\end{align}
defined in terms of the Hermite polynomials:
\begin{equation}
    H_n(u) = (-1)^n e^{u^2}\frac{d^n}{d u^n}e^{-u^2}
\end{equation}
From our discussion in Chapter 2, we know that the family $(h_n(u))_{n=0}^\infty$ forms a complete orthonormal basis for $L^2(\mathbb{R})$. This allows us to decompose all of the data in the problem into this basis:
\begin{align}
    g^2 =\sum_{i=0}^\infty a_ih_i && G^2 &=\sum_{i=0}^\infty b_ih_i
\end{align}
where the coefficients can be obtained by integrating:
\begin{align}
    a_i&=\braket{h_i|g^2}=\int_{-\infty}^\infty h_i(x)g^2(x)dx\\
    b_i&=\braket{h_i|G^2}=\int_{-\infty}^\infty h_i(x)G^2(x)dx
\end{align}
We can also decompose the unknown complex-valued function $f(u) =\sum_{i=0}^\infty c_ih_i(u)$, but of course we do not know complex-valued coefficients $c_i$ a priori. 
This allows us to reformulate the problem of phase generation \ref{prob:phase-generation} in the following way:

\begin{problem}
\label{prob:hermite-generation}
Define the family of real-valued symmetric operators indexed by $k$:
\begin{equation}
    P^{(k)}_{nm}\equiv\braket{h_k|h_n h_m}=\int h_k(x)h_n(x)h_m(x)dx
\end{equation}
and also related operators:
\begin{equation}
    Q^{(k)}_{nm}\equiv e^{in\pi/2}P^{(k)}_{nm}e^{-im\pi/2}
\end{equation}
Denote a set of $N$ real-valued observations of the operator $P^{(k)}$ by $a_k$ where $k\in\{1,\dots,N\}$. Similarly, let $b_k$ be observations of $Q^{(k)}$. Then consider the optimization problem is to find a quantum state $\ket{c}\in\mathbb{C}^N$ satisfying the observation constraints i.e.:
\begin{align}
    \bra{c}P^{(k)}\ket{c}&=a_k \quad \forall k\\
    \bra{c}Q^{(k)}\ket{c}&=b_k \quad \forall k
\end{align}
\end{problem}

If we relax this problem by allowing some error in the constraints, then this problem becomes related to the problem of phase generation \ref{prob:phase-generation}. Below is the sketch of the proof in the case of infinite $N$ and no error in the constraints. The version of the proof for finite $N$ and allowing for errors should be possible too, but we have not proven it yet.
\newpage

\begin{proof}
    ($\impliedby$) Suppose we start with a solution to the problem of phase generation in $L^2(\mathbb{R})$ that matches constraints $g^2$ and $G^2$ exactly. We start by decomposing our solution into Hermite-Gaussians:
    \begin{equation}
        f(u)=\sum_{i=0}^\infty c_ih_i(u)
    \end{equation}
    Then, as we said, constraints have to be satisfied. 
    \begin{align}
        g^2(u)&=|f(u)|^2\implies \sum_{k=0}^\infty a_kh_k=\left|\sum_{i=0}^\infty c_ih_i\right|^2\\
        G^2(u)&=|\mathcal{F}[f](u)|^2\implies \sum_{k=0}^\infty b_kh_k=\left|\sum_{i=0}^\infty c_ie^{-in\pi/2}h_i\right|^2
    \end{align}
    where we used the Fourier property of the Hermite basis \ref{eq:fourier-eigen}. Simplifying these constraints even further, we get:
    \begin{align}
    \label{eq:before-Pnmk}
        &\sum_{n=0}^\infty \sum_{m=0}^\infty c^*_n c_mh_nh_m=\sum_{k=0}^\infty a_k h_k \\
        &\sum_{n=0}^\infty \sum_{m=0}^\infty c^*_ne^{in\pi/2} c_me^{-im\pi/2}h_nh_m=\sum_{k=0}^\infty b_k h_k
    \end{align}
    but $h_nh_m$ is just another function in $L^2(\mathbb{R})$, which can be decomposed into Hermite Gaussian functions using $P^{(k)}_{nm}$ tensor:
    \begin{equation}
        h_nh_m=\sum P^{(k)}_{nm}h_k
    \end{equation}
    Plugging this back into (\ref{eq:before-Pnmk}) and equating the corresponding basis coefficients, we obtain the following equality for all $k\in\{0,1,\dots\}$:
    \begin{align}
        c_n^*P^{(k)}_{nm}c_m=a_k && c_n^*e^{in\pi/2}P^{(k)}_{nm}e^{-im\pi/2}c_m=b_k 
    \end{align}
    which are exactly the desired constraints, once we introduce a vectorized notation $\ket{c}$ and $P^{(k)}$.
    
    ($\implies$) Now, if we suppose that we know the quantum state that satisfies the observations, then we know the coefficients $\{c_i\}_{i=0}^\infty$ of the unknown signal $f$. So, we can obtain the required phase using:
    \begin{equation}
        \phi(u)=\text{Arg}\left[f(u)\right]=\text{Arg}\left[\sum_{i=0}^\infty c_ih_i(u)\right]
    \end{equation}
    To show that this indeed satisfies the constraints of the original phase generation problem is just a matter of repeating the argument we made above.
\end{proof}

\subsection{Interpretation}
\label{subsection:interpretation}

This theorem provides an elegant way to recast the problem of phase generation into the language of quantum learning. Instead of the unknown phase, we have an unknown quantum state $\ket{c}\in\mathbb{C}^N$. Instead of intensity measurements, we have real-valued measurements of operators $P^{(k)}$ and $Q^{(k)}$ for $k\in\{0,1,\dots\}$. 

The operator $P^{(k)}$ might seem mysterious at first, but it has a very nice physical interpretation. Its components $P^{(k)}_{nm}$ correspond to the following inner poduct:
\begin{equation}
    P^{(k)}_{nm}\equiv\braket{h_k|h_n h_m}=\int h_k(x)h_n(x)h_m(x)dx
\end{equation}
so, $P^{(k)}_{nm}$ is exactly a triple Hermite-Gaussian product. The interpretation here is that it captures how strongly the coefficients $c_i$ of the underlying signal $f$ affect individual modes of the intensity. Since we have to take the modulus squared of $f$, each doublet of coefficients $c_nc_m$ in principle can affect every $k$-th mode of the intensity. Another nice thing about this object is that it is analytically computable in terms of the hyper-geometric series (see eq. (6.8.3) in \cite{AndrewsAskeyRoy1999}), so one does not need to store a 3-index tensor in memory at any point, and instead can compute it on the fly.

\subsection{Fast State Tomography}
\label{subsection:fast-state-tomography}

The best-known approach for this type of problem is to use semi-definite programming (SDP) as explained in \cite{guta2018faststatetomographyoptimal}, which has well-defined optimal error bounds. An interesting point about the paper \cite{guta2018faststatetomographyoptimal} is that the problem is computationally much more difficult when the underlying quantum state is constrained to be a rank-1 object (i.e., a pure state), which is the case in our problem \ref{prob:hermite-generation}. The problem becomes much more tractable if we expand the space of optimization to the higher-rank density matrices.

This inspires an interesting new avenue for the problem of phase generation. Imagine that instead of a single laser beam, we have $N$ beams numbered $f_1,f_2,\dots,f_N$. The electric fields are additive, so they collectively produce a light field $f$, which has intensity in the position and momentum domains $g^2$ and $G^2$. As before, we can decompose the desired intensities into $a_k$ and $b_k$, and the $i$-th beam $f_i$ into $\ket{c^{(i)}}$ vector of coefficients. We can succinctly write the combined state as:
\begin{equation}
    \rho=\frac{1}{N}\ket{c^{(1)}} \bra{c^{(1)}}+\frac{1}{N}\ket{c^{(2)}} \bra{c^{(2)}}+\dots+\frac{1}{N}\ket{c^{(N)}} \bra{c^{(N)}}
\end{equation}
which must satisfy the following constraints for all $k$:
\begin{align}
    \Tr[P^{(k)}\rho]=a_k && \Tr[Q^{(k)}\rho]=b_k
\end{align}
Our intuition is that by promoting the quantum state to a rank-$N$ object, we can obtain a much better match in the intensity. Of course, this claim requires careful numerical investigation, which we leave as a future direction for this research.

%% file: sections/conclusions.tex
\chapter{Conclusions}
\label{chapter:conclusions}

In summary, this thesis provides a comprehensive mathematical framework for understanding modern problems in laser beam shaping. We employed the language of fractional Fourier transforms and Wigner distribution to arrive at new re-interpretations of the problems of phase generation and beam estimation.

Next, we established a deep theoretical connection between the ray-optics limit of the above problems and the mathematical theory of Optimal Transport. Using Optimal Transport, we arrived at state-of-the-art solutions to the problems of phase generation and beam estimation. 

Furthermore, we included a comprehensive summary of phase generation algorithms and demonstrated their advantages and drawbacks. To the best of our knowledge, our optimal transport solutions can improve convergence of most (if not all) iterative algorithms for phase generation.

Last but not least, we established a concrete theoretical connection between the problem of phase generation and quantum learning. We also outline a new approach to solving this problem using fast state tomography algorithms described in \cite{guta2018faststatetomographyoptimal}, which is one of the future directions for our work. 

All in all, this thesis paves the way for a new unprecedented spatial control of laser light! 

%% file: appendix/A.tex
\appendix
\chapter{Fourier Analysis}
\label{chapter:fourier-analysis}

Here, we define conventions and notation that we will use throughout this thesis. In our experience, careful definitions related to the Fourier transforms and coordinate grids are absolutely necessary for understanding and implementing proposed algorithms. 

\section{Conventions}
\label{section:appendixa-conventions}

\subsection{Variables}
\label{subsection:variables}

Throughout this thesis, we will refer to the electric field as a ``signal" to highlight the fact that this theory can be applied to any complex-valued function in an abstract separable Hilbert space. The small variables $x,y,z$ typically correspond to the position variables (units of length), and the capital variables $\sigma_x,\sigma_y,\sigma_z$ are the spatial frequencies (units of 1/length). We will also use $u$, $v$, $w$ for dimensionless position variables, and $\mu$, $\nu$, and $\eta$ for their dimensionless conjugate variables. Notice that for every physical system that we will describe, there will be a conversion factor $s$ (unit of length) that relates dimensional and dimensionless variables:
\begin{align}
    x/s = u && y/s=v && z/s = w \\
    \sigma_xs=\mu && \sigma_ys=\nu && \sigma_zs=\eta
\end{align}
This is the same notational convention as in \cite{ozaktas2001fractional}, and, in my humble opinion, it is one of the best ones I have seen.

\subsection{Dimensionless forms}
\label{sec:dimensionless-form}

Every physical signal $f$, whether an electric field or a quantum mechanical wavefunction, has its dimensional and dimensionless forms. To convert a function to its dimensionless form, we can divide the argument by the so-called scale parameter $s$ that has the same units as the input $x$. The exact choice of scale parameter depends on the details of the experimental set-up. For example, a convenient choice for optical set-ups with a single lens is $s=\sqrt{\lambda f_0}$, where $\lambda$ is the wavelength of light and $f_0$ is the focal distance. This same scale parameter $s$ can be used to convert the Fourier transform $F$ of the signal to its dimensionless form. More concretely, we define: \footnote{This definition and the following discussion are a concise summary of chapter 9.1 from \cite{ozaktas2001fractional}.}
\begin{align}
\label{eq:dimensionless}
\hat{f}(x)&\equiv \frac{1}{\sqrt{s}}f(x/s)\equiv\frac{1}{\sqrt{s}}f(u)\\
\hat{F}(\sigma_x)&\equiv \sqrt{s} F(s \sigma_x) \equiv \sqrt{s}F(\mu)
\end{align}
where $u\equiv x/s$ and $\mu\equiv sX$ are our new dimensionless variables, $s>0$ is the dimensional scale parameter, and hat designates that the function takes dimensional arguments. As before, we have that $F(\mu)=\int duf(u)\exp(-i2\pi\mu u)$ is the Fourier transform operating on dimensionless quantities. It is easy to check that our definition ensures that $\hat{F}(\sigma_x)=\int dx \hat{f}(x)\exp(-i2\pi x\sigma_x)$ is the Fourier transform of $\hat{f}(x)$. An important consequence of our definition is that all of the above functions satisfy Parallel's identity:
\begin{equation}
    \int|\hat{f}(x)|^2dx=\int|f(u)|^2du=\int|F(\mu)|^2d\mu =\int |\hat{F}(\sigma_x)|^2d\sigma_x
\end{equation}

\subsection{Application to Fractional Fourier Transforms}
\label{subsection:application-to-fractional-fourier-transforms}

One needs to be very careful considering the scaling factor $s$ in the context of the fractional Fourier transform of angle $\alpha$. Notice that if $\alpha=\pi/2$, the fractional Fourier transform coincides with the regular Fourier transform, and satisfies a property:
\begin{equation}
    \mathcal{F}_{\pi/2}[f(u/s)]=\sqrt{s}\mathcal{F}_{\pi/2}[f](su)
\end{equation}
This property is exactly what allowed for our dimensionless and dimensional forms of the function to ``play nicely" with the Fourier transform. In the case of a general fractional Fourier transform, the relevant equation is:
\begin{align}
    \mathcal{F}_{\alpha}[f(u/s)]&=|s|\sqrt{\frac{1-i\cot\alpha}{1-is^2\cot{\alpha}}} \exp\left[i\pi u^2\cot \alpha\left(1-\frac{\cos^2 \alpha'}{\cos^2 \alpha}\right)\right] \mathcal{F}_{\pi/2}[f]\left(\frac{su \sin \alpha'}{\sin \alpha}\right) \\
    \alpha' &\equiv \arctan(s^{-2} \tan \alpha) 
\end{align}
which is a nightmare to deal with. Thankfully, this pain can be avoided if we only apply operators to \textbf{dimensionless} version of the function. For example, consider a physical system that has inputs and outputs related by the fractional Fourier transform 
\begin{equation}
\label{eq:_dimensionless-fractional}
    g(u)=\mathcal{F_\alpha}[f](u)\equiv f_\alpha(u)
\end{equation}
in dimensionless arguments. To find the physical outputs $\hat{g}(x)$ when the physical input is $\hat{f}(x)$ we first translate $\hat{f}(x)$ to its dimensionless form using:
\begin{equation}
    \hat{f}(x)\equiv s^{-1/2}f(x/s)=s^{-1/2}f(u)
\end{equation}
then we apply the fractional Fourier transform (\ref{eq:_dimensionless-fractional}) to get $g(u)$ and then convert $g(u)$ to the physical $\hat{g}(x)$ using:
\begin{equation}
    \hat{g}(x)\equiv s^{-1/2}g(x/s)=s^{-1/2}g(u)
\end{equation}
Just to summarize, the mathematical mapping is the following:
\begin{equation}
    \hat{g}(x)=s^{-1/2}\mathcal{F_\alpha}[f](x/s)=s^{-1/2}f_a(x/s)=\hat{f}_\alpha(x)
\end{equation}
This phenomenon is not unique to the fractional Fourier transform, and it is always present when working with operators that map between domains with different units. In math literature, this issue is often overlooked since everything is dimensionless, but here we must be careful.

\section{Functional Spaces}
\label{section:functional-spaces}

We will be working in several functional spaces, so we will give a brief overview of the relevant definitions to avoid any possible confusion.

\subsection{Square Integrable Spaces}
\label{subsection:square-integrable-spaces}

In a continuous setting, we will be mostly working with a space of square-integrable functions \footnote{Here, the integral is in general a Lebesgue integral, but in practice all functions we will be working with are Riemann integrable, so we do not need to worry about this technical detail.}:
\begin{equation}
    L^2(\mathbb{R}^2) = \{f:\mathbb{R}^2\to\mathbb{C}|\int_{\mathbb{R}^2}|f|^2<\infty\}
\end{equation}
which is a Hilbert space with the following inner product:
\begin{equation}
    \braket{f|g}=\int_{-\infty}^\infty\int_{-\infty}^\infty f^*(u,v)g(u,v)dudv
\end{equation}
The inner product induces a norm $||f||_{L^2}=\sqrt{\braket{f|f}}$. This can be written as:
\begin{equation}
\label{eq:continuum-norm}
    ||f||_{L^2}=\int_{-\infty}^\infty\int_{-\infty}^\infty |f(u,v)|^2dudv
\end{equation}
where $|f(u,v)|^2\geq0$ is the squared amplitude. On some occasions, we might restrict the domain to a rectangular region around the origin $D =[-L_u/2,L_u/2]\times[-L_v/2,L_v/2]\subset\mathbb{R}^2$, where $L_u$ and $L_v$ are the width and height of our region. More formally, our space is then $L^2(D)$:
\begin{equation}
\label{eq:l2d-def}
    L^2(D)=\{f:D\to \mathbb{C} |\int_D |f|^2<\infty\}
\end{equation}
with the same notion of the inner product and norm as above. 

\subsection{Continuous Fourier Transform} 
\label{subsection:continuous-fourier-transform}

Pick any $f\in L^2(\mathbb{R})$, i.e. $f:\mathbb{R}^2\to\mathbb{C}$ is some complex-valued function defined on some position coordinates $x$ and $y$. For this function we define a continuous Fourier transform $F:\mathbb{R}^2\to\mathbb{C}$ using the following definition:
\begin{equation}
\label{eq:ft}
    F(\mu, \nu) = \mathcal{F}[f](\mu,\nu) =  \int_{-\infty}^{+\infty}\int_{-\infty}^{+\infty} f(u, v)\exp(-i 2\pi (\mu u + \nu v)dudv
\end{equation}
\begin{equation}
\label{eq:ift}
    f(u, v) = \mathcal{F}^{-1}[F](u,v) =  \int_{-\infty}^{+\infty}\int_{-\infty}^{+\infty} F(\mu, \nu)\exp(i 2\pi (\mu u + \nu v)d\mu d\nu
\end{equation}

Both $F$ and $f$ represent the same function, but they are expressed in different bases. More formally we say that each $F(\mu,\nu)$ is an element of a dual space $(L^2(\mathbb{R}^2))^*$, but luckily for us $L^2(\mathbb{R}^2)$ is self-dual, which means that $L^2(\mathbb{R}^2) \cong (L^2(\mathbb{R}^2))^*$. Practically speaking, this means that both the original function $f$ and its Fourier transform $F$ are in the same functional space $L^2(\mathbb{R}^2)$.

\subsection{Compact Fourier Transform}
\label{subsection:compact-fourier-transform}

Now consider a function $f\in L^2(D)$. Once we restrict the domain to a compact set, such as $D$, the Fourier basis becomes countable. For example, $L^2([0,1])$ has a countable orthonormal basis $\{\exp(i2\pi nu)\}_{n\in \mathbb{Z}}$ \cite{Young_1988}. Similarly, it can be proven that the following forms a countable orthonormal basis for $L^2(D)$:
\begin{equation}
    e_{n,m}(u,v)=\frac{1}{\sqrt{L_u}}\frac{1}{\sqrt{L_v}} \exp(i 2\pi n u / L_u)\exp(i 2\pi m v /L_v) \quad \quad \forall n,m\in\mathbb{Z}
\end{equation}
We can decompose our function $f$ into this countable basis:
\begin{equation}
    f(u,v)=\sum_{n\in\mathbb{Z}}\sum_{m\in\mathbb{Z}}c_{n,m}e_{n,m}(u,v)
\end{equation}
where $c_{n,m}$ coefficients can be expressed via an integral:
\begin{equation}
\label{eq:fourierCompact}
    c_{n,m}=\braket{e_{n,m}|f}=\frac{1}{\sqrt{L_uL_v}}\iint dudv f(u,v) \exp\left(-i2\pi \left(\frac{n}{L_u}u+\frac{m}{L_v}v\right)\right)
\end{equation}
Notice the difference between this compact ``Fourier transform" and the continuous case described above. In this case, the conjugate variable (momentum) takes discrete values $\mu = n/L_u$ and $\nu=m/L_v$ for discrete $n,m\in \mathbb{Z}$. So, it is perhaps not surprising that we can identify $L^2(D)$ with a space of infinite sequences $l^2(\mathbb{Z}^2)$.
\begin{equation}
    l^2(\mathbb{Z}^2)=\{(c_{n,m})_{n\in\mathbb{Z},m\in\mathbb{Z}}|\sum_{n=-\infty}^{\infty}\sum_{m=-\infty}^{\infty}|c_{n,m}|^2<\infty\}
\end{equation}
So, the equation (\ref{eq:fourierCompact}) exactly gives the isomorphism between the Hilbert spaces $L^2(D)\cong l^2(\mathbb{Z}^2)$. 

\subsection{Discrete Fourier Transform}
\label{subsection:discrete-fourier-transform}

Ultimately, we will be working with camera images, which are $N\times M$ arrays of numbers in both the position and momentum domains. The correct functional space in this case is $l^2([N]\times[M])$, where $[N]=\{1,2,...,N\}$.
\begin{equation}
    l^2([N]\times[M]) = \{f\in\mathbb{C}^{N \times M}|\sum_{i=1}^{N}\sum_{j=1}^{M}|f_{i,j}|^2<\infty \}
\end{equation}
which is really equivalent to the space of $N\times M$ complex valued matrices $\operatorname{Mat}_{N\times M}(\mathbb{C})$. This space inherits the usual $l^2$ inner product:
\begin{equation}
\label{eq:CNMinnerproduct}
    \braket{f|g}=\sum_{i=1}^{N}\sum_{j=1}^{M}f^*_{i,j}g_{i,j}=\Tr [f^\dagger g]
\end{equation}
where the last equality is just a convenient compact form of the usual inner product. Similarly to $L^2(\mathbb{R}^2)$, this functional space is self-dual, meaning that $l^2([N]\times[M]) \cong (l^2([N]\times[M]))^*$. That means that the Discrete Fourier transform (DFT) $F\in\mathbb{C}^{N \times M}$ defined below will be an element of $l^2([N]\times[M])$ as well.
\begin{equation}
\label{eq:dft-convention}
    F_{j'k'}= \frac{1}{\sqrt{N}}\frac{1}{\sqrt{M}}\sum_{j=1}^N\sum_{k=1}^M f_{jk}\exp{(-i2\pi (jj'/N+kk'/M))}
\end{equation}
\begin{equation}
    f_{j'k'}= \frac{1}{\sqrt{N}}\frac{1}{\sqrt{M}}\sum_{j=1}^N\sum_{k=1}^M F_{jk}\exp{(i2\pi (jj'/N+kk'/M))}
\end{equation}
 Notice that the Discrete Fourier Transform defined in (\ref{eq:dft-convention}) is just a direct discretization of (\ref{eq:fourierCompact}). To see the mapping, take $L_u=N\Delta u$ and $L_v=M\Delta v$ and evaluate the function $f$ on a grid of points $[N]\times[M]$. See Appendix B for the details of the numerical implementation. 

 \newpage

\subsection{Fourier Transform Properties}
\label{subsection:fourier-transform-properties}

Below, we define some useful properties of the Fourier transform. Notice that with our conventions, the convolution theorem and its corollaries (FT1 - FT4) have no factors of $\pi$, which is very convenient.
\begin{center}
\begin{varwidth}{\textwidth}
\begin{enumerate}[start=1,label={(FT\arabic*)}]
\item $\mathcal{F}[f \circledast g] = \mathcal{F}[f] \mathcal{F}[g] $ \label{FT1}
\item $\mathcal{F}[fg] = \mathcal{F}[f] \circledast \mathcal{F}[g]$  \label{FT2}
\item $\mathcal{F}^{-1}[F \circledast G] = \mathcal{F}^{-1}[f] \mathcal{F}^{-1}[g]$  \label{FT3}
\item $\mathcal{F}^{-1}[FG] = \mathcal{F}^{-1}[F] \circledast \mathcal{F}^{-1}[G]$ \label{FT4}
\item $\mathcal{F}[f(\alpha u, \beta v)] = \frac{1}{|\alpha \beta|}\mathcal{F}[f]\left(\frac{\mu}{\alpha}, \frac{}{\beta}\right)$  \label{FT5}
\item $\mathcal{F}^{-1}[F(\alpha \mu, \beta \nu)] = \frac{1}{|\alpha \beta|}\mathcal{F}^{-1}[F]\left(\frac{u}{\alpha}, \frac{v} {\beta}\right)$ \label{FT6}
\end{enumerate}
\end{varwidth}
\end{center}
These properties also apply to the discrete Fourier transformation \ref{eq:dft-convention}, but one has to be careful with the exact details of the discrete convolution operation, which might require some padding and cropping to keep the dimensions of the output the same as the input.

%% file: appendix/B.tex
\chapter{Numerical Implementation}
\label{chapter:numerical-implementation}

\section{Coordinate Lattices}
\label{section:coordinate-lattices}

To properly discretize a continuous function, we define a grid of points, which we call a \textit{lattice}:
\begin{align}
    L=\{(j\Delta u,k\Delta v):j\in\{-\floor{N/2},\dots,\floor{(N-1)/2}\},\\
    k\in\{-\floor{M/2},\dots,\floor{(M-1)/2}\}\}
\end{align}
Each lattice $L$ can be parametrized by discretization $\Delta u$, $\Delta v$, and the number of rows $M$ and columns $N$. Once we have the lattice defined, we can evaluate our continuous function $f$ on this grid to obtain a discretized function $f_{jk}:L\to \mathbb{C}$ as:
\begin{equation}
    f_{jk}=f(j\Delta x,k\Delta y)
\end{equation}
Now, we can use the DFT defined in (\ref{eq:dft-convention}) to compute $F_{j'k'}$, which lives on a dual lattice $L'$. The new lattice $L'$ has the discretization $\Delta \mu$ and $\Delta \nu$ and the same number of rows/columns $M$ and $N$. Discretization of the dual lattice is fixed via the following:
\begin{align}
    \Delta u \Delta \mu =\frac{1}{N} && \Delta v \Delta \nu = \frac{1}{M}
\end{align}

\subsection{Natural Lattice}
\label{subsection:natural-lattice}

One very convenient choice of the lattice is to pick $\Delta u=1/\sqrt{N}$ and $\Delta v=1/\sqrt{M}$. We call such a lattice a \textit{natural lattice}, $L_0$. Notice that the dual lattice of the natural lattice has the same discretization $\Delta \mu = 1/\sqrt{N}$ and  $\Delta \nu = 1/\sqrt{M}$. Thus, a natural lattice has a convenient property of being self-dual, meaning that $L_0'=L_0$. For most computations, we will use a \textit{square natural lattice}, which has $N=M$.

\subsection{FFT Convention}
\label{subsection:fft-convention}

We can specialize the DFT (\ref{eq:dft-convention}) to the square natural lattice $L_0$, which requires shifting the origin by $-\floor{N/2}$. We will also assume $N=M$ and $\Delta u=\Delta v$ for simplicity.
\begin{equation}
\label{eq:shifted-fft}
    F_{j'k'}= \frac{1}{N}\sum_{j=0}^{N-1}\sum_{k=0}^{N-1} f_{jk}\exp{(-i2\pi ((j-\floor{N/2})(j'-\floor{N/2})/N+(k-\floor{N/2})(k'-\floor{N/2})/N))}
\end{equation}
This can be compactly written using the following shifted DFT matrix $W\in\mathbb{C}^{N\times N}$, where $j, j'\in\{0,1,\dots,N-1\}$:
\begin{equation}
\label{eq:shifted-dft}
    W_{jj'} = \frac{1}{\sqrt{N}} \exp(-i2\pi(j-\floor{N/2})(j'-\floor{N/2})/N)
\end{equation}
Then the new signal is just the following tensor contraction:
\begin{equation}
    F_{j'k'}=\sum_{j,k}W_{jj'}f_{j,k}W_{kk'} 
\end{equation}

The matrix $W_{jj'}$ can be viewed as a change of basis matrix for $\mathbb{C}^{N\times N}$. In fact, it is unitary — $W^\dagger W=W W^\dagger=I$. Using this matrix is also very convenient, because it is easily implemented in code using \texttt{fftshift}, \texttt{ifftshift}. See code examples using Julia at the end of this section. 

\subsection{Visualizing Shifted DFT}
\label{subsection:visualizing-shifted-dft}

Let's elaborate on our use of the shifted DFT to avoid any possible confusion. The standard DFT matrix is defined for $j, j'\in\{0,1,\dots,N-1\}$:

\begin{equation}
\label{eq:normal-dft}
    DFT_{jj'} = \frac{1}{\sqrt{N}} \exp(-i2\pi jj'/N)
\end{equation}

\begin{figure}[!ht]
\centering
\includegraphics[width=1\columnwidth]{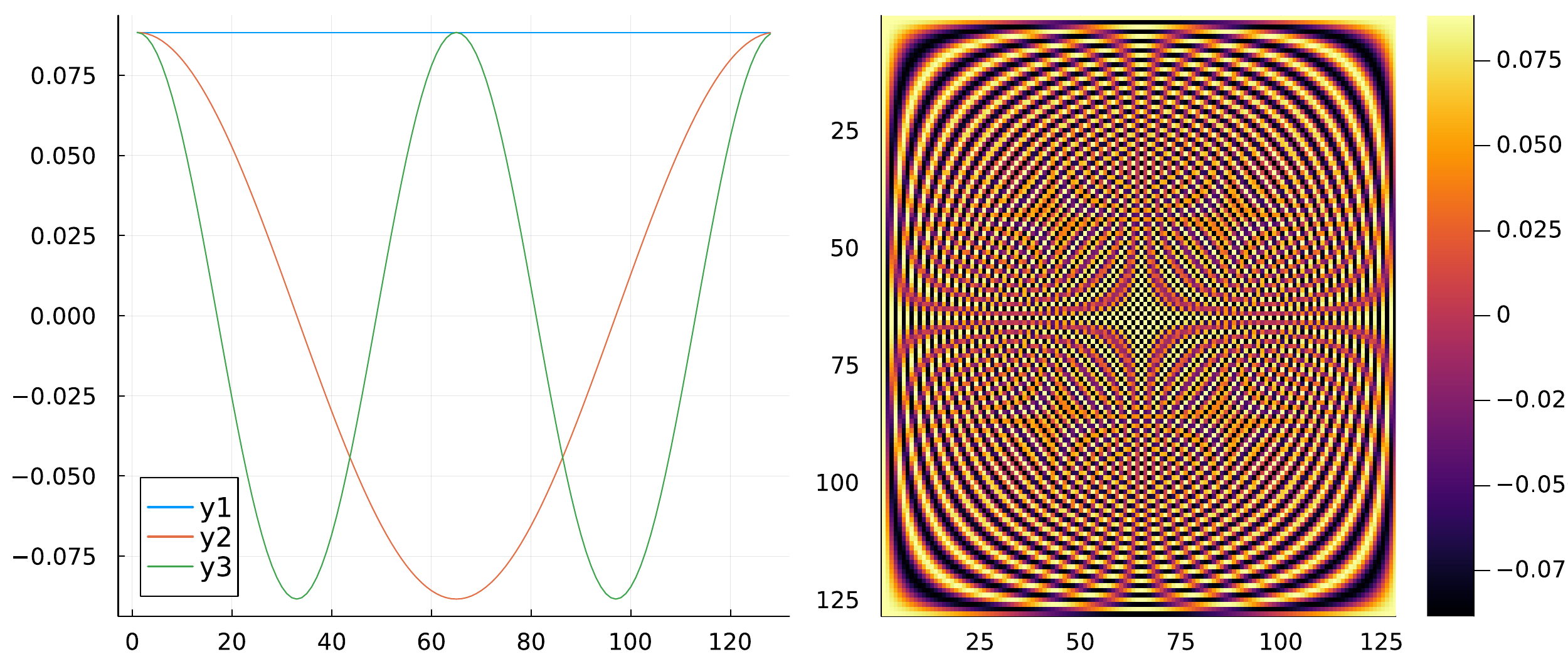}
\caption[Non-shifted DFT matrix]{Non-shifted DFT matrix. The left panel plots the first three columns. The right panel visualizes the real part of the DFT matrix.}
\label{fig:non-shifted-DFT}
\end{figure}

An important observation is that columns of the matrix correspond to the new basis elements, meaning that $e^{(1)}_j=DFT_{j1}$, $e^{(2)}_j=DFT_{j2}$, and $e^{(N)}_j=DFT_{jN}$. Notice that in the conventional non-shifted DFT, $e^{(1)}_j=1/\sqrt{N}$, $e^{(2)}_j=e^{-i2\pi j}/\sqrt{N}$, etc. Thus, the the low frequency modes are at the beginning $e^{(1)}, e^{(2)}, e^{(3)},\dots$ and at the end $e^{(N-2)}, e^{(N-1)}, e^{(N)}$, while middle range basis vectors correspond to high frequencies, with $e^{(\floor{N/2})}$ and $e^{(\floor{N/2}+1)}$ being the highest frequency, known as the so-called Nyquist frequency (see Figure \ref{fig:non-shifted-DFT}).

In the shifted DFT, which we call $W$ (\ref{eq:shifted-dft}), we have the opposite. The basis vectors $e^{(\floor{N/2})}$ and $e^{(\floor{N/2}+1)}$ are the lowest frequency, while the basis vectors at the beginning $e^{(1)}, e^{(2)}, e^{(3)},\dots$ and at the end $e^{(N-2)}, e^{(N-1)}, e^{(N)}$ are the highest frequencies (see Figure \ref{fig:non-shifted-DFT}). In other words, shifted DFT is nothing but a simple relabeling of the vectors in the basis.

\begin{figure}[!ht]
\centering
\includegraphics[width=1\columnwidth]{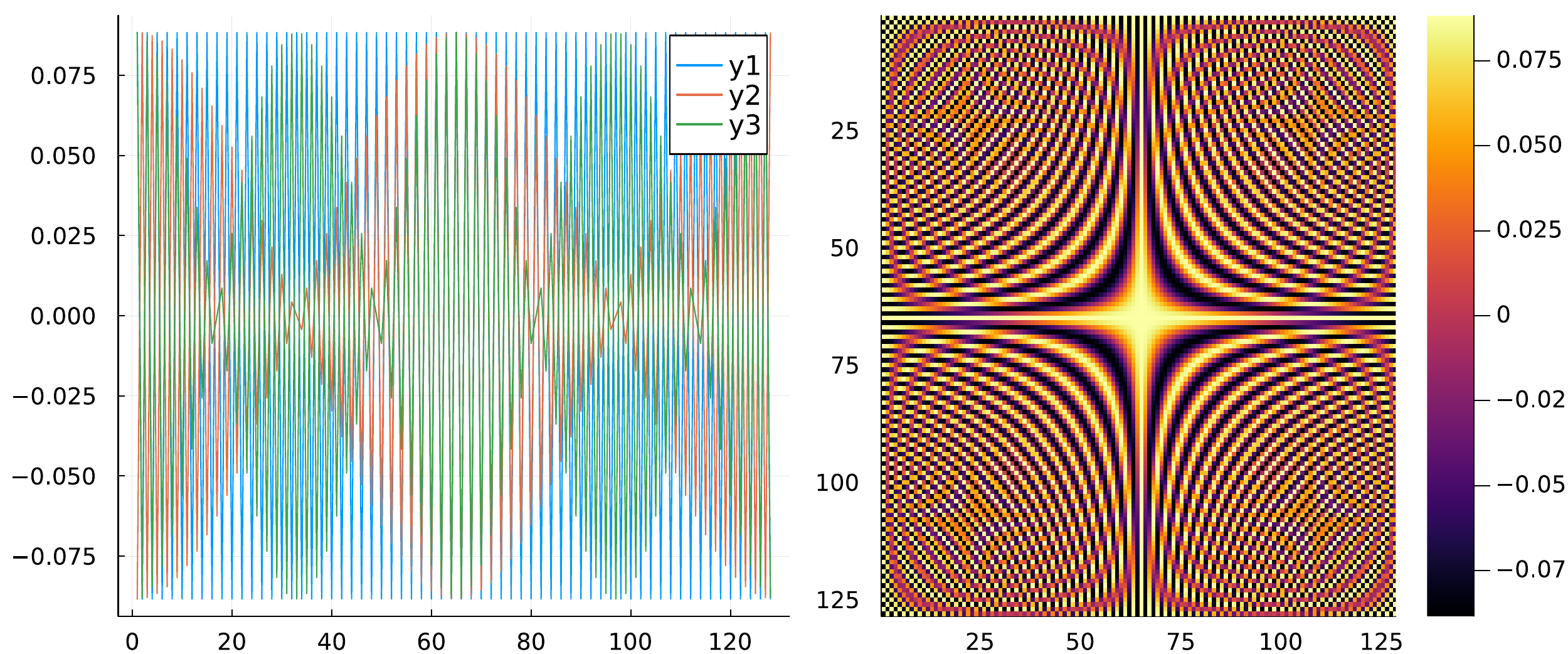}
\caption[Shifted DFT matrix]{Shifted DFT matrix. The left panel plots the first three columns. The right panel visualizes the real part of the shifted DFT matrix.}
\label{fig:shifted-DFT}
\end{figure}

The conventional FFT algorithm implements the non-shifted DFT matrix (\ref{eq:normal-dft}) (with slightly different normalization), but instead of performing a matrix multiplication directly, the FFT algorithm uses recursion to compute the output in $\mathcal{O}(N\log N)$. In order to use out-of-package FFT with the shifted-DFT convention (\ref{eq:shifted-dft}), one needs to insert \texttt{fftshift}, \texttt{ifftshift} in the appropriate places. Again, we refer interested readers to Section \ref{section:code-examples} for implementation details.

\subsection{Parseval's identity}
\label{subsection:parsevals-identity}

Our conventions are nice, because they obey Parseval's identity:
\begin{equation}
    \braket{f|f}=\braket{F|F}
\end{equation}
where $F$ is the discrete Fourier transform of $f$ defined by (\ref{eq:shifted-fft}) and $\braket{\cdot|\cdot}$ is the inner product defined in (\ref{eq:CNMinnerproduct}). To see this, use the trace definition:
\begin{equation}
    \braket{F|F} =\Tr[F^\dagger F]=\Tr[(WfW)^\dagger WfW]=\Tr[f^\dagger f] = \braket{f|f}
\end{equation}
where we used the cyclic property of trace and the fact that $WW^\dagger=W^\dagger W=I$. This identity is the reason why we included the factor of $1/\sqrt{N}$ in the definition of $W$.

\subsection{Discrete Hermite Gaussian Transform}
\label{subsection:discrete-hermite-gaussian-transform}

Now that we have the ability to compute the Fourier transform of the signal, we wish to have a similar matrix $H_{j,j'}$ to project an arbitrary signal onto the first $N$ Hermite Gaussians. We desire the following properties for this matrix:
\begin{enumerate}   
    \item \textbf{Orthonormality}: we want the matrix $H$ to have orthonormal rows, i.e. $\sum_{i=0}^{N-1} H^*_{i,n}H_{i,m}=\delta_{n,m}$. This can be compactly written as $H^\dagger H = I$.
    \item \textbf{Completeness}: we want to span the entire space, meaning that $H H^\dagger = I$. Together with the orthonormality condition, this ensures that $H^\dagger$ is exactly the inverse of $H$, so $H$ is unitary.
    \item \textbf{Eigenvalue property}: we want the columns of $H$ to be eigenfunctions of the shifted DFT matrix $W$ defined in (\ref{eq:shifted-dft}). Ideally, we want to see that $\sum_{j'=0}^{N-1}W_{jj'}h^{(n)}_{j'}=\exp(-in\pi/2)h^{(n)}_j$, where $h^{(n)}_{j}=H_{j,n}$ is the j-th column of the matrix.
    \item \textbf{Continuous limit}: we want to see that $h^{(n)}$ approaches continuous Hermite Gaussian functions defined in (\ref{eq:hermite-functions}) in the limit of $N\to \infty$.
\end{enumerate}
Finding a matrix that would satisfy all of the above is non-trivial, and in some ways impossible, since it's difficult to preserve the eigenvalue property while maintaining the continuous limit \cite{hermiteGaussianTransform}. Instead, we will relax the eigenvalue property and require that \textbf{most} columns have eigenvalues $\exp(-in\pi/2)$ while the high-frequency Hermite-Gaussian will be allowed to slightly deviate in their eigenvalues.

Notice also that direct diagonalization of the matrix $W$ will not work, because the eigenspaces are highly degenerate, i.e., every 4th Hermite-Gaussian has the same eigenvalue. The trick to avoid this (introduced in \cite{hermiteGaussianTransform}) is to diagonalize the Quantum Harmonic Oscillator Hamiltonian (QHO), which generates the Fourier transform.

We start by fixing a natural lattice $L$ on $N$ points \footnote{We multiply its values by $2\pi$ to adhere to the eigenvalue convention\ref{eq:qho-eigenvalues}.} and defining the position matrix $Q_{jj'}=L_j\delta_{j,j'}$. To obtain a momentum matrix, we simply conjugate $Q$ by the Fourier transform. The idea here is that the ``momentum" matrix looks like the position basis under the Fourier change of basis:
\begin{equation}
    P_{j,k} = (WQW^\dagger)_{jk}=\sum_{j',k'=0}^{N-1} W_{j,j'}Q_{j',k'}W^*_{k,k'}=\sum_{j'=0}^{N-1}W_{j,j'}L_{j'}W^*_{k,j'}
\end{equation}
Now that we have $P$ and $Q$, we can define a QHO Hamiltonian:
\begin{equation}
\label{eq:hamiltonian-to-be-diagonalized}
    \mathcal{H}_{j,k} = (P^\dagger P +Q^\dagger Q)_{j,k}
\end{equation}

\begin{figure}[!ht]
\centering
\includegraphics[width=1\columnwidth]{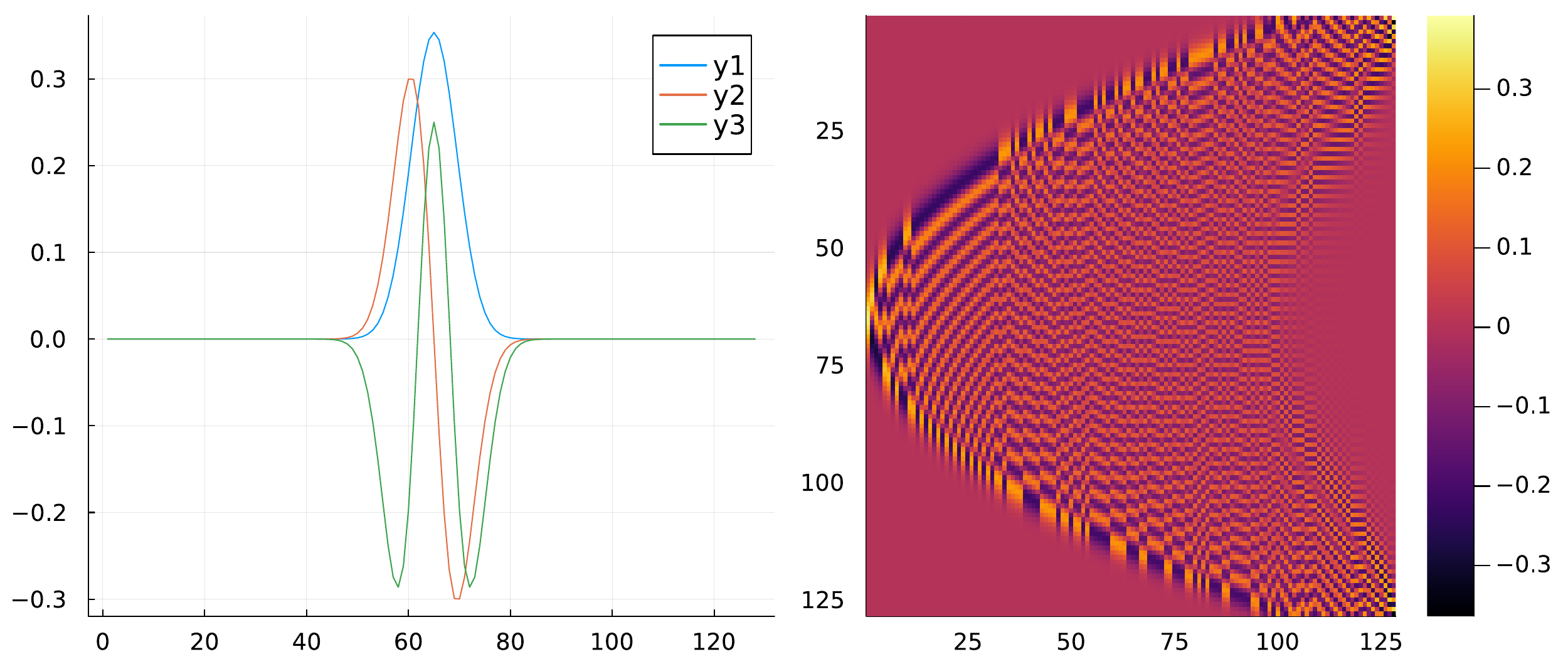}
\caption[Discrete Hermite Gaussian Transform H]{Discrete Hermite Gaussian Transform H. The right panel plots the first three columns. Left panel visualizes the matrix (it is real-valued by construction).}
\label{fig:DHT}
\end{figure}

One can prove that this matrix indeed commutes with $W$ \cite{hermiteGaussianTransform}, so by diagonalizing $\mathcal{H}$, we are also finding a simultaneous eigenbasis for $W$ (see Figure \ref{fig:DHT}). Moreover, the obtained Hermite-Gaussians approach their continuum limit as $N\to\infty$, as shown in \cite{hermiteGaussianTransform}. 

\begin{figure}[!ht]
\centering
\includegraphics[width=0.80\columnwidth]{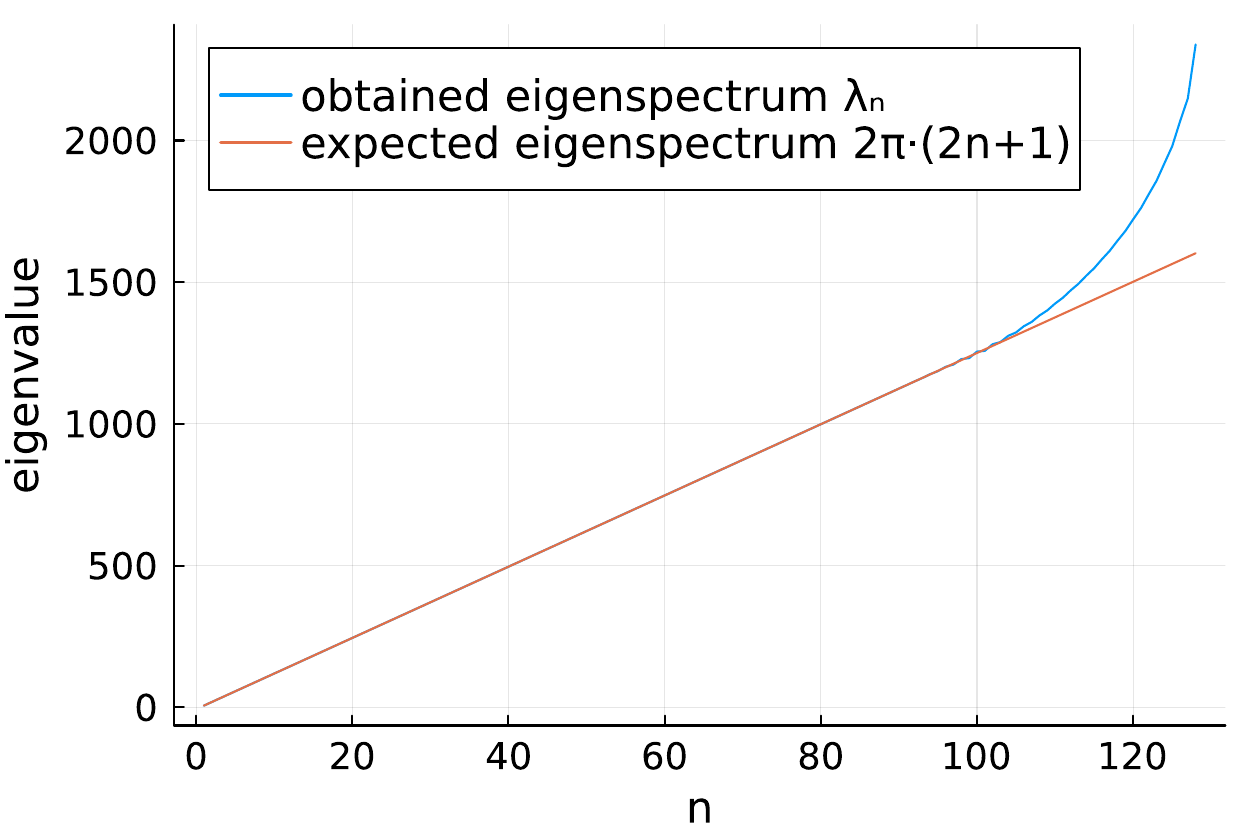}
\caption[Eigenspectrum from diagonalizing $\mathcal{H}$]{Eigenspectrum from diagonalizing $\mathcal{H}$ (defined in \ref{eq:hamiltonian-to-be-diagonalized}). The eigenspectrum deviates from the expected eigenspectrum for the continuous operator $\mathcal{H}$ (defined in (\ref{eq:qho-eigenvalues}))}
\label{fig:eigenvalue-chirp}
\end{figure}

The only downside of this approach is that the obtained eigenspectrum slightly ``chirps" from the expected $2\pi(2n+1)$ for large values of $n$ (see Figure \ref{fig:eigenvalue-chirp}). This can be explained by the fact that for large values of $n$, Hermite-Gaussians become over-aliased, and the chosen discretization can no longer capture the highly oscillatory behavior. That said, this issue is not big in practice, since most of the signals we are dealing with are well-approximated by the first $\floor{N/2}$ Hermite Gaussians. Hence, the coefficients next to the highly oscillating Hermite-Gaussians are usually close to 0.

\begin{figure}[!ht]
\centering
\includegraphics[width=0.80\columnwidth]{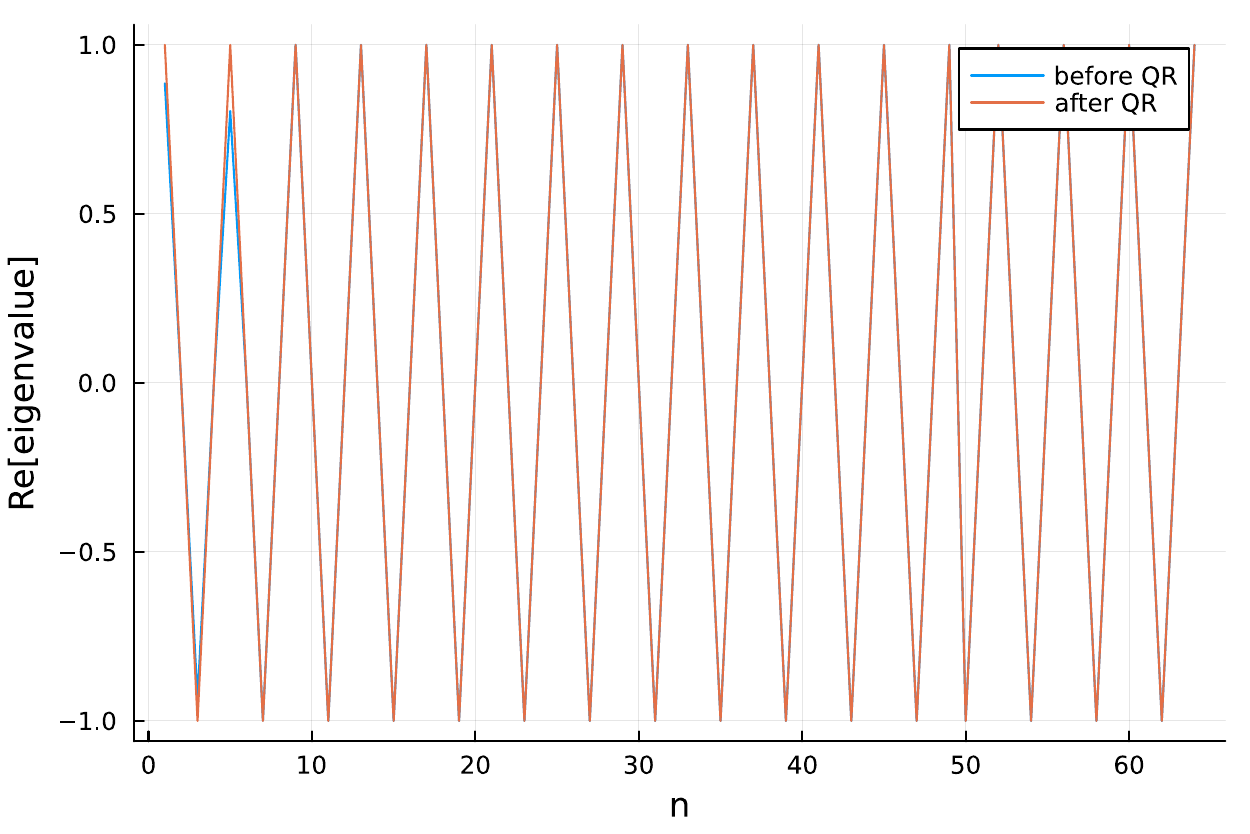}
\caption[Real part of eigenvalues obtained from $\tilde{H}^\dagger W\tilde{H}$ and $H^\dagger WH$]{Real part of eigenvalues obtained from $\tilde{H}^\dagger W\tilde{H}$ and $H^\dagger WH$ where $\tilde{H}$ is the complex-valued matrix obtained from diagonalizing $\mathcal{H}$ (defined in \ref{eq:hamiltonian-to-be-diagonalized}) and $H$ is a symmetrized real-valued matrix after QR decomposition. We can see a slight improvement for the first $\approx$ 10 eigenvalues}
\label{fig:fixed-eigenvalues}
\end{figure}

There is a small technicality that the diagonalization algorithm might output complex-valued Hermite Gaussians $\tilde{h}^{(n)}$. We can do a simple trick to make sure that all Hermite-Gaussians are real-valued. To do that, notice that both $\tilde{h}^{(n)}$ and $(\tilde{h}^{(n)})^*$ are eigenvectors for $H$ defined in (\ref{eq:hamiltonian-to-be-diagonalized}). Thus we can simply take $h^{(n)}=\tilde{h}^{(n)}+(\tilde{h}^{(n)})^*$ to ensure that $h^{(n)}$ is real-valued. In practice, this could mess up the orthonormality condition $H^\dagger H=I$, so we use the QR decomposition of the matrix to renormalize all vectors. One can show that this procedure does not mess up any of the desired properties outlined above. In fact, this symmetrization step, followed by QR decomposition, helps to slightly fix the eigenvalues of $H$ with respect to the shifted DFT $W$ (see Figure \ref{fig:fixed-eigenvalues}). To see the exact algorithm, see section (\ref{section:code-examples}).

\newpage

\subsection{Fractional Fourier Transform}
\label{subsection:fractional-fourier-transform}

\begin{figure}[!ht]
\centering
\includegraphics[width=\columnwidth]{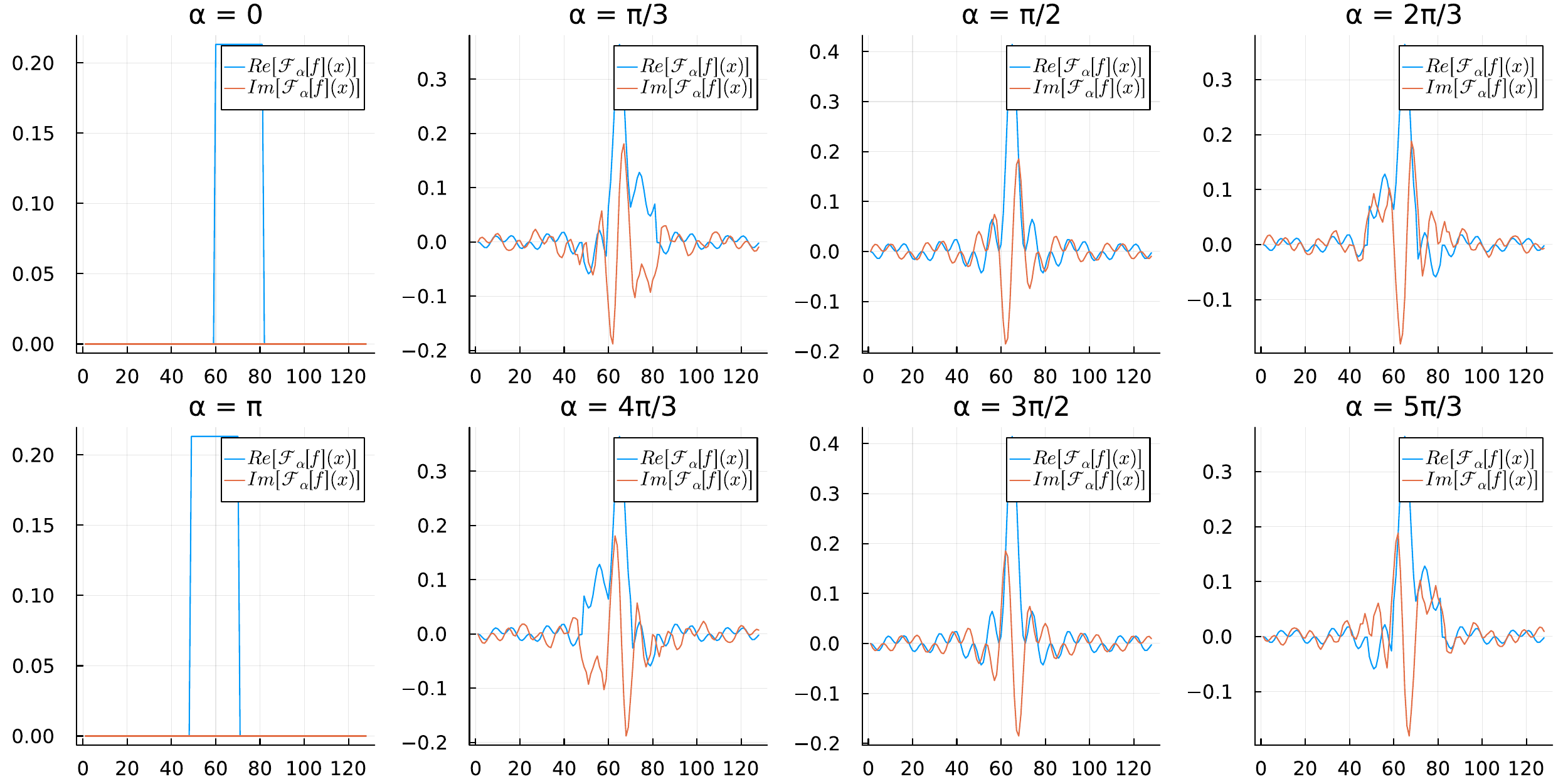}
\caption[Fractional Fourier transform of a rectangular signal for several values of $\alpha$]{Fractional Fourier transform of a rectangular signal for several values of $\alpha$.}
\label{fig:frft-signals}
\end{figure}

Once we obtain a Hermite-Gaussian Transform matrix $H$, it is very easy to construct a Fractional Fourier Transform matrix $W_\alpha$. To do so, consider a discretized version of the definition \ref{eq:frft-defintion}. 

\begin{equation}
    \mathcal{F_\alpha}[f] =\sum_{n=0}^{N-1}a_n e^{-in\alpha}h^{(n)}
\end{equation}

To implement a fractional Fourier transform, we simply have to project our signal onto the Hermite-Gaussian basis, multiply each coefficient by an associated exponent $e^{-in\alpha}$, and then convert the resulting expression into the original basis of the signal. 

\begin{equation}
    W_\alpha=H\Lambda^{2\alpha/\pi}H^T
\end{equation}
where $\Lambda$ are the eigenvalues of the shifted DFT matrix $W$:
\begin{equation}
    \Lambda = H^T W  H
\end{equation}
or in the component form:
\begin{equation}
    \Lambda_{i,j} =\braket{h^{(i)}|Wh^{(j)}}
\end{equation}
where the inner product is 0 whenever $i\neq j$ and extracts the eigenvalue of matrix $W$ associated with the eigenvector $h^{(i)}$  when $i=j$.

\begin{figure}[!ht]
\centering
\includegraphics[width=\columnwidth]{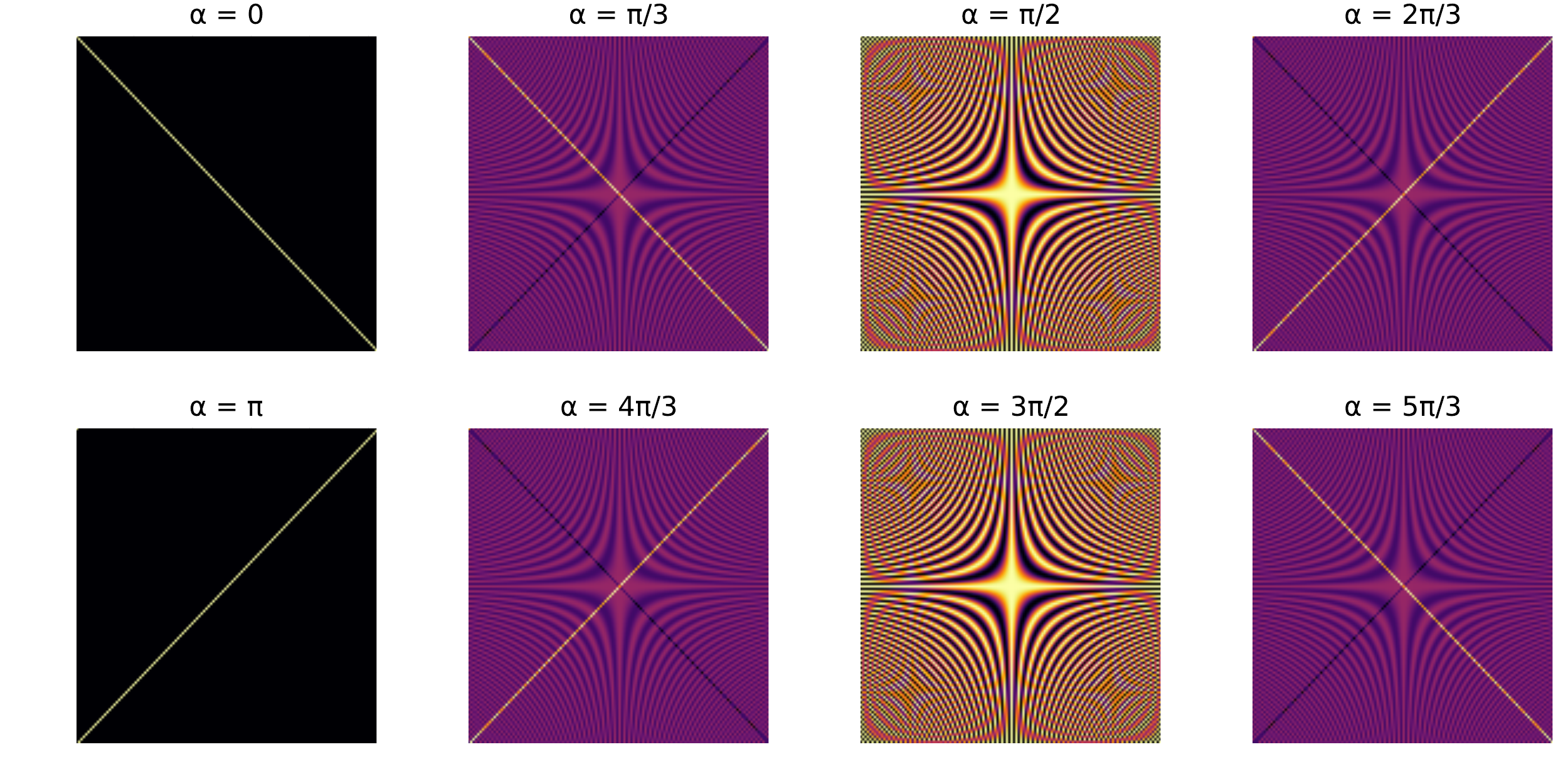}
\caption[Fractional Fourier transform matrix $W_\alpha$ (real part) for a several values of $\alpha$]{Fractional Fourier transform matrix $W_\alpha$ (real part) for a several values of $\alpha$.}
\label{fig:frft-matrix}
\end{figure}

In particular, when $\alpha=\pi/2$, we indeed recover that $W_\alpha=W$:
\begin{equation}
    W_{\pi/2}=H\Lambda H^T=HH^TWHH^T=IWI=W
\end{equation}
Now, it is also easy to show that $W_\pi =\mathcal{P}$ is the parity operator, and $W_{3\pi/2}=W^\dagger$ is the inverse Fourier, which is what we expect in the case of the continuous Fourier transform (see Figures \ref{fig:frft-signals} and \ref{fig:frft-matrix}). The implementation is straightforward and can be found in section (\ref{section:code-examples}).

\subsection{Wigner Distribution}
\label{subsection:wigner-distribution}

Recall the definition of the continuous Wigner distribution:
\begin{equation}
    W_f(u, \mu)=\int f(u+u'/2)f^*(u-u'/2)e^{-i2\pi u'\mu}du'
\end{equation}
It turns out that it is very difficult to discretize the expression above in such a way that it would satisfy all of the properties of the continuous Wigner distribution. One of the main challenges is the evaluation of the function $f$ on points $u\pm u'/2$ in the definition above, which forces us to have different lattices for $u$ and $u'$ variables. The different lattices make it difficult for the discrete Wigner implementation to satisfy the fractional Fourier transform property (\ref{fig:wigner-frft-commute}).

The authors of \cite{aliasWigner} have surveyed several methods and concluded that none of them satisfy all of the desired properties of the continuous Wigner distributions. They conjectured that there is no discrete implementation that would satisfy all of the properties. So, similarly to the discrete Hermite Gaussian transform, we must make some compromises. We insist on the following properties:
\begin{enumerate}   
    \item \textbf{Marginals}: we want the position and Fourier marginals to match the signal exactly. As a consequence, we want the Wigner distribution to sum up to the total energy, which is 1 for normalized distributions.
    \item \textbf{Compact support}: we want the Wigner distribution of $f$ to have the same support as the input function. For instance if $\text{supp} f=[-1,1]$ then $W(u,\mu)=0$ whenever $u$ is outside $[-1,1]$.
    \item \textbf{Continuous limit}: we want the Wigner Distribution to match the continuous limit at least for some simple, well-behaved signal, such as $\text{rect}(u)$.
\end{enumerate}
Notice that we do not require \textbf{all} of the marginals to be matched, but in practice, we will see that FrFT will roughly rotate our phase-space distribution.

There are two ways of implementing the discrete Wigner Distribution — using FFT and using Hermite-Gaussian decomposition of the Wigner Distribution \ref{eq:wigner-hermite-decompostion}, which leverages the connection to the Laguerre polynomials \ref{eq:wigner-laguerre-formula}. The FFT approach is $\mathcal{O}(N^2\log N)$ in time and $\mathcal{O}(N^2)$ in space. The Laguerre approach is $\mathcal{O}(N^4)$ in time and $\mathcal{O}(N^2)$ in space. Furthermore, we found a lot of numerical instabilities using the Laguerre method, so we highly recommend using the FFT approach.
\begin{figure}[!ht]
\centering
\includegraphics[width=\columnwidth]{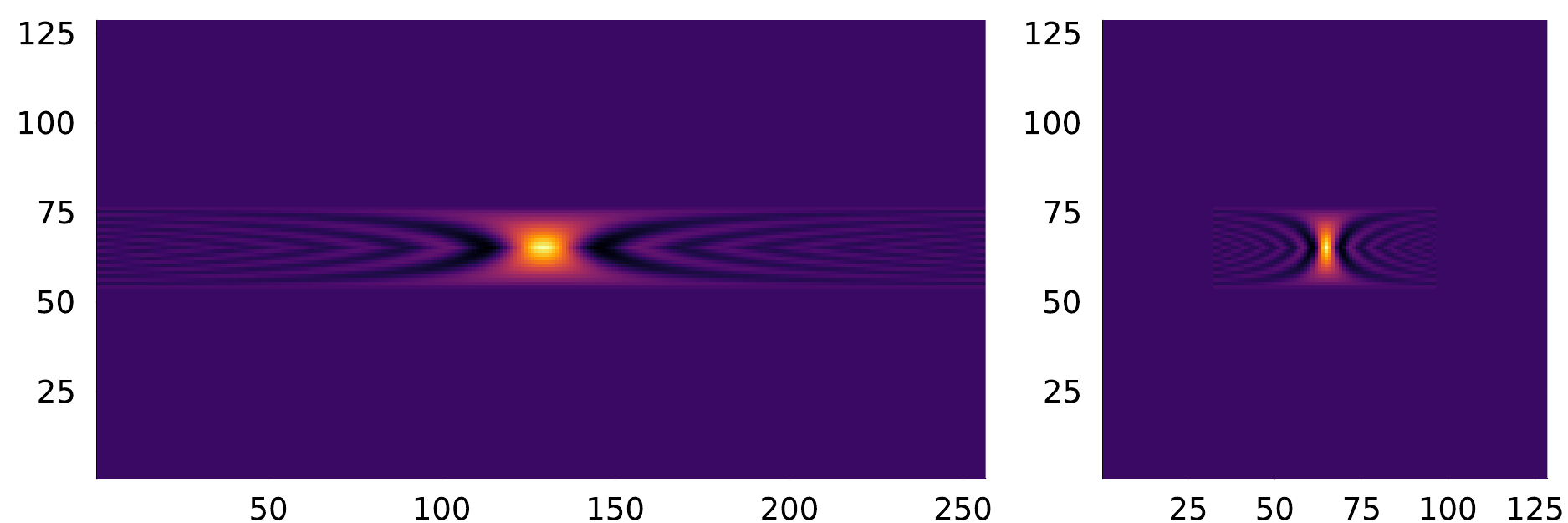}
\caption[Wigner distribution using FFT]{Direct output of the \texttt{wigner\_fft} (left) and the interpolated version onto a square natural lattice (right) for a signal $f(u)=\text{rect}(u)$.}
\label{fig:wigner-fft}
\end{figure}

\subsubsection{FFT Wigner}
\label{subsection:fft-wigner}

In this case, the input is a complex-valued signal $f$ and an associated natural lattice at points $N$, $L_0$. This approach involves precomputing $f(u+u'/2)$ and $f(u-u'/2)$ on a grid of points. Notice that we can use Topelitz matrices to efficiently store the data. Once we have a matrix of points $f(u+u'/2)f(u-u'/2)$, we simply perform a shifted FFT to get the desired Wigner distribution. 

\begin{figure}[!ht]
\centering
\includegraphics[width=\columnwidth]{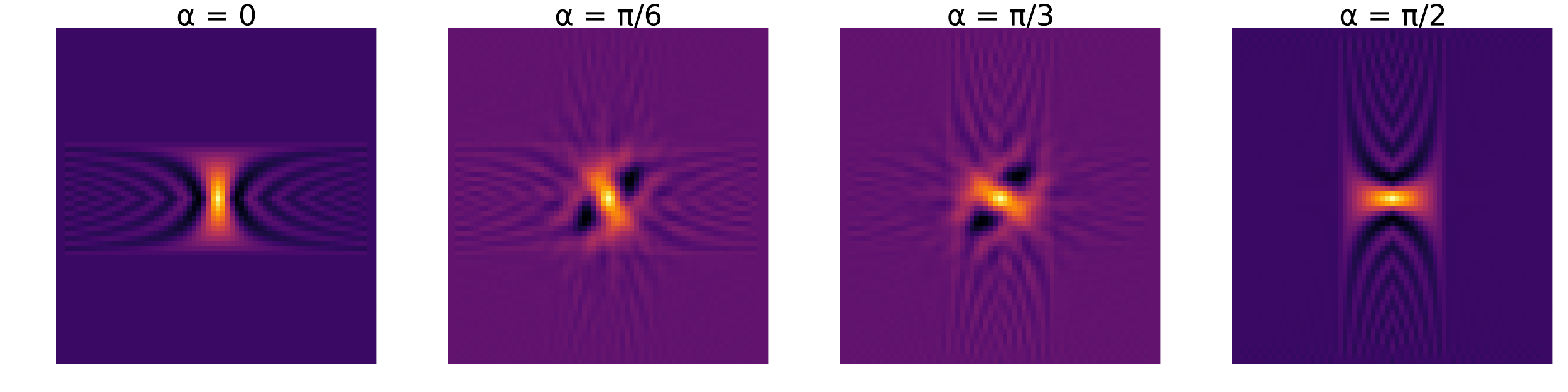}
\caption[Rotated Wigner distribution using FFT]{\texttt{wigner\_fft} applied to $\mathcal{F}_\alpha[f]$ where $f(u)=\text{rect}(u)$. We see a perfect match for $\alpha=0$ and $\alpha=\pi/2$ and a rough match in intermediate planes. The figure is $100\times100$ pixel cropped around the center.}
\label{fig:wigner-frft}
\end{figure}

Notice that because of $u'/2$, we have to perform padding in the input domain, which forces the momentum variable $\mu$ to live on a lattice with $2N$ points and spacing of $1/4\sqrt{N}$ (notice that this is not a natural lattice). The final step (if desired) is to perform the interpolation, such that both $x$ and $X$ live on a natural lattice with $N$ points. See the details in \ref{section:code-examples}. Also, see the example of our implementation applied to a rectangular signal (see Figure \ref{fig:wigner-fft}). We see that this method approaches the continuous limit (see Figure \ref{fig:wigner-rect}). Furthermore, this implementation roughly reproduces the marginals after a FrFT is applied to the signal — in other words, the FrFT operator rotates our Wigner distribution, as desired.

\subsection{Normalization}
\label{subsection:normalization}

There are several natural ways to normalize the beam depending on the application. For the majority of this thesis, we will use the unitary convention.

\subsubsection{Unitary convention}
We can normalize our function using the inner product (\ref{eq:CNMinnerproduct}):
\begin{align}
\label{eq:unitary-normalization}
    ||f||^2&=\sum_{ij}|f_{ij}|^2=\Tr[f^\dagger f]\\
    \tilde{f}_{ij} &= f_{ij}/||f||
\end{align}
This normalization ensures that both $f$ and its DFT pair $F$ have norm one (via the Parseval's identity). In addition, there is a nice interpretation of normalized $f\in \mathbb{C}^{N\times N}$ as a quantum state with $N^2$ components and outcome probabilities $|f_{ij}|^2$, which sum up to 1.

\subsubsection{Energy convention}
Another way to define a norm would be to explicitly include discretization factors $\Delta u$ and $\Delta v$.
\begin{align}
    ||f||^2=\sum_{ij}|f_{ij}|^2\Delta u\Delta v
\end{align}
This norm has an advantage by reproducing the continuum norm \ref{eq:continuum-norm} in the large $N$ limit (recall $\Delta u=1/\sqrt{N}$ for natural lattices). Also, in this interpretation, we can treat $|f_{ij}|^2$ as the energy density (i.e., energy per unit area) and the norm $||f||$ as the total energy of the beam. This is useful if we want to test efficiency or accuracy across different discretizations $\Delta u$, $\Delta v$.

\subsubsection{ROI convention}
Finally, in some cases, we will only care about the intensity on a subset of the lattice, which we call the region of interest (ROI). An example would be the MRAF algorithm, which optimizes intensity only within the given ROI. In this case, it is useful to normalize the norm to be 1 within the ROI. We do so by restricting indices of the summation.
\begin{align}
    ||f||^2&=\sum_{(i,j)\in\text{ROI}}|f_{ij}|^2\\
    \tilde{f}_{ij} &= f_{ij}/\sqrt{||f||}
\end{align}

\subsection{Loss Functions}
\label{subsection:loss-functions}
After the appropriate normalization is performed, we can measure the loss between the target amplitude $g$ and the reconstructed amplitude $f$.
\begin{equation}
    L_{\text{amp}}(f,g)=\sqrt{\sum_{i,j}|f_{i,j}-g_{i,j}|^2}
\end{equation}
Notice that this is exactly $||f-g||$ with the norm from $l^2$ induced from (\ref{eq:CNMinnerproduct}). This loss is measured between the \textit{amplitudes} of the beams, so ultimately it depends on both the real and imaginary parts of $f$ and $g$. 

There are some cases where we do not care about the phase of the resulting light-field $f$. Then, it makes more sense to compute the loss between \textit{intensities} of the beams as follows:
\begin{equation}
\label{eq:Lint-def}
    L_{\text{int}}(f,g)=\sum_{i,j}||f_{i,j}|^2-|g_{i,j}|^2|
\end{equation}
which can be viewed as an L1 norm between intensities $|f_{i,j}|^2$ and $|g_{i,j}|^2$. Notice that $L_\text{amp}$ penalizes phase mismatch, while $L_\text{int}$ ignores the phase completely. For normalized distributions, we can bound one loss in terms of the other. \footnote{This theorem was formulated and proven by my close collaborator, Hunter Swan.}
\newpage
\begin{theorem}
    Let $f,g\in \operatorname{Mat}_{N\times M}(\mathbb{C})$ be two normalized functions according to the unitary convention (\ref{eq:unitary-normalization}) i.e. $||f||_{l^2}=1$ and $||g||_{l^2}=1$. Then $L_{\text{amp}}$ and $L_{\text{int}}$ are related via the following inequalities:
    \begin{align}
        L_{\text{int}}(f,g) &\leq2L_{\text{amp}}(f,g) \\
        L_{\text{amp}}^2(|f|,|g|) &\leq L_{\text{int}} (f,g)
    \end{align}
\end{theorem}

\begin{proof}
We start with the proof from the showing the first inequality.
    \begin{align*}
        L_{\text{int}}(f,g) &= \sum_{i,j}||f_{i,j}|^2-|g_{i,j}|^2| \\
        &=\sum_{i,j}||f_{i,j}|-|g_{i,j}||\cdot||f_{i,j}|+|g_{i,j}|| \\
        &\leq \sqrt{\sum_{i,j}||f_{i,j}|-|g_{i,j}||^2}\sqrt{\sum_{i,j}||f_{i,j}|+|g_{i,j}||^2} \quad \quad \text{(by Cauchy-Schwartz)} \\
        &\leq \sqrt{\sum_{i,j}|f_{i,j}-g_{i,j}|^2}\sqrt{\sum_{i,j}|f_{i,j}+g_{i,j}|^2} \quad \quad \text{(by reverse } \triangle\text{-inequality)} \\
        &= L_{\text{amp}}(f,g)||f+g||_{l^2}\\
        &\leq L_{\text{amp}}(f,g)(||f||_{l^2}+||g||_{l^2}) \quad \quad \text{(by } \triangle\text{-inequality)}  \\
        &\leq 2 L_{\text{amp}}(f,g)
    \end{align*}
Now for the other direction:
\begin{align*}
        L^2_{\text{amp}}(|f|,|g|) &= \sum_{i,j} ||f_{i,j}| - |g_{i,j}||^2\\
        &=\sum_{i,j} ||f_{i,j}| - |g_{i,j}||\cdot ||f_{i,j}| - |g_{i,j}||\\
        &\leq \sum_{i,j} ||f_{i,j}| - |g_{i,j}||\cdot ||f_{i,j}| + |g_{i,j}||\\
        &=L_{\text{int}}(f,g)
\end{align*}
\end{proof}
We note that these are \textbf{not} equivalent distances in the topological sense because there is no constant $C$ for which  $L_{\text{amp}}(f,g) \leq C L_{\text{int}} (f,g)$. So, these two distances induce different topologies on the space of $\operatorname{Mat}_{N\times M}(\mathbb{C})$. What is true, however, is that the convergence with respect to the $L_\text{amp}$ distance implies convergence with respect to the  $L_\text{int}$ distance, but the other direction is false.

Another remark is that this argument can be trivially generalized to spaces $L^2(\mathbb{R}^2)$ and $L^2(D)$ if we simply replace summations with integrals.

\section{Code examples}
\label{section:code-examples}
Below, we provide code implementations in Julia for most of the numerical algorithms described in this thesis. We also include some code snippets for simple functions like shifted FFT and lattice coordinates to avoid any possible confusion with our conventions. We hope that this section proves useful for people who want to re-implement some of our algorithms for their own applications.

We would like to point out that all of this code (in a slightly more organized and general form) can be found in our open-source package \texttt{SLMTools} \cite{SLMTools}. If you find this code useful for your work, we kindly ask you to cite our work. 

\subsection{Julia code} 

\subsubsection{Lattice Coordinates}
Very often it's useful to retrieve the coordinates associated with the coordinate lattice. Using these coordinates makes it easy to discretize a continuous function over a lattice. See the following code snippet for an example.

\begin{minted}{julia}
function getCenteredCoordinates(N::Integer, M::Integer, dx::Real, dy::Real)
    """Generates Centered coordinates given lattice parameters"""
    x = range(-floor(N/2) * dx, stop=floor((N - 1)/2) * dx, length=N)
    y = range(-floor(N/2) * dy, stop=floor((N - 1)/2) * dy, length=M)
    Lx = repeat(x', N, 1)
    Ly = repeat(y, 1, M)
    return Lx, Ly
end

# example use
N = 128
M = 128
dx = 1 / sqrt(N)
dy = 1 / sqrt(M)
Lx, Ly = getCenteredCoordinates(N,M,dx,dy)
fgauss(x, y) = exp.(-(x.^2 .+ y.^2) ./ 2)
fij = fgauss(Lx, Ly)
\end{minted}

\subsubsection{Basis change}
Below we provide the code for generating shifted DFT, Hermite-Gaussian, and FrFT bases. We also show how to change bases for the \texttt{fij} matrix we defined above.

\newpage

\begin{minted}{julia}
using LinearAlgebra

function shiftedDFTBasis(N::Integer)
    """Generates shifted DFT basis"""
    shift = N ÷ 2
    W = [1/sqrt(N)*exp(-1im * 2 * π * (j-shift)*(k-shift)/N) for j in 0:N-1, k in 0:N-1];
    return W
end 

function hermiteBasis(N::Integer)
    """Generates discrete Hermite basis"""
    
    # scale L by 2pi (eigenvalue convention)
    L = range(-floor(N/2) * dx, stop=floor((N - 1)/2) * dx, length=N)
    L = L .* 2π

    # Generate a shifted DFT matrix
    W = shiftedDFTBasis(N)
    
    # Form a Hamiltonian matrix
    Q = diagm(L)
    P = W*Q*W'
    QHO = P'*P + Q'*Q

    # Perform the diagonalization
    λ, H1 = eigen(QHO);

    # QR decomposition to make them real-valued
    H1 = real((H1 + conj(H1))/2)
    H,_ = qr(H1)
    H = Matrix(H)
    
    return H, λ
end

function FrFTBasis(N::Integer, α::Real)
    """Generates FrFT basis"""
    
    # generates hermite basis
    W = shiftedDFTBasis(N)
    H, λ = hermiteBasis(N)
    
    # generates FrFT matrix
    λw = diag(H' * W * H)
    Λ = diagm(λw.^(2α/π))
    Wα = H * Λ * transpose(H)
    return Wα
end

# example use
N = 128
α = π/3

# basis change matrices
W = shiftedDFTBasis(N)
H = hermiteBasis(N)[1]
Wα = FrFTBasis(N, α)

# new representations 
# fij is defined above
F1ij = transpose(W) * fij * W
F2ij = transpose(H) * fij * H
F3ij = transpose(Wα) * fij * Wα
\end{minted}

\subsubsection{FFT implementation}

To change the basis to the Fourier domain, we recommend using the FFT approach over the direct matrix product with $W$. FFT scales as $\mathcal{O}(N \log N)$ as opposed to $\mathcal{O}(N^3)$ for the naive matrix multiplication algorithm shown above. We show that two approaches yield the same result.

\begin{minted}{julia}
using FFTW

function sft(fij::Matrix{T}) where {T<:Number}
    """Computes the shifted DFT (unitary convention) using FFT algorithm."""
    N = size(fij)[1]
    Fij = fftshift(fft(ifftshift(fij))) * 1/N
    return Fij
end

function isft(Fij::Matrix{T}) where {T<:Number}
    """Computes the inverse shifted DFT (unitary convention) using FFT algorithm."""
    N = size(Fij)[1]
    fij = ifftshift(ifft(fftshift(Fij))) * N
    return fij
end

# example use
Fij = sft(fij)
fij_ = isft(Fij)
# make sure it's unitary
@assert norm(fij.-fij_) < 1e-12
Ef = sum(fij .* conj.(fij))
EF = sum(Fij .* conj.(Fij))
@assert norm(Ef - EF) < 1e-12
# compare with W contraction
@assert norm(Fij - transpose(W) * fij * W) < 1e-12
\end{minted}

\subsubsection{Wigner Implementation}
Below you can find our FFT-based implementation of the Wigner distribution. This implementation is partially inspired by the \href{https://qutip.org/docs/4.0.2/modules/qutip/wigner.html}{implementation} from the QuTiP package. 
\begin{minted}{julia}
using ToeplitzMatrices

function wigner_fft(f::Vector{T}) where T<:Number
    """
    Computes the Wigner distribution using FFT method. The input 
    is any complex-valued function discretized on N points. 
    The output is an Nx2N array of real numbers. The first coordinate 
    is on the same lattice as the input data. The second coordinate 
    is on the lattice that has twice as many points, and is half as big 
    in the extent.
    """    
    # Reshape function
    ψ = reshape(f,1,:).|>ComplexF64
    N = size(f)[1]
    
    # Construct r1 and r2
    r1 = hcat([0], reverse(conj.(ψ)), zeros(1, N-1))
    r2 = hcat([0], ψ, zeros(1, N-1))
    
    # Create Toeplitz matrices
    w = Toeplitz(zeros(N), r1[:]) .* reverse(Toeplitz(zeros(N), r2[:]), dims=1)
    
    # compute FT 
    w = 1/(2N) * ifftshift(fft(fftshift(w, 2),2),2)
    w = real(w) # always real anyway
    return w
end

# example use
N = 128
rectf = [Int(abs(1/sqrt(N)*j) < 1) for j in -floor(N/2):floor((N - 1)/2)]
Wrect = wigner_fft(rectf)
\end{minted}

If one desires, it is possible to interpolate the final result to a natural lattice in both $x$ and $X$. As an example, we can use the package \href{https://github.com/hoganphysics/SLMTools}{SLMTools} \footnote{https://github.com/hoganphysics/SLMTools} \cite{SLMTools} to perform the interpolation step, assuming that the input function lives on a natural lattice.

\begin{minted}{julia}
using SLMTools

# define lattices
Lx = natlat(N)[1]
Lp = natlat(2*N)[1] ./ (2*sqrt(2))

# create a lattice field object
Wlf = LF{RealAmplitude}(Wrect, (Lx, Lp))

# downsample and multiply by 4 to keep the norm the same
Wlfdown = downsample(Wlf,(Lx, Lx))
Wlfdown = Wlfdown * 4
\end{minted}

We designed \texttt{SLMTools} to work with light field (\texttt{LF}) objects, which is a data structure that stores the array, along with the lattices that assign coordinates to the data stored in the array. It is very convenient to keep track of the coordinate grids as we often interpolate between different lattices depending on the application. For more details, see the documentation of \texttt{SLMTools} \cite{SLMTools}.

\subsubsection{Sinkhorn 2D}
\label{subsubsection:sinkhorn2D}
Here is our current implementation of the 2D Sinkhorn Algorithm that we described in Subsection \ref{subsection:efficient-2d-implementation} of Chapter \ref{chapter:algorithms}. We note that this algorithm is currently numerically unstable for values of $\epsilon$ less than $0.003$, but we are currently working on improving it further.
\newpage
\begin{minted}{julia}
function SinkhornConv2D(
    μ::Matrix{T}, 
    ν::Matrix{T}, 
    eps::Real, 
    max_iter::Integer
) where {T<:Real}
    """Our implementation of the Sinkhorn algorithm for 2D distributions. 
    μ and ν are the input and output intensities, eps is the regularization 
    parameter, and max_iter is the maximum number of iteration. We note 
    that the algorithm might be numerically unstable for eps<0.003"""
    
    # parameters
    N = size(μ)[1]
    u = ones(N, N)./N^2
    v = ones(N, N)./N^2 # this initialization doesn't matter

    # loss tracking
    loss = []
    prev = copy(u) 
    
    # create convolution matrix
    X, Y = natlat((N,N))
    A = exp.(-(X.^2 .+ Y'.^2) / (2 * N * eps))
    FA = sft(A)

    for i in range(1, max_iter)

        # row constraint
        row_sum = real.(isft(sft(u).*FA))
        v = ν ./ row_sum
        v = v ./ sum(v)
        
        # col constraint
        col_sum = real.(isft(sft(v).*FA))
        u = μ ./ col_sum
        u = u ./ sum(u)

        # log loss
        push!(loss, norm(u-prev))
        prev = copy(u)

    end
    
    # last iterarion (no normalization)
    row_sum = real.(isft(sft(u).*FA))
    v = ν ./ row_sum
    col_sum = real.(isft(sft(v).*FA))
    u = μ ./ col_sum
    return u, v, loss
end

function dualToGradients(
    u::Matrix{T}, 
    v::Matrix{T}, 
    μ::Matrix{S}, 
    eps::Real
) where {T<:Real, S<:Real}
    """Computes the gradients using the dual variables of the Sinkhorn algorithm"""
    # create convolution matrix
    N = size(u)[1]
    X, Y = natlat(N,N)
    A = exp.(-(X.^2 .+ Y'.^2) / (2 * N * eps))
    FA = sft(A)
    
    # moment calculation
    XV = X .* v
    YV = Y' .* v

    # convolution step
    AXV = real.(isft(sft(XV).*FA))
    AYV = real.(isft(sft(YV).*FA))
    
    # compute gradients
    gradphix = u .* AXV ./ μ
    gradphiy = u .* AYV ./ μ
    return gradphix, gradphiy
end
\end{minted}

Once the gradients are computed, it only remains to integrate them to obtain the unwrapped phase. In \texttt{SLMTools} we use a trapezoid method of integration, which works slightly better then naive \texttt{cumsum}. See the code below that shows how to compute the final optimal transport phase.

\begin{minted}{julia}
function otQuickPhase(
    g2::LF{Intensity,T,N}, G2::LF{Intensity,T,N}, 
    eps::Real, max_iter::Integer
) where {T<:Real,N}
    u, v, loss = SinkhornConv2D(g2.data, G2.data, eps, max_iter) 
    gradphix, gradphiy = dualToGradients(u, v, g2.data,  eps)
    phi = scalarPotentialN(cat(gradphix,gradphiy; dims=3), g2.L)
    return phi
end 
\end{minted}
Below is an example usage of this function using \texttt{SLMTools}.\footnote{Credit to Indra Periwal for the example code.}

\begin{minted}{julia}
# Generate an input grid and corresponding output grid
N = 1024
L0 = natlat((N,N))

# Generate an input beam and target output beam
inputBeam = lfGaussian(Intensity, L0, 4.0)
targetBeam = lfRing(Intensity, L0, 4, 1)#+1e-7
display(look(inputBeam, targetBeam))

# Use optimal transport to find an SLM phase to make an approximate output beam
phiOT = otQuickPhase(inputBeam,targetBeam,0.0035, 100)
outputOT = square(SLMTools.sft(sqrt(inputBeam) * phiOT))
look(targetBeam,outputOT)
\end{minted}

%% file: appendix/C.tex
\chapter{Primer on Convex Optimization}
\label{chapter:primer-on-convex-optimization}

The goal of this section is to introduce the reader to the basics of convex optimization and to introduce some basic notation and terminology, which we will use in the following chapters. Most of the notation is adopted from \cite{boyd2004convex}.

\section{Convex Basics}
\label{section:convex-basics}

\subsection{Standard Optimization Form}
\label{subsection:standard-optimization-form}

Assume we have the following optimization problem:
\begin{align}
\label{eq:general-optimization}
\min_{x} \quad &f_0(x)\\
\textrm{s.t.} \quad & f_i(x)\leq 0 \quad  i\in \{1,\dots, m\} \nonumber\\
\quad & h_i(x)= 0  \quad i\in \{1,\dots, p\} \nonumber
\end{align}
where $x\in \mathbb{R}^N$ is called the \textit{optimization variable}. The function $f_0:\Omega_0\to\mathbb{R}$ is the \textit{objective function}, defined on a domain $\Omega_0\subset\mathbb{R}^N$, which we are trying to minimize. The functions $f_i:\Omega_i\to\mathbb{R}$ are \textit{inequality constraints} defined on domains $\Omega_i\subset\mathbb{R}^N$ for $i\in \{1,\dots, m\}$. And, similarly, functions $h_i:\Omega_i\to\mathbb{R}$ are \textit{equality constraints} defined on domains $D_i\subset\mathbb{R}^N$. Next, we define \textit{feasibility domain} $\mathcal{D}$:
\begin{equation}
\label{eq:feasibility-domain}
\mathcal{D}=\left(\bigcap_{i=0}^m\Omega_i\right)\cap\left(\bigcap_{j=1}^pD_j\right)
\end{equation}
We will call a point $x\in\mathbb{R}^N$ \textit{feasible} if $x\in\mathcal{D}$. The goal of the optimization problem is to find a feasible point $x^\star$ that satisfies the constraints and minimizes the function $f_0$.

\subsection{Convex Optimization}
\label{subsection:convex-optimization}

Consider a special case of the optimization problem (\ref{eq:general-optimization}) such that:
\begin{enumerate}
    \item the objective $f_0$ is convex 
    \item inequality constraints $f_1,\dots,f_m$ are convex 
    \item equality constraints $h_1,\dots,h_p$ are affine
\end{enumerate}
When these conditions are met, we call such a problem \textit{convex optimization problem}. Convex problems have some nice properties that make them easy to solve. Practically speaking, convex problems do not have local minima where the optimization algorithm can get stuck. This does not mean that a solution is unique. Consider, for example, a trivial minimization problem where $f_0(x)=0$ with one additional constraint $f_1(x)=|x-1| \leq 0$. notice that this is a convex problem, where any $x^\star\in[-1,1]$ is a solution. What is important, however, is that the feasibility domain $\mathcal{D}$ and the optimal set of solutions are both convex sets, which makes it easy to optimize. 

Another important feature of this definition is that we did not require any of the convex functions $f_i$ to be differentiable. It turns out that even non-differentiable convex functions can be optimized using techniques of sub-differential calculus. In practice, convex optimization software requires you to compile the problem using a library of known convex functions, so that sub-gradients can be internally computed. 

\subsection{Change of Variables}
\label{subsection:change-of-variables}

The definition of the standard optimization problem may seem a bit restrictive, but it turns out that a lot of problems can be reduced to the problem (\ref{eq:general-optimization}) by some simple transformations. In general, we will call two optimization problems \textit{equivalent} if the solution to one can be easily obtained from the solution to the other (it is possible to make this definition precise \cite{boyd2004convex}). 

For instance, suppose we introduce a \emph{nonsingular affine} change of variables
$x = T y + c$ with $T \in \mathbb{R}^{N \times N}$ invertible and $c \in \mathbb{R}^N$.
Substituting into (\ref{eq:general-optimization}) yields

\begin{align}
\min_{y}\quad & f_0(Ty + c) \\
\text{s.t.}\quad &
      f_i(Ty + c) \le 0, \qquad i = 1,\dots,m, \\
&     h_j(Ty + c) = 0, \qquad j = 1,\dots,p.
\end{align}
Because $T$ is invertible, every feasible $x$ corresponds to exactly one $y$
and vice versa; thus, the two problems have the same optimal value,
and any optimizer satisfies $x^{\star} = T y^{\star} + c$.
Simple rescalings of the objective or constraints (e.g.\ multiplying all terms by a positive constant) are likewise equivalent. Such transformations let us rewrite a problem in whatever coordinates or units are most convenient while preserving solvability and optimality guarantees.

\subsection{Duality}
\label{subsection:duality}

There is a standard ``trick" for reformulating optimization problems, which is known as duality. It turns out that most\footnote{I say ``most" just in case some angry mathematician will find an optimization problem for which this is not possible} minimization optimization problems have an associated dual maximization problem. In general, it is always true that maximizing the dual gives the lower bound on the optimal value of the primal (original) optimization problem. The difference between the minimum of the primal objective and the maximum of the dual objective is known as the duality gap.

In a special case of convex problems, the duality gap is almost always zero\footnote{Rigorously, for this to hold, we need Slater's constraint qualification to be satisfied. In practice, this is always true for any ``real-world" convex optimization problem \cite{boyd2004convex}}. In fact, one can rigorously prove that this is always true for the optimal transport problem (\ref{eq:primal-OT}) and its entropic relaxation (\ref{eq:entopy-OT})\footnote{One can just observe that the outer product distribution of $\mu$, $\nu$ satisfies Slater's constraint qualification. So, strong duality holds, which means that the duality gap is 0.}. So, let us construct the dual maximization problem

There is a relatively standard procedure to convert a given optimization problem to its dual form. We refer the interested reader to Appendix C, which outlines this procedure, and also for a more in-depth discussion, we recommend chapter 5 from the textbook  \cite{boyd2004convex}.

\paragraph{Lagrangian.}
Introduce Lagrange multipliers
\[
\lambda\in\mathbb{R}^{m}_{\!+},\qquad
\nu\in\mathbb{R}^{p}
\]
corresponding to the inequality and equality constraints, respectively.  
The \emph{Lagrangian} is defined as
\begin{align}
\label{eq:general-lagrangian}
\mathcal{L}(x,\lambda,\nu)
   &=f_{0}(x)
     +\sum_{i=1}^{m}\lambda_{i}\,f_{i}(x)
     +\sum_{j=1}^{p}\nu_{j}\,h_{j}(x).
\end{align}
(The implicit domain of $\mathcal{L}$ is $x\in\mathcal{D}$ and
$\lambda\ge0$.)

\paragraph{Dual function.}  
The \emph{Lagrange dual} is obtained by minimizing the Lagrangian over
the primal variable:
\[
g(\lambda,\nu)
  \;=\;
  \inf_{x\in\mathcal{D}}\mathcal{L}(x,\lambda,\nu)
  \;=\;
  \inf_{x\in\mathcal{D}}
  \Bigl\{
     f_{0}(x)\;+\!\sum_{i=1}^{m}\lambda_{i}f_{i}(x)
     +\!\sum_{j=1}^{p}\nu_{j}h_{j}(x)
  \Bigr\}.
\]
For any fixed $(\lambda,\nu)$ with $\lambda\ge0$, $g(\lambda,\nu)$ is
the greatest lower bound that the Lagrangian can attain when $x$ varies
over the feasible domain.  If $\lambda\not\ge0$, the function takes the
value $-\infty$ by convention.

\paragraph{Lagrange dual problem.}
Maximizing the dual function gives the \textit{dual problem}:
\begin{align}
\label{eq:general-dual}
\max_{\lambda\in\mathbb{R}^{m}_{\!+},\,\nu\in\mathbb{R}^{p}}\quad
    &g(\lambda,\nu).
\end{align}

\section{Applications to Optimal Transport}
\label{section:application-to-optimal-transport}

We will apply the procedure outlined above to obtain dual reformulations of the optimal transport problem \ref{eq:primal-OT} and its entropic regularization \ref{eq:entopy-OT}. We start by formulating the usual definition of the discrete optimal transport problem.

\subsection{Discrete Optimal Transport}
\label{subsection:discrete-optimal-transport-appendix}

Suppose we are given discretized probability distributions $\mu \in \mathbb{R}^N$ and $\nu \in \mathbb{R}^M$. By definition, we assume that $\mu_i>0$, $\nu_i>0$ and $\sum \mu_i=1$ and $\sum \nu_i=1$. We are also given a positive cost matrix $C\in \mathbb{R}^{N\times M}$, such that $C_{ij}>0$. 

We can formulate the Kantorovich relaxation of the Optimal Transport as the following optimization problem:
\begin{align}
\label{eq:primal-OT}
\min_{\Gamma} \quad & \sum_{i,j}C_{ij}\Gamma_{ij}\\
\textrm{s.t.} \quad & \sum_j\Gamma_{ij}=\mu_i \nonumber \quad\sum_i\Gamma_{ij}=\nu_j \nonumber \quad \Gamma_{ij} \geq 0
\end{align}
where $\Gamma\in\mathbb{R}^{N\times N}$ is the optimization variable. Here, optimization function, equality, and inequality constraints are all affine, so the problem is convex, specifically, it is a linear program (LP) \cite{boyd2004convex}. So, we can use any convex optimization software to find $\Gamma$.

\subsection{Trace Reformulation}
\label{subsection:trace-reformulation}

A very convenient reformulation of the optimal transport optimization problem is the following:
\begin{align}
\label{eq:matrix-OT}
\min_{\Gamma} \quad & \Tr[C \Gamma^T]\\
\textrm{s.t.} \quad & \Gamma \mathds{1}_M=\mu \nonumber \quad \mathds{1}^T_N\Gamma=\nu^T \nonumber \quad \forall_{ij}\Gamma_{ij} \geq 0
\end{align}
where $\mathds{1}_N$ and $\mathds{1}_M$ are column vectors of all ones of size $N$ and $M$ respectively. The equality conditions are obvious, and to see how to get the trace in the objective function, observe:

\begin{equation}
    \sum_{ij}C_{ij}\Gamma_{ij} =\sum_{ij}C_{ij}\Gamma^T_{ji} =\Tr \left[C \Gamma^T\right] 
\end{equation}
This reformulation is nice because it has fewer floating indices and it invokes a natural Frobenius inner product between $C$ and $\Gamma$.

\subsection{Kantorovich Dual}
\label{subsection:kantorovich-dual}

Next, we apply the general duality construction procedure to the discrete optimal transport problem defined in \ref{eq:primal-OT}.

\paragraph{Feasibility domain.} We start by considering the feasibility of the problem, which in our case is a space of all real-valued $N\times N$ matrices:
\begin{equation}
    \mathcal{D}=\{\Gamma\in \mathbb{R}^{N\times N}\}
\end{equation}

\paragraph{Lagrangian.}
Introduce Lagrange multipliers
\begin{align}
    \phi\in\mathbb{R}^N,\qquad 
    \psi\in\mathbb{R}^M,\qquad
    \Lambda\in\mathbb{R}^{N\times M}_{\!+}
\end{align}
for the row, column, and non-negativity constraints, respectively.  
The Lagrangian is
\begin{align}
\label{eq:lin-OT-lagrangian}
\mathcal{L}(\Gamma,\phi,\psi,\Lambda)
	&=\sum_{i,j} C_{ij}\Gamma_{ij}
	  \;+\;\sum_{i}\phi_i\!\left(\mu_i-\sum_{j}\Gamma_{ij}\right)
	  \;+\;\sum_{j}\psi_j\!\left(\nu_j-\sum_{i}\Gamma_{ij}\right)
	  \;-\;\sum_{i,j}\Lambda_{ij}\Gamma_{ij} \nonumber\\[2pt]
	&=\sum_{i,j}\Gamma_{ij}\bigl(C_{ij}-\phi_i-\psi_j-\Lambda_{ij}\bigr)
		\;+\;\sum_{i}\phi_i\mu_i
		\;+\;\sum_{j}\psi_j\nu_j.
\end{align}

\paragraph{Dual function.}
The Lagrange dual $g(\phi,\psi,\Lambda)$ is obtained by minimizing
$\mathcal{L}$ over $\Gamma\in\mathcal{D}$:
\begin{equation}
    g(\phi,\psi,\Lambda)
  \;=\;
  \inf_{\Gamma\in\mathcal{D}}\!
  \Bigl\{
       \sum_{i,j}\Gamma_{ij}\bigl(C_{ij}-\phi_i-\psi_j-\Lambda_{ij}\bigr)
       +\sum_{i}\phi_i\mu_i+\sum_{j}\psi_j\nu_j
  \Bigr\}.
\end{equation}

Now consider two cases. First, suppose there exists a single pair of indecies $i$ and $j$, for which $K_{ij}\equiv(C_{ij}-\phi_i-\psi_j-\Lambda_{ij})\neq0$. Then set $\Gamma_{ij}=-tK_{ij}$ for that $i,j$ and $0$ otherwise. It is easy to see that by taking $t\to\infty$, we have $g\to-\infty$. 

The other case is that for each $i,j$ we have $K_{ij}=0$, which corresponds to:
\begin{equation}
    C_{ij}-\phi_i-\psi_j=\Lambda_{ij}\quad\forall ij
\end{equation}
In other words, we can express $\Lambda$ in terms of other variables. To summarize, we get:
\begin{align}
    g(\phi,\psi,\Lambda)
  =\begin{cases}
\sum_{i}\phi_i\mu_i+\sum_{j}\psi_j\nu_j \quad & \text{if } K_{ij}=0 \quad \forall ij\\
-\infty & \text{if } \exists ij.K_{ij}=0
  \end{cases}
\end{align}
Recall that we want to maximize over $g$, so we can put $K_{ij}=0$\quad$\forall ij$ as the explicit constraint. Notice that this corresponds to setting $C_{ij}-\phi_i-\psi_j\geq0$, since $\Lambda_{ij}\geq0$ by definition. This results in the following maximization problem:

\paragraph{Kantorovich dual.} 
\begin{align}
\label{eq:dual-OT}
\max_{\phi\in\mathbb{R}^N,\;\psi\in\mathbb{R}^M}\quad
	&\sum_{i}\phi_i\mu_i \;+\;\sum_{j}\psi_j\nu_j \\[2pt]
\text{s.t.}\quad
	&\phi_i+\psi_j \;\le\; C_{ij},
	\qquad\forall i,j.\nonumber
\end{align}

\subsection{Entropy Regularized Optimal Transport}
\label{subsection:entropy-regularized-optimal-transport}

In entropic relaxation of optimal transport, there is an additional term $\epsilon\Omega(\Gamma)$ added to the loss function, where $\Omega(\Gamma)=\sum_{ij}\Gamma_{ij}\log(\Gamma_{ij})$ is the negative entropy of the transport plan. Notice that entropy has an implicit constraint of $\Gamma_{ij}\geq0$ because of the logarithm. Thus, the problem becomes:
\begin{align}
\label{eq:entopy-OT}
\min_{\Gamma} \quad & \sum_{i,j}C_{ij}\Gamma_{ij} +\epsilon \sum_{ij}\Gamma_{ij}\log(\Gamma_{ij})\\
\textrm{s.t.} \quad & \sum_j\Gamma_{ij}=\mu_i \nonumber \quad\sum_i\Gamma_{ij}=\nu_j \nonumber 
\end{align}
where the \textit{feasibility domain} implicitly includes the non-negativity constraint:
\begin{equation}
    \mathcal{D}=\{\Gamma\in\mathbb{R}^{N\times M}:\Gamma_{ij}\geq0\}
\end{equation}

\subsection{Sinkhorn Dual}
\label{subsection:sinkhorn-dual}
Now, let us compute the dual function to \ref{eq:entopy-OT}.

\paragraph{Lagrangian} of the problem above:
\begin{align}
\label{eq:sinkhorn-lagrangian}
    \mathcal{L}(\Gamma, \phi,\psi) &= \sum_{i,j}C_{ij}\Gamma_{ij} +\epsilon \sum_{ij}\Gamma_{ij}\log(\Gamma_{ij}) + \sum_i \phi_i\left(\mu_i-\sum_{j}\Gamma_{ij}\right) + \sum_j \psi_j\left(\nu_j-\sum_{i}\Gamma_{ij}\right) \nonumber\\
    &= \sum_{i,j} \Gamma_{i,j}\left(C_{i,j}+\epsilon\log(\Gamma_{i,j})-\phi_i-\psi_j\right) + \sum_i\phi_i\mu_i + \sum_i\psi_i\nu_i
\end{align}
\paragraph{Lagrange dual} is defined as a minimum over $\Gamma$:
\begin{align}
    g(\phi,\psi)=\inf_{\Gamma\in\mathcal{D}}\mathcal{L}(\Gamma, \phi,\psi)
\end{align}
The minimum must satisfy $\partial_{\Gamma_{ij}}\mathcal{L}(\Gamma, \phi,\psi)=0 \quad\forall ij$.
\begin{equation}
    \partial_{\Gamma_{ij}}\mathcal{L}(\Gamma, \phi,\psi)=C_{i,j}+\epsilon\log(\Gamma_{i,j})-\phi_i-\psi_j + \epsilon=0 
\end{equation}
From which immediately follows that:
\begin{equation}
    \Gamma_{i,j} = \exp((\phi_i+\psi_j - \epsilon -C_{i,j})/\epsilon)
\end{equation}
Plugging this back into (\ref{eq:sinkhorn-lagrangian}) we obtain the Lagrange dual:
\begin{equation}
    g(\phi,\psi) = \sum_i\phi_i\mu_i + \sum_i\psi_i\nu_i -\sum_{i,j}\epsilon \exp((\phi_i+\psi_j - \epsilon -C_{i,j})/\epsilon)
\end{equation}
Thus, we can equivalently solve the following optimization problem:
\begin{align}
\label{eq:sinkhorn-OT-dual}
\max_{\phi,\psi} \quad & \sum_i\phi_i\mu_i + \sum_i\psi_i\nu_i -\sum_{i,j}\epsilon \exp((\phi_i+\psi_j - \epsilon -C_{i,j})/\epsilon)
\end{align}

\section{Change of Basis in Optimal Transport}
\label{section:change-of-basis-in-optimal-transport}

\subsection{Defining a Basis}
\label{subsection:defining-a-basis}

Let $v^{(1)},v^{(2)},\dots v^{(J)} \in \mathbb{R}^N$ be a set of orthogonal vectors in $\mathbb{R}^N$ and $w^{(1)},w^{(2)},\dots w^{(K)} \in \mathbb{R}^M$ be a set of orthogonal vectors in $\mathbb{R}^M$. Form a matrix $V\in \mathbb{R}^{N\times n}$ using columns $v^{(1)},v^{(2)},\dots v^{(J)}$, and matrix $W = w^{(1)},w^{(2)},\dots w^{(K)}$. The orthogonality condition of vectors implies that:

\begin{equation}
\label{eq:orthov}
V = \begin{bmatrix} 
    | & | &  & | \\
    v^{(1)} & v^{(2)} & \cdots & v^{(J)} \\
    | & | &  & | 
\end{bmatrix} \implies V^TV=\mathds{1}_J
\end{equation}

\begin{equation}
\label{eq:orthow}
W = \begin{bmatrix} 
    | & | &  & | \\
    w^{(1)} & w^{(2)} & \cdots & w^{(K)} \\
    | & | &  & | 
\end{bmatrix} \implies W^TW=\mathds{1}_K
\end{equation}

For the respective collections of vectors to be called a basis, they must span all of $\mathbb{R}^N$ and $\mathbb{R}^M$. Since they are all orthogonal and, thus, independent, we just need to require that $J=N$ and $K=M$. Then, in addition to $\ref{eq:orthov}$ and $\ref{eq:orthow}$, we have the completeness relation, which states that $VV^T=\mathds{1}_N$ and $WW^T=\mathds{1}_N$. The theorem below is agnostic to the exact choice of the basis sets $V$ and $W$, but one should consider picking a basis that is well suited for the problem of interest — e.g., Hermite-Gaussian polynomials or Laguerre polynomials for the sake of beam-shaping.

\subsection{Basis Change Theorem}
\label{subsection:basis-change-theorem}

\begin{theorem}[Basis change in OT]
\label{theorem:basis-change}
Let $V \in\mathbb{R}^{N\times N}$ and $W \in\mathbb{R}^{M\times M}$ be matrices with orthogonal rows, $V^TV=I$ and $W^TW=I$, and orthogonal columns, $VV^T=I$ and $WW^T=I$. Then, Kantorovich relaxation of OT for distributions $\nu$, $\mu$, cost $C$, for the transportation plan $\Gamma$ is equivalent to the following optimization problem:
\begin{align}
\label{eq:matrix-basis-OT}
\min_{\gamma} \quad & \Tr[c \gamma^T]\\
\textrm{s.t.} \quad & \gamma \bar{w}=a \nonumber \quad \bar{v}^T\gamma=b^T \nonumber \quad \forall_{ij}(V\gamma W^T)_{ij} \geq 0
\end{align}
where $\gamma =V^T\Gamma W$, $c=V^TCW$, $a=V^T\mu$, and $b = W^T\nu$. Also, $\bar{v}=V^T1_N$ and $\bar{w}=W^T1_M$.
\end{theorem}

\begin{proof}
We want to show $\Gamma$ is the solution to the original optimization problem (\ref{eq:matrix-OT}) iff $\gamma=V^T\Gamma W$ is the solution to (\ref{eq:matrix-basis-OT}). First, observe that the objective functions are the same:
\begin{align}
    \Tr \left[ c \gamma^T \right]  &=\Tr \left[V^TCW (V^T\Gamma W)^T\right] \\
    &=\Tr \left[V^TCW W^T\Gamma^T V\right] \\
    &=\Tr \left[VV^TC W W^T \Gamma^T \right] \\
    &=\Tr \left[ C \Gamma^T \right] 
\end{align}
where we used the cyclic property of trace and completeness of the basis $VV^T=I$ and $WW^T=I$. Now, to see that equality conditions are equivalent, observe:
\begin{align}
     a &= V^T\mu = V^T\Gamma1_M =V^T(V\gamma W^T) 1_M=\gamma \bar{w} \\
     b^T &= (W^T\nu)^T =1_N^T\Gamma W=1_N^T(V\gamma W^T)W=\bar{v}^T\gamma
\end{align}
where we used equality conditions $\mu = \Gamma1_M $ and $\nu^T=1_N^T\Gamma$ and basis orthogonality $V^TV=I$ and $W^TW=I$. Finally, to enforce the inequality constraint notice that $\gamma =V^T\Gamma W$ implies that  $\Gamma =V\gamma W^T$ if $VV^T=I$ and $WW^T=I$. So, we get that:
\begin{equation}
    \forall_{ij}\Gamma_{ij} \geq 0\iff \forall_{ij}(V\gamma W^T)_{ij}\geq0
\end{equation}
Thus, if $\Gamma^\star$ minimizes (\ref{eq:matrix-OT}) then $\gamma^\star=V^T\Gamma^*W$ minimizes (\ref{eq:matrix-basis-OT}), and if $\gamma^\star$ minimizes (\ref{eq:matrix-basis-OT}) then $\Gamma^\star=V\gamma^*W^T$ minimizes (\ref{eq:matrix-OT})
\end{proof}

\subsection{Relaxing Completeness Condition}
\label{subsection:relaxing-completeness-condition}

It turns out that if we only keep the top $J<N$ vectors in the basis for $V$ and the top $K<M$ vectors in the basis for $W$, then we can still find an approximate transport plan between the two distributions.

Notice that the new truncated bases $V_\text{trunc}\in \mathbb{R}^{N\times J}$ and $W_\text{trunc}\in \mathbb{R}^{N\times K}$ still satisfy the orthogonality condition, meaning that $V_\text{trunc}^TV_\text{trunc}=\mathds{1}_J$ and $W_\text{trunc}^TW_\text{trunc}=\mathds{1}_K$, where $\mathds{1}_J$ and $\mathds{1}_K$ are identity matrices on $\mathbb{R}^J$ and $\mathbb{R}^K$.

However, the completeness condition is no longer true. To see that write:
\begin{align}
V_\text{trunc}V_\text{trunc}^T&=\sum_{i=1}^{J}\ket{v^{(i)}}\bra{v^{(i)}}=\mathds{1}_N - 
\sum_{i=J+1}^{N}\ket{v^{(i)}}\bra{v^{(i)}} \\
W_\text{trunc}W_\text{trunc}^T&=\sum_{i=1}^{K}\ket{v^{(i)}}\bra{v^{(i)}}=\mathds{1}_N - 
\sum_{i=K+1}^{N}\ket{v^{(i)}}\bra{v^{(i)}}
\end{align}
So, we are missing exactly the remaining $N-J$ vectors in the $V$ basis and $N-K$ vectors in the $W$ basis. Nevertheless, our preliminary numerical experiments show that it is still possible to retrieve an approximate $\tilde{\Gamma}$, which looks like a blurred out approximation of the original $\Gamma$. Because we did not have enough time to investigate this further, we leave this is as a direction for future research.

%% file: appendix/D.tex
\chapter{Deep Learning Experiments}
\label{chapter:deep-learning-experiments}

We also amended this thesis with my work on deep learning for the problem of phase retrieval. This work was produced as a final project for a class, CS231N: Deep Learning for Computer Vision, that I took in the spring of 2024 at Stanford. I would like to acknowledge my close friend and co-author of this work, John Wang, without whom results below would not be possible.

Notice that conventions for the variables and the loss slightly differ from the ones used in this thesis, but nevertheless qualitative results can still be interpreted clearly — the combination of the Optimal Transport, polished by the Gerchberg-Saxton algorithm, is very difficult (if not impossible) to beat with a deep neural network. This matches our intuition outlined in Chapter 4.


\section*{Supplementary material}

# Optimal Transport Informed Phase Retrieval Using Deep Learning

Andrii Torchlyo
Department of Physics and Mathematics
Stanford University
torchlyo@stanford.edu

John Wang
Department of Computer Science
Stanford University
jwang003@stanford.edu

## Abstract

*The Phase Retrieval problem is the task of estimating the phase a the complex-valued function given two amplitude constraints. This problem arises in many fields of science that study propagation of waves. There have been many attempts to solve Phase Retrieval using classical algorithms like the Gerchberg-Saxton (GS) algorithm [4]. Recently, neural network architectures for computer vision have gained popularity in Phase Retrieval [11]. In this work, we present a new paradigm for understanding and solving the phase retrieval problem by incorporating an Optimal Transport-based initialization into current algorithmic approaches. We show both theoretically and experimentally that Optimal Transport guarantees an improved solution over existing methods. Furthermore, we introduce a method inspired by Perturbation Theory to incorporate Optimal Transport solutions into deep learning models by modifying the Phase Retrieval task. Finally, we show that an adaptive regularization schedule improves model performance and stabilizes training. These advancements result in improvements of over 4.5 times the current publicly available Phase Retrieval algorithms. Code for the project is available at $https://github.com/jwang307/phase-retrieval$.*

## 1. Introduction

The computational problem of phase retrieval is central to many areas of science, including atomic physics, biology, and astronomy. In this paper, we will motivate the importance of phase retrieval by discussing the specific example of laser beam shaping, which is a common and significant challenge in atomic physics.

Suppose we want to characterize laser light as a function of position. If we assume that the laser is monochromatic and uniformly polarized, we can represent the laser light as a complex-valued function $E : \mathbb{R}^3 \to \mathbb{C}$, which can be expressed as:

$$E(x, y, z) = A(x, y, z)e^{i\phi(x,y,z)} \tag{1}$$

where $A : \mathbb{R} \to \mathbb{R}_{\geq 0}$ is called the amplitude of the field, and $\phi : \mathbb{R}^3 \to [0, 2\pi]$ is called the phase of the field. Intuitively, the square of the amplitude represents how much energy is present in a spatial location, while the gradient of the phase points the direction in which this energy will flow.

To fully characterize the field, it suffices to specify the amplitude and the phase at some plane $z = z_1$. Then, using Maxwell's equations, it is possible to determine the electric field at any other plane. In particular, if the other plane at $z = z_2$ is very far (Fraunhofer far field limit) then the equation becomes:

$$E(X, Y, z_2) = \iint E(x, y, z_1)e^{2\pi i(xX + yY)} \tag{2}$$

$$= \mathcal{F}[E(x, y, z_1)](X, Y). \tag{3}$$

So, up to the rescaling of coordinates and the overall scale factor, the electric fields are related by a 2D Fourier Transform.

In most experimental settings, it is simple to measure the amplitude of the light field by placing a camera at some plane $z = z_1$. However, it is usually very difficult to obtain a direct measurement for the phase of the light field. A common workaround is to make two amplitude measurements — at the input plane $z = z_1$ and at the far field (Fourier plane) $z = z_2$. Then, the task is to computationally find the phase $\phi(x, y, z_1)$ which will match the target amplitude at the Fourier plane.

In the next section we formally describe the optimization problem to solve.

### 1.1. Mathematical Formulation

Given two $N \times N$ amplitude images $A_1, A_2 \in \mathbb{R}^{N \times N}$, we want to find a phase $\phi \in [0, 2\pi]^{N \times N}$ such that predicted amplitude $\hat{A}_2 = |\mathcal{F}\{A_1 e^{i\phi}\}|$ is close to $A_2$ constraint. Here $\mathcal{F}$ refers to a 2D discrete fourier transform, and $|\cdot|$ is the amplitude operation. Specifically, we are interested in finding the phase satisfying:

$$\phi = \arg\min_\phi d\left(|\mathcal{F}\{A_1 e^{i\phi}\}|, A_2\right) \tag{4}$$

1

In the equation above $d(A, B)$ refers to a suitable distance metric between images. For most experiments we will be using $L^2$ norm, defined as $d(A, B) = ||A - B||_2$.

## 2. Related Work

Existing algorithms for Phase Retrieval can be divided into classical algorithms and deep learning based methods.

### 2.1. Gerchberg-Saxton Algorithm

The industry standard for solving the Phase Retrieval problem is the Gerchberg-Saxton (GS) algorithm [4], which works by iteratively applying Fourier and inverse Fourier transforms while enforcing amplitude constraints on every iteration:

---

**Algorithm 1. Gerchberg-Saxton Algorithm**

1: $\phi \leftarrow \text{Angle}(\mathcal{F}[A_2])$     ▷ Phase initialization
2: $i \leftarrow 0$
3: **while** $i \leq N$ **do**
4:     $E_2 \leftarrow \mathcal{F}[A_1 e^{i\phi}]$    ▷ Estimate Fourier plane field
5:     $\phi \leftarrow \text{Angle}(E_2)$         ▷ Discard amplitude
6:     $E_1 \leftarrow \mathcal{F}^{-1}[A_2 e^{i\phi}]$   ▷ Estimate input plane field
7:     $\phi \leftarrow \text{Angle}(E_1)$         ▷ Discard amplitude
8: **end while**

---

It is mathematically possible to prove that GS algorithm is guaranteed to reduce the L2 loss on every iteration [3]. However, the issue with the GS algorithm is that despite a monotonically decreasing loss function, the algorithm often converges on a sub-optimal solution. The primary reason for this is because of "phase vortices," which are points in the predicted phase map where the phase contour terminates (we say that phase "wraps" from 0 to $2\pi$ around a phase vortex). A visual of this is presented in the results section, where we use the GS algorithm as a baseline. Phase vortices result in a pixel-scale black dots on the output intensity, which is known to be the main obstacle that prevents GS algorithm from converging on a better solution [5].

### 2.2. Deep Learning Methods

Recently, deep learning methods have been developed for solving Phase Retrieval. Most solutions can be placed into one of two main approaches: (1) an untrained, iterative scheme and (2) a data driven, trained scheme [10]:

In both approaches, a neural network typically receives an input amplitude and a target amplitude with the task of predicting the phase that maps between the two images. Existing approaches to do this leverage computer vision architectures from simple CNNs to UNets [11]. However, the landscape of existing solutions is quite sparse: the current state of the art solution, PhysenNet, leverages a UNet architecture with 4

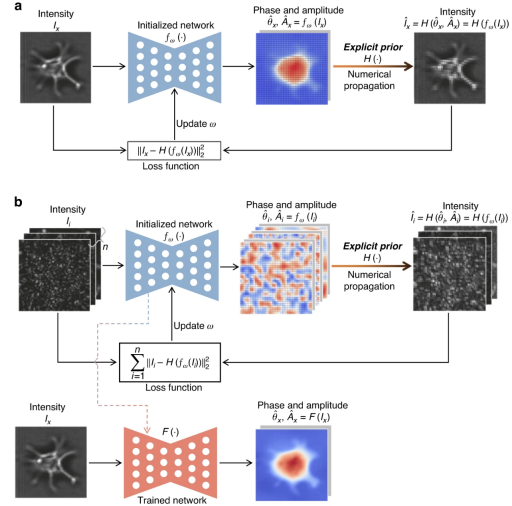

Figure 1. Overview of Deep Learning Based Approaches. 1a shows approach (1) and 1b shows approach (2). Figure taken from [11].

downsampling layers, a bottleneck, and 4 upsampling layers [10].

While approach (2) is the typical set up for most machine learning approaches, we are interested in approach (1), where the neural network acts as an iterative optimizer. This approach does not require a training dataset and typically provides more accurate phase predictions because it iterates continuously over the same input target pair, while the trained network must predict the target phase with one forward pass. The primary disadvantage of approach (1) is the time needed to reach convergence. In practice, this takes around 5 minutes per image pair on an M1 MacBook and around 20 seconds on an H100 GPU. Because we typically don't need to screen massive amounts of image pairs for laser beam shaping, this runtime is acceptable.

## 3. Methods

We compare deep learning methods initialized by optimal transport against several baseline models, including the GS algorithm and PhysenNet. In the following section, we describe the optimal transport algorithm and the deep learning architectures developed. We also outline strategies for training the deep learning models while incorporating the optimal transport solution.

### 3.1. Optimal Transport Algorithm

A novel approach to approximating the phase is to use Optimal Transport [1]. Empirically, the GS algorithm improves by an order of magnitude when initialized with the Optimal Transport phase solution. We begin with a short derivation.

---

[1]This is a new idea that has not been published before. It was discovered by Hunter Swan and first implemented by me (Andrii Torchylo) as a part of summer research in the Hogan lab at Stanford

2

Denote the input plane coordinates with $\vec{x}$ and output coordinates with $\vec{X}$. Then, a perfect phase solution $\phi(\vec{x})$ will satisfy the following equation.

$$A_2(\vec{X}) = |\mathcal{F}[A_1(\vec{x})e^{i\phi(\vec{x})}](\vec{X})| \tag{5}$$

$$= \left| \int A_1(\vec{x})e^{i\phi(\vec{x})}e^{-i2\pi\vec{x}\cdot\vec{X}}d\vec{x} \right| \tag{6}$$

Mathematically solving for $\phi(\vec{x})$ from equation above is close to impossible (although one can derive an analytic solution in 1D). Under suitable convexity and smoothness assumptions about the phase $\phi$, (6) can be approximated to the following differential equation (See Appendix A for the derivation and intuition):

$$A_2^2\left(\nabla\phi(\vec{x})\right) = \frac{A_1^2(\vec{x})}{\det \nabla^2\phi(x)} \tag{7}$$

This is a second order non-linear partial differential equation for a phase $\phi$, which is known as Monge-Ampere differential equation. One can try to solve this differential equation for the unknown phase $\phi(\vec{x})$ using finite difference methods. However, in practice it is very difficult due to numerical stability issues.

We will bypass solving this differential equation using a very useful fact from Optimal Transport theory: finding the gradient of the phase, $\phi$, is equivalent to finding an optimal transport plan between probability distributions $A_1^2$ and $A_2^2$ using a quadratic cost function.

In our experiments, we use Sinkhorn regularized Optimal Transport ($\lambda = 0.001$) to extract gradient of the phase, after which we perform numerical integration to obtain the phase itself.

In practice, Optimal Transport performs well when target features are not sharp. A known downside of Optimal Transport is the smoothing of sharp features, which is shown in our results section. This is because an assumption when deriving (4) is that the phase does not vary rapidly.

### 3.2. UNet and ResUNet

Out of existing deep learning architectures for Phase Retrieval, UNets are typically considered state of the art [11]. This intuitively follows because the phase retrieval problem is inherently a reconstruction task. We base our deep learning models on the UNet architecture. While we tested several configurations of downsampling and upsampling layers, we settled on the following configuration for our UNet:

- Input Convolution: Input Image to 32 channels

- Four downsampling layers: 32 to 512 channels, kernel size=3 and padding=1

- Four upsampling layers: 512 to 32 channels, kernel size=3 and padding=1

- Output Convolution: 32 channels to phase prediction

The input convolutions, output convolutions, and down and upsampling layers utilize the Double Convolution layer outlines below:

| Algorithm 2. Double Convolution Sequence |
| --- |
| 1: $conv1 \leftarrow$ Conv2d($in, mid, 3, 1$, False) |
| 2: $batchnorm1 \leftarrow$ BatchNorm2d($mid$) |
| 3: $leakyrelu1 \leftarrow$ LeakyReLU() |
| 4: $conv2 \leftarrow$ Conv2d($mid, out, 3, 1$, False) |
| 5: $batchnorm2 \leftarrow$ BatchNorm2d($out$) |
| 6: $leakyrelu2 \leftarrow$ LeakyReLU() |

Specifically, the downsampling layer is as follows:

| Algorithm 3. Downsampling Sequence |
| --- |
| 1: $doubleconv \leftarrow$ DoubleConv($in, out$) |
| 2: $maxpool \leftarrow$ MaxPool2D($2$) |

While the upsampling layer is detailed below:

| Algorithm 4. Upsampling Sequence |
| --- |
| 1: $conv_t \leftarrow$ ConvTranspose($in, in/2, kernel = 2$) |
| 2: $doubleconv \leftarrow$ DoubleConv($in, out$) |

In addition to testing the UNet archiecture, we also look at other popular image reconstruction methods. Namely, the ResUNet architecture, which has been shown to improve on UNets in several segmentation tasks [12]. We implement the ResUNet architecture with the same overall layer details. The primary difference between architecture is that instead of concatenating the corresponding downsampling tensor during the upsampling process, ResUNet treats the upsampling procedure as a residual connection, performing an addition operation instead. Due to this, the double convolution during the upsampling process takes $in/2$ channels instead.

#### 3.2.1 Phase Correction

In the GS algorithm, it is straightforward to initialize the algorithm with an informed phase guess. However, when training a neural network to predict the phase, it is less clear how to "initialize" it with the optimal transport solution. Here, we incorporate the optimal transport phase prediction by modifying the UNet task: instead of predicting the phase, the UNet is tasked with predicting the correction on the optimal transport solution. Specifically, we train UNet to find a phase $\phi'$ such that the combined phase:

$$\phi = \phi_{\text{OT}} + \lambda_r\phi' \tag{8}$$

minimizes the distance metric outlined in equation (4). In the equation above, $\lambda_r$ is the regularization term of the phase guess, which controls the size of the correction. Empirically, we find that in the first few iterations the phase correction guess is usually quite large due to a random initialization of the weights, so a small regularization term allows the model to stay within the region of the optimal transport solution.

This approach was inspired by perturbation theory, which is commonly used to approximate solutions to complicated Hamiltoninans in Quantum Mechanics. Consider the expansion of the perfect phase solution with respect to some order parameter :

$$\phi = \phi_0 + \epsilon\phi_1 + \epsilon^2\phi_2 + ...  \qquad (9)$$

Then we can interpret the Optimal Transport phase $\phi_{OT}$ as a leading term contribution $\phi_0$, while the output of the neural network generates a first order correction $\phi_1$. The perturbation parameter $\epsilon$ is controlled by $\lambda_r$.

#### 3.2.2 Adaptive Regularization

In our preliminary experiments, we noticed that small values of the regularization parameter $\lambda_r$ led to a very slow training process with sub-optimal convergence. On the other hand, large values of $\lambda_r$ led to instable training within the first couple of iterations. Ideally, regularization should start small but then adiabatically increase as the training progress.

To this end, we explored the effect of adaptive regularization using a regularization scheduler. We test two regularization schedules: a linear schedule with a constant decay rate in each iteration, and an exponential schedule with a decay rate calculated by the difference between the desired start and end $\lambda_r$ terms. Specifically, given $a$ as the initial regularization term, $b$ as the final regularization term, $i$ as the current iteration, and $n$ as the number of iterations, we have:

$$\lambda_r = a + (b - a) \cdot \frac{i}{n}  \qquad (10)$$

for the linear scheduler. For the exponential scheduler, we have:

$$\lambda_r = a \cdot e^{r \cdot i} \quad \text{where} \quad r = \frac{log(\frac{b}{a})}{n}  \qquad (11)$$

#### 3.2.3 Smoothness Regularization

An alternative idea to the Optimal Transport is to enforce smoothness constraint of the generated phase via smoothness regularizaiton. We adopt the regularization term from Mahendran et al. that used a finite difference approximation of the total variation regularizer [6]. Specifically we consider the effect of adding a regularization term to the loss function, given by:

$$\mathcal{R}_V^\beta = \sum_{i,j} \left( (\phi_{i,j+1} - \phi_{i,j})^2 + (\phi_{i+1,j} - \phi_{i,j})^2 \right)^{\beta/2}  \qquad (12)$$

where $\phi_{i,j}$ refers to the value of the $i,j$-th pixel of the generated phase. For our experiments, we set $\beta = 1$. Also, we weight the contribution of the smoothness regularization via $\lambda_s$, which is a hyperparameter that we tune during grid search.

### 4. Data

Because our approach to Phase Retrieval uses the neural network as an iterative optimizer, we do not require a training dataset. Instead, we only need the input image and the target output to test the predicted phase against at every iteration.

We create the synthetic input and output 128x128 images that resemble a typical laser beam shaping task. The input is a mixture of Gaussians, one centered at $\sigma = 25$ pixels with another smaller Gaussian at $\sigma = 12$, which is offset by $(12, -12)$ pixels. The target is a Gaussian ring, which is a one pixel circle of radius $R = 25$ convolved with a Gaussian with $\sigma = 12$. In addition, we add a small Gaussian with $\sigma = 12$, which is offset by $(12, 12)$ pixels.

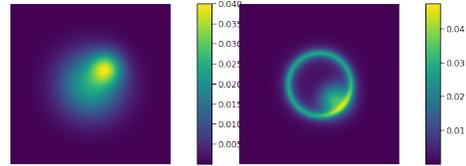

Figure 2. Input and output amplitudes for our experiment

Models are trained to predict the phase that maps between the two amplitudes. Therefore, both images in Fig. 2 are input into the model. Since the images are grayscale, the input is a $2 \times 128 \times 128$ tensor where each element is from 0 to 255 denoting the magnitude of each pixel. There are two channels, one for the input amplitude and one for the target amplitude. The output of the model is the predicted phase $\phi$, which is a grayscale image $1 \times \times 128 \times 128$.

We test the performance of the Phase Retrieval algorithms by comparing the L2 norm between the target amplitude $A_2$ and the predicted amplitude $\tilde{A}_2 = |\mathcal{F}[A_1 e^{i\phi}]|$ — where $\phi$ is the phase solution and $A_1$ is the input amplitude.

We additionally, compute Schroff Error between the predicted and the target amplitude, which is a common metric for a task of phase retrieval [7, 8]. Specifically, we use the definition of the Shroff Error from [9] (future publication):

$$d(A, B) = \sqrt{\frac{1}{|U_{1/2}|} \int_{U_{1/2}} \frac{(A(\vec{x})^2 - B(\vec{x})^2)^2}{A(\vec{x})^4} d\vec{x}},  \qquad (13)$$

where $U_{1/2} := \{\vec{x} \in \mathbb{R}^2 \mid A^2(\vec{x}) \geq \frac{1}{2} \max_{\vec{x}} A(\vec{x})^2\}$ is the region where the target output beam intensity $A^2$ is at least half its maximum value, and where $|U_{1/2}|$ denotes the area of $U_{1/2}$.

# 5. Experiments

We ran experiments to obtain performance on baseline models and evaluate the effectiveness of our proposed regularization techniques and deep model architectures.

For our baselines, we ran the GS algorithm and OT algorithm individually. We also ran the GS algorithm initialized with the OT solution. Each experiment was ran to convergence, which we empirically measured at or before 5000 iterations.

For our deep learning baseline, we ran a UNet with no prior guess. We ran each UNet configuration to 5000 iterations. Similarly, we observed convergence at or before 5000 iterations. We performed a hyperparameter sweep over $\eta, \lambda_r$, and weight decay. The exact test configurations are all possible combinations of the following:

- $\eta : 1e-6, 1e-5, 1e-4, 1e-3, 1e-2, 0.05, 0.1$
- $\lambda_r : 1e-2, 0.1, 0.25, 0.5$
- $w_d : 1e-8, 1e-6, 1e-4, 1e-2$

In addition to the UNet baseline, we also ran baselines on the UNet architecture with adaptive regularization and smoothness regularization to compare against OT initialized UNet performance in those runs. For adaptive regularization, we performed a hyperparameter sweep of $\eta$, the beginning $\lambda_{ra}$, and the end $\lambda_{rb}$. We sweep over $\eta$ again because we empirically observe a different optimal learning rate when using the scheduler. However, all other hyperparameters remain optimal

- $\eta : 1e-5, 1e-4, 1e-3, 1e-2$
- $\lambda_{ra} : 1e-8, 1e-6, 1e-4, 1e-2$
- $\lambda_{rb} : 0.1, 0.25, 0.5, 2$

Similarly, for smoothness regularization, we keep all optimal hyperparameters except $\eta$ and sweep over the smoothness weight $\lambda_s$. The tested configurations are:

- $\eta : 1e-5, 1e-4, 1e-3, 1e-2$
- $\lambda_s : 1e-8, 1e-6, 1e-4, 1e-2$

For each of these configurations, we train the model to 5000 iterations. Runs were performed on a single H100 GPU with 80 GB RAM.

# 6. Results and Discussion

We show the phase and target intensities predicted by each model.

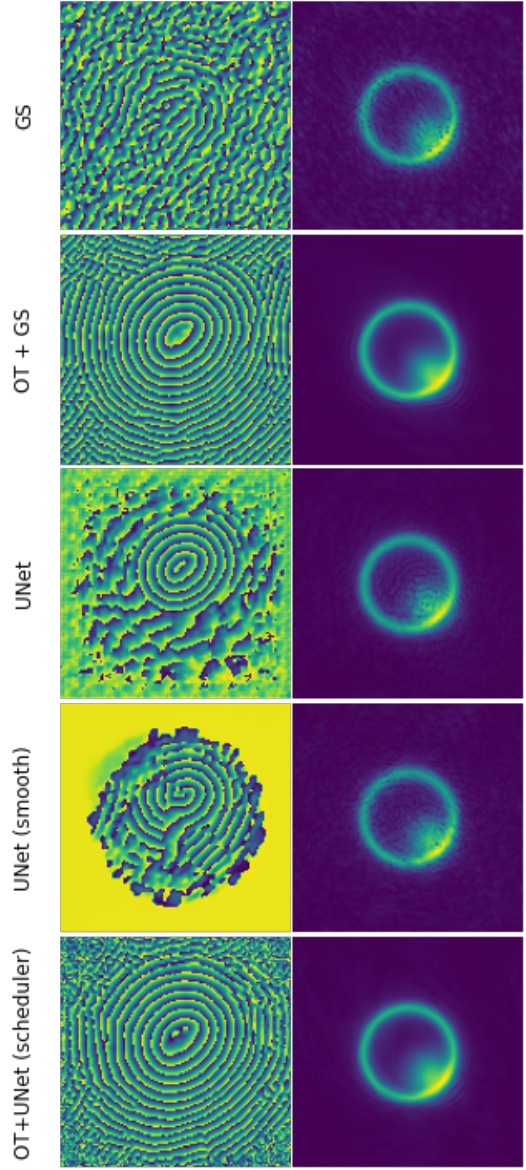

Figure 3. Phase maps and amplitudes for selected methods. Each method corresponds to a row of the figure, where image on the left is the predicted phase $\phi$ and image on the right is the predicted amplitude $\hat{A}_2 = |\mathcal{F}\{A_1 e^{i\phi}\}|$. Note that all phases were modded by $2\pi$ before plotting.

We observe an improvement over all baselines when using the optimal transport initialized UNet model. We also observe improvements when using adaptive regularization, and a marginal improvement in Schroff error when using a ResUNet. However, smoothness regularization did not improve the model and worsened performance in some cases. The complete results are shown in Table 1.

5

| Method | L2 Loss | Schroff Error |
|---|---|---|
| GS | 0.051935 | 0.175288 |
| UNet | 0.013398 | 0.092572 |
| OT + GS | 0.003182 | 0.056895 |
| UNet (smooth) | 0.024730 | 0.137531 |
| OT + UNet | 0.003113 | 0.054462 |
| OT + UNet (smooth) | 0.003502 | 0.048023 |
| OT + UNet (scheduler) | **0.002853** | 0.044446 |
| OT + ResUNet (scheduler) | 0.003073 | **0.042387** |

Table 1. L2 Loss and Schroff Error Comparison for all Methods.

## 6.1. Baselines

We consider the GS algorithm as the classical baseline since it is the most simple and popular approach to phase retrieval. The UNet architecture described above was adopted from PhysenNet, which we consider our deep learning baseline. It is notable that among available options for Phase Retrieval, these two methods are generally considered to be among the best.

Furthermore, we consider another baseline: OT initialized GS algorithm. Here, we initialize the GS algorithm with the OT phase prediction. In this approach, we see an improvement in the L2 error of the predicted amplitude by over an order of magnitude (Fig. 3). Notably, this method itself is still not published. However, since we focus on an OT initialized deep learning method in this project, we consider OT + GS as a clear state-of-the-art goal to outperform.

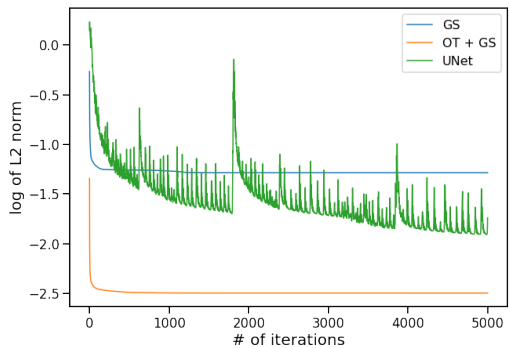

Figure 4. L2 Distance of Predicted Amplitudes for all Baseline Methods

We observe relatively quick convergence of the GS algorithm and OT + GS algorithm during training. Notably, since the GS algorithm guarantees a monotonic decreasing error, the training is much more stable than the UNet baseline. However, the GS algorithm with no initialization performs

the worst among all baselines, converging around 0.0519 after about 500 iterations (see GS curve on Figure 4). We attribute this sub-optimal convergence due to the formation of the phase vortices (see the first row of Figure 3), which are the points where the phase contours terminate. Notice that each phase vortex results in a black dot on the output amplitude, which dominates L2 loss for the GS algorithm.

Contrast the phase generated GS algorithm to OT+GS approach (second row of Figure 3). We can see that OT+GS is almost entirely vortex-free, which is reflected in the better amplitude prediction and almost an order of magnitude lower L2 loss around 0.0032.

While the UNet baseline achieves a better solution than the GS algorithm without initialization, we observe unstable training. In this paradigm, training instability doesn't affect performance as one can just save the minimum error achieved during the iterative optimization; regardless stable training is obviously preferred, and is a shortcoming of the baseline UNet. Additionally, the UNet performs significantly worse than the OT + GS baseline, and the visual quality of the prediction retains the phase vortices seen in the GS solution. However, an interesting feature of the UNet solution is that it still attempts to produce a smooth geometric solution in the middle part of the phase map similar to the OT+GS solution (see third row of the Figure 3). This is surprising given that UNet baseline was not informed by the OT solution in any way.

## 6.2. Deep Learning and Regularization Methods

We refer to our primary deep learning approach OT + UNet, which uses the UNet to predict the phase correction on top of the OT phase guess. In addition, we observe an $\approx 8\%$ improvement in L2 norm when incorporating an adaptive regularization schedule, which we term OT + UNet (scheduler). Compared to the baseline models, both these models outperform all baselines in L2 norm of the predicted amplitude and Schroff error.

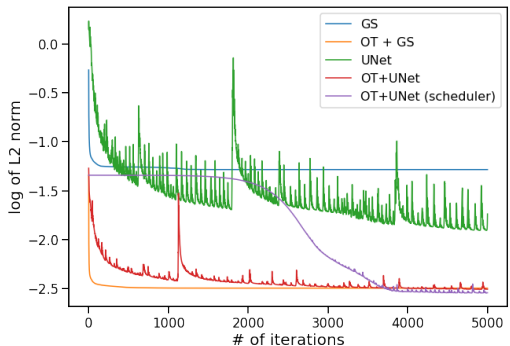

Figure 5. L2 Norm of Primary Deep Learning Approaches Compared to Baselines

We observe nearly identical predictions between OT + UNet and OT + GS. However, with the regularization scheduler, we observe a substantial decrease in L2 norm. Additionally, the exponential scheduler results in much more stable training, without periodic spikes in error. We observe similar spikes when using a linear regularization schedule, while an exponential schedule consistently stabilizes the training (Fig. 7, Appendix). While we are unsure exactly why this is the case, we believe that an exponential regularization schedule gives the model more time to calibrate the correct phase correction scale since $\lambda_r$ increases much more slowly than a linear schedule.

Throughout all experiments with alternate methods (ResUNet, smoothness, adaptive regularization), we find that only adaptive regularization substantially improves model performance over OT + UNet. However, all OT initialized models vastly outperform uninitialized UNet baselines.

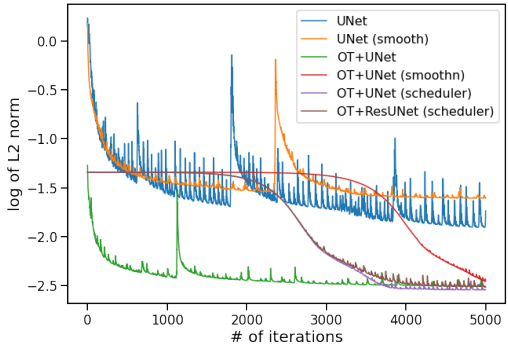

Figure 6. L2 Norm of all Deep Learning Methods

While the ResUNet does not substantially alter performance over a base UNet, we observe a longer period to convergence. For completeness, we ran OT + ResUNet (scheduler) to 8000 iterations and the converged L2 norm was nearly identical to the OT + UNet (scheduler) solution, at 0.0029. This is consistent with past results which show none to marginal improvements over UNet, an example of which is the ResUNet paper itself [12].

Furthermore, we observed that smoothness regularization stabilizes the training and reduces the noise on the edges of the phase map. However, smoothness regularization alone is not sufficient to capture the geometry of the problem. The generated phase map for UNet with smoothness regularizaiton (see row 4 of Figure 4) contains phase vortecies, which significantly increases the L2 loss.

## 7. Conclusion and Future Work

In this work, we introduce three main improvements to existing Phase Retrieval solutions:

1. Optimal transport as an effective initial guess for algorithms to improve on.

2. Incorporating OT phase predictions into deep learning models by modifying the model task to predict the phase correction on the initial guess.

3. Introducing an exponential regularization schedule on phase correction which lowers L2 norm while stabilizing training.

Our experiments show that Optimal Transport is an optimal method (pun intended) of initializing iterative solvers for phase retrieval, which is substantially better than a smoothness regularization approach. We observed that it allows models to converge on an order of magnitude better than baseline methods, prevents the formation of phase vortices, and stabilizes training.

In the future, we hope to explore more deep learning architectures. While the UNet remains a popular model for image segmentation, recent transformer based models such as Swin-UNet could improve on these results with an OT-based initialization [1]. The scope of this project remained limited to UNet based models, but foundational models such as the Vision Transformer could also be trained to predict the phase with an output projection head [2].

Additionally, a necessary improvement to our experiment (that we didn't implement due to the limited compute) would be to train and evaluate each model on a batch of synthetically generated data. In this project, we use one synthetic data point that is representative of most laser beam patterns. However, a comprehensive evaluation of the model should take place on a variety of input and target intensities.

Finally, the primary drawback of this iterative approach is that each image pair takes several minutes to run with a CPU. While we were fortunate enough to have access to an H100, we recognize that many research institutions, especially those in the physical scientific domains this model would be most beneficial for, do not possess extensive compute resources. Given more time, we hope to explore ways to improve time to convergence and overall prediction speed for iterative models.

## 8. Appendix

### 8.1. Derivations

Recall that we had the following equation for the unknown phase $\phi$

$$A_2(\vec{X}) = |\mathcal{F}[A_1(\vec{x})e^{i\phi(\vec{x})}](\vec{X})| \qquad (14)$$

$$= \left| \int A_1(\vec{x})e^{i\phi(\vec{x})}e^{-i2\pi\vec{x}\cdot\vec{X}}d\vec{x} \right| \qquad (15)$$

The idea is to Taylor expand the exponent and perform integral analytically. So, let's start with an approximation. We will assume that the function

$$\psi(\vec{x}) = \phi(\vec{x}) - 2\pi\vec{x}\cdot\vec{X} \qquad (16)$$

is (1) convex for each $\vec{X}$ and (2) doesn't vary rapidly. From assumption (1), we know that for each value of $\vec{X}$ there will be a global minimum where gradient of phase vanishes. Call this point $x_0$. Then we have:

$$\nabla\psi(\vec{x}_0) = 0 \implies \nabla\phi(\vec{x}_0) = 2\pi\vec{X} \qquad (17)$$

So, then we see that $\vec{x}_0$ maps to a unique point on the output plane $\vec{X}_0$, via the map $\nabla\phi$. This is the essence of the ray optics approximation to the problem of phase retrieval. In a sense, $\nabla\phi$ can be interpreted as a transport map that moves intensity from the input plane to the output plane. Let's explore this idea further by doing a taylor expansion of $\psi$ up to the second order:

$$\psi(\vec{x}) = \psi(\vec{x}_0) + (\vec{x} - \vec{x}_0)^T \nabla\psi(\vec{x}_0) + \qquad (18)$$

$$+ (\vec{x} - \vec{x}_0)^T \nabla^2\psi(\vec{x}_0)(\vec{x} - \vec{x}_0) \qquad (19)$$

Recall that the middle term vanishes at $\vec{x}_0$. So, now let's look at the second term:

$$\nabla^2\psi(\vec{x_0}) = \nabla^2\phi(\vec{x_0}) - \nabla^2(2\pi\vec{x}\cdot\vec{X}) = \nabla^2\phi(\vec{x_0}) \qquad (20)$$

Furthermore, we will also approximate $A_1(x) \approx A_1(x_0)$. So, putting everything back into the integral equation (3) we obtain:

$$A_2(\vec{X}) = |\int A_1(\vec{x_0})e^{i(-2\pi\vec{x}_0\cdot\vec{X})} \cdot \qquad (21)$$

$$\cdot e^{i(\phi(\vec{x}_0) + (\vec{x} - \vec{x}_0)^T\nabla^2\psi(\vec{x}_0)(\vec{x} - \vec{x}_0))}d\vec{x}| \qquad (22)$$

$$= \left| e^{i(\phi(\vec{x}_0) - 2\pi\vec{x}_0\cdot\vec{X})} \right| \cdot \qquad (23)$$

$$\cdot \left| A_1(\vec{x}_0) \int e^{i(\vec{x} - \vec{x}_0)^T\nabla^2\psi(\vec{x}_0)(\vec{x} - \vec{x}_0)}d\vec{x} \right| \qquad (24)$$

$$= A_1(\vec{x}_0) \left| \int e^{i(\vec{x} - \vec{x}_0)^T\nabla^2\psi(\vec{x}_0)(\vec{x} - \vec{x}_0)}d\vec{x} \right| \qquad (25)$$

Notice that it seems like all of the $\vec{X}$ dependence vanished from the right hand side, but it's not the case. Recall that the we chose $\vec{x}_0$ such that $\nabla\phi(\vec{x}_0) = 2\pi\vec{X}$. So the right hand side is implicitly a function of $\vec{X}$. The integral is a 2d gaussian integral and it can be computed analytically. The value of the integral is simply $\sqrt{2\pi/\det\nabla^2\phi(x_0)}$. So, we obtain an equation:

$$A_2\left(\frac{\nabla\phi(\vec{x}_0)}{2\pi}\right) = A_1(\vec{x}_0)\sqrt{\frac{2\pi}{\det\nabla^2\phi(x_0)}} \qquad (26)$$

Squaring both sides and doing a change of variables results in the following differential equation:

$$A_2^2(\nabla\phi(\vec{x})) = \frac{A_1^2(\vec{x})}{\det\nabla^2\phi(x)} \qquad (27)$$

### 8.2. Plots

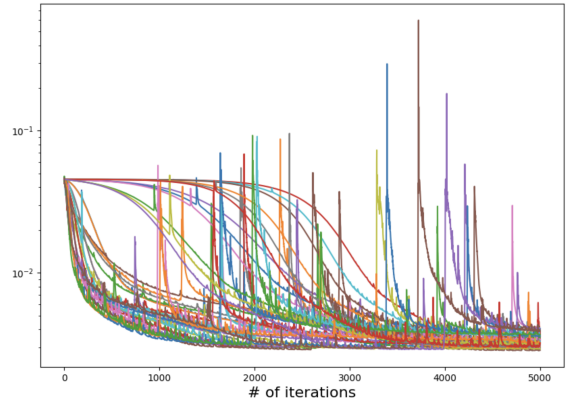

Figure 7. All L2 norm curves in Hyperparameter Sweep of OT + UNet (scheduler)

## 9. Contribution and Acknowledgements

A.T. formulated the project task. J.W. explored existing deep learning solutions. A.T. and J.W. implemented models. J.W. performed experiments. A.T. generated metrics. A.T. and J.W. performed evaluations and wrote the paper.

We acknowledge Hunter Swan and the Jason Hogan Lab as a whole for developing the idea of using Optimal Transport for the task of Phase Retrieval. J.W. is a member of the Arc Institute, which provided compute for this project.

 